\documentclass[12pt,twoside]{amsbook}
\usepackage[b5paper,top=2.45cm,bottom=2.45cm,left=2.55cm,right=1.25cm]{geometry}

\usepackage{amsfonts}
\usepackage{amssymb}
\usepackage{amsthm}
\usepackage{graphicx}

\usepackage{amsmath}
	\numberwithin{equation}{chapter}

\usepackage[caption=false]{subfig}
\usepackage{algpseudocode}
\usepackage[acronym]{glossaries}
\usepackage{hyphenat}

\usepackage{fancyhdr}

\numberwithin{figure}{chapter}
\numberwithin{table}{chapter}
	
\newcommand{\ket}[1]{ | \, #1 \rangle}
\newcommand{\bra}[1]{ \langle #1 \, |}
\newcommand{\proj}[1]{\ket{#1}\bra{#1}}

\newcommand{\be}{\begin{equation}}
\newcommand{\ee}{\end{equation}}

\newcommand{\ba}{\begin{aligned}}
\newcommand{\ea}{\end{aligned}}

\newcommand{\zeroOp}{\mathbb{O}}
\newcommand{\idOp}{\mathbb{I}}

\newtheorem{trm}{Theorem}[chapter]
\newtheorem{lem}[trm]{Lemma}
\newtheorem{definition}[trm]{Definition}

\fancypagestyle{plain}{\fancyhead{} }

\makeatletter
\newcommand\RedeclareMathOperator{%
  \@ifstar{\def\rmo@s{m}\rmo@redeclare}{\def\rmo@s{o}\rmo@redeclare}%
}
\newcommand\rmo@redeclare[2]{%
  \begingroup \escapechar\m@ne\xdef\@gtempa{{\string#1}}\endgroup
  \expandafter\@ifundefined\@gtempa
     {\@latex@error{\noexpand#1undefined}\@ehc}%
     \relax
  \expandafter\rmo@declmathop\rmo@s{#1}{#2}}
\newcommand\rmo@declmathop[3]{%
  \DeclareRobustCommand{#2}{\qopname\newmcodes@#1{#3}}%
}
\@onlypreamble\RedeclareMathOperator
\makeatother

\DeclareRobustCommand{\gobblefour}[4]{}
\newcommand*{\SkipTocEntry}{\addtocontents{toc}{\gobblefour}}

\DeclareMathOperator{\Tr}{Tr}

\DeclareMathOperator{\Interior}{Interior}

\DeclareMathOperator{\Mat}{Mat}
\RedeclareMathOperator{\vec}{vec}
\DeclareMathOperator{\eig}{eig}
\DeclareMathOperator{\Ext}{Ext}

\makeindex
\makeglossaries

\theoremstyle{plain}

\newglossaryentry{Re}
{
  name={\ensuremath{Re}},
  description={the real part},
}

\newglossaryentry{Im}
{
  name={\ensuremath{Im}},
  description={the imaginary part},
}

\newglossaryentry{negation}
{
  name={\ensuremath{\neg}},
  description={the negation of a bit},
}

\newglossaryentry{realMatrix}
{
  name={\ensuremath{\mathbb{R}^{k \times l}}},
  description={the set of real $k \times l$ matrices},
}

\newglossaryentry{symmetricMatrix}
{
  name={\ensuremath{\mathbb{S}^{n \times n}}},
  description={the set of real symmetric $n \times n$ matrices},
}

\newglossaryentry{hermMatrix}
{
  name={\ensuremath{\mathbb{H}^{n \times n}}},
  description={the set of hermitian $n \times n$ matrices},
}

\newglossaryentry{realPlus}
{
  name={\ensuremath{\mathbb{R}_{+}}},
  description={the set of real non-negative numbers},
}

\newglossaryentry{real}
{
  name={\ensuremath{\mathbb{R}}},
  description={the set of real numbers},
}

\newglossaryentry{complex}
{
  name={\ensuremath{\mathbb{C}}},
  description={the set of complex numbers},
}

\newglossaryentry{succ}
{
  name={\ensuremath{\succ}},
  description={the strict L\"{o}wner's partial order of real symmetric matrices},
}

\newglossaryentry{succeq}
{
  name={\ensuremath{\succeq}},
  description={the non-strict L\"{o}wner's partial order of real symmetric matrices},
}

\newglossaryentry{T}
{
  name={\ensuremath{^{T}}},
  description={the matrix transposition},
}

\newglossaryentry{herm}
{
  name={\ensuremath{^{\dagger}}},
  description={the matrix hermitian conjugate},
}

\newglossaryentry{conj}
{
  name={\ensuremath{^{*}}},
  description={the complex conjugate of a number},
}

\newglossaryentry{mT}
{
  name={\ensuremath{^{-T}}},
  description={the matrix inverse and transposition},
}

\newglossaryentry{trace}
{
  name={\ensuremath{\Tr}},
  description={the trace of a matrix},
}

\newglossaryentry{Interior}
{
  name={\ensuremath{\Interior}},
  description={the interior of a set},
}

\newglossaryentry{Mat}
{
  name={\ensuremath{\Mat}},
  description={the matrix created out of a vector (in column-wise order)},
}

\newglossaryentry{vec}
{
  name={\ensuremath{\vec}},
  description={the vector created out of a matrix (in column-wise order)},
}

\newglossaryentry{eig}
{
  name={\ensuremath{\eig}},
  description={the set of eigenvalues of a matrix},
}

\newglossaryentry{zeroOp}
{
  name={\ensuremath{\zeroOp}},
  description={the null operator or the zero matrix},
}

\newglossaryentry{idOp}
{
  name={\ensuremath{\idOp}},
  description={the identity operator or the unit matrix},
}

\newglossaryentry{Pabxy}
{
  name={\ensuremath{P(a,b|x,y)}},
  description={the joint probability of outcomes $a$ and $b$ given settings $x$ and $y$},
}

\newglossaryentry{PABXY}
{
  name={\ensuremath{\mathbb{P}(A,B|X,Y)}},
  description={the joint probability distribution for outcomes in sets $A$ and $B$, and settings from sets $X$ and $Y$},
}

\newglossaryentry{quantumSet}
{
  name={\ensuremath{\mathcal{Q}}},
  description={the set of all quantum probability distributions},
}

\newglossaryentry{QkSet}
{
  name={\ensuremath{\mathcal{Q}_k(A,B|X,Y)}},
  description={the set of probability distributions in $k$-th level of NPA for given scenario},
}

\newglossaryentry{AQset}
{
  name={\ensuremath{\mathcal{Q}_{1+AB}}},
  description={the Almost Quantum set of probability distributions},
}

\newglossaryentry{localSet}
{
  name={\ensuremath{\mathcal{L}}},
  description={the set of local probability distributions},
}

\newglossaryentry{noSignalSet}
{
  name={\ensuremath{\mathcal{N}}},
  description={the set of no-signaling probability distributions},
}

\newglossaryentry{PAaxy}
{
  name={\ensuremath{P_A(a|x,y)}},
  description={the probability of the outcome $a$ for Alice for settings $x$ and $y$},
}

\newglossaryentry{PAax}
{
  name={\ensuremath{P_A(a|x)}},
  description={the marginal probability of the outcome $a$ for Alice for the setting $x$},
}

\newglossaryentry{PBbxy}
{
  name={\ensuremath{P_B(b|x,y)}},
  description={the probability of the outcome $b$ for Bob for settings $x$ and $y$},
}

\newglossaryentry{PBby}
{
  name={\ensuremath{P_B(b|y)}},
  description={the marginal probability of the outcome $b$ for Bob for the setting $y$},
}

\newglossaryentry{barB}
{
  name={\ensuremath{\bar{B}}},
  description={the set of outcomes of Bob in the prepare-and-measure scenario},
}

\newglossaryentry{barX}
{
  name={\ensuremath{\bar{X}}},
  description={the set of settings of Alice in the prepare-and-measure scenario},
}

\newglossaryentry{barY}
{
  name={\ensuremath{\bar{Y}}},
  description={the set of settings of Bob in the prepare-and-measure scenario},
}

\newglossaryentry{HilbertSpace}
{
  name={\ensuremath{\mathbb{H}}},
  description={a Hilbert space of a finite dimension},
}

\newglossaryentry{PdBXY}
{
  name={\ensuremath{\mathbb{P}_d (\bar{B}|\bar{X},\bar{Y})}},
  description={a conditional probability distribution on a Hilbert space of dimension $d$},
}

\newglossaryentry{PPXY}
{
  name={\ensuremath{\mathcal{P}^{(P)}(\bar{X},\bar{Y})}},
  description={a certain subset of binary probability distribution},
}

\newglossaryentry{PpmdBXY}
{
  name={\ensuremath{\mathcal{P}_d (\bar{B}|\bar{X},\bar{Y})}},
  description={a set of all prepare-and-measure probability distributions with restriction on the dimension},
}

\newglossaryentry{PpmBXY}
{
  name={\ensuremath{\mathcal{P} (\bar{B}|\bar{X},\bar{Y})}},
  description={the set of all prepare-and-measure probability distributions},
}

\newglossaryentry{dimWit}
{
  name={\ensuremath{W (\bar{B}, \bar{X}, \bar{Y}, \{\beta_{b,x,y}\}, C_W)}},
  description={a dimension witness},
}

\newglossaryentry{noiseParameter}
{
  name={\ensuremath{\eta}},
  description={a white noise parameter},
}

\newglossaryentry{purity}
{
  name={\ensuremath{p}},
  description={the purity of a state},
}

\newglossaryentry{A}
{
  name={\ensuremath{A}},
  description={the set of measurement outcomes of the first party},
}

\newglossaryentry{Ais}
{
  name={\ensuremath{\{A_i\}}},
  description={linear constraint matrices (real symmetric matrices)},
}

\newglossaryentry{b}
{
  name={\ensuremath{b}},
  description={the RHS linear constraint (a real vector)},
}

\newglossaryentry{B}
{
  name={\ensuremath{B}},
  description={the set of measurement outcomes of the second party},
}

\newglossaryentry{C}
{
  name={\ensuremath{C}},
  description={the linear coefficient (a real symmetric matrix)},
}

\newglossaryentry{X}
{
  name={\ensuremath{X}},
  description={the primal variable (a real positive definite matrix) \textbf{or} the set of measurement settings of the first party},
}

\newglossaryentry{y}
{
  name={\ensuremath{y}},
  description={the dual variable (a real vector)},
}

\newglossaryentry{Y}
{
  name={\ensuremath{Y}},
  description={the set of measurement settings of the second party},
}

\newglossaryentry{Z}
{
  name={\ensuremath{Z}},
  description={the dual slack variable (a real positive definite matrix)},
}

\newglossaryentry{rp}
{
  name={\ensuremath{r_p}},
  description={the primal residual (a real vector)},
}

\newglossaryentry{rd}
{
  name={\ensuremath{r_d}},
  description={the dual residual (a real vector)},
}

\newglossaryentry{normRp}
{
  name={\ensuremath{\epsilon_P}},
  description={the primal residual norm (a real number)},
}

\newglossaryentry{normRd}
{
  name={\ensuremath{\epsilon_D}},
  description={the dual residual norm (a real number)},
}

\newglossaryentry{HpM}
{
  name={\ensuremath{H_P(M)}},
  description={the Monteiro-Zhang symmetrization (a square matrix to symmetric matrix function)},
}

\newglossaryentry{mathcalE}
{
  name={\ensuremath{\mathcal{E}}},
  description={the derivative of a symmetrization with respect to the primal variable},
}

\newglossaryentry{mathcalF}
{
  name={\ensuremath{\mathcal{F}}},
  description={the derivative of a symmetrization with respect to the dual slack variable},
}

\newglossaryentry{n}
{
  name={\ensuremath{n}},
  description={the size of the primal variable $X$ and the dual slack variable $Z$},
}

\newglossaryentry{m}
{
  name={\ensuremath{m}},
  description={the size of the dual variable $y$, \textit{i.e.} the number of linear constraint matrices},
}

\newglossaryentry{FrobNorm}
{
	name={\ensuremath{{| \cdot |_F}}},
	description={the Frobenius norm of a matrix},
}

\newglossaryentry{BellI}
{
	name={\ensuremath{I(A,B,X,Y,\{\alpha_{a,b,x,y}\},C_I)[\mathbb{P}(A,B|X,Y)]}},
	description={a Bell operator},
}

\newglossaryentry{BellIhat}
{
	name={\ensuremath{\hat{I}(X, Y, \{\alpha_{x,y}\}, \hat{C}_I)[\mathbb{P}(\{o_1,o_2\},\{o_1,o_2\}|X,Y)]}},
	description={a Bell operator on in correlation form},
}

\newglossaryentry{mathcalPABXY}
{
	name={\ensuremath{\mathcal{P}(A,B|X,Y)}},
	description={the set of all probability distributions for scenario given by $A$, $B$, $X$ and $Y$ allowed by considered theory},
}

\newglossaryentry{kron}
{
	name={\ensuremath{\otimes}},
	description={the Kronecker product of matrices},
}

\newglossaryentry{kronSym}
{
	name={\ensuremath{\otimes_S}},
	description={the symmetric Kronecker product of matrices},
}

\newglossaryentry{kronDelta}
{
	name={\ensuremath{\delta_{a,b}}},
	description={the Kronecker delta ($\delta_{a,b}=1$ for $a=b$, and $0$ otherwise)},
}

\newacronym{AHO}{AHO}{Alizadeh, Haeberly and Overton search direction}
\newacronym{DI}{DI}{device-independent}
\newacronym{HKM}{HKM}{Helmberg, Rendl, Vanderbei, Wolkowicz / Kojima, Shindoh, Hara / Monteiro search direction}
\newacronym{iid}{\textit{i.i.d.}}{independent and identically distributed}
\newacronym{IPM}{IPM}{interior point method}
\newacronym{LHS}{LHS}{left hand side}
\newacronym{LP}{LP}{linear programming}
\newacronym{MEX}{MEX}{MATLAB Executable}
\newacronym{MPA}{MPA}{Masanes-Pironio-Ac\'in quantum key distribution protocol}
\newacronym{NPA}{NPA}{Navascues-Pironio-Ac\'in method}
\newacronym{NT}{NT}{Nesterov-Todd search direction}
\newacronym{P-C}{P-C}{predictor-corrector}
\newacronym{PD}{PD}{positive definite}
\newacronym{PM}{PM}{projective measurement}
\newacronym{P-M}{P-M}{prepare-and-measure}
\newacronym{POVM}{POVM}{positive operator valued measure}
\newacronym{PSD}{PSD}{positive semi-definite}
\newacronym{RHS}{RHS}{right hand side}
\newacronym{RNG}{RNG}{random number generator}
\newacronym{QI}{QI}{Quantum Information}
\newacronym{QKD}{QKD}{quantum key distribution}
\newacronym{QRNG}{QRNG}{quantum random number generator}
\newacronym{PRNG}{PRNG}{pseudo-random number generator}
\newacronym{SDI}{SDI}{semi-device-independent}
\newacronym{SDP}{SDP}{semi-definite programming}
\newacronym{SV}{SV}{Santha-Vazirani}
\newacronym{SVD}{SVD}{singular value decomposition}

\begin{document}
\frontmatter
\title[SDP in quantum information]{Applications of semi-definite optimization in quantum information protocols}
\author[Piotr Mironowicz]{Piotr Mironowicz}

\date{\today{}}

\begin{abstract}

This work is concerned with the issue of applications of the semi-definite programming (SDP) in the field of quantum information science. Our results of the analysis of certain quantum information protocols using this optimization technique are presented, and an implementation of a relevant numerical tool is introduced. The key method used is NPA discovered by Navascues et al. [Phys. Rev. Lett. 98, 010401 (2007)].

In chapter~1 a brief overview of mathematical methods used in this work is presented. In chapter~2 an introduction to quantum information science is given. Chapter~3 concerns the device-independent (DI) and semi-device-independent (SDI) approaches in quantum cryptography. In chapter~4 our results regarding quantum information protocols which we developed using SDP are described. The results include a new type of quantum key distribution protocol based on Hardy's paradox; several protocols for quantum random number expansion; a method of formulation of SDP relaxations of SDI protocols; and a DI Santha-Vazirani source of randomness amplification protocol. In chapter~5 an introduction to interior point methods is given. In chapter~6 the implementation of our numerical tool dedicated to problems occurring in chapter~4 are discussed and compared with other solutions.

\end{abstract}

\thanks
{
I would like to thank my supervisor prof. Krzysztof Giaro for all his advice and friendly and patient help in many situations, including those in which the main problem were my personal character flaws.
}
\thanks
{
I would also like to thank my ancillary supervisor, dr Marcin Paw{\l}owski, for showing me that you can still be a kid at the age of 37, and that it is possible to get serious results without serious work. Now I hope he will show me how to do this.
}
\thanks
{
Well, I have to admit, that he also gave me a scientific help. A really big help.
}
\thanks
{
I am also grateful to Jadwiga and Ryszard Horodecki for help and forbearance at the time when I needed it.
}
\thanks
{
I thank Ravishankar Ramanathan for his critical opinions and for keeping me company at work when other people were sleeping.
}
\thanks
{
I also thank Irena Moszczy\'nska-Janicka for her language corrections and editorial help.
}
\thanks
{
Obviously, I would also like to thank my wife, Aleksandra, for all possible means of help I could expect.
}
\thanks
{
Some results of this work were supported by the grant from the Ministry of Science and Higher Education of the Republic of Poland IDEAS PLUS (IdP2011 000361), the National Science Centre (NCN) grant 2013/08/M/ST2/00626, the National Science Centre project Maestro DEC-2011/02/A/ST2/00305, FNP TEAM and the InterPhD scholarship from POKL.04.01.01-00-368/09 project.
}
\thanks
{
Parts of this work were written at Gda\'{n}sk University of Technology, the National Quantum Information Centre in Gda\'{n}sk, WiMBP Agency 65 and in the forests of Sopot.
}
\thanks
{
The usage of OCTAVE 3.8.1 \cite{octave}, SeDuMi \cite{SeDuMi} and SDPT3 \cite{TTT12} is acknowledged.
}

\hyphenation{Lagrangian OCTAVE MATLAB SeDuMi Kronecker}

\maketitle
\glsaddall
\tableofcontents

\printglossaries

\chapter*{Preface}

This work concerns the interdisciplinary issue of applications of the semi-definite programming in the field of quantum information science. In particular the results cover the analysis of certain quantum information protocols using this optimization technique. The key tool we use is the \gls{NPA} method discovered by Miguel Navascues, Stefano Pironio and Antonio Ac\'in \cite{NPA07}. Contrary to the majority of works employing this method, our aim is to deal not only with the results obtained with it, but also to investigate some aspects of the implementation issues.

In general the NPA method may be viewed as a definition of certain sets of probability distributions which are interesting on their own \cite{AQ}, and are possible to be formulated as solutions to semi-definite problems. If the classical computer science can be viewed as considerations of capabilities of Turing machines supplied with certain probability distributions, the quantum information tries to answer the question what happens if we consider a wider family of probabilities. It reveals that semi-definite programming is an excellent tool for this analysis.

The task of this work is two-fold, namely we use the semi-definite programming model of a certain set of probability distributions to investigate some aspects of random number generation and cryptography, and we develop a suitable numerical tool for this task. To be more precise, we
\begin{itemize}
	\item Formulate models based on semi-definite programming for analysis of the reliability of quantum random number generation and quantum key distribution protocols.
	\item Implement some variants of interior point method for semi-definite programming and compare their performance in the inquiry of quantum probability distributions.
\end{itemize}

The considered types of protocols are currently of particular interest. Although the efforts in construction of quantum computers \cite{IBMQuantumComputing15} find more interest among laity, quantum cryptography is the field which is more probable to find applications in everyday's life in the coming years. There already exist commercial devices implementing quantum randomness generation, \textit{e.g.} the device \textit{Quantis} produced by \textit{id~Quantique}\cite{IDQ}, or \textit{qStream} by Quintessence Labs\cite{QLabs}. Such devices are available by PCI-express or USB interface. Also quantum key distribution devices are present on the market, \textit{e.g.} Cerberis QKD by \textit{id~Quantique}, or solutions by MagiQ \cite{magiq}.

The aim of the SECOQC project (2004-2008) was ``\textit{evolving quantum cryptography into an instrument that can be operated in an economic environment}'' \cite{secoqc}. In October 2008 they performed the first live demonstration of a working network with quantum key distribution in a metropolitan environment. Other institutions involved in the development of quantum key distribution include NIST \cite{NISTQKD} and Toshiba \cite{ToshibaQKD}. A collaborative work by NEC, Mitsubishi Electric, NTT, Toshiba and other institutions reported in 2011 \cite{Tokio11} a first implementation of TV conference secured with quantum cryptography over a distance of 45~kilometers. A spectacular implementation milestone was an experiment with quantum teleportation \cite{Canary12} and entanglement swapping \cite{Canary14} at a distance of 143~kilometers over Canary Islands of La Palma and Tenerife.

The work is organized as follows. The two main chapters of this work are chapter four and six. These chapters contain our contribution to the investigation of quantum protocols and implementation of semi-definite programming method tools involved.

In the first chapter we start with a brief overview of mathematical methods which will be useful in this work. Next we introduce some basic notions regarding semi-definite programming. In particular we formulate the primal and dual problems, show some of their properties and give examples not related to the main topic to demonstrate that this optimization technique finds a wide range of applications.

In second and third chapters we give a brief introduction to quantum information science which is needed to understand our results. The former deals with general methods of this discipline, and the latter concentrates on the so-called device-independent approach in quantum cryptography.

In chapter four we describe our results regarding quantum information protocols which we developed using semi-definite programming. We start with a discussion of a new type of quantum key distribution protocol which uses the so-called Hardy's paradox in order to certify its security. Then we move to the task of certification of the credibility of random numbers obtained with several quantum protocols. This analysis is conducted with minimal assumptions about the reliability of devices used for randomness generation. Such an approach is called device-independent when we do not assume anything about the internal working of the device, or semi-device-independent when we limit a communication between different parts of the device. We propose a number of randomness certification protocols of both of these kinds. We finish this chapter with a discussion of a quantum protocol performing the so-called randomness amplification, \textit{i.e.} a kind of improvement of the quality of some biased source of randomness. We show how these protocols can be investigated using semi-definite programming relaxations of the set of quantum probability distributions.

The fifth chapter deals with interior point methods. This is one of the methods of solving semi-definite programming problems. Other methods include the spectral bundle method \cite{Bundle00} and the augmented Lagrangian method \cite{SDPNAL10}. We have chosen to concentrate on this particular approach for several reasons. First of all, interior point method had an enormous impact on the field optimization \cite{Wright05}. Secondly, most of the state of the art implementations use this method, including solvers already popular in the quantum information community. Thirdly, interior point methods seem to have the most extensive literature.

In chapter six we discuss the issues related to the implementation of interior point methods suited for the form of problems occurring in chapter four. First we investigate the performance of two popular implementations of semi-definite programming solvers, SDPT3\index{SDPT3} and SeDuMi\index{SeDuMi}, for these problems. Then we introduce a new method of calculation of the Schur complement equation occurring in each iteration of interior point method. Next we discuss a warm-start strategies, \textit{i.e.} methods of finding a starting iterate for interior point method. Then we analyze a number of strategies for perturbing the iterates in order to prevent a semi-definite programming solver from failure. We finish with our proposal of semi-definite programming solver dedicated for the problems from chapter five. We compare its performance with other solvers.

To sum up, the premises of this thesis are the following:
\begin{enumerate}
	\item Semi-definite programming can be employed to evaluate the amount of randomness generated in quantum protocols, and the security of a new kind of quantum key distribution protocols.
	\item The form of problems occurring in the analysis of the discussed quantum protocols can be exploited in order to improve the performance of dedicated solvers.
\end{enumerate}

Readers will notice that the content of this work covers many different fields. The scope includes quantum information, which itself employs mathematical theories of, among others, Hilbert spaces, probability theory, theory of information, and cryptography. We look at the themes from the point of view of numerical optimization. Thus our toolbox requires also elements of matrix analysis and numerical optimization.

We tried to make this work self-contained to the greatest possible extent. Nonetheless, it seems to be impossible to cover this range of topics in an elaborate and concise way. For this reason some issues are addressed only briefly, to provide general background to explain why some tasks were met in a particular manner. On the other hand, readers not interested in details may skip several more technical sections, like sec.~\ref{sec:extractors} describing the topic of randomness extraction, or sec.~\ref{sec:estimation} dealing with experimental estimation of physical quantities. We assume the knowledge of the basics of linear programming and familiarity with the bra-ket notation of Hilbert spaces. Nonetheless, we review some basic notions of these fields below.

This work is not dealing with physical aspects of the considered protocols. In particular we do not consider the issues related to physical realizations of the protocols.

\begin{flushright}
May-July 2015
\end{flushright}

This work contains the results of the following papers:
\begin{enumerate}
	\item Li, H. W., \textbf{Mironowicz, P.}, Paw{\l}owski, M., Yin, Z. Q., Wu, Y. C., Wang, S., Chen, W., Hu, H.-G., Guo, G.-C., Han, Z. F. (2013). \textit{Relationship between semi-and fully-device-independent protocols}, Physical Review A, 87(2), 020302. \cite{HWL13}
	\item \textbf{Mironowicz, P.}, Gallego, R., Paw{\l}owski, M. (2015). \textit{Robust amplification of Santha-Vazirani sources with three devices}, Physical Review A, 91(3), 032317. \cite{MP13}
	\item \textbf{Mironowicz, P.}, Li, H. W., Paw{\l}owski, M. (2014). \textit{Properties of dimension witnesses and their semi-definite programming relaxations}, Physical Review A, 90(2), 022322. \cite{HWL14}
	\item \textbf{Mironowicz, P.}, Paw{\l}owski, M. (2013). \textit{Robustness of quantum\hyp{}randomness expansion protocols in the presence of noise}, Physical Review A, 88(3), 032319. \cite{LubiePlacki}
	\item Rahaman, R., Parker, M. G., \textbf{Mironowicz, P.}, Paw{\l}owski, M. (2013). \textit{Device-independent quantum key distribution based on Hardy's paradox}, Physical Review A 92, 062304 (2015). \cite{Ramij-our}
\end{enumerate}
Their results are referred to mainly in chapter~\ref{chap:quantumProtocols}.

Chapter~\ref{chap:solver} contains results not announced previously.

\mainmatter

\chapter{Introduction}

In this chapter we give a concise introduction to several topics used in further parts of this work. We start with some mathematical background regarding matrix analysis. Then we formulate the task of semi-definite programming (\acrshort{SDP}), discuss some of its properties, and give examples of applications. We finish this chapter with the discussion of the Newton's iterative method of finding roots of functions, and theory of information.

\section{Preliminaries}

In this section we first state the notation. Then we briefly overview some mathematical background we will be using further in the work. This includes the notion of the symmetric Kronecker product and discussion of selected properties of positive semi-definite matrices.

\subsection{Notation}


\gls{real} is the set of real number, and \gls{complex} is the set of complex numbers.

Sets of integer numbers are denoted with upper case letters, \textit{e.g.} $A$, $B$, $X$, $Y$. Sets of other objects are denoted with calligraphy upper case letters, \textit{e.g.} $\mathcal{S}$. Multidimensional array, like joint probability distributions are denoted with blackboard bold upper case letters, \textit{e.g.} $\gls{PABXY}$.

Matrices are denoted with upper case letters, \textit{e.g.} $A,C,M,X,Z$, whereas vectors with lower case letters like $b,v,x,y,z$. $M_{k,l}$ refers to the element in $k$-th row and $l$-th column. For a vector $v$ its $k$-th element is $v_k$. Vectors are represented by one column matrices. Matrix elements are numbered from $1$.

The set of real $k \times l$ matrices is denoted by \gls{realMatrix}, and of real $n \times n$ symmetric matrices by \gls{symmetricMatrix}. We refer to the pair of value $k$ and $l$ (for matrices of arbitrary size) or $n$ (for square matrices) as \textbf{size}\index{matrix!size} of the matrix.

The sets of real and complex vectors with $k$ elements are denoted by $\mathbb{R}^k$ and $\mathbb{C}^k$, respectively.

The function $\vec(\cdot)$ defines a vector containing the elements of the given matrix in the column-wise order. $\Mat(\cdot)$ is the inverse of this function. This conversion will be used further in this work very often. For example we have
\be
	\nonumber
	\vec \left( \begin{bmatrix} a & c \\ b & d \end{bmatrix} \right) = \begin{bmatrix} a \\ b \\ c \\ d \end{bmatrix}.
\ee

We will also use the following standard convention in which upper-case letters denote matrices, and lower-case letters denote vectors of elements of the matrices, \textit{e.g.}
\be
	\label{eq:vecMat}
	\ba
		& x = \gls{vec}(X) \in \mathbb{R}^{n^2}, \\
		& X = \gls{Mat}(x) \in \mathbb{R}^{n \times n}.
	\ea
\ee

For two matrices $A, B \in \mathbb{R}^{m \times n}$ we define the relation $A \leq B$ to hold if and only if
\be
	\nonumber
	\forall_{i = 1, \dots, m} \forall_{j = 1, \dots, n} A_{i,j} \leq B_{i,j}.
\ee
We define relations $A < B$, $A \geq B$ and $A > B$ in an analogous way.

We denote the real part of a complex number of matrix by $\gls{Re}$, and the imaginary part by $\gls{Im}$.

For bits we define the negation $\neg 0 \equiv 1$ and $\neg 1 \equiv 0$.

The symbol $\cdot$ is used to separate expressions in those places in which it (in our opinion) improves readability of formulas, and has no special mathematical meaning.

\subsection{Matrix products}

We will now introduce two matrix products, namely the symmetric Kronecker product and the Frobenius product. The result of the former is a matrix, and of the latter it is a number.

\subsubsection{The symmetric Kronecker product}

The \textbf{symmetric Kronecker product}\index{symmetric Kronecker product} has been introduced in \cite{AHO98} (see also \cite{TTT98})\footnote{We note here that \cite{AHO98,TTT98,TTT12} use a different notion of symmetrized Kronecker product, which allows to represent symmetric matrices with only the unique entries. We do not use this form of the Kronecker product in this work, since it would not allow for the techniques from sec.~\ref{sec:Schur}.}. Below we show its several properties.

Let $A,B,C \in \mathbb{R}^{k \times k}$.

We define the operation $\gls{kronSym}$ by the equation
\be
	\label{eq:symKron}
	A \otimes_S B \equiv \frac{1}{2} \left( A \otimes B + B \otimes A \right),
\ee
where $\otimes$ denotes the ordinary Kronecker product.

It is easy to check that
\be
	\nonumber
	\left( A \otimes B \right) \vec(C) = \Vec{\left( B C A^T \right)}.
\ee
From this we get
\be
	\nonumber
	\left( A \otimes_S B \right) \vec(C) = \frac{1}{2} \Vec{\left( B C A^T + A C B^T \right)}.
\ee

We also have
\be
	\nonumber
	\left( A \otimes_S B \right) \left( C \otimes_S C \right) = AC \otimes_S BC.
\ee

Obviously $A \otimes_S A = A \otimes A$ and $A \otimes_S B = B \otimes_S A$.

\subsubsection{The Frobenius product}

The \textbf{Frobenius product}\index{Frobenius product} of two real matrices, $A, B \in \mathbb{R}^{k \times l}$ is defined as $\Tr (A^T B)$.

It can be easily shown that
\be
	\label{eq:Frob}
	\Tr (A^T B) = \Tr (A B^T) = \sum_{i = 1, \dots, k} \sum_{j = 1, \dots, l} A_{i,j} B_{i,j} = \vec(A)^T \vec{B}.
\ee
Thus the Frobenius product is the sum of the elements of the element-wise product of entries of two matrices.

The symmetry of this product comes directly from~\eqref{eq:Frob}. One can also show that for symmetric $A$ and antisymmetric $B$ we have
\be
	\nonumber
	\Tr(A^T B) = \Tr(A B) = 0.
\ee

The Frobenius product introduces a \textbf{Frobenius norm}\index{Frobenius norm} of a matrix, \gls{FrobNorm}. We define
\be
	\nonumber
	|A|_F = \sqrt{\Tr (A^T A)}.
\ee
This norm is called also a Hilbert-Schmidt norm. This is a direct generalization of the vector Euclidean norm, as can be seen from~\eqref{eq:Frob}.

\subsection{Properties of positive matrices}
\label{sec:propPSD}

In order to formulate semi-definite programming problems let us introduce the notion of positive-definite (\acrshort{PD}) and positive semi-definite (\acrshort{PSD}) matrices. A symmetric or hermitian matrix $M$ is called PD (PSD), denoted $M \succ 0$ ($M \succeq 0$), if all its eigenvalues are positive (non-negative). Equivalently, a symmetric matrix $M \in \mathbb{R}^{n \times n}$ is PD (PSD) if and only if
\be
	\nonumber
	\forall_{\substack{x \in \mathbb{R}^n \\ x \neq 0}} x^T M x > 0 \left( \forall_{\substack{x \in \mathbb{R}^n \\ x \neq 0}} x^T M x \geq 0) \right),
\ee
and a hermitian matrix $M \in \mathbb{C}^{n \times n}$ is PD (PSD) if and only if
\be
	\label{eq:hermitianPSD}
	\forall_{\substack{x \in \mathbb{C}^n \\ x \neq 0}} x^\dagger M x > 0 \left( \forall_{\substack{x \in \mathbb{C}^n \\ x \neq 0}} x^\dagger M x \geq 0 \right).
\ee
In our opinion the former definition is more intuitive, but the latter is more common in the literature on the subject. A more detailed treatment on the properties of PD and PSD matrices may be found in \cite{matrixAnalysis,matrixAnalysis2}. We note that a real PD (PSD) matrix satisfies \eqref{eq:hermitianPSD}, and thus is a complex PD (PSD) matrix. On the other hand, for a complex PD (PSD) $M$ we have that
\be
	\nonumber
	\forall_{\substack{x \in \mathbb{C}^n \\ x \neq 0}} x^\dagger \frac{1}{2} (M + M^{\dagger}) x = x^\dagger Re(M) x \geq 0,
\ee
and thus the matrix $Re(M)$ is a real PD (PSD) matrix.

The relation $\succeq$ introduces the so-called L\"{o}wner's partial order of semi-definite matrices. For two symmetric or hermitian matrices $A$ and $B$ we have $A \succeq B$, if $A - B \succeq 0$.

It can be easily shown that if $A, B \succeq 0$, then $A + B \succeq 0$. It is also easy to see that if we multiply a PSD matrix by a non-negative constant, we get another PSD matrix. Thus the set of PSD matrices forms a pointed convex cone. It also follows that $\Tr(A B) \geq 0$, for $A, B \succeq 0$, and that $A^{\frac{1}{2}}$ exists and is PSD.

One may prove the following theorem \cite{V13,matrixAnalysis2}:
\begin{trm}
	For a symmetric matrix $M \in \mathbb{R}^{n \times n}$ we have that $M \succeq 0$ is equivalent to each of the following statements:
	\begin{itemize}
		\item For all $x \in \mathbb{R}^{n}$ we have $x^{T} M x \geq 0$.
		\item There exists $L \in \mathbb{R}^{n \times n}$ such, that $M = L^{T} L$ ($L$ is the Cholesky decomposition of M).
		\item There exists ${v_{1}, \ldots, v_{n}}$, $v_{i} \in \mathbb{R}^{n}$ such, that $M_{i, j} = v_{i}^{T} \cdot v_{j}$.
	\end{itemize}
\end{trm}

Let $B \in \mathbb{C}^{n \times n}$ be a hermitian matrix, $B^R$ and $B^I$ its real and imaginary parts, respectively. Then $B \succeq 0$ if, and only if
\be
	\label{eq:hermitianToSymmetric}
	\begin{bmatrix} B^R & -B^I \\ B^I & B^R \end{bmatrix} \succeq 0.
\ee
Indeed, for any complex vector $w = u + i v \in \mathbb{C}^n$ we have
\be
	\nonumber
	\ba
		& w^{\dagger} B w = \left( u^T - i v^T \right) \left( B^R + i B^I \right) (u + i v) = \\
		& = \left( u^T B^R u + v^T B^R v - u^T B^I v + v^T B^I u \right) + \\
		& + i (u^T B^R v - v^T B^R u + u^T B^I u + v^T B^I v) \geq 0
	\ea
\ee
if, and only if
\be
	\nonumber
	\ba
		\begin{bmatrix} u^T & v^T \end{bmatrix}
		\begin{bmatrix} B^R & -B^I \\ B^I & B^R \end{bmatrix}
		\begin{bmatrix} u \\ v \end{bmatrix}
		\succeq 0.
	\ea
\ee
This is because $u^T B^I u = v^T B^I v = 0$ and $u^T B^R v = v^T B^R u$. Thus any SDP problem defined in terms of complex vectors and hermitian matrices can be stated as a problem involving only real vectors with symmetric matrices.

One may prove a very useful result, namely the Sylvester’s Criterion\index{Sylvester’s Criterion}. This criterion says that a real symmetric matrix is PD if and only if all its \textbf{leading} principal minors are positive. It is PSD if and only if all the principal minors are non-negative.

Further in this work we will omit the explicit statement that a matrix is (real) symmetric, if the notion of $\succ$ or $\succeq$ is used.

\section{Basics of semi-definite programming}

In this section we discuss some basic notions related to SDP. We start with a discussion of the formulation of the problem, and compare it to LP. Then we briefly mention norms of the so-called infeasibility which are used further in this work. We finish with a discussion of the duality of SDP.

\subsection{Real linear- and semi-definite problems}
\label{sec:SDPformulation}

Let $m$ and $n$ be positive integers, $m \leq \frac{n(n+1)}{2}$.

A semi-definite programming problem in a \textbf{primal}\index{primal problem} form is the following optimization task in a variable $X \in \mathbb{S}^{n \times n}$:
\begin{align}
	\label{SDP-primal}
	\begin{split}
		\text{minimize } &\null \Tr(C X) \\
		\text{subject to } &\null \Tr(A_i X) = b_i, \text{ for } i = 1, \cdots, m \\
		&\null X \succeq 0, \\
	\end{split}
\end{align}
where $C \in \mathbb{S}^{n \times n}$ and $A_1, \cdots A_m \in \mathbb{S}^{n \times n}$ are symmetric\footnote{The fact that these matrices are symmetric is not restrictive. For a symmetric matrix $X$ and a matrix $C$ we have $\Tr(C X) = \Tr \left( \frac{1}{2}(C + C^T) X \right)$, and thus we may always take a symmetric matrix $\frac{1}{2}(C + C^T)$ instead of $C$.} matrices. The matrices $A_i$, $C$ and vector $b \in \mathbb{R}^m$ define the SDP problem. We assume that $A_1, \cdots A_m$ are linearly independent (otherwise we can reduce this set).

Recall that the primal form of linear programming (\acrshort{LP}) problems is the following optimization task in variable $x$:
\begin{align}
	\label{LP-primal}
	\begin{split}
		\text{minimize } &\null c^{T} \cdot x \\
		\text{subject to } &\null A^T x = b, x \geq 0,
	\end{split}
\end{align}
where $A \in \mathbb{R}^{n \times m}$, $b \in \mathbb{R}^m$, $x, c \in \mathbb{R}^n$.

Obviously LP problem may be written in the form of SDP, if $X$ is constrained to be a diagonal matrix, with the diagonal entries used as the $x$ variable. Thus LP can be considered as a particular case of SDP.

A \textbf{dual}\index{dual problem} SDP problem for \eqref{SDP-primal} is the optimization task in variables $y \in \mathbb{R}^m$ and $Z \in \mathbb{S}^{n \times n}$ of the following form
\begin{align}
	\label{SDP-dual}
	\begin{split}
		\text{maximize } &\null b^{T} \cdot y \\
		\text{subject to } &\null C - \sum_{i=1}^{m} y_i A_i = Z. \\
		&\null Z \succeq 0,
	\end{split}
\end{align}
In case of LP the dual problem is
\begin{align}
	\label{LP-dual}
	\begin{split}
		\text{maximize } &\null b^{T} \cdot y \\
		\text{subject to } &\null c - A y = z, \\
		&\null z \geq 0.
	\end{split}
\end{align}

In the above problems, the variable \gls{X} or $x$ is called the \textbf{primal variable}\index{primal variable}, \gls{y} the \textbf{dual variable}\index{dual variable}, \gls{Z} the \textbf{dual slack variable}\index{dual slack variable}, $A$ and \gls{Ais} are \textbf{linear constraint matrices}\index{linear constraint matrices}, \gls{b} is the \textbf{RHS linear constraint}\index{RHS linear constraint}, and \gls{C} or $c$ is the \textbf{linear coefficient}\index{linear coefficient}.

If $X, Z \in \mathbb{R}^{n \times n}$ and $y \in \mathbb{R}^{m}$ satisfies conditions specified by~\eqref{SDP-primal} and~\eqref{SDP-dual}, then they are called a feasible solution\index{feasible solution}. Feasible variable $X$ is called a \textbf{primal solution}\index{solution!primal}, and feasible variables $Z$ and $y$ a \textbf{dual solution}\index{solution!dual}. An optimal solution is expected to be feasible. The values of $\Tr(C X)$ and $b^{T} \cdot y$ are called the values of the primal and dual solutions, respectively\index{value of a problem}. We have $\Tr(C X) \geq b^{T} \cdot y$. Usually an SDP solver is expected to find both primal and dual solutions.

The fact that primal formulation refers to minimization, and dual to maximization problems, is not restrictive. We can always change the sign of the matrix $C$ or the vector $b$ to get the desired optimization problem fitting into the standard form in~\eqref{SDP-primal} and~\eqref{SDP-dual}.

What is more, a problem formulated in one of the forms given by~\eqref{SDP-primal} and~\eqref{SDP-dual} may be reformulated in the other one. The issue of choosing the proper formulation is not always obvious, and can have a very significant impact of the difficulty of the problem to a solver \cite{dualizeIt}. This can be illustrated by the example in tab.~\ref{tab:CHSHsizes} showing the sizes of some SDP problems in dual and primal formulations. See sec.~\ref{sec:formulationsGamma} for a discussion of this issue in the context of problems considered in this work. One should choose the formulation which leads to a smaller number of constraints, given by the number $m$ (unless the structure of the problem can be exploited in the other formulation).

If either $C = \zeroOp$ or $b = 0$, then such a problem is called \textbf{feasibility problem}\index{feasibility problem} and refers to finding whether \textit{any} solution of given, primal or dual, problem exist.

The key property of SDP problems is the fact that they may be efficiently solved numerically using interior point algorithm described further in chapter~\ref{chap:IPM}.

Further in this work we often use the following notation for SDP problems. Let us introduce the matrix
\be
	\label{eq:mathcalA}
	\mathcal{A} \equiv [a_1; \dots; a_m] \in \mathbb{R}^{n^2 \times m},
\ee
where
\be
	\nonumber
	a_i = \vec(A_i).
\ee
Thus $a_i$ is $i$-th column of $\mathcal{A}$. Then we have
\be
	\nonumber
	\ba
		& \vec \left( \sum_{i=1}^{m} y_i A_i \right) = \mathcal{A} y, \\
		& \Tr(A_i X) = (\mathcal{A}^T x)_i,
	\ea
\ee
where $(\mathcal{A}^T x)_i$ is the $i$-th element of the vector $\mathcal{A}^T x$ and $x = \vec(X)$. These expressions allow reformulation of the problems given in~\eqref{SDP-primal} and~\eqref{SDP-dual} in a form similar to LP formulations in~\eqref{LP-primal} and~\eqref{LP-dual}.

\subsubsection{Primal and dual infeasibility norms}

We note here that practical implementations of SDP solver usually find solutions which are not feasible in a strict sense. Instead, the solutions satisfy the condition from~\eqref{SDP-primal} and~\eqref{SDP-dual} only with some accuracy. Here we discuss the expressions we use further in this work to evaluate primal and dual infeasibility. See \cite{Mittelmann12} for more details on the issue of infeasibility norms.

Let $c \equiv \vec(C)$, $x \equiv \vec(X)$ and $z \equiv \vec(Z)$. Let us define the following terms, \textit{viz.} the residuals for feasibility conditions in~\eqref{SDP-primal} and~\eqref{SDP-dual} (\textit{cf.}~\eqref{eq:primalFeas} and~\eqref{eq:dualFeas} below)
\begin{subequations}
	\be
		\label{eq:rp}
		\gls{rp} \equiv b - \mathcal{A}^T x \in \mathbb{R}^m,
	\ee
	\be
		\label{eq:rd}
		\gls{rd} \equiv c - \mathcal{A} y - z \in \mathbb{R}^{n^2}.
	\ee
\end{subequations}

The primal infeasibility norm we use in this work is given by
\be
	\label{eq:rpNorm}
	\gls{normRp} \equiv \frac{1}{1 + |b|_F} |b - \mathcal{A}^T x|_F.
\ee
The dual infeasibility norm is defined in this work as
\be
	\label{eq:rdNorm}
	\gls{normRd} \equiv \frac{1}{1 + |c|_F} |c - \mathcal{A} y - z|_F.
\ee

\subsubsection{Complex semi-definite problems}
\label{sec:complexSDP}

One can also consider problems \eqref{SDP-primal} and \eqref{SDP-dual} in complex variables. In such case $C, \{A_i\} \in \mathbb{H}^{n \times n}$, and $b \in \mathbb{R}^m$. Then $X$ and $Z$ are complex PSD matrices, and $y$ is a real vector.

It is easy to see that if $C \in \mathbb{S}^{n \times n}$, and $X \in \mathbb{H}^{n \times n}$, $X^R \equiv Re(X)$ and $X^I \equiv Im(X)$, then
\be
	\nonumber
	\Tr (C X) = \Tr (C X^R) + i \Tr(C X^I) = \Tr (C X^R),
\ee
since $X^I$ is antisymmetric, and the Frobenius product of symmetric and antisymmetric matrix is always equal to $0$. From the above considerations it follows that if $C$ is real and we are interested only in finding the value of the solution, then we can neglect the imaginary part occurring in the problem\footnote{If $C$ is hermitian, we still may reformulate the complex problem as a real problem with the construction in~\eqref{eq:hermitianToSymmetric}.}.

\subsection{A note on mixed linear and semi-definite problems}
\label{sec:mixed}

We briefly note that one often considers the so-called mixed cone. The primal problem in variables $(x_L,X_S) \in \mathbb{R}^{n_L} \times \mathbb{S}^{n_S \times n_S}$ is the following:
\begin{align}
	\nonumber
	\begin{split}
		\text{minimize } &\null c_L^{T} \cdot x_L + \Tr(C_S X_S) \\
		\text{subject to } &\null (A_L^T)_{i,:} \cdot x_L + \Tr({A_S}_i X_S) = b_i, \text{ for } i = 1, \cdots, m \\
		&\null x_L \geq 0, \\
		&\null X_S \succeq 0.
	\end{split}
\end{align}
where $A_L \in \mathbb{R}^{n_L \times m}$, $(A_L^T)_{i,:}$ denotes $i$-th row of the matrix $A_L^T$, $b \in \mathbb{R}^m$, $x_L, c_L \in \mathbb{R}^{n_L}$, $C_S \in \mathbb{S}^{n_S \times n_S}$ and ${A_S}_1, \cdots {A_S}_m \in \mathbb{S}^{n_S \times n_S}$.

The dual mixed problem in variables $(y,z_L,Z_S) \in \mathbb{R}^m \times \mathbb{R}^{n_L} \times \mathbb{S}^{n_S \times n_S}$ is of the following form
\begin{align}
	\nonumber
	\begin{split}
		\text{maximize } &\null b^{T} \cdot y \\
		\text{subject to } &\null c_L - A_L y = z_L, \\
		&\null C_S - \sum_{i=1}^{m} y_i {A_S}_i = Z_S, \\
		&\null z_L \geq 0, \\
		&\null Z_S \succeq 0.
	\end{split}
\end{align}

Since any LP can be reformulated as SDP, the mixed problems are not more general than the SDP problems. It suffices to place linear variables on the diagonal of an SDP variable of size $n_L + n_S$. The reason why mixed problems are considered is that the numerical methods needed to solve SDP are more expensive in terms of computational effort that LP. If a problem is stated in the mixed form, then it is possible to reduce this complexity.

\subsection{Duality of semi-definite problems}
\label{sec:duality}

An important property of primal and dual formulations is the fact, that the solution of the primal problem is an upper bound on the solution for the dual problem:
\be
	\nonumber
	\begin{aligned}
		\Tr(C X) - b^{T} \cdot y &= \Tr((Z + \sum_{i=1}^{m} y_i A_i) X) - b^{T} \cdot y = \\
		& \Tr(Z X) + \sum_{i=1}^{m} y_i \Tr(A_i X) - y^{T} \cdot b = \Tr(Z X) \geq 0.
	\end{aligned}
\ee
This property is called a \textit{weak duality}\index{duality!weak}. In the case of LP, the values of primal and dual solutions are always equal, if the solution exists, such property is called a \textit{strong duality}\index{duality!strong}.

Let $p^{*}$ be the value of the optimal solution of the primal problem,~\eqref{SDP-primal}, and $d^{*}$ the value of the optimal solution of the dual problem,~\eqref{SDP-dual}. It can be shown \cite{NN94,Rockafellar,SDP} that for $p^{*} = d^{*}$ to hold, it is sufficient if one of the following conditions is satisfied:
\begin{itemize}
	\item There exists $X \succ 0$ such that $\Tr(A_i X) = b_i$, for $i = 1, \cdots m$ (strict primal feasibility).
	\item There exists $Z \succ 0$ such that $C - \sum_{i=1}^{m} y_i A_i = Z$ (strict dual feasibility).
\end{itemize}
If both conditions hold, then the optimal values can be obtained for both primal and dual problems, \textit{i.e.} both primal and dual solutions exist. Their values are then equal and finite. Thus these conditions are sufficient for strong duality to hold for an SDP problem.

Further the expression $\Tr(Z X)$ is referred to as the \textit{gap}. Note that strong duality of an SDP problem implies that the optimal primal and dual variables are orthogonal, \textit{i.e.}
\be
	\label{eq:strongDuality}
	\Tr(Z X) = 0,
\ee
meaning that the gap is equal to $0$.

\section[Examples of semi-definite problems]{Examples and overview of applications of semi-definite programming}

In this section we show a few examples of SDP problems. The aim of the first of them is to give an illustration of how they are formulated. The remaining examples were chosen because of their importance.

A more comprehensive overview of applications of SDP may be found, \textit{e.g.} in \cite{SDP,Boyd04}. These include a famous MAX-CUT and MAX-k-SAT relaxations by Goemans and Williamson \cite{maxcut}, maximum eigenvalue, matrix norm minimization, and combinatorial optimization problems \cite{GLS84,A91,Overton92,MoharPoljak93,A95,Goemans97,B00,BYZ00}.

\subsection{Primal and dual formulation}

Let us consider the following problem
\begin{align}
	\nonumber
	\begin{split}
		\text{minimize } &\null \tilde{x} \\
		\text{subject to } &\null \begin{bmatrix} \tilde{x} & 1 \\ 1 & \tilde{x} \end{bmatrix} \succeq 0. \\
	\end{split}
\end{align}
Using the Sylvester's Criterion one may infer that the solution reads $\tilde{x} = 1$.

Beginning with the primal formulation, we have
\begin{align}
	\nonumber
	\begin{split}
		\text{minimize } &\null x_{1 1} \\
		\text{subject to } &\null x_{1 1} = x_{2 2}, 2x_{1 2} = 2, \\
		&\null X = \begin{bmatrix} x_{1 1} & x_{1 2} \\ x_{1 2} & x_{2 2} \end{bmatrix} \succeq 0
	\end{split}
\end{align}
In this case $C = \begin{bmatrix} 1 & 0 \\ 0 & 0 \end{bmatrix}$, $A_1 = \begin{bmatrix} 1 & 0 \\ 0 & -1 \end{bmatrix}$, $A_2 = \begin{bmatrix} 0 & 1 \\ 1 & 0 \end{bmatrix}$, and $b = \begin{bmatrix} 0 \\ 2 \end{bmatrix}$. Indeed, these matrices give the following primal problem equivalent to the initial one:
\begin{align}
	\nonumber
	\begin{split}
		\text{minimize } &\null \Tr(C X) = \Tr \left( \begin{bmatrix} 1 & 0 \\ 0 & 0 \end{bmatrix} \begin{bmatrix} x_{1 1} & x_{1 2} \\ x_{1 2} & x_{2 2} \end{bmatrix} \right) = x_{1 1} \\
		\text{subject to } &\null 0 = \Tr (A_1 X) = \Tr \left( \begin{bmatrix} 1 & 0 \\ 0 & -1 \end{bmatrix} \begin{bmatrix} x_{1 1} & x_{1 2} \\ x_{1 2} & x_{2 2} \end{bmatrix} \right) = x_{1 1} - x_{2 2} \\
		&\null 2 = \Tr (A_2 X) = \Tr \left( \begin{bmatrix} 0 & 1 \\ 1 & 0 \end{bmatrix} \begin{bmatrix} x_{1 1} & x_{1 2} \\ x_{1 2} & x_{2 2} \end{bmatrix} \right) = 2 x_{1 2} \\
		&\null X \succeq 0.
	\end{split}
\end{align}
It is less obvious that the dual of this problem, \textit{viz.}
\begin{align}
	\begin{split}
		\text{maximize } &\null b^T \cdot y = 2 y_2 \\
		\text{subject to } &\null Z \equiv \begin{bmatrix} 1 - y_1 & -y_2 \\ -y_2 & y_1 \end{bmatrix} \succeq 0
	\end{split}
\end{align}
gives the same value.

The dual formulation of the initial problem gives $C = \begin{bmatrix} 0 & 1 \\ 1 & 0 \end{bmatrix}$, $A_1 = \begin{bmatrix} -1 & 0 \\ 0 & -1 \end{bmatrix}$, and $b = [-1]$, namely
\begin{align}
	\nonumber
	\begin{split}
		\text{maximize } &\null b^T \cdot y = -y_1 \\
		\text{subject to } &\null Z \equiv C - y_1 A_1 = \begin{bmatrix} y_1 & 1 \\ 1 & y_1 \end{bmatrix} \succeq 0,
	\end{split}
\end{align}
but since we have replaced minimization with maximization of an expression of the opposite sign, we have to negate the sign of the obtained result. Writing this problem in the primal form we get
\begin{align}
	\nonumber
	\begin{split}
		\text{minimize } &\null \Tr(CX) = 2x_{1 2} \\
		\text{subject to } &\null \Tr(A_1 X) = -x_{1 1} - x_{2 2} = b_1 = -1 \\
		&\null X \succeq 0.
	\end{split}
\end{align}
This formulation gives the result $-1$, \textit{i.e.} exactly the negation of the desired solution.

From the above example we see that both formulations give the same result. Moreover, if we start with a primal formulation, and take its dual, we get the same value. Similarly if we start with a dual formulation, and take its primal, we also get the same result. Thus, we have four formulations of the same problem. This illustrates also the importance of a careful choice of the formulation. For this case the primal seems to be more natural, but the dual results in a smaller number of constraints.

\subsection{Eigenvalues of matrices}
\label{sec:eigenvaluesSDP}

The problem minimization of maximal eigenvalue of a linearly constrained matrix can be formulated as SDP. More detailed treatment of this topic can be found in \cite{MoharPoljak93}.

Let the set of matrices $\mathcal{S}$ be parametrized by $y_i$s, \textit{i.e.}
\be
	\nonumber
	\mathcal{S} = \left\{ M(y_2, \cdots, y_m) \equiv -C + \sum_{y=2}^m A_i y_i : y_i \in \mathbb{R} \right\},
\ee
where $C$ and $A_i$ are symmetric matrices. Then the problem of minimizing the maximal eigenvalue over $y_i$s can be written as the following dual SDP problem:
\begin{align}
	\nonumber
	\begin{split}
		\text{minimize } &\null y_1 \\
		\text{subject to } &\null y_1 \idOp - M(y_2, \cdots, y_m) \succeq 0.
	\end{split}
\end{align}
We have changed the maximization of $-y_1$ to minimization of $y_1$ in this form and set $A_1 = \idOp$.

Indeed, when the optimum is attained, the variables $y_2, \cdots, y_m$ parametrize such a matrix which can be subtracted from $y_1 \idOp$, and the result is PSD. Thus $y_1$ represents the maximal eigenvalue of $M(y_2, \cdots, y_m)$, with $y_2, \cdots, y_m$ chosen in such a way that this maximal eigenvalue is as small as possible within the set $\mathcal{S}$.

\subsection{Shannon capacity of a graph}

The notion of a capacity of a channel represented by a graph was introduced by Shannon in \cite{Shannon56}, which is defined below. Unfortunately this entity is difficult to be calculated. In \cite{Lovasz79} Lovasz formulated an SDP relaxation of this problem called Lovasz $\theta$ function\index{$\theta$ function}. This function had a strong impact both on classical and quantum information theories \cite{citeLovasz1,citeLovasz2,citeLovaszQuantum}, and also on other disciplines, like graph theory \cite{Goemans98,KMS98}.

For a pair of graphs, $G$ and $H$, let us define $G \cdot H$ in the following way. Let $V(G \cdot H) = V(G) \times V(H)$, where $V(\cdot)$ is the set of vertices of a graph. The vertex $(x_1, y_1)$ is adjacent to $(x_2, y_2)$ if and only if one of the following holds:
\begin{itemize}
	\item $x_1$ is adjacent to $x_2$ in $G$, and $y_1$ is adjacent to $y_2$ in $H$, or\footnote{The original paper \cite{Lovasz79} states explicitly only this condition.}
	\item $x_1=x_2$, and $y_1$ is adjacent to $y_2$ in $H$, or
	\item $x_1$ is adjacent to $x_2$ in $G$, and $y_1=y_2$.
\end{itemize}
This is the so-called strong product of graphs\index{strong product} \cite{Sabidussi60}. Using the notion of the strong product we define $G^1 = G$ and $G^{k+1} = G^k \cdot G$.

Let us consider an $n$ letter alphabet, and a graph $G$ with vertices labeled with the letters, and edges between the letters which are possible to be confused with each other for a given model of communication \textit{via} a channel. Obviously the number of one letter messages which are impossible to be confused is equal to the size of the largest independent set of the graph, denoted $\alpha(G)$.

It is easy to see that the number of $k$-letter messages which are possible to be send without confusion is $\alpha(G^{k}) \geq \alpha(G)^{k}$, \textit{viz.} if using one letter message we are able to formulate $l = \alpha(G)$ different messages impossible to be confused, then with $k$ letters we can encode at least $l^k$ different messages without the risk of confusion. For example we have $\alpha(C_{5}) = 2$, and $\alpha(C_{5}^{2}) = 5$.

The Shannon capacity of a graph $G$ is defined by
\be
	\nonumber
	\Theta(G) = \sup_{k} \alpha(G^{k})^{\frac{1}{k}}
\ee

Lovasz's relaxation has the property that $\Theta(G) \leq \theta(G)$. The $\theta(G)$ is defined as follows.

Let us consider a set $\mathcal{S}$ of all symmetric matrices $A$ satisfying the following condition. For two nodes $i$ and $j$ of $G$, if $i=j$ or $i$ and $j$ are not adjacent in $G$, then $A_{i j} = 1$. Other entries of these matrices are unconstrained. $\theta(G)$ is defined to be the minimum of largest eigenvalue of matrices from $S$. The problem can be formulated as SDP with the method described in sec.~\ref{sec:eigenvaluesSDP}.

One may show that $\theta(G \cdot H) \leq \theta(G) \theta(H)$.

\section{The Newton's method}
\index{Newton"'s method}

The Newton's method, called also the Newton\hyp{}Raphson method, is a technique of finding approximations roots of differentiable functions. Here we only sketch this method in an intuitive way. Readers interested in a more rigorous treatment are referred to the wide range of literature on the subject, \textit{e.g.} \cite{Hildebrand87,Boyd04}.

Let $f(x) = [f_1(x) \cdots f_k(x)]^T$ be a real vector valued differentiable function of many variables, $x = (x_1, \cdots, x_l)$. The Jacobian matrix is the following $k \times l$ real matrix containing all first order derivatives of the function:
\be
	J(x) \equiv
	\begin{bmatrix}
		\frac{\partial f_1}{\partial x_1}(x) & \cdots & \frac{\partial f_1}{\partial x_l}(x) \\
		\vdots & \ddots & \vdots \\
		\frac{\partial f_k}{\partial x_1}(x) & \cdots & \frac{\partial f_k}{\partial x_l}(x)
	\end{bmatrix}.
\ee

The method is iterative. We start at some point $x^{(0)}$. The Newton step $\Delta x$ is calculated with the following equation:
\be
	\nonumber
	J \left( x^{(i)} \right) \Delta x^{(i)} = -f \left( x^{(i)} \right).
\ee
Afterward we iterate with
\be
	\nonumber
	x^{(i+1)} = x^{(i)} + \alpha^{(i)} \Delta x^{(i)}
\ee
for some $(\alpha^{(i)})_i$, till the desired accuracy is attained. The sequence $(\alpha^{(i)})_i$ depends on the variant of the method. In many cases one takes $\alpha^{(i)} = 1$ for all iterations (\textit{cf.} sec.~\ref{sec:step-lengths}).

The convergence of this method can be proved under certain assumptions \cite{Rheinboldt70,Rheinboldt74}. In particular the sequence $\left(x^{(i)}\right)_i$ converges to a root (if a root exists) for convex functions with the properly chosen $x^{(0)}$ \cite{Spivak}.

The intuition behind this method is that we have
\be
	\nonumber
	f(x + \Delta x) \approx f(x) + J(x) \Delta x
\ee
and thus we expect that $f(x + \Delta x) \approx 0$.

\section{Measures of information}
\label{sec:entropy}
\label{sec:min-entropy}\index{entropy}

One of the basic notions of the theory of information is entropy. It formalizes the notion of a measure of information.

Let $\mathbb{P}_X = (p_1, \ldots, p_N)$ be a discrete probability distribution of a random variable $X$.

The most commonly used entropy is \textbf{Shannon entropy}\index{entropy!Shannon}, or simply \textbf{entropy}, introduced by Shannon in 1948 in a revolutionary paper \textit{A Mathematical Theory of Communication} \cite{Shannon48}. For a random variable $X$ it is defined by
\be
	\label{eq:entropy}
	H(X) \equiv -\sum_{i=1}^N p_i \log_b{p_i},
\ee
with $b=2$, and $0 \log 0 = 0$. A unit of this entropy is called the bit\index{bit} (or, rarely, the shannon). If $b=e$ then the unit is called nat\index{nat}, and for $b=10$ it is called hartley\index{hartley}. These units are defined in an international standard IEC 80000-13. One usually omits $b$ if it is equal to $2$, or if its value is obvious from the context.

A commonly used measure of randomness in cryptographic context is \textbf{min-entropy}\index{entropy!min-entropy} \cite{OperMinEn,NIST800632,ColPHD,CK11}, denoted $H_{\infty}$. It is defined as
\be
	\nonumber
	H_{\infty}(X) \equiv - \log_{2} \left( \max_{i} p_i \right).
\ee
Min-entropy is directly related to the guessing probability of the value of a particular variable with distribution $\mathbb{P}$ with the strategy in which one guesses the most probable result. As specified by the National Institute of Standards and Technology, in the context of guessing cryptographic keys, min-entropy is a measure of the difficulty of guessing the easiest single key in a given distribution of keys \cite{NIST800632}.

Entropy is related to the widely understood \textit{uniformity} of a probability distribution. The maximal value of both entropies is attained by uniform probability distributions, and the minimal value by deterministic variables (with $p_i = 1$ for a certain $i$).

These two entropies are additive, \textit{i.e} for a random variable
\be
	\nonumber
	X = (X_1, \dots, X_k),
\ee
where $X_i$ are independent random variables, we have
\be
	\nonumber
	H(X) = \sum_i H(X_i),
\ee
and similarly for min-entropy.

Both Shannon entropy and min-entropy are particular cases of a more general concept of \textbf{Renyi entropy}\index{entropy!Renyi} \cite{Renyi61} defined by
\be
	\nonumber
	H_{\alpha}(X) \equiv \frac{1}{1-\alpha} \log_b{\left( \sum_i p_i^{\alpha} \right)}.
\ee
The Shannon entropy is obtained when we take the limit of $\alpha$ tending to $1$, and the min-entropy with $\alpha$ tending to infinity.

Another important quantity is the \textbf{conditional entropy}\index{entropy!conditional}, $H \left( \mathbb{P}_X|\mathbb{P}_Y \right)$. This is defined as
\be
	\nonumber
	H (X|Y) \equiv - \sum_{x,y} P_{X,Y}(x,y) \log_2{P_{X|Y}(x|y)},
\ee
where $\mathbb{P}_{X,Y}$ is the joint probability distribution of random variables $X$ and $Y$, and $\mathbb{P}_{X|Y}$ is their conditional probability distribution.

The conditional min-entropy is given by
\be
	\nonumber
	H_{\infty} \left( \mathbb{P}_X|\mathbb{P}_Y \right) \equiv -\log_2 \left( \sum_y P_Y(y) \max_x P_X|Y(x|y) \right).
\ee

\chapter{Basics of quantum information science}
\label{chap:basicsQI}

In this chapter we provide a brief introduction and overview of a few topics in quantum information science (\acrshort{QI}). QI is an interdisciplinary field. It concerns the topics of computer science, the theory of information in particular, like data communication or processing. This field provides a broader view on these topics, since it covers not only the tasks possible to be treated by machines governed by the law of classical physics, but it considers devices which are allowed by the laws of the quantum mechanics.

From the point of view of this work, a fundamental paper by John Bell from 1964, \textit{On the Einstein-Podolsky-Rosen paradox} \cite{Bell64}, which explicitly stated the notion of the later called Bell inequalities, is of particular importance. Bell considered there a problem stated by Einstein, Podolsky and Rosen in their revolutionary paper, \textit{Can Quantum-Mechanical Description of Physical Reality be Considered Complete?}, published in 1935 \cite{EPR35}.

The field of QI covers, among others, the issues of construction of the so-called quantum computers \cite{QComp,QComp98}, possibility of quantum dense coding \cite{denseCoding}, quantum computational complexity \cite{qComplexity93,qComplexity00}, quantum communication complexity \cite{quantumCC03,quantumCC04}, and quantum error correction \cite{quantumErrorCorrection96,quantumErrorCorrection13}. Recently the idea of quantum internet has emerged \cite{QuantumInternet08}. The QI partially conveys also issues related to physical realizations of quantum devices, like quantum logical gates and registers \cite{SaffmanWalker10}, quantum routers \cite{routers10}, or quantum repeaters \cite{repeaters11}.

The most spectacular successes of quantum information science are in the field of quantum cryptography \cite{qCrypto02,QKDreview}. These include famous BB84 \cite{BB84} and E91 \cite{E91} quantum key distribution protocols. Further in this work we will deal with a modern approach to quantum protocols, called \textit{device-independent}, which lessens the assumptions on the internal workings of cryptographic devices significantly.

The key role in our numerical calculations will play the \textit{NPA} method introduced by Navascues, Pironio and Ac\'in in paper \cite{NPA07, NPA08}. These authors define an infinite hierarchy of conditions which are satisfied by any quantum probability distribution (see the definition \eqref{def:quantum} below). Each level of this hierarchy determines a semi-definite optimization problem. We give a more detailed description of this method in sec.~\ref{sec:NPA}.

We introduce the formalism of Hilbert spaces for quantum mechanics in sec.~\ref{sec:Hilbert} only to define the set of quantum probability distributions, $\gls{quantumSet}$, in the definition~\ref{def:quantum} and to justify the SDP relaxation in sec.~\ref{sec:NPA} regarding the NPA method. Readers familiar with the notion of quantum information or uninterested in the details may skip sec.~\ref{sec:Hilbert}, and only remember that the NPA method provides a relaxation of some set $\gls{quantumSet}$ describing quantum devices. In our further considerations we will employ the so-called device-independent approach (described in details in chapter~\ref{chap:DI}) which abstracts from physical realizations and considers only sets of probability distributions.

\section[Hilbert spaces and probability distributions]{Hilbert space formalism and multipartite probability distributions}

Since the details of physical aspects of QI are far beyond the scope of this work, we give in this section only a short review of topics which are necessary for our further considerations. Readers interested in details of the formalism of QI should refer to other works. A standard textbook in QI is a work of Nielsen and Chuang \cite{NC10}. A more concise overview is contained in a paper by Keyl \cite{Keyl02}. Another review, with more emphasis on physical realizations, is \cite{Zukowski12}.

\subsection{Hilbert space formalism}
\label{sec:Hilbert}

The formulation of quantum mechanics is based on a formalism of Hilbert spaces developed by von~Neumann in 1930s \cite{vN55}. Further we restrict our considerations to systems described by finite dimensional Hilbert spaces. This is justified because all systems we are dealing with can be considered as finite state systems. 

Formally a \textbf{Hilbert space}\index{Hilbert space} is an inner product space (\textit{i.e.} a vector space with a defined inner product), such that the metric space with a norm induced by the inner product is complete. A simple example of a Hilbert space is a complex linear vector space $\mathbb{C}^d$ with inner product of vectors $x$ and $y$ defined by $x^{\dagger} \cdot y$, where $x^{\dagger}$ is the Hermitian conjugate of $x$. In quantum mechanics and QI one usually denotes a vector by $\ket{x}$ and its conjugate by $\bra{x}$. This is the bra-ket notation\footnote{Strictly speaking the conjugate vectors $\{\bra{x}\}$ belong to the \textbf{dual} Hilbert space.}.

The \textbf{computational basis}\index{basis!computational}, or simply basis, of a Hilbert space of dimension $d$ is denoted by kets $\ket{0}, \cdots, \ket{d-1}$. We may identify $\ket{i} = \begin{bmatrix} \underbrace{0 \dots 0}_{\times i - 1} 1 \underbrace{0 \dots 0}_{\times d - i} \end{bmatrix}^T$ and use the standard basis as the computational basis.

For two vectors $\ket{\psi}$, $\ket{\phi}$ we can write their inner product as
\be
	\nonumber
	\bra{\phi} \ket{\psi} = \left( \bra{\psi} \ket{\phi} \right)^{*}.
\ee
The vectors are orthogonal if $\bra{\phi} \ket{\psi} = 0$. We can also consider $\ket{\phi} \bra{\psi}$, which defines a linear operator $\mathbb{C}^d \rightarrow \mathbb{C}^d$ of rank one. Such an operator is represented by a square, real or complex, matrix.	

In this work we treat the quantum mechanics as a statistical theory. This means that we are interested only in probabilistic descriptions of the physical systems, in particular of the quantum devices employed for solving the tasks of computer science. This statistical nature of the theory means that the \textbf{behavior}\index{behavior} of the considered device can be observed only in a sequence of experiments starting with the same state and observing the same property of the device. This leads us to the problem of defining the \textbf{quantum state}\index{quantum state} and the \textbf{observable}\index{observable}. We discuss this below.

The state (pure) is represented by a unit vector on a Hilbert space called a \textbf{state space}. This space depends on the physical system we want to describe. A state is denoted as a ket vector, \textit{e.g.} $\ket{\psi}$.

A second basic component of the formalism is the idea of \textbf{observables}, \textit{i.e.} measurable operators, which are represented by self-adjoint linear operator acting on the space. The behavior of the device is the result of performing measurement on a state. A particular case of measurement is \textbf{projective measurement}\index{projective measurement} (\acrshort{PM}) which is represented by projectors. A \textbf{projector}\index{projector} is an idempotent observable, \textit{i.e.} for the projector $P$ we have $P P = P$. Projectors are represented by matrices with spectrum contained in the set $\{0,1\}$.

A set of projectors $\{E^i\}_i$ is called PM if they sum to the unit matrix, $\sum_i E^i = \idOp$. It follows that for a PM we have the orthogonality property, $E^i E^j = \delta_{i,j} E^i$. The probability of a given result $i$ of a measurement $\{E^i\}_i$ on a state $\ket{\psi}$ is given by
\be
	\label{eq:probMeas}
	P(i) = \bra{\psi} E^i \ket{\psi} = \Tr \left( E^i \proj{\psi} \right).
\ee
The indices $i$ of operators within a PM refers to relevant possible results of a performed experiment. We may consider many different experimental setups given by PMs, $\left\{ \{E^i_j\}_i \right\}_j$. In such a case $j$ refers to the setup, and $i$ to a measurement result of the setup. Each set $\{E^i_j\}_i$ for given $j$ is a PM.

We say that a \textbf{projection succeeded}\index{projection succeeded} if PM is given by $\{E_0,E_1\}$, with $E_0$ of rank one, and the result of the measurement is $0$.

Statistical mixtures of pure states are represented by the so-called \textit{density matrices}, which are self-adjoint operators trace $1$ operators on the space. They are represented by Hermitian PSD matrices. The mixture of states $\{\ket{\psi_i}\}$ with probabilities of occurrence $p_i$, $\sum_i p_i = 1$ is represented by a density matrix
\be
	\nonumber
	\rho = \sum_i p_i \proj{\psi_i}.
\ee
The probability of the result $i$ is for this mixture given by
\be
	\label{eq:probMeasMixed}
	P(i) = \Tr \left( E_i \rho \right) = \sum_j p_j \Tr \left( E_i \proj{\psi_j} \right).
\ee

A \textbf{Hadamard basis}\index{basis!Hadamard} for a two dimensional computational basis is defined by the following two vectors:
\be
	\nonumber
	\ba
		\ket{+} \equiv \frac{\sqrt{2}}{2} (\ket{0} + \ket{1}), \\
		\ket{-} \equiv \frac{\sqrt{2}}{2} (\ket{0} - \ket{1}).
	\ea
\ee
A key property of this basis is that if one measures the state $\ket{0}$ or $\ket{1}$, with PM defined by projectors on the vectors $\ket{+}$ and $\ket{-}$ (and \textit{vice versa}), then the probability of each result is $0.5$, \textit{e.g.} for $E_{+}=\proj{+}$ and $\rho=\proj{0}$ we have $\Tr(E_{+} \rho) = \frac{1}{2}$.

General measurement is described by the so-called \acrshort{POVM}s (positive operator valued measure). These are represented by sets $\{E^i\}_i$ of PSD Hermitian matrices, not necessarily idempotent (and thus not orthogonal), which sum to the unit matrix. The probabilities of different results of a given POVM are given by the same formulas, \eqref{eq:probMeas} and \eqref{eq:probMeasMixed}, as for PMs. Any POVM may be equivalently replaced with a PM on a space of higher dimension using the so-called Gelfand--Naimark--Segal construction \cite{GelfandNaimark43,Segal47}.

Quantum mechanics considers also the evolution of quantum states in time. The evolution is a unitary transformation of quantum states and is described by operators called Hamiltonians\footnote{The evolution operator, or propagator, $U$ after time $t$ for Hamiltonian $H$ is given by $U = \exp(-i t H)$.}. These issues will not be considered in this work.

We briefly mention an important property of quantum states, comprising of more that one part, called \textit{entanglement}. Without going into details, entangled states are these states which reveal some non-local properties. The most prominent examples are the so-called \textbf{maximally entangled states}\index{{maximally entangled states}|see {singlet}}, or \textbf{singlets}\index{singlet}. Such states comprise of two parts. For a given dimension $d$ of the Hilbert space, these states are defined by
\be
	\nonumber
	\frac{1}{\sqrt{d}} \sum_{i=0}^{d-1} \ket{i} \otimes \ket{i}.
\ee
Eq.~\eqref{eq:singlet2d} shows the singlet state on a product of two Hilbert spaces of dimension $2$ in density matrix form.

For the sake of completeness, we also mention that states in a space of dimension $2$ are of particular interest, especially in relation with computer science applications. Such states are called \textbf{qubits}, as a direct quantum generalization of bits. Qubits are usually parametrized using the so-called Bloch sphere. The parametrization is given by
\be
	\label{eq:Bloch}
	\ket{\psi} = \cos \left( \frac{\theta}{2} \right) \ket{0} + \exp{i \phi} \sin \left( \frac{\theta}{2} \right) \ket{1}.
\ee
We note that if the above formula is multiplied by a phase factor $\exp{i \omega}$, $\omega \in \mathbb{R}$, then the probabilities observed on the state will not change.

\subsection{Multipartite probability distributions}
\label{sec:multiProbDistr}

Let us consider a case in which two parties, Alice and Bob, are separated (\textit{e.g.} spatially) and conduct an experiment on two subsystems which had previously interacted.

\begin{figure}[htbp]
	\centering
		\includegraphics[width=0.5\textwidth]{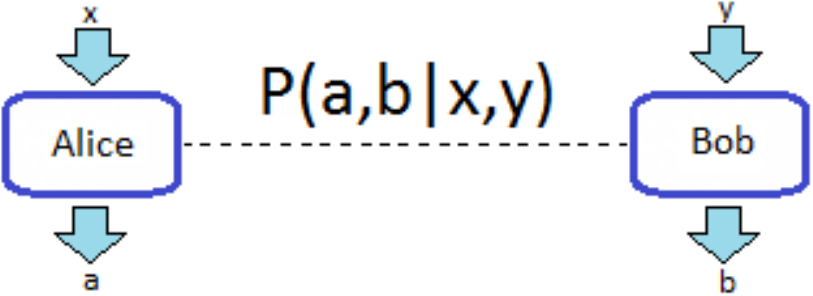}
		\caption{A measurement scenario with two parties.}
	\label{fig:AliceBob}
\end{figure}

Suppose that Alice in her part performs a measurement $x$ from a set $X$ of her possible measurement settings (\textit{e.g.} possible positions of a knob on her apparatus), and obtains a result labeled as $a$ from a set of possible results $A$ (\textit{e.g.} a set of characters on the display). Similarly Bob performs a measurement $y \in Y$ with a result $b \in B$. The assumption of separation means in particular that the information of the choice of $x$ does not reach Bob, nor $y$ to Alice. The situation is depicted in Fig.~\ref{fig:AliceBob}. In this work we consider only the case when these sets, $A$, $B$, $X$ and $Y$ are finite and non-empty. We refer to those four sets as an \textbf{experimental scenario}\index{scenario}, or simply \textbf{scenario}.

Below we denote by $\gls{Pabxy}$ a joint probability of outcomes $a$ and $b$ for settings $x$ and $y$. We define $\gls{PAaxy}$ as the probability that Alice gets outcome $a$ if she chooses the setting $x$, and similarly we define $\gls{PBbxy}$ for Bob:
\be
	\nonumber
	\ba
		& \gls{PAaxy} \equiv \sum_{b \in B} P(a,b|x,y), \\
		& \gls{PBbxy} \equiv \sum_{a \in A} P(a,b|x,y).
	\ea
\ee
The probability distribution
\be
	\nonumber
	\mathbb{P}(A,B|X,Y) = \{P(a,b|x,y)\}_{\substack{a \in A, b \in B \\ x \in X, y \in Y}}
\ee
gives the complete characterization of the behavior\index{behavior} of the quantum device.

In general we may consider $\mathbb{P}(A,B|X,Y)$ as a vector in $\mathbb{R}_{+}^{|A| \cdot |B| \cdot |X| \cdot |Y|}$ satisfying standard Kolmogorovian conditions, and possibly some other constraints. One may also ask a question which joint probability distributions for \textit{fixed scenario under interest} are allowed by physical theories? In other words, we ask about the set,
\be
	\nonumber
	\mathcal{P}(A,B|X,Y) \subset \mathbb{R}_{+}^{|A| \cdot |B| \cdot |X| \cdot |Y|},
\ee
containing \textit{all allowed probability distributions} satisfying constraints imposed by some theory, \textit{cf.} sets $\gls{localSet} \subset \gls{quantumSet} \subset \gls{noSignalSet}$ defined below. Below we consider several theories, for each of them the set of all allowed probability distributions is a convex set\footnote{In general, this does not need to be true.}, \textit{i.e.} the convex combination of probabilities in a given set belongs to that set.

A common-sense, \textit{i.e.} classical, approach to the description of the state and measurement has the following properties:
\begin{itemize}
	\item The whole system at the beginning is in one of possible internal states $\lambda \in \Lambda$ with probability $P(\lambda)$ given by distribution $\mathbb{P}(\Lambda)$. This state is shared by both subsystems.
	\item $a \in A$ is some (possibly random) function of $x \in X$ and $\lambda \in \Lambda$, and similarly $b \in B$ depends on $y \in Y$ and $\lambda \in \Lambda$. Thus we have probability distributions $\mathbb{P}_A(A|X,\Lambda)$ and $\mathbb{P}_B(B|Y,\Lambda)$. The measurements are performed locally.
\end{itemize}
Motivated by these considerations we introduce the following definition:
\begin{definition}
	\label{def:local}
	The joint probability distribution $\mathbb{P}(A,B|X,Y)$ for which there exist conditional probability distributions
	\be
		\nonumber
		\mathbb{P}(\Lambda), \mathbb{P}_A(A|X,\Lambda), \mathbb{P}_B(B|Y,\Lambda)
	\ee
	satisfying 
	\be
		\label{eq:Lprob}
		P(a,b|x,y) = \sum_{\lambda} P(\lambda) \cdot P(a|x,\lambda) \cdot P(b|y,\lambda)
	\ee
	is called a \textbf{local}\index{distribution!local} or \textbf{classical}\index{distribution!classical} probability distribution.
\end{definition}
The intuition behind the local probability distributions is the following. The device consists of two parts which possibly were initially connected. The parts share some random variable $\lambda$. When the experiment is performed on this device, the behavior of both parts is completely local, meaning that the measurement and its result on one part does not influence the result in the second part.

The set of all local distributions (for fixed scenario) is denoted by $\gls{localSet}$. This set of probabilities is of interest of the ``classical'' computer science. As noted in the preface, the main step leading from computer science to QI is in taking into account a wider class of probability distributions.


One may consider the so-called no-signaling probability distributions. These are bounded by a condition that an immediate communication is forbidden. This is formalized by the no-signaling principle, which is commonly considered as a fundamental property of Nature. The principle imposes the following definition.
\begin{definition}
	\label{def:no-signal}
	We say that a joint probability distribution
	\be
		\nonumber
		\mathbb{P}(A,B|X,Y)
	\ee
	is \textbf{no-signaling}\index{distribution!no-signaling} if and only if there exist marginal probability distributions $\mathbb{P}_A(A|X)$ and $\mathbb{P}_B(B|Y)$ satisfying the following conditions for all $a$, $b$, $x$ and $y$:
	\be
		\label{eq:marginals}
		\ba
			& \gls{PAaxy} = \gls{PAax}, \\
			& \gls{PBbxy} = \gls{PBby}.
		\ea
	\ee
\end{definition}
These constraints are formulated in abstraction to any underlying physical theory. The set of such probability distributions (for fixed scenario) is denoted by \gls{noSignalSet}. Thus for no-signaling probability distribution we may define the marginal probability distributions $\mathbb{P}_A(A|X)$ and $\mathbb{P}_B(B|Y)$. The intuitive description of this condition is that the setting on one part of the device does not influence the second part of the device (or that the information cannot be sent immediately over the distance).

For our further considerations it is crucial to define a set of probability distributions which are allowed by quantum mechanics \cite{NPA08}. In this definition the projective measurements are given by operators $\{E^a_x\}$ for Alice, and $\{F^b_y\}$ for Bob.
\begin{definition}
	\label{def:quantum}
	We say that a joint probability distribution
	\be
		\nonumber
		\mathbb{P}(A,B|X,Y)
	\ee
	is \textbf{quantum}\index{distribution!quantum} if and only if there exists a Hilbert space $\gls{HilbertSpace}$, on which it is possible to define a state (a unit vector) $\ket{\psi}$, and a set of projective measurement operators
	\be
		\label{eq:realizationPMs}
		\{E^a_x, F^b_y\}_{a \in A, b \in B, x \in X, y \in Y},
	\ee
	for which the following conditions are fulfilled:
	\begin{enumerate}
		\item $E^a_x = {E^a_x}^{\dagger}$ and $F^b_y = {F^b_y}^{\dagger}$ (operators are Hermitian),
		\item $E^a_x E^{a^{\prime}}_x = 0$ and $F^b_y F^{b^{\prime}}_y = 0$ for $a \neq a^{\prime}$ and $b \neq b^{\prime}$ (orthogonality of different results with the same setting),
		\item $\sum_{a \in A} E^a_x = \mathbb{I}$ and $\sum_{b \in B} F^b_y = \mathbb{I}$ (sum of probabilities of all outcomes for each setting is equal to $1$),
		\item $[E^a_x, F^b_y] = 0$ (measurements of Alice and Bob commute, \textit{i.e.} their results do not depend on the time order in which they were performed, which is justified by the no-signaling principle),
		\item $P(a,b|x,y) = \bra{\psi} E^a_x F^b_y \ket{\psi} = \Tr \left( \proj{\psi} E^a_x F^b_y \right)$ (the probability distribution is given by measurements on the quantum state).
	\end{enumerate}
	We say that $\ket{\psi}$ and $\{E^a_x, F^b_y\}_{a,b,x,y}$ \textbf{realize} $\mathbb{P}(A,B|X,Y)$.

	We denote this probability by
	\be
		\nonumber
		\mathbb{P} \left[ \{\{E^a_x\}_{a \in A}\}_{x \in X}, \{\{F^b_y\}_{b \in B}\}_{y \in Y} \right].
	\ee
	
	We call $\ket{\psi}$ together with the set in~\eqref{eq:realizationPMs} a \textbf{quantum realization}\index{quantum realization}, or realization, of a quantum probability distribution.

	If $A = B = \{0,1\}$, then $\mathbb{P}(A,B|X,Y)$ is called \textbf{binary}.

	The set of \textit{all quantum probability distribution (for fixed scenario)} is denoted as $\gls{quantumSet}$.
\end{definition}
This definition formalizes the intuition that all bipartite quantum probability distributions are realized by a physical scenario in which Alice and Bob share some quantum state and perform independent measurements on it.

From the conditions $\sum_{a \in A} E^a_x = \mathbb{I}$ and $\sum_{b \in B} F^b_y = \mathbb{I}$ we see that for $a_0 \in A$, and $\tilde{A} \equiv A \setminus \{a_0\}$ we have
\be
	\nonumber
	E^{a_0}_x = \mathbb{I} - \sum_{a \in \tilde{A}} E^a_x,
\ee
and for $b_0 \in B$, $\tilde{B} \equiv B \setminus \{b_0\}$
\be
	\nonumber
	E^{b_0}_y = \mathbb{I} - \sum_{b \in \tilde{B}} F^b_y.
\ee
Thus to define a realization it is enough to specify the set
\be
	\nonumber
	\{E^a_x, F^b_y\}_{a \in \tilde{A}, b \in \tilde{B}, x \in X, y \in Y}.
\ee
The sets $\tilde{A}$ and $\tilde{B}$ are called \textbf{reduced outcome sets}\index{reduced outcome set}.

\begin{figure}[htbp]
	\centering
		\includegraphics[width=0.5\textwidth]{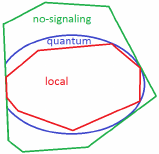}
	\caption{Relation between sets of probability distributions allowed by different theories.}
	\label{fig:probabilities_sets}
\end{figure}

We have the following inclusion relation $\gls{localSet} \subset \gls{quantumSet} \subset \gls{noSignalSet}$. These inclusions are strict. This is schematically depicted in Fig.~\ref{fig:probabilities_sets}. All the above considerations are easily generalized for cases with more than two parties.

If we abstract from the physical implementation of some scenario, and consider only the probability distribution $\mathbb{P}(A,B|X,Y)$ generated by it, then we call this probability distribution a \textbf{behavior}\index{behavior}, and refer to the physical object as a \textbf{box}\index{box} or \textbf{device}\index{device}.

For joint probability distributions one often introduces the correlation function. In considered distributions the correlation is usually defined for outcomes if given settings are chosen. For the case with two outcomes, \textit{i.e.} $A=B=\{o_1,o_2\}$, we define the following expression as the correlation of outcomes for settings $x \in X$ and $y \in Y$:
\be
	\label{eq:correlation}
	C(x,y) \equiv P(o_1,o_1|x,y) - P(o_1,o_2|x,y) - P(o_2,o_1|x,y) + P(o_2,o_2|x,y).
\ee
In particular for binary outcomes this expression reads
\be
	\nonumber
	C(x,y) \equiv P(0,0|x,y) - P(0,1|x,y) - P(1,0|x,y) + P(1,1|x,y).
\ee

\subsubsection{Examples of deterministic and quantum boxes}

We will illustrate the above definitions by an explicit example of a scenario with two settings and two outcomes on Alice and Bob side, \textit{i.e.} the case $A=B=X=Y=\{1,2\}$.

A simple box from the set $\mathcal{L}$ is shown in tab.~\ref{tab:exampleDeterministicBox}. This box is obtained when both Alice and Bob always give the result $1$. We contrast it with an example of a box from the set $\mathcal{Q}$ shown in tab.~\ref{tab:exampleQuantumBox}. In sec.~\ref{sec:CHSH} we give an example of a game which can be won with the deterministic box with probability $0.75$ and with the quantum box with probability about $0.851$.

\begin{table}[htbp]
	\caption{An example of a deterministic box.}
	\begin{tabular}{|l|l|r|r|r|r|}
		\hline
		 &  & \multicolumn{2}{c|}{$y=1$}  & \multicolumn{2}{c|}{$y=2$} \\ \hline
		 &  & \multicolumn{1}{l|}{$b=1$} & \multicolumn{1}{l|}{$b=2$} & \multicolumn{1}{l|}{$b=1$} & \multicolumn{1}{l|}{$b=2$} \\ \hline
		$x=1$ & $a=1$ & $1$ & $0$ & $1$ & $0$ \\ \hline
		 & $a=2$    & $0$ & $0$ & $0$ & $0$ \\ \hline
		$x=2$ & $a=1$ & $1$ & $0$ & $1$ & $0$ \\ \hline
		 & $a=2$    & $0$ & $0$ & $0$ & $0$ \\ \hline
	\end{tabular}
	\label{tab:exampleDeterministicBox}
\end{table}

\begin{table}[htbp]
	\caption[An example of a quantum box.]{An example of a quantum box. In the table we use $q = \frac{1}{4}\left(1+\frac{\sqrt{2}}{2}\right) \approx 0.427$. These probabilities are obtained when Alice and Bob share a two dimensional singlet state, Alice performs measurement in the basis given by \eqref{eq:Bloch} with angles $\phi=0$ for $x=1$ and $\phi=\frac{\pi}{2}$ for $x=2$, and Bob measures in the basis specified by angles $\frac{\pi}{4}$ and $-\frac{\pi}{4}$ for $y=1$ and $y=2$, respectively.}
	\begin{tabular}{|l|l|r|r|r|r|}
		\hline
		 &  & \multicolumn{2}{c|}{$y=1$}  & \multicolumn{2}{c|}{$y=2$} \\ \hline
		 &  & \multicolumn{1}{l|}{$b=1$} & \multicolumn{1}{l|}{$b=2$} & \multicolumn{1}{l|}{$b=1$} & \multicolumn{1}{l|}{$b=2$} \\ \hline
		$x=1$ & $a=1$ & $q$ & $\frac{1}{2}-q$ & $q$ & $\frac{1}{2}-q$ \\ \hline
		 & $a=2$ & $\frac{1}{2}-q$ & $q$ & $\frac{1}{2}-q$ & $q$ \\ \hline
		$x=2$ & $a=1$ & $q$ & $\frac{1}{2}-q$ & $\frac{1}{2}-q$ & $q$ \\ \hline
		 & $a=2$ & $\frac{1}{2}-q$ & $q$ & $q$ & $\frac{1}{2}-q$ \\ \hline
	\end{tabular}
	\label{tab:exampleQuantumBox}
\end{table}

\subsection{White noise}

We note here that in practical realization of experiments some noise occurs. We model this with the so-called \textbf{white noise}\index{white noise} parametrized by $\gls{noiseParameter}$. Let us consider a probability distribution $\mathbb{P}(A,B|X,Y)$. In this model the experimentally observed probabilities are
\be
	\nonumber
	(1-\eta) \mathbb{P}(A,B|X,Y) + \eta \mathbb{P}_U(A,B|X,Y),
\ee
where $\mathbb{P}_U(A,B|X,Y)$ is the uniform probability distribution for each $x \in X$ and $y \in Y$.

Another useful parameter is \textbf{purity}, denoted by $\gls{purity}$. For the purpose of this work we define it to be equivalent to white noise with $\eta=1-p$. Thus the value $p = 1$ refers to the case when there is no noise.

From this we get that, for binary outcomes, the correlations in this case are given by $p \cdot C(x,y)$, where $C(x,y)$ are the correlations in the noiseless case.

\section{Bell operators and inequalities}
\label{sec:BI}

In this section we give a short introduction to the topic of Bell inequalities and Bell operators \cite{Bell64}. Intuitively, a Bell inequality\index{Bell inequality} is the limit on some expression of probabilities not possible to be violated in a world governed by classical physics, but possible in the quantum mechanics.

A violation of Bell inequalities may be observed experimentally, and give a conclusive evidence that the world is not governed by the laws of the classical physics. The famous experiments were conducted in 1980s by Aspect \textit{et al.} \cite{Aspect81,Aspect82a,Aspect82b}. A \textbf{Bell experiment} involves at least two separated parties who share a quantum state and perform subsequent measurements with different settings without any communication between them. After series of such measurements, the collected data are used to estimate the joint probabilities of the outcomes conditioned on the settings.

In fact further in this work we will look at Bell experiments in a more general way. Namely we do not make any assumptions on the experimental setup, apart from the condition that parts of the setup are separated. Thus we consider only the information theoretic property, \textit{i.e.} the probability distribution generated in the setup, and ignore the physical realization\footnote{To be strict, we retain a hidden assumption that the setup obeys the laws of quantum physics, or some of its relaxations described in sec.~\ref{sec:NPA}.}.

The quantum maximum of a Bell operator is called the Tsirelson bound\index{Tsirelson!bound}. This notion has been introduced by a Russian-Israeli mathematician, Boris Semyonovich Tsirelson\index{Tsirelson}, in 1980 in the revolutionary paper \cite{Tsirelson80}.

For the purpose of this work we define a \textbf{Bell operator}\index{Bell operator} as a linear functional defined on a multipartite probability distribution $\mathbb{P}$. More formally, a bipartite Bell operator
\be
	\nonumber
	\gls{BellI}
\ee
is a linear function defined, in particular, for probability distributions
\be
	\nonumber
	\mathbb{P}(A,B|X,Y).
\ee
It is of the form:
\be
	\label{BI}
	\ba
		I & (A,B,X,Y,\{\alpha_{a,b,x,y}\},C_I)[\mathbb{P}(A,B|X,Y)] \equiv \\
		& = \sum_{a \in A} \sum_{b \in B} \sum_{x \in X} \sum_{y \in Y} \alpha_{a,b,x,y} P(a,b|x,y) + C_I,
	\ea
\ee
where $\alpha_{a, b, x, y}, C_I \in \mathbb{R}$. Further in this work we omit $\mathbb{P}$ if it is obvious which probability distribution is considered. A Bell operator of a particular interest, \gls{BellIhat}, is of the following \textit{correlation} form for binary probability distribution $\mathbb{P}(A,B|X,Y)$ with $A=B=\{o_1,o_2\}$:
\be
    \label{BIhat}
    \begin{aligned}
        \hat{I} & (X, Y, \{\alpha_{x,y}\}, \hat{C}_I)[\mathbb{P}(\{o_1,o_2\},\{o_1,o_2\}|X,Y)] \equiv \\
        & = \sum_{x \in X} \sum_{y \in Y} \hat{\alpha}_{x,y} C(x,y) + \hat{C}_I,
    \end{aligned}
\ee
with $\hat{\alpha}_{x,y},\hat{C}_I \in \mathbb{R}$, and $C(x,y)$ defined by~\eqref{eq:correlation}. Obviously, the form (\ref{BI}) conforms the form (\ref{BIhat}) if and only if
\be
	\nonumber
	\alpha_{o_1,o_1,x,y} = \alpha_{o_2,o_2,x,y} = -\alpha_{o_1,o_2,x,y} = -\alpha_{o_2,o_1,x,y} = \hat{\alpha}_{x,y}.
\ee

A Bell operator $I$ may possess a property that its maximal value allowed on the set $\gls{quantumSet}$, $I_Q$, is strictly larger than its maximal value on the set $\gls{localSet}$, $I_L$. The fact that such operators exist is a result of the Bell's theorem \cite{Bell87}. A \textit{Bell inequality} is the relation $B[\mathbb{P}] \leq I_L$ stating the limit on the value of this operator under local theories. We say that a Bell's inequality is violated if given quantum probability distribution $\mathbb{P}_Q$ we have $I[\mathbb{P}_Q] > I_L$.

The constant term $C_I$ in a Bell inequality does not change its properties. Still, we retain this general form, both for Bell inequalities, and dimension witnesses in sec.~\ref{sec:DW}.

Below we give a short overview of Bell operators which are important for the further discussion of our results. In most of the cases we are not interested in the relevant Bell inequality, since it is not needed in the analysis of the protocols further in this work.

\subsection{CHSH}
\label{sec:CHSH}

The most prominent example of Bell inequalities is the so-called CHSH\index{Bell operator!CHSH} inequality, which stands for the names of its authors, John Clauser, Michael Horne, Abner Shimony, and Richard Holt \cite{CHSH}.

In this scenario we have $A=B=X=Y=\{1,2\}$ The CHSH operator has the correlation form,~\eqref{BIhat}:
\be
	\label{eq:CHSH}
	C(1, 1) + C(1, 2) + C(2, 1) - C(2, 2).
\ee
The maximal value of this operator in $\gls{localSet}$ is $2$, whereas in $\gls{quantumSet}$ it is $2 \sqrt{2}$.

There are many different formulations of this operator. Another important one is a game formulation. Let us consider Alice and Bob playing a cooperative game in which they are not allowed to communicate. We take $A=B=X=Y=\{0,1\}$, Alice gets bit $x$, and Bob gets bit $y$ (they do not know the bit of the other party). They win the game if the bits they return, $a$ and $b$, satisfy
\be
	\label{eq:CHSHgame}
	a \oplus b = x \cdot y.
\ee
The probability distribution $\mathbb{P}(A,B|X,Y)$ determines the strategy of Alice and Bob which is allowed in the given theory. The maximal success probability if $\mathbb{P}(A,B|X,Y)$ is in the set $\gls{localSet}$, $\gls{quantumSet}$ and $\gls{noSignalSet}$, is $0.75$, $\frac{1}{2}\left(1+\frac{1}{\sqrt{2}} \right) \approx 0.851$ and $1$, respectively.

\subsection{I3322}
\label{sec:I3322}

The I3322\index{Bell operator!I3322} stands for: three settings for the first and second party, and two outcomes for the first and second party. The inequality related to the I3322 operator is sometimes called the Froissard inequality \cite{Froissart81}. It has been reintroduced in \cite{CG04}.

The I3322 operator is given by the following formula
\be
	\label{eq:I3322}
	\begin{aligned}
		I3322 & \equiv -P_A(1|1) - 2 P_B(1|1) - P_B(1|2) + P(1,1|1,1) + P(1,1|1,2) \\
		& + P(1,1|1,3) + P(1,1|2,1) + P(1,1|2,2) - P(1,1|2,3) \\
		& + P(1,1|3,1) - P(1,1|3,2),
	\end{aligned}
\ee
where $\gls{PAax}$ and $\gls{PBby}$ are marginal probability distributions of Alice and Bob, see~Eq.~\eqref{eq:marginals}. The relevant Bell inequality is $I3322 \leq 0$.

The exact Tsirelson bound on this operator is not known precisely. Using SDP (the NPA method discussed further in sec.~\ref{sec:NPA}, and the so-called see-saw \cite{seesaw10} SDP method) one may get that this bound is close to $0.25089$.

\subsection{Braunstein-Caves family of chained Bell operators}
\label{prot:BC}

The Braunstein-Caves\index{Bell operator!Braunstein-Caves} family of chained Bell operators was introduced in 1988 in the paper \cite{BC88}. The members of this family are parametrized by an odd number $n \geq 2$. This parameter gives the number of binary measurement settings for two parties. For these operators the sets of settings are $X=Y=\{1,\cdots,n\}$. Operators from this family consist of chains of operators of correlation between subsequent measurement settings of Alice and Bob,. They fit into the correlation form,~\eqref{BIhat}.

The most commonly used member of this family is the Braunstein-Caves operator with three settings for each of the two parties, BC3. BC3 is defined as
\be
	\label{BC3BI}
	BC3 \equiv C(1, 1) + C(1, 2) + C(2, 2) + C(2, 3) + C(3, 3) - C(3, 1),
\ee

The general formula for $n^{th}$ Braunstein-Caves operator is
\be
	\label{BCn}
	\ba
		BCn & = C(1, 1) + C(1, 2) + C(2, 2) + C(2, 3) + C(3, 3) \\
		& + C(3, 4) + \ldots + C(n-1, n-1) + C(n-1, n) \\
		& + C(n, n)- C(n, 1)
	\ea
\ee

The Tsirelson bound for $n$-th Braunstein-Caves operator is given by the formula \cite{W06}
\be
	\nonumber
	2 n \cos \left( \frac{\pi}{2 n} \right).
\ee
In particular BC3 is limited by $6 \cos \left( \frac{\pi}{6} \right) \approx 5.19$.

\subsection{Modified CHSH}
\label{prot:modCHSH}\index{Bell operator!modified CHSH}

Now let us consider the following Bell operator, which we have considered in the paper \cite{MP13}. We have called it a \textit{modified CHSH} because it is a CHSH operator with one additional correlation term. Four terms of this operator form a CHSH operator. The properties of this operator have been further investigated in \cite{Scarani14}.

Modified CHSH is an operator of the following form:
\be
	\label{modCHSH}
		C(1, 2) + C(1, 3) + C(2, 1) + C(2, 2) - C(2, 3)
\ee
Its Tsirelson bound is $1 + 2 \sqrt{2}$.

\subsection{Tn Bell operators}
\label{sec:T3}

In the paper \cite{HWL13} we have introduced a family of Bell operators, $Tn$, with $n \in \{2, 3, \dots\}$. They are obtained from a family of dimension witnesses introduced in \cite{HWL12} using our method described in sec.~\ref{sec:sdi}. Both of these families are denoted by $Tn$.

This family of operators is given by the following formula
\be
	\label{eq:Tn}
	Tn \equiv \sum_{x,y} (-1)^{x_y} C(x,y),
\ee
where $x = (x_1, x_2, \cdots, x_n)$, with $x_2,\cdots,x_n \in \{0,1\}$ and $x_1=0$ (\textit{i.e.} $|X| = 2^{n-1}$), and $y \in Y = \{1, \cdots, n\}$. Their Tsirelson bounds are $\sqrt{n} 2^{n-1}$.

The member of this family $T3$ is of particular interest further in this work. The Bell operator in this case is
\be
	\label{T3BI}
	\ba
		T3 = & C(1,1)+C(2,1)+C(3,1)+C(4,1) +C(1,2) +C(1,3) \\
		&  +C(2,2) -C(2,3) -C(3,2) +C(3,3) -C(4,2) -C(4,3),
	\ea
\ee
where we have used labels $000 \equiv 1$, $001 \equiv 2$, $010 \equiv 3$ and $011 \equiv 4$ for the setting of Alice.

\subsection{CGLMP}
\label{sub:CGLMP}\index{Bell operator!CGLMP}

The CGLMP operator has been introduced by Collins, Gisin, Linden, Massar and Popescu in \cite{CGLMP}. In this case both Alice and Bob have two measurement settings with three outcomes,
\be
	\nonumber
	A=B=\{0,1,2\},
\ee
and
\be
	\nonumber
	X=Y=\{1,2\}.
\ee
The operator has the following form:
\be
	\label{eq:CGLMP} 
	\ba
		& CGLMP \equiv \\
		& P(0,0|1,1)-P(0,2|1,1)+P(0,0|1,2)-P(0,2|1,2) \\
		& -P(1,0|1,1)+P(1,1|1,1)-P(1,0|1,2)+P(1,1|1,2) \\
		& -P(2,1|1,1)+P(2,2|1,1)-P(2,1|1,2)+P(2,2|1,2) \\
		& -P(0,0|2,1)+P(0,1|2,1)+P(0,0|2,2)-P(0,2|2,2) \\
		& -P(1,1|2,1)+P(1,2|2,1)-P(1,0|2,2)+P(1,1|2,2) \\
		& +P(2,0|2,1)-P(2,2|2,1)-P(2,1|2,2)+P(2,2|2,2).
	\ea
\ee
This operator in $\gls{localSet}$ can obtain at most value $3$, giving the Bell inequality
\be
	\nonumber
	CGLMP \leq 3,
\ee
which may be violated in $\gls{quantumSet}$.

\subsection{Other inequalities}
\label{prot:other}

Now we describe two Bell operators which we have introduced in \cite{HWL13}. These two were taken as examples from wider groups of inequalities, which use at most $4$ measurement settings of Alice and $3$ of Bob, and consist of $7$ and respectively $10$ correlations. Operators from the first group are similar to these from Braunstein-Caves family.

The operators are defined by the following equations:
\be
\label{eq:I1}
	\ba
		I_1 \equiv & C(1, 2) - C(1,3) - C(2, 1) - C(2, 2)+ \\
		& C(3, 1) + C(3, 3) + C(4, 1)
	\ea
\ee
with Tsirelson bound $1 + 6 \cdot \cos \left( \frac{\pi}{6} \right) \approx 6.19$, and
\be
\label{eq:I2}
 \ba
	I_2 & \equiv -C(1,2) + C(1, 3) + C(2, 1) + C(2, 2) \\
		& + C(2, 3) + C(3, 2) - C(3, 3) \\
		& + C(4, 1) + C(4, 2) + C(4, 3)
 \ea
\ee
with Tsirelson bound $2 + 4 \cdot \sqrt{2} \approx 7.66$.

The maximal values of the operators occurring in the inequalities are possible to be obtained with qubits.

\subsection{Mermin game}
\label{sec:Mermin}\index{Bell operator!Mermin}

The last important Bell operator considered in this work is the one introduced by Mermin in 1990 \cite{Mermin}. This scenario differs from the previous cases in that it involves three parties, instead of two.

The standard formulation of this scenario is the following game. Alice, Bob and Charlie each receive one input bit, $x,y$ and $z$, respectively. There is a promise that
\be
	\nonumber
	x \oplus y \oplus z = 1.
\ee
Each of the players returns a single bit denoted $a,b$ and $c$. They win if $a \oplus b \oplus c = xyz$, or, using the promise, $a \oplus b \oplus c = xy$. In the unbiased (\textit{i.e.} with a uniform distribution of $x$ and $y$) version of this game the classical success probability is $\frac{3}{4}$, whereas quantum mechanics allows to reach 1.

\section{Navascues-Pironio-Ac\'in method}
\label{sec:NPA}

From definitions~\ref{def:local} and \ref{def:no-signal} of local and no-signaling probability distributions, it follows that for given $A,B,X,Y$ these two sets of probability distributions, $\gls{localSet}$ and $\gls{noSignalSet}$, are described by a finite set of linear constraints. In consequence the task of optimization of expressions linear in probabilities, \textit{e.g.} Bell operators, is solved by LP, and thus is easy.

Indeed, the no-signaling constraints in the definition~\ref{def:no-signal} are explicitly linear constraints. If we treat joint probability distributions as non-negative vectors with proper normalization constraints, then it is easy to see that we get an LP problem. The LP formulation of $\gls{localSet}$ takes advantage of the fact that any local joint probability distribution can be written as a convex combination of deterministic probability distributions \cite{FHSW10}, \textit{i.e.} such distribution for which $a \in A$ is a (deterministic) function of $x \in X$, and $b \in B$ is a deterministic function of $y \in Y$.

The task of optimization of such expressions, in particular Bell operators, over $\gls{quantumSet}$ is NP-hard, as shown by Kempe \textit{et al.} in 2008 at FOCS \cite{Kempe08}. Fortunately, as mentioned above, in 2006 Navascues, Pironio and Ac\'in proposed a method of relaxation which allows to formulate the task of optimization with expressions linear in probabilities as an SDP.

The NPA allows to analyze the scenario shown in Fig.~\ref{fig:AliceBob}, \textit{i.e.} the situation in which two parties, Alice and Bob, perform some measurements on a possibly entangled system and obtain some joint probability distribution. For the sake of simplicity we assume that each measurement setting gives outcomes from the same set, namely Alice gets a result $a \in A$, and Bob a result $b \in B$, for some sets $A$ and $B$.

The \textbf{sequence of operators}\index{sequence of operators}, or sequence, is a concatenation of the projective measurement operators, \textit{e.g.} $S = E^1_2 E^3_2 F^2_1 E^1_1$ is a sequence of four operators. Using the fact that any operator of Alice, $E^a_x$, commutes with any operator of Bob, $F^b_y$, we can reorder this sequence to get
\be
	\nonumber
	E^1_2 E^3_2 F^2_1 E^1_1 = E^1_2 E^3_2 E^1_1 F^2_1.
\ee
Recall that $E^a_x E^{a^{\prime}}_x = 0$ and $F^b_y F^{b^{\prime}}_y = 0$ for $a \neq a^{\prime}$ and $b \neq b^{\prime}$. Using this and the commutation property we get for example
\be
	\nonumber
	E^2_1 F^3_3 E^1_1 = E^2_1 E^1_1 F^3_3 = 0
\ee
since $E^2_1$ and $E^1_1$ are orthogonal. From the fact that $E^a_x$ are projectors, we have $(E^a_x)^k = E^a_x$ for any $k \geq 1$, and similarly for $F^b_y$. The \textbf{length}\index{sequence of operators!length} of a sequence of operators $S$ is the minimal number of projectors needed to formulate it. A \textbf{null sequence}\index{sequence of operators!null sequence} is the identity operator, $\idOp$. We define its length to be $0$.

The idea of this method is as follows. Suppose that a given joint probability distribution $\mathbb{P}(A,B|X,Y)$ is quantum, which by definition~\ref{def:quantum} means that there exists a realization with a state $\ket{\psi}$ and projective measurements $\{ E^a_x, F^b_y \}$ such that, for all settings $x \in X$ and $y \in Y$ and outcomes $a \in A$ and $b \in B$, it satisfies
\be
	\label{eq:P-NPA}
	P(a,b|x,y) = \bra{\psi} E^a_x F^b_y \ket{\psi}.
\ee

Let us consider an $n$-element set $\mathcal{S}$ of sequences of operators, \textit{e.g.}
\be
	\label{eq:AQseqence}
	\mathcal{S}_{1+AB} = \left\{ \idOp, E^a_x, F^b_y, E^a_x F^b_y \right\}_{\substack{a \in A, b \in B \\ x \in X, y \in Y}}.
\ee
For $O_i, O_j \in \mathcal{S}$ let us define
\be
	\label{eq:gamma}
	\Gamma_{O_i, O_j} \equiv \bra{\psi} O_i^{\dagger} O_j \ket{\psi}.
\ee
By this equation we have that $\Gamma_{E^a_x, E^b_y} = P(a,b|x,y)$, and $\Gamma_{\idOp, \idOp} = 1$. This equation defines a $n \times n$ matrix, in which rows and columns are indexed by elements of $\mathcal{S}$.

Now we show that $\Gamma \succeq 0$. Let $v \in \mathbb{C}^n$. For $V = \sum_j v_j O_j$ we have
\be
	\begin{aligned}
		v^\dagger \Gamma v & = \sum_{i,j} v_i^{*} \Gamma_{i,j} v_j = \sum_{i,j} v_i^{*} \bra{\psi} O_i^{\dagger} O_j \ket{\psi} v_j \\
		& = \bra{\psi} V^{\dagger} V \ket{\psi} = | V \ket{\psi} |^2 \geq 0,
	\end{aligned}
\ee
and thus $\Gamma \succeq 0$. Moreover it is easy to see that the elements of the matrix $\Gamma$ satisfy the following linear constraints
\be
	\label{eq:gammaConds}
	\ba
		\Gamma_{i,j} = \Gamma_{k,l} \Leftarrow O_i^{\dagger} O_j = O_k^{\dagger} O_l, \\
		O_i^{\dagger} O_j = \zeroOp \Rightarrow \Gamma_{i,j} = 0.
	\ea
\ee
Further in this work we are interested in optimizing the values of expressions which are linear functions of the elements of $\Gamma$. The discussion in sec.~\ref{sec:complexSDP} allows us to consider only the real part of the matrix $\Gamma$ for such kind of problems. Thus further in this work all SDP problems are in real variables.

With the above formulation, the conditions of the definition~\ref{def:no-signal} also have to be imposed as additional linear constraints. There exists a method which allows to assure that the distributions will be no-signaling. This is done by taking into account only the reduced outcome sets\index{reduced outcome set}. Then, for example, in the binary case we consider only sequences with operators $E^0_x$ and $F^0_y$ (since the outcomes ``$1$'' can be recovered with $E^1_x=\idOp-E^0_x$ and $F^1_y=\idOp-F^0_y$). In consequence the $\Gamma$ matrix contains only probabilities $P_A(0|x)$, $P_B(0|y)$ and $P(0,0|x,y)$ stated explicitly (see~Eq.~\eqref{eq:P-NPA}). We use the normalization and no-signaling of probabilities to get
\be
	\label{eq:probReducedGamma}
	\ba
		& P_A(1|x) = 1 - P_A(0|x), \\
		& P_B(1|y) = 1 - P_B(0|y), \\
		& P(1,0|x,y) = P_B(0|y) - P(0,0|x,y), \\
		& P(0,1|x,y) = P_A(0|x) - P(0,0|x,y), \\
		& P(1,1|x,y) = 1 - P(0,0|x,y) - P(0,1|x,y) - P(1,0|x,y).
	\ea
\ee
This gives the \textit{reduced} $\Gamma$, which we further refer to simply as $\Gamma$\index{$\Gamma$ matrix}.

From this, we have some relaxation of the conditions on the probability distribution $\mathbb{P}(A,B|X,Y)$ which are quantum. Instead of assuming that the state $\ket{\psi}$ and the proper measurement operators exist, we check if it is possible to construct an PSD matrix $\Gamma$ which satisfies conditions \eqref{eq:gammaConds} together with no-signaling constraints. It is easy to see that the family of matrices satisfying conditions in~\eqref{eq:gammaConds} is convex.

To sum up, we consider sequences of operators $\mathcal{S}$ of a certain length with $a \in \tilde{A}$, $b \in \tilde{B}$, $x \in X$ and $y \in Y$. We use them to define a set of PSD matrices with constraints \eqref{eq:gammaConds}. We relate certain entries of these matrices with probabilities, either directly or using \eqref{eq:probReducedGamma}. The optimization task linear in entries of matrices from this set is an SDP optimization problem.

We note that the problem of checking whether given probability distribution is quantum can be modeled as an SDP feasibility problem with relevant entries of matrices determined by a given distribution.

With the NPA method it is possible to construct a hierarchy of relaxations by taking different choices of the set of sequences $\mathcal{S}$. One defines a set $\mathcal{S}_k$ as a set of all sequences of operators $\{E^a_x, E^b_y\}$ on length at most $k$. We have that $\mathcal{S}_{1+AB} = \mathcal{S}_1 \cup \{ E^a_x E^b_y \}$, where $\mathcal{S}_{1+AB}$ is defined in~\eqref{eq:AQseqence}.

Note that the relaxations allow for a wider class of joint probability distributions than the quantum mechanics. The sets allowed by $\mathcal{S}_k$ are denoted by $\mathcal{Q}_k$. The larger set $\mathcal{S}$ we choose, the more restrictive is the relaxation. In \cite{NPA08} it was shown, that the hierarchy of such relaxations converges to quantum mechanics in a sense stated in~\eqref{eq:QnRelations}.

Hierarchy of level $\mathcal{Q}_{2}$ means that the set $\mathcal{S}$ consists of all sequences of measurement operators of length $2$, whereas in level $\gls{AQset}$, $\mathcal{S}$ is a set of all sequences of length $1$ and sequences with one operator of Alice and one of Bob. Obviously $\mathcal{Q}_{2}$ is defined by a longer $\mathcal{S}$, than $\gls{AQset}$. To sum up we have
\be
	\label{eq:QnRelations}
	\ba
		& \mathcal{S}_1 \subset \mathcal{S}_2 \cdots \subset \mathcal{S}_{\infty}, \\
		& \mathcal{Q}_1 \supset \mathcal{Q}_2 \cdots \supset \mathcal{Q}, \\
		& \mathcal{Q} = \bigcap_{k=1}^{\infty} Q_k.
	\ea
\ee
The last equality, \textit{i.e.} convergence to the quantum set, is proven in \cite{NPA08} and is crucial for this work. The sizes of sets $\mathcal{S}_k$, and thus the sizes of $\Gamma$ matrices, grow exponentially as
\be
	\nonumber
	O \left( (|A|\cdot|X| + |B|\cdot|Y|)^k \right).
\ee
In practice one rarely uses sets beyond $Q_2$, and $Q_{1+AB}$ is sufficient for most purposes.

We mention that in the first level of the hierarchy, $\mathcal{Q}_1$, the probabilities occurring in $\Gamma$ can possibly be negative. For this reason for this hierarchy level one has to add the constraint of non-negativity of probabilities explicitly as linear constraints. The hierarchy $\gls{AQset}$, and the subsequent levels does satisfy the no-signaling principle in a natural way, without additional constraints.

Further in this work we refer to the $k$-th NPA hierarchy with settings $A$ and $B$, and outcomes $X$ and $Y$, for Alice and Bob respectively, as
\be
	\nonumber
	\gls{QkSet}.
\ee
Because the settings and outcomes can be numbered, we will usually refer only to the sizes of the relevant sets, and write $\mathcal{Q}_k(|A|,|B| | |X|,|Y|)$. The $\Gamma$ matrix is sometimes called a \textbf{certificate}\index{certificate!of quantum probability distribution}, since it gives a clue that the related joint probability distribution,~\eqref{eq:P-NPA}, possibly is quantum.

We also note that the hierarchy can be in a natural way defined for scenarios involving more than two parties \cite{AQ}.

An example of an SDP matrix defining an NPA problem is given in appendix~\ref{app:exampleNPA}. This kind of matrices is discussed in more details in sec.~\ref{sec:NPAmatrices}.

\subsubsection{The Almost quantum set}

The above relaxations are very efficient. The work \cite{AQ} considers the possibility that correlations which are observed in the physical world are in fact described not by quantum mechanics, but by hypothetical theory which allows the set $\gls{AQset}$. We will say a few words on this problem.

The problem of deriving the laws of physics out of some simple postulates is not new. The special relativity theory is build on postulates that the laws of physics are invariant in all inertial systems, and that the speed of light in a vacuum is the same for all observers.

There were many efforts to find relevant postulates for quantum theory \cite{PR94,NTCC06,NANC07,ML09,LO13}. It reveals that the set $\gls{AQset}$ satisfies all of them. Only the relation between this set and the postulate of Information Causality \cite{IC09} is currently unknown. For these reasons this set is called an \textbf{Almost Quantum}\index{distribution!Almost Quantum} set of probability distributions. For most of practical application this set is close enough to the quantum set. The situations in which higher levels of the NPA hierarchy are needed are rare.

\subsection[Formulations and complexity of $\Gamma$]{Primal and dual formulations and the complexity of the $\Gamma$ matrix}
\label{sec:formulationsGamma}

As we mentioned in sec.~\ref{sec:SDPformulation}, an SDP problem can be stated either in a primal or in a dual form. We will now discuss the issue of formulation of the SDP problems involving the $\Gamma$ matrix.

The problem we wish to consider is the following: optimize some linear expression $F(\Gamma)$ under linear constraints of the form $G_i(\Gamma) \geq g_i$ and $H_j(\Gamma) = h_j$, with $\Gamma$ satisfying also linear constraints stated in~\eqref{eq:gammaConds}.

We start with a primal formulation, which is suggested in the original NPA paper \cite{NPA08}. In this case $\Gamma = X$ is the primal variable, and constraints $H_j(\Gamma) = h_j$ with the defining constraint from~\eqref{eq:gammaConds} are expressed with the linear constraint matrices $\{A_i\}_{i=1}^{m}$. Each equality constraint from~\eqref{eq:gammaConds} is imposed by one matrix $A_i$. The constraints $G_i(\Gamma) \geq g_i$ are easily stated in the mixed form (see sec.~\ref{sec:mixed}), or by adding additional diagonal terms in an SDP matrix (the latter method is more expensive). The problem with primal formulation is that the number of linear constraints, $m$, is large, see tab.~\ref{tab:CHSHsizes} for the example of the scenario with two parties, with two settings and two outcomes.

In the dual formulation we identify all entries of the $\Gamma$ matrix which are constrained to be equal with one of the values in the dual variable $y$. Recall that $\Gamma_{\idOp,\idOp} = 1$, and this value is put into the $C$ matrix. Similarly, like in the primal formulation, the constraints $G_i(\Gamma) \geq g_i$ are expressed either as diagonal terms or in the mixed form. The constraints $H_j(\Gamma) = h_j$ can be imposed in two ways. One way is to use the dependencies stated by them to reduce the number of degree of freedom of the dual variable $y$. The other way is to replace each constraint by two inequalities,
\be
	\nonumber
	\ba
		H_j(\Gamma) \geq h_j, \\
		H_j(\Gamma) \leq h_j.
	\ea
\ee
Although the latter method slightly increases the difficulty of the problem, it allows to employ the method of sec.~\ref{sec:sparsityNPA} to improve the performance of a solver.

The sizes of problems (without constraints $G_i$ and $H_j$) are determined by the size of the $\Gamma$, $n$, and the number of the linear constraint matrices $\{A_i\}_{i=1}^{m}$, $m$. Tab.~\ref{tab:CHSHsizes} shows examples of sizes of problems in the scenario involving two parties, each with two settings and two outcomes in both primal and dual formulations. It is easy to see that the dual formulation seems to be more suitable. Further in this work we state the problems in the dual form.

\begin{table}[htbp]
	\caption[Sizes of different hierarchy levels for CHSH.]{Sizes of SDP problems occurring in different hierarchy levels of the NPA method for CHSH problem, \textit{i.e.} for the quantum scenario involving two parties, each with two settings and two outcomes on their measurement devices. The column n refers to the size of the $\Gamma$ matrix, and thus the size of SDP variables. Columns m-primal and m-dual contain the number of linear constraint matrices\index{linear constraint matrices}, $\{A_i\}$ (\textit{cf.}~\eqref{SDP-primal} and~\eqref{SDP-dual}) in primal and dual formulation of problem constrained with conditions from~\eqref{eq:gammaConds}. The average density is given by $\frac{\text{m-dual}}{n^2}$, and express the sparsity of the linear constraint matrices in the dual formulation.}
	\begin{tabular}{|l|r|r|r||r|}
	\hline
		hierarchy level & \multicolumn{1}{l|}{n} & \multicolumn{1}{l|}{m-dual} & \multicolumn{1}{l|}{average density (dual)} & \multicolumn{1}{l|}{m-primal} \\ \hline
		$\mathcal{Q}_2$ & 13 & 31 & 0.183 & 137 \\ \hline
		$\mathcal{Q}_3$ & 25 & 61 & 0.098 & 563 \\ \hline
		$\mathcal{Q}_4$ & 41 & 101 & 0.060 & 1579 \\ \hline
		$\mathcal{Q}_5$ & 61 & 151 & 0.041 & 3569 \\ \hline
		$\mathcal{Q}_6$ & 85 & 211 & 0.029 & 7013 \\ \hline
		$\mathcal{Q}_7$ & 113 & 281 & 0.022 & 12487 \\ \hline
		$\mathcal{Q}_8$ & 145 & 361 & 0.017 & 20663 \\ \hline
		$\mathcal{Q}_9$ & 181 & 451 & 0.014 & 32309 \\ \hline
		$\mathcal{Q}_{10}$ & 221 & 551 & 0.011 & 48289 \\ \hline
		$\mathcal{Q}_{11}$ & 265 & 661 & 0.009 & 69563 \\ \hline
		$\mathcal{Q}_{12}$ & 313 & 781 & 0.008 & 97187 \\ \hline
		$\mathcal{Q}_{13}$ & 365 & 911 & 0.007 & 132313 \\ \hline
		$\mathcal{Q}_{14}$ & 421 & 1051 & 0.006 & 176189 \\ \hline
		$\mathcal{Q}_{15}$ & 481 & 1201 & 0.005 & 230159 \\ \hline
	\end{tabular}
	\label{tab:CHSHsizes}
\end{table}

\subsection{Example of Navascues-Pironio-Ac\'in problem}
\label{sec:exampleNPA}

We will now give a simple example of the method. We consider a scenario $\mathcal{Q}_1(2,2|2,2)$. This is a case with two parties, each having two setting, $X=Y=\{1,2\}$, and obtaining two outcomes, $A=B=\{0,1\}$. We consider the reduced outcomes set, which in this case is $\{E^0_1, E^0_2, F^0_1, F^0_2\}$. The sequence of operators of length at most $1$ is
\be
	\nonumber
	\mathcal{S}_1 = \{\idOp, E^0_1, E^0_2, F^0_1, F^0_2\} \equiv \{O_1, O_2, O_3, O_4, O_5\}.
\ee
This five sequences will label the rows and columns of the certificate, and thus this leads to a $5 \times 5$ hermitian PSD matrix:
\be
	\nonumber
	\Gamma^{(complex)} \equiv
	\begin{bmatrix}
		1 & \chi_1 & \chi_2 & \chi_4 & \chi_7 \\
		\chi_1^{*} & \chi_1 & \chi_3 & \chi_5 & \chi_8 \\
		\chi_2^{*} & \chi_3^{*} & \chi_2 & \chi_6 & \chi_9 \\
		\chi_4^{*} & \chi_5^{*} & \chi_6^{*} & \chi_4 & \chi_{10} \\
		\chi_7^{*} & \chi_8^{*} & \chi_9^{*} & \chi_{10}^{*} & \chi_7.
	\end{bmatrix}
\ee
This comes directly from the definition \eqref{eq:gamma}. For example, we have for some state $\ket{\psi}$ and operators $\{E^0_1, E^0_2, F^0_1, F^0_2\}$:
\be
	\nonumber
	\ba
	\Gamma^{(complex)}_{1,2} & = \bra{\psi} O_1^{\dagger} O_2 \ket{\psi} = \bra{\psi} E^0_1 \ket{\psi} \equiv \chi_1 \\
	& = \bra{\psi} E^0_1 E^0_1 \ket{\psi} = \Gamma^{(complex)}_{2,2} = \Gamma^{(complex)}_{2,1} = \chi_1^{*}.
	\ea
\ee
Thus $\chi_1$ is a real number. We can repeat this reasoning for other $\chi_i$, apart from $\chi_3$ and $\chi_{10}$\footnote{This is because $\chi_3 \equiv \bra{\psi} E^0_1 E^0_2 \ket{\psi}$ in general is not equal to $\bra{\psi} E^0_2 E^0_1 \ket{\psi}$, and similarly for $\chi_{10}$.}, which are thus complex numbers. On the other hand, as explained above, we take as the certificate the real part of this matrix. Thus $\Gamma$ will be a $5 \times 5$ real matrix:
\be
	\nonumber
	\Gamma \equiv
	\begin{bmatrix}
		1 & y_1 & y_2 & y_4 & y_7 \\
		y_1 & y_1 & y_3 & y_5 & y_8 \\
		y_2 & y_3 & y_2 & y_6 & y_9 \\
		y_4 & y_5 & y_6 & y_4 & y_{10} \\
		y_7 & y_8 & y_9 & y_{10} & y_7
	\end{bmatrix} \succeq 0,
\ee
$y_i \in \mathbb{R}$. It is easy to see that this matrix satisfies conditions \eqref{eq:gammaConds}. In the dual formulation of SDP this matrix can be parametrized in the following way:
\be
	\nonumber
	\ba
		& C - \sum_i y_i A_i = \\
		& \begin{bmatrix}
			1 & 0 & 0 & 0 & 0 \\
			0 & 0 & 0 & 0 & 0 \\
			0 & 0 & 0 & 0 & 0 \\
			0 & 0 & 0 & 0 & 0 \\
			0 & 0 & 0 & 0 & 0
		\end{bmatrix}
		- y_1
		\begin{bmatrix}
			0 & -1 & 0 & 0 & 0 \\
			-1 & -1 & 0 & 0 & 0 \\
			0 & 0 & 0 & 0 & 0 \\
			0 & 0 & 0 & 0 & 0 \\
			0 & 0 & 0 & 0 & 0
		\end{bmatrix} 
		\dots - y_{10} 
		\begin{bmatrix}
			0 & 0 & 0 & 0 & 0 \\
			0 & 0 & 0 & 0 & 0 \\
			0 & 0 & 0 & 0 & 0 \\
			0 & 0 & 0 & 0 & -1 \\
			0 & 0 & 0 & -1 & 0
		\end{bmatrix} \succeq 0.
	\ea
\ee
If we use the notation from \eqref{eq:mathcalA}, then the matrices $A_i$ are columns of a matrix $\mathcal{A}$ given in appendix~\ref{app:exampleNPA}.

We have
\be
	\nonumber
	\ba
		& P_A(0|0) = \Gamma_{\idOp,E^0_0} = y_1, \\
		& P_A(0|1) = \Gamma_{\idOp,E^0_1} = y_2, \\
		& P_B(0|0) = \Gamma_{\idOp,F^0_0} = y_4, \\
		& P_B(0|1) = \Gamma_{\idOp,F^0_1} = y_7, \\
		& P(0,0|0,0) = \Gamma_{E^0_0,F^0_0} = y_5, \\
		& P(0,0|1,0) = \Gamma_{E^0_1,F^0_0} = y_6, \\
		& P(0,0|0,1) = \Gamma_{E^0_0,F^0_1} = y_8, \\
		& P(0,0|1,1) = \Gamma_{E^0_1,F^0_1} = y_9.
	\ea
\ee
Only these probabilities occur directly in the $\Gamma$, because we have used the reduced outcome set\index{reduced outcome set}. We can recover the remaining probabilities using formulas from~\eqref{eq:probReducedGamma}.

Note that $\Gamma_{E^0_0,E^0_1} = y_3$ and $\Gamma_{F^0_0,F^0_1} = y_{10}$ are not probabilities. They refer to $\bra{\psi} E^0_0 E^0_1 \ket{\psi}$ and $\bra{\psi} F^0_0 F^0_1 \ket{\psi}$, which have no physical meaning, because Alice and Bob perform their measurements only once.

Let us denote by $M_{F[\mathbb{P}(A,B|X,Y)]}$ a matrix which under Frobenius product operation with $\Gamma$ gives the value of a linear function of probabilities, $F[\mathbb{P}(A,B|X,Y)]$. For example with $F[\mathbb{P}(A,B|X,Y)] = P_B(0|1) - P(0,0|1,0)$ we construct
\be
	\nonumber
	M_{P_B(0|1) - P(0,0|1,0)} \equiv
	\begin{bmatrix}
		0 & 0 & 0 & 0 & \frac{1}{2} \\
		0 & 0 & 0 & 0 & 0 \\
		0 & 0 & 0 & -\frac{1}{2} & 0 \\
		0 & 0 & -\frac{1}{2} & 0 & 0 \\
		\frac{1}{2} & 0 & 0 & 0 & 0
	\end{bmatrix}.
\ee
Then we have
\be
	\nonumber
	\Tr \left( \Gamma M_{P_B(0|1) - P(0,0|1,0)} \right) = y_7 - y_6 = P_B(0|1) - P(0,0|1,0).
\ee
We use matrices $M_{F[\mathbb{P}(A,B|X,Y)]}$ in primal formulation of NPA. For dual formulation we define $v_{F[\mathbb{P}(A,B|X,Y)]}$ with
\be
	\nonumber
	v_i \equiv \Tr \left( A_i M_{F[\mathbb{P}(A,B|X,Y)]} \right).
\ee
To formulate maximization of the expression $F[\mathbb{P}(A,B|X,Y)]$ as primal SDP we use $C = - M_{F[\mathbb{P}(A,B|X,Y)]}$ and explicitly state constraints \eqref{eq:gammaConds} with $A_i$ matrices and vector $b$. In dual formulation we use $b = v_{F[\mathbb{P}(A,B|X,Y)]}$ as the target. Usually we will not use canonical forms of SDP in this work, but formulate problems in the most intuitive way.

\subsection{Maximization of Bell operators}

In this section we give two examples of SDP problems using NPA method. We will use these optimization problems further in chapter~\ref{chap:solver} as performance testing cases for different SDP solvers. This examples will also explain the notation we use further in this work.

The first example is the CHSH maximization in Almost Quantum\index{distribution!Almost Quantum} level:
\begin{align}
	\label{problem-CHSH}
	\begin{split}
		\text{maximize } & \Tr \left( \Gamma M_{CHSH} \right) \\
		\text{subject to } &\null \Gamma \in \mathcal{Q}_{1+AB}(2,2|2,2)
	\end{split}
\end{align}
where
\be
	\nonumber
	M_{CHSH} \equiv M_{C(1,1)} + M_{C(1,2)} + M_{C(2,1)} - M_{C(2,2)}.
\ee
This problem gives the exact value, $2 \sqrt{2}$, up to numerical precision. This illustrates how the NPA method can be used to calculate an approximation of the Tsirelson bound.

The second example is the I3322\index{Bell operator!I3322} maximization in the same level of the hierarchy formulated as SDP in the following way, \textit{cf.}~Eq.~\eqref{eq:I3322}:
\begin{align}
	\label{problem-I3322}
	\begin{split}
		\text{maximize } & \Tr \left( \Gamma M_{I3322} \right) \\
		\text{subject to } &\null \Gamma \in \mathcal{Q}_{1+AB}(2,2|3,3)
	\end{split}
\end{align}
where
\be
	\nonumber
	\ba
		M_{I3322} & \equiv M_{P_A(1|1)} - 2 M_{P_B(1|1)} - M_{P_B(1|2)} +  M_{P(1,1|1,1)} + M_{P(1,1|1,2)} \\ 
		& + M_{P(1,1|1,3)} + M_{P(1,1|2,1)} + M_{P(1,1|2,2)} - M_{P(1,1|2,3)} \\
		& + M_{P(1,1|3,1)} - M_{P(1,1|3,2)}.
	\ea
\ee

\section{Dimension witnesses}
\label{sec:DW}\index{dimension witness}

In this section we give a brief overview of topics related to the so-called dimension witnesses \cite{HWL11,DW4,DimWit,PB11,DW5}.

\textbf{Dimension witnesses}\index{dimension witness} are defined in a scenario with two parties, Alice and Bob. Let $\bar{X}$ and $\bar{Y}$ be sets labeling the settings of Alice and Bob, and let $\bar{B}$ be a set of the outcomes that Bob can obtain. Both parties get some random inputs, $x \in \bar{X}$ and $y \in \bar{Y}$, for some sets $\bar{X}$ and $\bar{Y}$. Afterward Alice sends a message to Bob depending on her input. The message can either be a classical message (a number) or a quantum state of some dimension $d$. In the latter case, Bob after receiving the quantum state of dimension $d$ performs some measurement on it and obtains a result $b \in \bar{B}$. In the former (classical) case Bob calculates his result using some (possibly random) function of the message and his input, $y$. The message is then one digit, $\{0, \dots, d-1\}$, which can be modeled as sending a state $\phi \in \{\ket{0}, \dots, \ket{d-1}\}$.

\begin{figure}[htbp]
	\centering
		\includegraphics[width=0.5\textwidth]{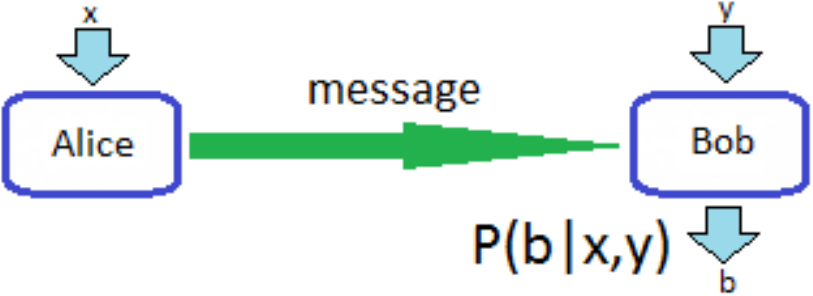}
		\caption{Prepare-and-measure scenario.}
	\label{fig:AliceBobDW}
\end{figure}

These considerations lead to some conditional probability distribution $\mathbb{P}(\bar{B}|\bar{X},\bar{Y})$ in a \textbf{prepare-and-measure}\index{prepare-and-measure} scheme (\acrshort{P-M}), depicted in Fig.~\ref{fig:AliceBobDW}. We formalize the conditions for this distribution in the following definition~\ref{probSDI}. Roughly speaking, a probability distribution in P-M is realized by a setup in which Alice prepares and sends states from a Hilbert space $\gls{HilbertSpace}$ of dimension $d$ to Bob who, after receiving, performs on them some operations leading him to his outcome.
We assume that there is no entanglement between Alice and Bob.

\begin{definition}
	\label{probSDI}
	Let $\bar{B}$, $\bar{X}$, and $\bar{Y}$ be sets, and $\gls{HilbertSpace}$ be a Hilbert space of a finite dimension $d$.

	A \textit{probability distribution in P-M scheme} is a conditional probability distribution $\gls{PdBXY}$ such that for $b \in \bar{B}$, $x \in \bar{X}$ and $y \in \bar{Y}$ we have
	\be
		\nonumber
		P(b|x,y) = \Tr \left(\rho_{x} M^b_y\right),
	\ee
	where $\{\rho_{x}\}_{x \in \bar{X}}$ is a set of density matrices on the space $\gls{HilbertSpace}$, and $\{M^b_y\}_{b \in \bar{B}}$ are POVMs on this space for all $y \in \bar{Y}$.

	We say that $\mathbb{P}_d$ is \textbf{realized}\index{probability distribution!realization} by sets $\{\rho_x\}_{x \in \bar{X}}$ and $\left\{ \{M^b_y\}_{b \in \bar{B}} \right\}_{y \in \bar{Y}}$, and denote it
	\be
		\nonumber
		\mathbb{P}_d \left[ \{\rho_x\}_{x \in \bar{X}}, \{\{M^b_y\}_{b \in \bar{B}}\}_{y \in \bar{Y}} \right].
	\ee

	If $\bar{B} = \{0,1\}$, then $\mathbb{P}_d(\bar{B}|\bar{X},\bar{Y})$ is called a \textit{binary probability distribution}\index{probability distribution!binary}. Then $\bar{B}$ can be omitted, $\mathbb{P}_d(\bar{X},\bar{Y})$.

	The set of all P-M probability distributions for given $d$, $\bar{B}$, $\bar{X}$ and $\bar{Y}$ is denoted by $\gls{PpmdBXY}$.
	
	The set of all P-M probability distributions for given $\bar{B}$, $\bar{X}$ and $\bar{Y}$, but without restriction on the dimension is denoted by $\gls{PpmBXY}$, \textit{viz.}
	\be
		\nonumber
		\gls{PpmBXY} = \bigcup_{d=2}^{\infty} \gls{PpmdBXY}.
	\ee
	
	We denote by $\gls{PPXY}$ the subset of binary probability distribution with $\{M^b_y\}_{y \in \bar{Y}}$ being projective measurements with restrictions that $d = 2$ and
	\be
		\nonumber
		\forall_{b \in \{0,1\}} \forall_{y \in \bar{Y}} \Tr M^b_y = 1,
	\ee 
	meaning that for any $y \in \bar{Y}$ both $M^0_y$ and $M^1_y$ project on different subspace of dimension $1$.
\end{definition}

Let us note that we have
\be
	\label{eq:dimInrease}
	\mathcal{P}_d (\bar{B}|\bar{X},\bar{Y}) \subseteq \mathcal{P}_{d+1} (\bar{B}|\bar{X},\bar{Y}),
\ee
since increasing the dimension of communicated state we can send at most the same amount of data. The key property of dimension witnesses is that they allow to distinguish the dimensions for which inclusion in \eqref{eq:dimInrease} is strict.

Using the definition of probability distributions in P-M scheme, we may introduce a notion of dimension witnesses which is analogous to the concept of Bell operators.

\begin{definition}
	\label{def:binaryZerosummming}
	Let $\bar{B}$, $\bar{X}$, and $\bar{Y}$ be sets, $\mathbb{P}(\bar{B}|\bar{X},\bar{Y}) \in \gls{PpmBXY}$, $\beta_{b,x,y} \in \mathbb{R}$ for all $b \in \bar{B}$, $x \in \bar{X}$ and $y \in \bar{Y}$, and $C_W \in \mathbb{R}$. A dimension witness $W$ is a linear function of conditional probability distributions, \textit{i.e.} it has the following form:
	\be
		\label{DW}
		\gls{dimWit} [\mathbb{P}(\bar{B}|\bar{X},\bar{Y})] \equiv \sum_{b \in \bar{B}} \sum_{x \in \bar{X}} \sum_{y \in \bar{Y}} \beta_{b,x,y} P(b|x,y) + C_W.
	\ee

	If $\bar{B} = \{0,1\}$, then the dimension witness is called \textbf{binary}\index{dimension witness!binary}.
	
	If $\forall_{b \in \bar{B}} \forall_{y \in \bar{Y}} \sum_{x \in \bar{X}} \beta_{b,x,y} = 0$, then the dimension witness is called \textbf{zero-summing}\index{dimension witness!zero-summing}.
\end{definition}

From the above definition it follows that dimension witnesses are linear functionals on $\gls{PpmBXY}$.

A dimension witness states a maximal value of a certain linear combination of these probabilities that is possible to be obtained with a given dimension of the state space. Since, in contrary to the case of Bell inequalities, the limits possible to be reached with given dimensions of the communicated system in classical and quantum cases interlace, there is no direct analogy of the Bell inequality. Instead the limits allowed for each particular case of the dimension of the communicated system are considered. Thus the dimension witnesses can be used to lower bound the size of communication needed to reproduce some probability distribution $\mathbb{P} (\bar{B}|\bar{X},\bar{Y})$ \cite{DimWit,PB11,DW4}.

\chapter[Device-independent protocols]{Device-independent and semi-device-independent protocols}
\label{chap:DI}\index{device-independent}\index{semi-device-independent}

Nowadays information has become one of the most important resources, which is illustrated by an enormous success of computer science and information technology. These disciplines deal in particular with the information processing. Another important investigated issue is the reliability of the information. The purpose of this chapter is to discuss how this problem is addressed by QI.

In 1998 at FOCS Mayers and Yao \cite{MY98} proposed a new approach to attaining a credibility of information. This approach, called \textit{device\hyp{}independent} (\acrshort{DI}), allows to gain a trust to a device in the case in which we \textbf{do not trust the vendor of our devices}, and even do not know how the device is constructed. No matter how unbelievable this claim sounds, it reveals that in some cases one may draw conclusions only from the observed results which the device returns, or in other words, by performing some self-testing instead of investigating the internal working of the device.

It may sound a little paradoxical that we can gain some certainty about a device without the knowledge of its construction, especially if the device has been constructed by our enemy wanting to mislead us. One thing we can be certain about is the fact that the device is governed by the laws of physics, so its working is limited by what is allowed by quantum mechanics. Without looking inside the device we also may easily state that it comprises of several separate parts. DI approach shows that these two facts may in many cases be enough to make sure that the untrusted device works as expected. As we will describe in sec.~\ref{sec:randExpansion}, one of the ways to perform self-tests on a certain class of devices is to conduct a Bell experiment.

Still, Bell experiments are very difficult to perform, since they require a high degree of precision and extremely high efficiency of the detectors used. For this reason a slightly different approach has been introduced \cite{PB11}, which assumes some knowledge about the device. Since we have to know something about the construction of the device, this approach is called \textit{semi-device-independent} \cite{SDI,SDI_effects} (\acrshort{SDI}). It offers a good compromise between security and experimental feasibility.

In SDI we usually work in the prepare-and-measure\index{prepare-and-measure} scheme, see sec.~\ref{sec:DW}. In this framework dimension witnesses, instead of	 Bell inequalities, are used. We still do not make any assumption of the internal working of the devices, but we assume that the communication between parties is bounded to the extent known to us and that there is no entanglement between parts of the device. This approach is much more convenient in practical realization.

Both for prepare-and-measure protocols in SDI, and for correlation protocols in DI, we would like to define a value that measures how reliable its particular realization is. As this value we take the expectation value of the relevant dimension witness or Bell operator, respectively, attained in the relevant protocol. This value is called a \textbf{security parameter}\index{security parameter}. It is possible to use more than one security parameter. In general a security parameter does not have to be the expectation value of a dimension witness or Bell operator, but any linear function of probabilities occurring in a given setup.

Below we discuss some particular applications of DI and SDI. Readers not interested in details of the methods used in DI may skip sec.~\ref{sec:extractors} which gives an overview of the topic of the randomness extraction. Similarly, sec.~\ref{sec:estimation}, in which we briefly discuss the issue of estimation of parameters in experiments, may be skipped. The only aim of that section is to illustrate how the values of expressions like the values of Bell operators are estimated in the experiment. Readers should remember from these sections that the protocols described further in chapter~\ref{chap:quantumProtocols} are indeed possible to conduct, and the randomness they generate can be quantified and used for generic purposes.

We note that the NPA method has become a basic tool for investigation of DI scenarios \cite{RNGCBT,LubiePlacki}. In our works \cite{HWL13,HWL14} we developed a method to apply the NPA for SDI scenario. Other methods for optimization in SDI are \cite{NTV,MiguelVertesi}.

\section{Randomness expansion}
\label{sec:randExpansion}\index{randomness!expansion}

Random number generation is an important issue in computer science. The applications of random numbers are ubiquitous. They occur in such topics as authentication \cite{NIST800632}, cryptography, gambling and system modeling.

Nonetheless, most of the random number generators (\acrshort{RNG}s) are based on a purely algebraical manipulation on the initial seed. Since the series of numbers produced by such generators are created in a deterministic manner, these RNGs are called pseudo-random number generators (\acrshort{PRNG}s). However, it is very difficult to develop a reliable pseudo-random number generation method. On the other hand, in many cases safe RNG method is crucial for the security of computer systems. A prominent example of systems whose security was undermined by PRNG is the Secure Socket Layer encryption protocol in early versions of Netscape web browser which used a seed of a small entropy (the time of a day, the process and the parent process ids) \cite{Netscape96}. In 2008 it was discovered that the results of PRNG of Debian's OpenSSL package are easy to predict \cite{Debian08}. The vulnerability of the PRNG of Windows~2000 operating system has been investigated in \cite{DGP09}.

Although there exist tests \cite{NIST80022} which allow checking whether a given sequence of numbers is reliable, or determine whether it conforms to a particular probability distribution, we can never be sure its security without the knowledge of how the sequence was generated. These tests can never give a guarantee that the numbers were indeed trustworthy, that is they cannot be predicted by an adversary. If we know the pseudo-random generating algorithm and the initial seed (or some long enough sequence of generated numbers), then we can predict every sequence of number that will ever be obtained by PRNG. All classical PRNGs have this significant \cite{Mitnick} drawback.

There also exist RNGs based on some chaotic classical physical processes such as electric or atmospheric noise or on estimating the entropy of hardware interrupts (for example in $/dev/random$ RNG on Linux systems), but they either need a third party supplier (the former case), or are not able to attain a satisfactory rate (the latter cases).

On the contrary, one of the most striking properties of quantum mechanics is that it is intrinsically random. The randomness is present in the deepest level of quantum physics, and is unavoidable. We know that some processes are intrinsically random, if only the no-signaling principle is valid, which as we have said is one of the most fundamental assumptions of modern physics (see~sec~\ref{sec:Hilbert}).

In this situation it is natural to try to use properties of quantum mechanics in order to generate entirely random sequences of numbers. There were efforts to make use of such quantum processes like nuclear decay (\textit{e.g.} HotBits \cite{HotBits}), or photons hitting a semi-transparent mirror (for example, \textit{id~Quantique}'s RNGs \cite{IDQ}). 

However, when using one of the commercially available quantum random number generators (\acrshort{QRNG}s), we still have to trust the vendor of the device. This assumption is crucial for QRNGs which produce randomness, \textit{e.g.} by performing measurements in Hadamard basis with states from computational basis. Such a device works only if it indeed implements the declared states and measurements. Unfortunately, in cryptographic context one does not always trust his devices.

Suppose there is an honest constructor who wants to produce and sell QRNGs. His problem is the lack of trust among his potential customers. Since he does not want to cheat his clients, he can make the design of his device open. But still some parties may distrust that the device is constructed in the declared manner. This way the idea of the quantum randomness certification emerged \cite{RNGCBT}. As mentioned briefly in the introduction to this chapter if we want to be sure that a device does really produce random numbers, we can perform a self-test with a Bell experiment.

One of the achievements of DI which is recently developing rapidly is the usage of the violation of certain Bell inequalities as a \textbf{certificate}\index{certificate!of security}\footnote{We note that \textbf{certificate} is a homonym, used both as a term for the existence of a relevant $\Gamma$\index{certificate!of quantum probability distribution} matrix in NPA method, and for a warranty of reliability in context of DI and SDI\index{certificate!of security}. The meaning is clear from the context.} of randomness for series of numbers from QRNGs constructed in a suitable way \cite{ColPHD,RNGCBT,CK11,HWL13}. In \cite{ColPHD,RNGCBT} as a certificate of randomness the violation of the CHSH Bell inequality\index{Bell operator!CHSH} \cite{CHSH} was used, while in \cite{CK11} the GHZ correlations were used instead. In our work \cite{LubiePlacki} we have investigated a few other ways to certify the randomness within the DI scheme, see chapter~\ref{chap:quantumProtocols}.

The QRNGs which need some initial randomness for the purpose of certification\footnote{If the QRNG is certified to be reliable, one does not need any randomness to generate subsequent numbers. On the other hand, in sec.~\ref{sec:estimation} we describe methods which interweave the process of certification and generation.} are called \textit{randomness expanders}. This is the case in all discussed protocols.

We would like to emphasize that the certification with Bell experiments works independently of the internal construction of the device used. If the value obtained in the experiment attains a certain threshold, we are sure that the generated results are indeed random, even if the device has been prepared by our foe. A lower bound on the amount of the obtained secure randomness can then be quantified by means of min-entropy using, \textit{e.g.} the methods discussed in this work. 

We refer to the total value of min-entropy generated by a working randomness expansion protocol as \textit{amount of randomness}\index{randomness!amount of}.

As already mentioned, SDI approach is more feasible in realization than DI. It reveals that this scheme also allows an efficient randomness certification \cite{HWL11}. For this reason in our papers \cite{HWL13,HWL14} we have investigated the relation between random number expansion protocols based on correlations occurring in the Bell scenario, and on protocols relying on the prepare-and-measure scheme. Further in sec.~\ref{sec:sdi}, we discuss this relation and the issue of SDI certification in more details.

A typical scheme of a DI randomness certification protocol is the following. One chooses an expression, which can be a Bell operator or set of operators, which will be used to certify the amount of randomness generated in the protocol, and a pair of settings, $x_0$ and $y_0$, used to produce randomness. A protocol runs in a series of iterations and involves typically two parties, Alice and Bob, who share a series of entangled states, one per iteration. In a single iteration they randomly choose their measurement settings using private trusted RNG\footnote{As mentioned previously, the protocols are randomness expanders\index{randomness!expansion}, since they require some initial randomness. One may decide to perform certification first using PRNG, and then always to choose $x_0$ and $y_0$. Otherwise one interweaves the certification with the proper randomness generation. Because in a vast majority of the iterations the settings $x_0$ and $y_0$ are chosen, the initial randomness is significantly less that the generated randomness.}. After the measurements they reveal their measurement settings and outcomes to perform estimation of the value(s) of operator(s) used as certificates\index{certificate!of security}, using for example a method described in sec.~\ref{sec:Azuma}. This value (values) is used to quantify the amount of certified randomness contained in the pairs of outcomes of Alice and Bob generated in iterations with settings $x_0$ and $y_0$.

An SDI randomness certification protocol works in a similar way. This time the parties do not share entangled states, but instead in each iteration Alice prepares a quantum state depending on her setting and sends it to Bob. Random numbers are obtained from the outcomes of Bob for the selected settings $x_0$ and $y_0$. One uses dimension witnesses for certification instead of Bell operators.

Before we move to the discussion of other applications of DI, we develop some issues related to DI analysis of two types of probability distributions.

\subsection{Joint probability distributions}

Let us take two sets, $X$ and $Y$, which label the measurement settings of Alice and Bob in DI scheme, and two sets, $A$ and $B$, which label their respective outcomes.

In sec.~\ref{sec:min-entropy} we defined the min-entropy of a probability distribution. We will now use this notion in order to quantify the randomness contained in joint probability distributions. The \textit{global} randomness is the one which is contained in the pair of outcomes of Alice and Bob. For the joint probability distribution $\mathbb{P}(A,B|X,Y)$ the global min-entropy given $x \in X$ and $y \in Y$ is
\be
	\nonumber
	- \log_{2} \left( \max_{a \in A, b \in B} P(a,b|x,y) \right).
\ee
We refer to the randomness of an outcome of a single party as a \textbf{local} randomness\index{randomness!local}.

Throughout this work we use min-entropy as a measure of the efficiency of the protocols, see sec.~\ref{sec:entropy}.

For given $A$, $B$, $X$, $Y$, $x_0 \in X$, $y_0 \in Y$, a Bell inequality $I$ and $s \in \mathbb{R}$ we define as the probability of guessing outcomes the following terms:
\be
		P_{guess}(\mathbb{P}(A,B|X,Y),x_0,y_0) \equiv \max_{a \in A, b \in B} P(a,b|x_0,y_0) \nonumber
\ee
This is motivated by intuition that if one tries to guess the outcomes, he will guess the most probable, \textit{cf.}~above discussion of min-entropy. Thus we have the following expression of min-entropy of outcomes (see~sec.~\ref{sec:min-entropy}):
\be
	\nonumber
	H_{\infty} (\mathbb{P}(A,B|X,Y),x_0,y_0) \equiv -\log_2 \left( P_{guess}(\mathbb{P}(A,B|X,Y),x_0,y_0) \right)
\ee

We define the certified randomness with the value $s$ of a Bell operator $I$ by
\be
	\label{eq:certHAB}
	\ba
		H_{\infty}^{cert} & (I,x_0,y_0,s) \equiv \\
		& \min_{\mathbb{P}(A,B|X,Y) \in \mathcal{Q}} H_{\infty}(\mathbb{P}(A,B|X,Y),x_0,y_0), \\
		& \text{subject to } I[\mathbb{P}(A,B|X,Y)] \geq s.
	\ea
\ee
$H_{\infty}^{cert}(B,x_0,y_0,s)$ is called the min-entropy certified by the value $s$ of $B$. Note that DI independent problem defined above in~\eqref{eq:certHAB} can be relaxed to an NPA problem if we replace $\mathcal{Q}$ with $\mathcal{Q}_n(A,B|X,Y)$ for some hierarchy level $n$, \textit{e.g.} $\gls{quantumSet}$.

\subsection{Prepare-and-measure probability distributions}

As we mentioned above, if we can bound the dimension of the communicated system, we may use the \textit{prepare and measure scheme} and certify the randomness \cite{HWL11}.

For given $\bar{B}$, $\bar{X}$, $\bar{Y}$, $x_0 \in \bar{X}$, $y_0 \in \bar{Y}$, a dimension witness $W$ (see~sec.~\ref{sec:DW}), $s \in \mathbb{R}$ and $d \geq 2$ as the probability of guessing we define the following term:
\be
	\label{Pguess}
	P_{guess}(\mathbb{P}_d(\bar{B}|\bar{X},\bar{Y}),x_0,y_0) \equiv \max_{b \in \bar{B}} P(b|x_0,y_0),
\ee
with min-entropy given by
\be
	\nonumber
	H_{\infty}(\mathbb{P}_d(\bar{B}|\bar{X},\bar{Y}),x_0,y_0) \equiv -\log_2 \left( P_{guess}(\mathbb{P}_d(\bar{B}|\bar{X},\bar{Y}),x_0,y_0) \right).
\ee

In analogy to~\eqref{eq:certHAB}, we define
\be
	\label{PguessCert}
	\ba
		P_{guess}^{cert} & (W,x_0,y_0,s,d) \equiv \max_{\mathbb{P}_d(\bar{B}|\bar{X},\bar{Y}) \in \gls{PpmdBXY}} \max_{b \in \bar{B}} P(b|x_0,y_0), \\
		& \text{subject to } W[\mathbb{P}_d(\bar{B}|\bar{X},\bar{Y})] \geq s,
	\ea
\ee
and
\be
	H_{\infty}^{cert}(W,x_0,y_0,s,d) \equiv -\log_2 \left( P_{guess}^{cert}(W,x_0,y_0,s,d) \right) \nonumber
\ee
for the case when randomness is certified with a dimension witness with communication limited by $d$ (see the definition~\ref{probSDI}), which is the min-entropy certified by the value $s$ of $W$ (for the dimension $d$).

We also introduce the particular case of certified randomness, \textit{viz.}
\be
	\label{PguessCertP}
	\begin{aligned}
		P_{guess}^{cert(P)} & (W,x_0,y_0,s) \equiv \\
		& \max_{\mathbb{P}_2(\bar{B}|\bar{X},\bar{Y}) \in \gls{PPXY}} \max_{b \in \bar{B}} P(b|x_0,y_0). \\
		& \text{subject to } W[\mathbb{P}_2(\bar{B}|\bar{X},\bar{Y})] \geq s,
	\end{aligned}
\ee
where $\gls{PPXY}$ is defined in definition~\ref{probSDI}. Note that in \eqref{PguessCert} and \eqref{PguessCertP} maximum is taken at two stages: we consider that probability distribution for which the most probable result has the highest probability among all other probability distributions.

The meaning of the above equations is as follows. \eqref{Pguess} expresses the probability that the eavesdropper correctly guesses a single outcome generated on the side of Bob. As usually, the strategy used by the eavesdropper is to guess the most probable result for a given distribution of outcomes $\mathbb{P}_d(\bar{B}|\bar{X},\bar{Y})$. \eqref{PguessCert}) refers to the maximal guessing with all possible probability distributions of outcomes being in accordance with the observed security parameter\index{security parameter}. \eqref{PguessCertP} gives this probability with a further restriction to the case with the dimension two and projective measurements.

\subsection{Randomness extractors}
\label{sec:extractors}\index{randomness!extraction}

Having some string of characters from the source with a given min-entropy per character it is possible to \textit{extract} its randomness, that is to create a shorter string with a higher min-entropy per character \cite{Trevisan01,DPVR09,TRSS10}. Now we will give a short overview of the randomness extraction. This section is not necessary to understand the remaining part of this work, and can be omitted by uninterested readers.

The task of randomness extraction is the following \cite{DPVR09,rndexp6}. One has the access to a source of uniformly random bits which is considered to be expensive, and thus one would like to reduce its usage. This source produces a seed $S$ of length $d$, $S \in \{0,1\}^d$. One also has the access to a source of random bits which are not uniform, but one knows the min-entropy of the source. This source produces a string $R$ of bits of length $n$, $R \in \{0,1\}^n$, of min-entropy $H_{\infty}(R) \geq k$. Both sources are assumed to be independent. One is interested in obtaining a sequence $\Ext(R,S) \in \{0,1\}^m$ of $m$ bits which are in some sense close to be uniformly random. We say that a function
\be
	\nonumber
	\Ext: \{0,1\}^n \times \{0,1\}^d \rightarrow \{0,1\}^m
\ee
is a $(m,k,\delta)$ extractor\index{randomness extractor} with uniform seed if under the above assumptions we have
\be
	\nonumber
	\frac{1}{2} \sum_{r_m \in \{0,1\}^m} \left| P \left[ Ext(R,S) = r_m \right]  - 2^{-m} \right| \leq \delta,
\ee
where term $2^{-m}$ comes from uniform probability distributions.

It is important to ask, how long the seed should be so that a $(m,k,\delta)$-strong extractor exists. One may show \cite{DPVR09} that for any $k \in [0,n]$ and $\delta > 0$ there exists such an extractor with $m = k - 4 \log{\frac{1}{\delta}} - O(1)$ which requires a seed of length $d = O \left( \log^2\left(\frac{n}{\delta}\right) \log(m) \right)$.

In short, it is possible to obtain a probability distribution close to the uniform distribution on $m$ bits with a source of min-entropy close to $m$ and a seed of length depending logarithmically on $m$.

\section{Randomness amplification}
\label{sec:randAmplify} \index{randomness!amplification}

As mentioned in chapter~\ref{chap:basicsQI}, the application of the laws of quantum mechanics allows to perform tasks impossible in classical information theory, like secure cryptography, or randomness certification. Recently, another area where QI theory makes new things possible has emerged. It is the amplification of weak randomness \cite{Amp1}.

Let us consider a source generating a sequence of bits,
\be
	\nonumber
	\mathbf{s} =  (s_0, s_1, \cdots, s_{N-1}).
\ee
This source is described by a constant $\epsilon$. The bits from the sequence may be correlated with each other, and also with some variable $e$, with a constraint saying that these correlations are limited by the following condition
\be
	\label{svcon}
	\forall_i \quad \frac{1}{2} - \epsilon \leq P(s_i=0 | s_0, \dots, s_{i - 1}, e) \leq \frac{1}{2} + \epsilon.
\ee
The source of randomness described by this constraint is usually referred to as a \textit{Santha-Vazirani source} (\acrshort{SV}).

If the constraints in~\eqref{svcon} hold, we say that the sequence $\mathbf{s}$, or the generating source, is $\epsilon$-free. The parameter $\epsilon$ gives a bound on the bias of the generated bits from the perfect uniformly distributed \acrshort{iid} sequence, and for this reason is called the \textbf{bias}\index{bias} of the source. Obviously $\epsilon=0$ corresponds to the case where the output of the source is perfectly random. On the other hand, if $\epsilon=\frac{1}{2}$, then the sequence can be even deterministic. We assume that the variable $e$ is known by a treacherous party, but it is unknown to us.

The task of \textbf{randomness amplification} is to use some post\hyp{}processing of the generated sequence $\mathbf{s}$ to obtain another, possibly shorter, sequence $\mathbf{y}$ which is $\epsilon'$-free with $\epsilon'<\epsilon$. The question of the possibility of randomness amplification using classical (\textit{i.e.} not quantum) algorithms has been answered negatively in a famous paper by Santha and Vazirani, presented at FOCS in 1984 \cite{SV}. They have shown that for any deterministic function $f: \{0,1\}^N \rightarrow \{0,1\}$ there exists an $\epsilon$-free SV such that the bias of $f(\mathbf{s})$ is at least $\epsilon$.

Colbeck and Renner in their groundbreaking paper \cite{Amp1} proved that, contrary to classical amplification, the procedure of amplification is possible in the quantum case. Their idea is based on performing a Bell experiment and applying a hashing function to the measurement outcomes. However, the protocol that they presented works only if the source of randomness is $\epsilon$-free with $\epsilon$ less than about $0.086$.

It is important to distinguish the task of amplification and the task of randomness expansion discussed in the previous section. The randomness expanders generate sequences of random bits, for which the \textit{average} min-entropy is known, but do not guarantee anything about a particular bit. In randomness amplification we are interested in generating a bit with bias less than the initial one.

The randomness amplification not only has practical applications, but it also sheds light on fundamental issues of quantum mechanics. One is that if there exist only slightly random processes in the Universe, then also processes with arbitrary high randomness do \cite{GMTDAA12}. So far the possibility of randomness amplification has been demonstrated only under very restrictive conditions, nonetheless it is a very active area of research \cite{Amp1,GMTDAA12,Amp3,Amp4,Amp5,Amp6,Amp8}.

In \cite{MP13} we have introduced the first randomness amplification protocol which allows amplification of arbitrary weak randomness and is possible to be realized in practice. We do so by presenting an amplification protocol which is based on the Mermin game (see~sec.~\ref{sec:Mermin}). It works for any $\epsilon<\frac{1}{2}$ and can tolerate a finite amount of noise and experimental imperfections. We describe this amplification protocol in sec.~\ref{sec:amplification}.

\section{Cryptography}
\label{sec:cryptography}\index{cryptography}

The basic task of cryptography is to establish a mean of a secret communication between distant parties.  This problem is at least as old as humanity. Cryptography uses methods involving some data manipulation \textit{i.e.} encoding. In contrast to, \textit{e.g.}, dispatching a guarded convoy with just an envelope containing a letter, this way of contact keeps the message even if it is known to an eavesdropper.

One of the most famous and important historical cryptographic protocols is the Caesar cipher\index{cryptography!protocols!Caesar}, an example of a substitution cipher\index{cryptography!substitution cipher} in which letters of the alphabet in the message are rearranged. It is easy to break this kind of ciphers.

Let us note that the Caesar cipher employs a \textbf{cryptographic key}\index{cryptographic key} which determines the method of rearranging the alphabet. The cipher requires both parties to know the key used to encode and decode the message. The cryptography which requires both parties to know the same key is called \textit{symmetric}\index{cryptography!symmetric}. In modern cryptography one of the basic requirements for a protocol is the Kerckhoffs's principle stated in 1883 saying that the protocol should be secure even if its algorithm is publicly known, or, as formulated by Shannon \cite{Shannon49} in 1949, \textit{the enemy knows the system}. Thus the more difficult to guess the key is, the more secure the message is, but we assume that the eavesdropper knows our protocol.

The \textbf{asymmetric}, or public key, cryptography\index{cryptography!asymmetric} distinguishes the key used to encode the message, and the key which has to be used to decode it. In this kind of encoding one of the parties makes his public key available to everyone. A party who wants to send the message uses a relevant algorithm to encode it and sends it to the owner of the key. The owner then uses his private key to decode the message. The security of this form of cryptography is based on the difficulty of certain mathematical tasks like computing the discrete logarithm, or integer factorization. Still, at least in principle, it is possible to break this kind of ciphers. For this reason a notion of \textbf{information-theoretic security}\index{cryptography!information-theoretic security} has been introduced. This refers to such cryptographic protocols which are secure against an eavesdropper with unbounded computational power \cite{Shannon49}.

In this work we are interested in methods of establishing a key used for symmetric cryptography. The considered scenarios involve two parties considered as honest, Alice and Bob, who want to communicate with each other, and an eavesdropper, called Eve, who wants to learn the content of their messages. Alice and Bob agree on the protocol they want to use, and the protocol is known to Eve. Their task is to establish an information-theoretic secure key.

We always assume that the data which have not been published anywhere are secure. In particular the eavesdropper does not have the \textit{direct} access to the generated key either on Alice's, or on Bob's side, since in this case the cryptography is pointless.

\subsection{Quantum cryptography}
\index{quantum key distribution}

Quantum key distribution (\acrshort{QKD}) is a significant achievement in the field of cryptography. Its techniques allow Alice and Bob to meet the task of establishing an information-theoretic secure key using a quantum device, public communication, and secure private RNG. We refer readers interested in the history of QKD to the work of Brassard \cite{Brassard05}.

The first quantum cryptographic protocol is BB84\index{cryptography!protocols!BB84}, announced in 1984 by Charles Bennett and Gilles Brassard in the groundbreaking paper \cite{BB84}. The protocol is the following. Alice uses her private secure RNG to generate two $N$-bit strings, $x, b \in \{0,1\}^N$. Then she sends a sequence of $N$ quantum states. The state sent in $i$-th iteration depends on the $i$-th bits of these strings, $x_i$ and $b_i$. Bob uses his private secure RNG to generate an $N$-bit string $y \in \{0,1\}^N$. He receives the series of the quantum states, and performs measurements depending on his bits, $y_i$ for $i$-th iteration. The prepared states and measurements are chosen in such a way that the conditional probability distribution $\mathbb{P}(b|x,y)$ is given by\footnote{To be more precise the states are the following. If $x_i=0$ then Alice sends a qubit from the computational basis\index{basis!computational}, $\ket{b_i}$. If $x_i=1$ then she sends a qubit from the Hadamard basis\index{basis!Hadamard}, $\ket{+}$ for $b_i=0$, and $\ket{-}$ for $b_i=1$. Bob measures in the basis depending on his input $y_i$: in the computational basis for $y_i=0$ and in the Hadamard basis for $y_i=1$.}
\be
	\nonumber
	\ba
		P(b_i|x_i,x_i)=1, P(\neg b_i|x_i, x_i)=0, \\
		P(b_i|x_i, \neg x_i) = P(\neg b_i|x_i, \neg x_i) = \frac{1}{2}.
	\ea
\ee
This means that in iterations in which $x_i=y_i$ are the same, Bob recovers the full information on $b_i$, and if $x_i \neq y_i$, then he gets no information. \textit{After} his measurements Bob announces his string $y$, and Alice announces her string $x$ over a public channel. Parties compare their strings, and create the key using these bits in $b$ for which $x_i=y_i$. Thus both parties share the same string of length on average $\frac{N}{2}$. One may see that a potential man-in-the-middle attack would be to intercept some of the states, perform some measurement on it, and send another state to Bob instead of the previous one. It can be shown that such kind of eavesdropping can always be detected statistically by Alice and Bob.

Another important QKD protocol was proposed by a Polish physicist Ekert in 1991\index{cryptography!protocols!E91} \cite{E91}. In this protocol parties share a maximally entangled state of two qubits and each of them performs a measurement in one of three basis selected in a way which allows them both to establish a shared key, and calculate a security parameter to detect potential eavesdropping. This protocol is similar to DI protocols in that it employs the security parameter to prove confidence. Nonetheless, we stress that this is not a DI protocol, since it assumes the implementation of the shared quantum state and measurements, thus requires a trusted device.

\subsection{Device-independent quantum key distribution}
\index{quantum key distribution}

The task of secure cryptography is particularly exposed to the threat of an untrusted vendor. For this reason it is natural to use the DI approach for further improvements of reliability of the cryptographic protocol. As mentioned in the introduction to this chapter, with DI approach to the analysis of quantum protocols, it is possible to cope with the situation when the vendor of the cryptographic device is not trusted. This lack of credit concerns the source of particles, the communication channel, and their measuring apparatuses \textit{etc}. We usually assume that the source of randomness needed by a QKD protocol is trusted. We discuss this issue in sec.~\ref{sec:biased}.

Some examples of early partially-DI QKD protocols are considered in \cite{BHK05,AGM06,SGBMP06}, but these were vulnerable to some particular types of attacks and needed additional assumptions. In 2007, Ac\'in \textit{et al.} \cite{ABGSPS07}, introduced the first device independent QKD protocol secure against the so-called collective attacks. Another important step in the development of DI-QKD protocols was the work by Masanes \textit{et al.} \cite{MPA11}. In that paper they provided a more general security scheme based on causally independent measurement processes. 

Common properties of DI-QKD protocols include:
\begin{enumerate}
	\item A protocol runs series of iterations.
	\item The parties share some entangled state (unknown to the parties).
	\item The parties possess private secure RNG, which generate private sequences of numbers.
	\item The random numbers are used to select measurement settings for each of the iterations.
	\item After a series of iterations parties publish some of their data and use them to estimate one or more security parameters.
	\item The unrevealed data are used by each of the parties to calculate a \textbf{raw key}\index{raw key}.
	\item The parties perform the \textbf{key reconciliation}\index{key reconciliation}, which requires them to publish some parts of their raw keys to obtain a shared value of the key.
\end{enumerate}
Most of DI-QKD protocols assume that their private RNGs in p.~3 generate unbiased random numbers. The protocols usually use only a single security parameter in p.~4. The data published in p.~5 are usually the measurement settings, and the raw key is simply some selected subset of outcomes of the parties. In sec.~\ref{sec:HardyQKD} we describe a protocol which differs in these points.

As mentioned, a protocol works in iterations. The \textbf{key rate}\index{key rate} refers to the average number of secure shared bits obtained in a single iteration of a protocol.

A method for estimation of security parameters in p.~4 is given in sec.~\ref{sec:estimationSymmetric}. The key reconciliation \cite{reconciliation94} in p.~7 is a standard procedure, which allows to obtain exactly the same strings of bits for all parties, even if they start with slightly different strings. The amount of communication depends on the conditional min-entropies between the initial strings.

\subsubsection{Masanes\hyp{}Pironio\hyp{}Ac\'in protocol}
\label{sec:MPAprotocol}

We will briefly discuss the quantum key distribution protocol given by Masanes, Pironio and Ac\'in (\acrshort{MPA})\index{cryptography!protocols!MPA} \cite{MPA11}. This protocol assumes that the state of the cryptographic device does not depend on the previous iterations (\textit{i.e.} the device does not use any internal memory). The parties can use any Bell operator to certify the key rate. The paper \cite{MPA11} in particular considers the following Bell scenario.

Let $N \in \{2,3\}$. The parties obtain binary outcomes of measurements, thus $A=B=\{0,1\}$. The set of settings for Alice is $X = \{0, \dots, N-1\}$, and for Bob $Y = \{-1,0, \dots, N-1\}$. Let $\delta(y)$ be equal to $1$ when $y=-1$, and $0$ otherwise. The Bell operator used to certify the security of the protocol is
\be
	\nonumber
	\sum_{a \in A, b \in B} \sum_{x=0}^{N-1} \sum_{y=x-1}^x (-1)^{a+b+\delta(y)} P(a,b|x,y).
\ee
For $N=2$ the above expression contains a CHSH\index{Bell operator!CHSH} operator. The authors of \cite{MPA11} show that when the Bell operator attains its maximal quantum value, then
\be
	\nonumber
	P_A(0|N-1)=P_A(1|N-1)=\frac{1}{2}.
\ee
They also calculate the min-entropy $H_{\infty}(N-1,s)$ of the outcomes of Alice when the setting $x$ is $N-1$ as a function of the security parameter $s$.

For majority of iterations Alice fixes her setting to $N-1$, and Bob to $N$. The protocol also assumes that the conditional entropy of the setting of Bob given the setting of Alice for this pair of settings is low, $0 \leq H(b|a,N-1,N) \ll 1$.

After performing all measurements, both parties publish their measurement settings. They use these iterations in which Alice has chosen the setting $N-1$ and Bob $N$ to create their raw keys. Afterward they perform the reconciliation, requiring them to publish at average $H(b|a,N-1,N)$ bits per bit of the raw key. Thus the key rate of the protocol when the value of the security parameter is $s$, is given by the expression
\be
	\label{eq:keyrateMPA}
	P(x=N-1,y=N) \cdot \left( H_{\infty}(N-1,s) - H(b|a,N-1,N) \right).
\ee

\section{Experimental estimation of parameters}
\label{sec:estimation}

We will now briefly discuss the problem of experimental estimation of physical parameters. This issue goes beyond the main topic of this work, and we will only sketch it. This section is not necessary to understand the rest of this work.

The protocols considered in this work in chapter~\ref{chap:quantumProtocols}, secs~\ref{sec:expansion}, \ref{sec:sdi} and~\ref{sec:amplification}, have in common several properties important from the point of view of experimental estimation. All these protocols need to estimate the average value of at least one expression (\textit{e.g.} Bell operators or dimension witnesses), and involve more than one party. On the other hand, the task of these protocols, \textit{i.e.} the generation of randomness, allows the parties to communicate with each other, which is not the case in QKD protocols, like the one described in sec.~\ref{sec:HardyQKD}.

QKD protocols, including the protocol from sec.~\ref{sec:HardyQKD}, usually involve (at least) two parties which can communicate only \textit{via} a public channel, and thus the messages they exchange cannot be considered as unknown to a potential eavesdropper, in particular if their private data are used to generate a secret key. On the other hand, the parties have to estimate some parameters concerning both of them, and thus requires them to send some of their internal data to the other party. The parties want to have a strong confidence in their estimation and to publish not more information than it is necessary.

As we will see, this requires us to employ a different method for each of these two tasks.

Currently there are two popular methods of estimation of parameters: the first is using martingales\footnote{Martingales\index{martingale} are stochastic processes with the property that the conditional expectation of any of their future values is equal to their current value. They formalize the intuition of a gain in a fair gambling game.} and the second uses symmetric variables. The former is suitable for the task randomness generation, whereas the latter should be used for QKD.

In the analysis of QKD protocols we restrict here our considerations to protocols secure against \textit{eavesdroppers without memory}. This assumption means in particular that the state of the device does not depend on the iteration. This assumption is standard, see~\cite{MPA11} for a discussion. The issue of attaining the security against a general eavesdropper remained unsolved for many years, and has been recently addressed by Arnon-Friedman, Renner and Vidick in \cite{AFRV14}\footnote{The security against an eavesdropper with memory has been previously proved for several protocols, \textit{e.g.} in \cite{BCK12,VV12}, and in our work \cite{Ramij-our} for the case without noise.}.

\subsection{Estimation with martingales and Azuma theorem}
\label{sec:Azuma}

The method of estimation involving martingales was introduced to the field of quantum information science in 2001 by Richard D. Gill \cite{Accardi}. Further it was employed in a seminal paper by Pironio \textit{et al.} \cite{RNGCBT} in the context of randomness certification.

We discuss here the estimation of the average value of a single Bell operator. This is easily applied to the case with dimension witnesses or to the case with more expressions to estimate.

Recall~\eqref{BI} with a general form of a Bell expression (we omit $C_I$, since one does not need to estimate a known constant value):
\be
	\nonumber
	I = \sum_{a \in A} \sum_{b \in B} \sum_{x \in X} \sum_{y \in Y} \alpha_{a,b,x,y} P(a,b|x,y).
\ee
In each iteration $i = 1, \dots, N$ the behavior of the box is described by some probability distribution $\mathbb{P}_i(A,B|X,Y)$. Since this probability is determined by the internal working of the device, which is unknown to us, we do not know this distribution. Because in each iteration this distribution may be different, we even cannot estimate this distribution by repeating the experiment many times. Nonetheless, the device \textit{has} some behavior, and one can think of the value of the Bell operator in $i$-th round. Let us denote it by $I_i$.

In the considered scenario we run a sequence of $N$ measurements on a quantum box. We are interested in estimating the average value of the Bell operator, $\bar{I}$, \textit{i.e.}
\be
	\nonumber
	\bar{I} \equiv \frac{1}{N} \sum_{i=1}^N I_i.
\ee

In $i$-th iteration of the setting for Alice is $x_i$, and the setting for Bob is $y_i$. 

Let $\chi_i(a,b|x,y)$ be an indicator, \textit{i.e.} a binary function equal $1$ if in $i$-th iteration of the performed experiment the settings are $x$ and $y$, and the outcomes are $a$ and $b$ for Alice and Bob, and equal to $0$ otherwise. Let us define a random variable:
\be
	\label{eq:IEst}
	\hat{I}_i = \sum_{a \in A} \sum_{b \in B} \sum_{x \in X} \sum_{y \in Y} \alpha_{a,b,x,y} \frac{\chi_i(a,b|x,y)}{P(x,y)},
\ee
where $\mathbb{P}(X,Y)$ is the distribution of settings. Then
\be
	\nonumber
	\hat{I} \equiv \frac{1}{N} \sum_{i=1}^N \hat{I}_i
\ee
is an estimator of $\bar{I}$.

Using the Azuma-Hoeffding inequality\index{Azuma-Hoeffding inequality} \cite{Azuma67,Hoeffding63} one may show \cite{RNGCBT,rndexp5} that for $\epsilon > 0$, $p_{\min} \equiv \min_{x \in X, y \in Y} P(x,y)$, $\alpha_{\max} \equiv \max(\alpha_{a,b,x,y})$, and $I_{\max}$ being the Tsirelson bound of the Bell operator, we have
\be
	\nonumber
	P(\bar{I} \leq \hat{I} - \epsilon) \leq \exp \left( - \frac{\epsilon^2 N}{2 \left( \frac{\alpha_{\max}}{p_{\min}} + I_{\max} \right)} \right).
\ee
This means that the probability that the estimator $\bar{I}$ differs significantly from the actual value $\hat{I}$ decreases exponentially with the number of runs.

In \cite{rndexp5,rndexp6} it was shown how the above estimation can be applied in the context of randomness certification. In the original paper \cite{RNGCBT} the author used this method improperly.

Let $x_i$ and $y_i$ denote the settings used by Alice and Bob in the $i$-th iteration of the experiment, and $N_{x,y}$ the number of iterations in which the pair of settings $x$ and $y$ occurred. We have $\sum_{x,y} N_{x,y} = N$.

In \cite{rndexp5} the authors divide the interval $[I_{\min},I_{\max}]$, where $I_{\min}$ is the minimal quantum value of the Bell operator, into $m$ disjoint parts,
\be
	\nonumber
	\{\Omega_l = [J_l,J_{l+1})\}_{i = 1, \dots, m},
\ee
and define the so-called ``good'' even $\mathfrak{G}$. We will not describe their method in details and just state that they prove the following theorem:
\begin{trm}
	For any $\epsilon,\delta>0$ there exists a ``good'' even $\mathfrak{G}$ with probability
	\be
		\nonumber
		P[\mathfrak{G}] \geq 1 - m \cdot 2^{-\delta N} - 3 \cdot 2^{-c p_{\min} \epsilon^2 \cdot N}
	\ee
	such that if $\mathfrak{G}$ occurs, and $\hat{I} \in \Omega_l$, then
	\be
		\nonumber
		\sum_{i=1}^N H_{\infty} (\mathbb{P}_i(A,B|X,Y),x_i,y_i) \geq \left( \sum_{x,y} N_{x,y} \cdot H_{\infty}^{cert}(I,x,y,J_{l-1}) \right) - \delta N - 1.
	\ee
\end{trm}

The intuitive sense of this theorem is that in a long series of iterations of the randomness generating device the probability that the generated min-entropy is on average less than expected decreases exponentially with the number of iterations.

Further in this work, in sec.~\ref{sec:expansion}, we calculate $H_{\infty}^{cert}(I,x,y,s)$ for a particular choice of $x = x_0$ and $y = y_0$ and $s \in [I_{\min},I_{\max}]$. From the above theorem it follows that we may use a distribution of settings $\mathbb{P}(X,Y)$ which prefers this particular pair of settings with probability arbitrarily close to $1$, and certify with probability of failure decreasing exponentially with $N$ that the device produced at least
\be
	\nonumber
	N \cdot \left(P(x_0,y_0) \cdot H_{\infty}^{cert}(I,x_0,y_0, \hat{I} - \epsilon) - \delta \right) - 1
\ee
bits of randomness in the sequence of outcomes $(a_i,b_i)_i$ from iterations with settings $x_i = x_0$ and $y_i = y_0$.

\subsection{Symmetric probability distributions and the task of estimation}
\label{sec:estimationSymmetric}

In the scenario considered in the previous section, the parties could reveal all their results to each other. In the case of estimation in QKD, they do not have such a possibility, because of the lack of a secure private channel (since the aim of QKD is to establish such a channel). Thus, the parties want to publish only some subset of their data in order to estimate their security parameters. In \cite{MRCWB09} the following method which can be used in this scenario was introduced.

Let $\mathcal{V} \subset \mathbb{R}$ be a finite alphabet, and $\mathbf{V} = (V_1, \cdots, V_N)$ be a random variable with values in $\mathcal{V}^N$, and probability distribution $\mathbf{P}_{\mathbf{V}}$. We say that $\mathbf{V}$ is \textbf{symmetrically distributed}\index{symmetrically distributed random variable} if for any permutation
\be
	\nonumber
	\pi: \{1, \cdots, N\} \rightarrow \{1, \cdots, N\}
\ee
we have
\be
	\nonumber
	\mathbf{P}_{\mathbf{V}} (v_1, \dots, v_N) = \mathbf{P}_{\mathbf{V}} (v_{\pi(1)}, \cdots, v_{\pi(N)}).
\ee
Let us denote
\be
	\nonumber
	v_{+} \equiv \max \{|v| \in \mathcal{V}\}.
\ee

In \cite{MRCWB09}, in lemma~5, the following property of symmetric random variables is stated:
\begin{lem}
	\label{lem:MRCWB09}
	Let $\mathbf{V} = (V_1, \cdots, V_N)$ be a symmetrically distributed random variable, let $N_1, N_2$ be positive integers satisfying $N_1+N_2=N$, and $V_{est}$ be a random variable defined as
	\be
		\label{eq:Vest}
		V_{est} \equiv \frac{1}{N_2} \sum_{i=N_1+1}^N V_i.
	\ee
	Then the probability that
	\be
		\nonumber
		\sum_{\substack{v_1 \in \mathcal{V} \\ \dots \\ v_N \in \mathcal{V}}} \mathbf{P}_{\mathbf{V}} (v_1, \dots, v_N) \cdot (v_1 \cdot v_2 \cdot \dots \cdot v_{N_1}) \geq \left( V_{est} + N_2^{-\frac{1}{4}} \right)^{N_1}
	\ee
	is not greater than
	\be
		\nonumber
		2 (N+1)^{|\mathcal{V}|-1} \exp \left( -\frac{\sqrt{N_2}}{4 v_{+}^2} \right).
	\ee
\end{lem}

This lemma gives the following method of estimation which is possible to be used in QKD.

We are interested in estimating the number of the generated bits of the secure key in a QKD scenario, \textit{i.e.} the min-entropy or, equivalently, the guessing probability of a sequence obtained in given QKD protocol.

Let $s_1, \cdots, s_K$ be a set of security parameters\footnote{Recall that security parameters\index{security parameter} are all linear functions of probabilities occurring in the problem.}. We assume that we know a convex function $P_{guess}(s_1, \cdots, s_K)$ giving an upper bound on the average guessing probability of a \textit{single} digit (bit or, in general, dit) of a key.

Lemma~\ref{lem:MRCWB09} gives a method to estimate the average guessing probability of a digit in a long key. We start with expressing $P_{guess}$ as a function of probabilities occurring in the problem, $\mathbb{P}$. Since $P_{guess}$ is convex, it is possible to construct a function $\tilde{P}_{guess}$ linear in $\mathbb{P}$, which upper bounds $P_{guess}$.

For a given iteration $i$ we may express an estimator of the value of $\tilde{P}_{guess}$, $\hat{P}_{guess,i}$, in a similar way as we did for a Bell operator in~\eqref{eq:IEst}. In order to calculate the average value of $\tilde{P}_{guess}$ over long series of iterations, we have to make their distribution symmetric.

As previously mentioned, we restrict to the case where the device does not have memory. This means in particular that the values of $\hat{P}_{guess,i}$ for different $i$ are independent, still not necessarily equal. Using initial private randomness, one of the honest parties involved in the QKD, say Alice, after the series of all measurements, may choose a random permutation $\pi: \{1, \dots, N\} \rightarrow \{1, \dots, N\}$, and announce it. Let us define
\be
	\nonumber
	V_i \equiv \hat{P}_{guess,\pi(i)},
\ee
\textit{i.e.} $\mathbf{V} = (V_1, \cdots, V_N)$ permute the estimators of iterations. It is now straightforward to use lemma~\ref{lem:MRCWB09} to have a method of estimation of the value of $\tilde{P}_{guess}$, and thus get an upper bound on the probability that one can guess the value of a bit of the key, $P_{guess}$. This simply requires parties to publish the settings and measurements in $N_2$ iterations to calculate the value of~\eqref{eq:Vest}.

This method has been used in QKD protocols, \textit{e.g.} in \cite{MPA11}, and in our work \cite{Ramij-our}.

\chapter{Quantum protocols}
\label{chap:quantumProtocols}

This chapter covers the main topic of this work, namely the presentation of our results obtained using SDP in the analysis of quantum protocols. In each section we describe some results of our previous works \cite{HWL13,MP13,LubiePlacki,Ramij-our,HWL14}. All the figures in this chapter are taken from these works. The organization of this chapter is as follows.

The results of \cite{Ramij-our} are contained in sec.~\ref{sec:HardyQKD}, where we provide an analysis of a DI-QKD protocol belonging to a family we introduced. The protocol has several novel features distinguishing it from the QKD protocols discussed in sec.~\ref{sec:cryptography}.

The main advantage of the protocol is the smaller vulnerability to imperfect RNGs used for generation of the settings. This way we show that using our method it is possible to construct a QKD protocol which retains its security even if the source of randomness used by communicating parties is strongly biased.

We analyze the robustness of the protocol using SDP. We also introduce post-processing methods. The post-processing has a paradoxical property that rejecting a random part of the private data can increase the key rate of the protocol.

The paper \cite{LubiePlacki} concerning DI randomness expansion is discussed in sec.~\ref{sec:expansion}. We use the methods of SDP to calculate the min-entropy certified by a number of DI-QRNG protocols.

The issue of analysis of SDI protocols \cite{HWL13,HWL14} is divided in two sections. In sec.~\ref{sec:propDW} we discuss some properties of dimension witnesses, and then use these properties in sec.~\ref{sec:sdi} to develop a method to solve SDI problems relaxations with the NPA.

In sec.~\ref{sec:amplification} we refer our result of \cite{MP13} that amplification of arbitrarily weak randomness is possible using quantum resources. We present a protocol that uses the Mermin game for randomness amplification, and calculate its parameters. We derive the necessary bounds on the violation of the inequality as a function of the initial randomness quality.

\section{Quantum key distribution using Hardy's paradox}
\label{sec:HardyQKD}\index{cryptography!protocols!Hardy}\index{device-independent}\index{Hardy"'s paradox}\index{quantum key distribution}

In this section we generalize the results of \cite{MPA11} stating that a condition imposed on a single Bell operator may certify the randomness of the outcomes. We consider the case when instead of a Bell operator, the Hardy's paradox is used, and the key is formed from the measurement settings with the outcomes made public. The last property is shared with the non-DI prepare-and-measure protocol SARG04\index{cryptography!protocols!SARG04} \cite{sarg}. These properties distinguish our proposal from other QKD protocols which use a single Bell operator, publish their settings, and use outcomes to generate the key.

The proposed protocol has an important advantage over other solutions. The other QKD protocols, in which the key is obtained from the outcomes, are known \cite{Honza,badRNGqkd} to posses a drawback that even small imperfectness in private RNGs significantly decreases the key rate. In contrast, we demonstrate that they are not so dangerous for our protocol. More precisely, we take a standard DI-QKD described in sec.~\ref{sec:MPAprotocol}, which uses CHSH, and show that it fails if the bias of RNGs is greater than 0.1. On the other hand, our protocol allows positive key rates far beyond this point.

We use SDP in order to evaluate the key rate of the discussed protocol.

Before we start the discussion of the protocol, we have to state the Hardy's paradox.

\subsection{The Hardy's paradox and the Hardy's state}
\label{sec:HardyState}\index{Hardy"'s paradox}

The Hardy's paradox is another phenomenon which is not possible in the set of classical probabilities, $\mathcal{L}$, but is allowed in the quantum set, $\mathcal{Q}$. The paradox has been introduced in 1992 by Lucien Hardy \cite{Har92}.

Let us consider a physical system consisting of two subsystems shared between two distant parties, Alice and Bob. Each of the parties have access to one of the subsystems. Both of them can choose one of two binary measurement settings, and get binary outcomes. The Hardy's paradox introduces the following set of conditions on the probability distribution $\mathbb{P}(A,B|X,Y)$ (in this case $A=B=X=Y=\{0,1\}$):
\be
	\label{eq:Hardy2q}
	\ba
		& P(0,0|0,0) = q > 0, \\
		& P(0,0|1,0) = 0, \\
		& P(0,0|0,1) = 0, \\
		& P(1,1|1,1) = 0,
	\ea
\ee
where
\be
	\nonumber
	q \equiv \frac{5\sqrt{5}-11}{2}.
\ee

One may show \cite{Jor94,Kar97,Ramij-our,RZS12} that the state and measurement which realize a quantum probability distribution which satisfies conditions \eqref{eq:Hardy2q} must be a unique, up to a local unitary, pure two-qubit entangled state.

\subsection{The idea of the protocol}
\label{sec:HardyProtocol}

We consider a scenario in which two distant parties, Alice and Bob, want to generate a secure key. They are allowed to use public classical communication.

We discuss the case when private sources of randomness are biased in sec.~\ref{sec:biased}. For the rest of this section we assume the randomness to be unbiased. We consider two distributions: uniform and the one described in the section \ref{sec:nonuniform}, further referred to as \textit{nonuniform}. Note that one of the benefits of using nonuniform distribution is the fact that it requires less randomness (in terms of min-entropy).

The QKD protocol proceeds in the following way.

We assume that both parties share parts of a sequence of entangled states. Each part of an entangled pair is called a subsystem. The pairs can be obtained from an untrusted device. The protocol runs in an iterative way. One entangled pair is needed for each iteration.

In a single iteration $i$, Alice and Bob choose randomly their settings, $x_i \in \{0,1\}$ and $y_i \in \{0,1\}$ with their private sources of randomness. Then they put these settings into untrusted measurement devices and get results $a_i$ and $b_i$. Both parties keep their settings $x_i$ and $y_i$ private, and publish their outcomes, $a_i$ and $b_i$.

We briefly mention that in order to certify the key rate the parties have to reveal some subset of their settings. These are needed to estimate the values of probabilities occurring in the statement of the Hardy's paradox. In this section we explicitly state which part of the joint probability concerns settings, and which outcomes, \textit{e.g.} we write $P(a=0,b=0|x=1,y=0)$ instead of $P(0,0|1,0)$. With this notation the Hardy's conditions are defined in the following way by, \textit{cf.}~Eq.~\eqref{eq:Hardy2q}:
\be
	\nonumber
	\ba
		& P(a=0,b=0|x=0,y=0) = q, \\
		& P(a=0,b=0|x=1,y=0) = 0, \\
		& P(a=0,b=0|x=0,y=1) = 0, \\
		& P(a=1,b=1|x=1,y=1) = 0.
	\ea
\ee
The idea of the protocol emerges from Hardy's conditions in the following way. The second and third conditions state that if both Alice and Bob get outcomes $0$, then their settings have to be the same, or, in other words, $P(x \neq y|a=b=0) = 0$. The first condition states that the probability of getting both outcomes $0$ is positive. From the fourth condition it follows that the device cannot be described by a local probability distribution; otherwise a malevolent constructor would be able to predict the behavior of the device. The probability of guessing of the settings when both outcomes were $0$ depends on the ratio between $P(a=0,b=0|x=0,y=0)$ and $P(a=0,b=0|x=1,y=1)$. The analysis of this ratio is the main topic of this section, since it is directly related to the key rate of the protocol.

In order to generate the key, Alice and Bob select those iterations for which $a_i=b_i=0$. From the definition of the Hardy's paradox,~\eqref{eq:Hardy2q}, it is easy to see that in the noiseless case they have $x_i=y_i$, and thus may generate the key simply by taking their local settings, which are secret and equal for both parties.

Below we use SDP to analyze the noisy case, in which the probabilities only approximate the one of the noiseless case. In this noisy case the key reconciliation\index{key reconciliation} has to be performed. This requires parties to reveal some values of their secret bits.

The quantum probability distribution satisfying the condition of the noiseless Hardy case,~\eqref{eq:Hardy2q}, is unique. In this case we have
\be
	\label{eq:noiselessHardy}
	\ba
		0 & < P(a=0,b=0|x=0,y=0) = \frac{5 \sqrt{5} - 11}{2}  \\
		& < P(a=0,b=0|x=1,y=1) = \sqrt{5} - 2.
	\ea
\ee
In the noisy case instead of the ideal Hardy's probabilities we take
\be
	\label{eq:HardyNoisy}
	\begin{aligned}
		& P(a=0,b=0|x=0,y=0) = h_1, \\
		& P(a=0,b=0|x=1,y=0) = h_2, \\
		& P(a=0,b=0|x=0,y=1) = h_3, \\
		& P(a=1,b=1|x=1,y=1) = h_4.
	\end{aligned}
\ee
Further in this work we will calculate the guessing probability of a key for a grid of values $h_i$. This will allow us to approximate a function
\be
	\nonumber
	P_{guess}[\mathbb{P}(2,2|2,2)]
\ee
giving an upper bound on the probability of guessing a single bit of the key by an eavesdropper. In sec.~\ref{sec:estimationSymmetric} it has been discussed how such a function is used in an experimental estimation of the security of a protocol.

Below, for the analysis of the noisy case, we decided to use the model of white noise. In particular we are interested in a projection of that grid along the line $h_i$ defined by 
\be
	\label{h_noise}
	\ba
		h_1 = p \cdot q + \frac{1-p}{4}, \\
		h_2 = h_3 = h_4 = \frac{1-p}{4},
	\ea
\ee
since these are the probabilities observed in the case with white noise.

\subsection{Analysis of the protocol}

In order to analyze the protocol we have to introduce some basic notions. We start with a definition of distribution of settings. Then we move to the discussion of the guessing probability of a setting which differs significantly from the guessing probability of an outcome. We finish with a description of the method allowing to express the guessing probability of a setting as the NPA problem.

\subsubsection{Distribution of settings}

We say that the distribution of settings is fair, or \textbf{unbiased}\index{probability distribution!unbiased}, if all settings are \acrshort{iid} with a probability distribution defined by
\be
	\label{perfectDistribution}
	\ba
		\mathbb{P}&_{unbiased}(X,Y) = (P(0,0) = p_{A} \cdot p_{B}, P(0,1) = p_{A} \cdot (1 - p_{B}), \\
			&P(1,0) = (1 - p_{A}) \cdot p_{B}, P(1,1) = (1 - p_{A}) \cdot (1 - p_{B})).
	\ea
\ee

In the analysis of the Hardy's protocol the following two distributions are of particular interest. The first probability distribution is defined by
\be
	\nonumber
	p_{A} = p_{B} = \frac{1}{2}.
\ee
Further, in sec.~\ref{sec:nonuniform}, we consider a \textit{nonuniform} probability distribution defined by
\be
	\nonumber
	p_{A} = p_{B} = \frac{1}{2} (\sqrt{5} - 1).
\ee

\subsubsection{The guessing probability of a setting}
\label{sec:guessing}

We will now consider conditional probabilities of Alice's \textit{settings} $x$ when we know that both Alice and Bob got the outcome $0$, namely $P(x|a=0,b=0)$. These probabilities may be expressed in terms of $\mathbb{P}(A,B|X,Y)$ with the use of Bayes rule, as
\be
	\label{guessingBayes0}
	P(x=0|a=0,b=0) = \frac{\sigma}{\sigma+\nu}
\ee
and
\be
	\label{guessingBayes1}
	P(x=1|a=0,b=0) = \frac{\nu}{\sigma+\nu}.
\ee
Here $\sigma$ and $\nu$ are defined as
\be
	\label{guessingBayesX}
	\ba
		\sigma \equiv & P(a=0,b=0|x=0,y=0) P(x=0,y=0) + \\
		& P(a=0,b=0|x=0,y=1) P(x=0,y=1)
	\ea
\ee
\be
	\label{guessingBayesY}
	\ba
		\nu \equiv & P(a=0,b=0|x=1,y=0) P(x=1,y=0) + \\
		& P(a=0,b=0|x=1,y=1) P(x=1,y=1).
	\ea
\ee

\begin{figure}[!htbp]
	\includegraphics[width=0.88\textwidth]{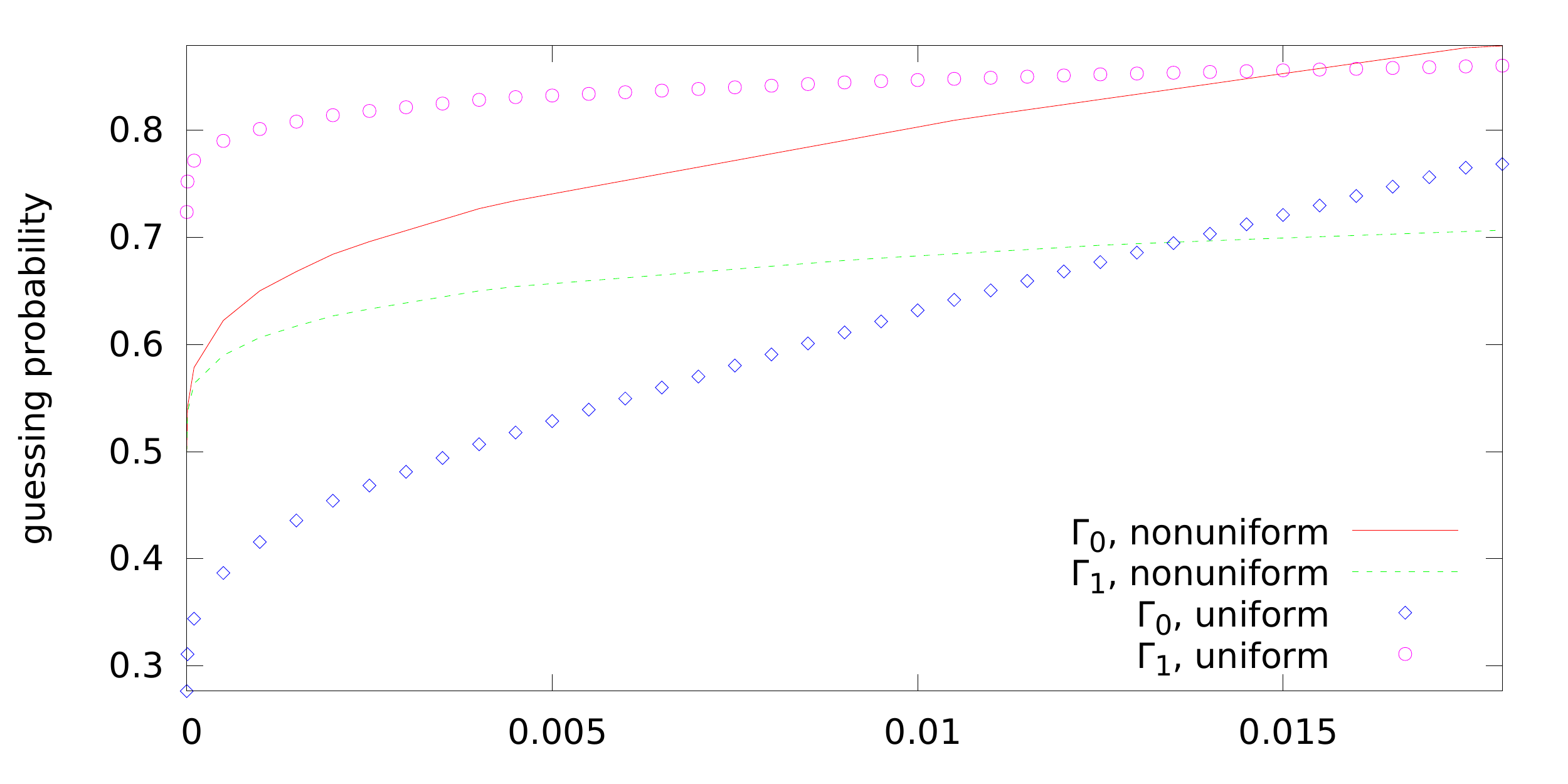}
	\caption[Upper bounds on guessing probabilities of settings in Hardy's QKD.]{Functions $\Gamma_0(\mathbf{h})$ and $\Gamma_1(\mathbf{h})$ for uniform and nonuniform (see~sec.~\ref{sec:nonuniform}) distribution of settings. These functions give upper bounds for values of $P(x=0|a=0,b=0)$ and $P(x=1|a=0,b=0)$ for $\mathbf{h}$ given by \eqref{h_noise} with $\eta = 1-p$.\label{fig:gammas}}
\end{figure}

Let us consider a vector $\mathbf{h}$ expressing the assumption of the probabilities occurring in the Hardy's paradox, see~\eqref{eq:HardyNoisy}. We introduce two concave functions, $\Gamma_0(\mathbf{h})$ and $\Gamma_1(\mathbf{h})$, that give upper bounds for values of $P(x=0|a=0,b=0)$ and $P(x=1|a=0,b=0)$, respectively, allowed by quantum mechanics or one of its NPA relaxations. Note, that these functions do not make any assumptions about the shared state and the measurements of Alice and Bob, so they give DI bounds. Examples of these functions for $\mathbf{h}$ given by~\eqref{h_noise} are shown in fig.~\ref{fig:gammas}.

\subsubsection{Semi-definite programming relaxation of the guessing probability}
\label{sec:sdpRelax}

This section describes how to use SDP in order to evaluate upper bounding functions $\Gamma_0(\mathbf{h})$ and $\Gamma_1(\mathbf{h})$.

To use the NPA method, we introduce functions $\tilde{\Gamma}_0(\mathbf{h})$, and $\tilde{\Gamma}_1(\mathbf{h})$, which give the relevant upper bounds on probability $P(x|a=0,b=0)$ assuming that the state under consideration is pure. Then it is easy to see that the functions $\Gamma_0(\mathbf{h})$ and $\Gamma_1(\mathbf{h})$ are concave hulls of $\tilde{\Gamma}_0(\mathbf{h})$ and $\tilde{\Gamma}_1(\mathbf{h})$, respectively.

Let us consider a particular entangled pair used in $i$-th iteration of the protocol. Without loss of generality, using no signaling principle, we may assume that an eavesdropper performs his measurement with result $e$ before Alice and Bob start the protocol. What is more, in order to consider probability distributions allowed for a particular subsystem $l$, we may ignore (``trace out'') other subsystems and perform optimization over bipartite states.

In \cite{MPA11} the authors have been able to use the NPA in a straightforward way in order to find the upper bounds for the case they considered. It was possible because they were interested in the probability of guessing the outcome if the setting is known, $\gls{PAax}$, which appears directly in NPA problems as a variable. In our case there is no variable corresponding to $P(x|a,b)$, and thus we cannot use the NPA directly. To be more precise, the expressions in \eqref{guessingBayes0} and~\eqref{guessingBayes1} are not linear in variables occurring in the NPA, and thus they cannot be used either as a target, or as a constraint.

In order to overcome this difficulty let us note that for given $\mathbf{h}$ using~\eqref{eq:HardyNoisy} we get
\be
	\label{eq:sigmaHardyNoisy}
	\sigma = h_1 \cdot P(x=0,y=0) + h_3 \cdot P(x=0,y=1),
\ee
which is a constant, and
\be
	\label{eq:nuHardyNoisy}
	\nu = h_2 \cdot P(x=1,y=0) + P(a=0,b=0|x=1,y=1) \cdot P(x=1,y=1).
\ee
Note that the probabilities $P(x,y)$ depend on the random number generator, and not on the internal construction of the device, and we assume that they are under control of Alice and Bob. Thus the only variable occurring in the problem is the probability $P(a=0,b=0|x=1,y=1)$.

It is easy to see that the expressions \eqref{guessingBayes0} and~\eqref{guessingBayes1} achieve their maximal values as functions of~\eqref{eq:sigmaHardyNoisy} and~\eqref{eq:nuHardyNoisy}, if $\nu$ gets its minimal or maximal value, respectively. Thus we only need to use SDP optimization to establish the range of possible values of $\nu$ for a given $\mathbf{h}$. This can be done using the NPA directly.

Obviously, calculating the function for all possible values of continuous parameters $\mathbf{h}$ is impossible. Instead we calculate it only for some values. Now, if we represent the function with set of vectors, each containing the coordinates of a single point together with the value of the function, then the problem of linearly constrained optimization over this function can be solved with linear programming. Examples of such problems are problems stated in~\eqref{program1} and~\eqref{program2} further in this section.

\subsection{Post-processing strategies}

We will now discuss methods of post-processing strategies which can be applied in order to improve the key rate obtained with the described protocol.

The analysis of the case without noise helps to understand the reason why the use of the dropping strategy can increase the key rate. It also explains the role of nonuniform distribution of settings.

\subsubsection{A dropping strategy}

As already mentioned, in the noiseless case with uniform distribution of settings we have
\be
	\nonumber
	P(a=0,b=0|x=0,y=0) = \frac{5 \sqrt{5} - 11}{2} \approx 0.090167,
\ee
and
\be
	\nonumber
	P(a=0,b=0|x=1,y=1) = \sqrt{5} - 2 \approx 0.236068,
\ee
so that
\be
	\nonumber
	P(a=0,b=0|x=0,y=0) < P(a=0,b=0|x=1,y=1).
\ee
From this we get that the guessing probability for an eavesdropper is higher than $\frac{1}{2}$, since the probability that the settings (used in key generation) are equal to $1$ is greater that the probability that they are equal to $0$. Below we give a strategy which allows to get the guessing probability equal to $\frac{1}{2}$.

With the dropping strategy, after performing her measurements, Alice randomly selects only the following fraction of iterations with $a_i=b_i=0$ and $x_i=1$ (in the perfect case we then have also that $y_i=1$):
\be
	\nonumber
	\frac{P(a=0,b=0|x=0,y=0)}{P(a=0,b=0|x=1,y=1)}.
\ee
Afterward Alice sends the list of selected iterations to Bob \textit{via} an authenticated public channel.

In this reduced list of iterations Alice has an approximately equal number of settings $0$ and $1$. (Obviously in the noiseless case they correspond to the same values on the side of Bob.) Thus the guessing probability for an eavesdropper is now exactly equal to $\frac{1}{2}$.

Now, we have
\be
	\nonumber
	\ba
		& P_\text{not dropped}(a=0,b=0|x=0,y=0) \\
		& + P_\text{not dropped}(a=0,b=0|x=1,y=1) = 5 \sqrt{5} - 11 \approx 0.180334.
	\ea
\ee
To get the actual ratio of the total outcomes that are contained in the key, this should be multiplied by
\be
	P(x=y=0) = P(x=y=1) = \frac{1}{4}. \nonumber
\ee
Thus the key rate is equal to $\frac{5 \sqrt{5} - 11}{4} \approx 0.04508$.

\subsubsection{Nonuniform distribution of settings}
\label{sec:nonuniform}

Instead of choosing measurement settings with equal probabilities, both Alice and Bob may choose the measurement settings $x=0$ (resp. $y=0$) and $x=1$ (resp. $y=1$) with a ratio $r : 1-r$, for some $r=p_A=p_B$, where the joint probability distribution $\mathbb{P}(X,Y)$ is given by~\eqref{perfectDistribution}.

In order to obtain the guessing probability equal to $\frac{1}{2}$, meaning that
\be
	\nonumber
	P(a=0,b=0,x=0,y=0)=P(a=0,b=0,x=1,y=1),
\ee
the condition for $r$ reads
\be
	\label{eq:Hardy_rCond}
	P(a=0,b=0|x=0,y=0) r^2 = P(a=0,b=0|x=1,y=1) (1-r)^2,
\ee
or, equivalently
\be
	\nonumber
	r = \frac{\sqrt{P(a=0,b=0|x=1,y=1)}}{\sqrt{P(a=0,b=0|x=0,y=0)}+\sqrt{P(a=0,b=0|x=1,y=1)}}.
\ee
Thus we have
\be
	\label{eq:Hardy_rNonuniform}
	p_A = p_B = r = \frac{1}{2} \left(\sqrt{5} - 1\right) \approx 0.61803.
\ee

The key rate in the noiseless case, where there is no need for reconciliation, is thus
\be
	\label{eq:keyrateR}
	\ba
		& P(a=0,b=0|x=0,y=0) \cdot P(x=0,y=0) + \\
		& P(a=0,b=0|x=1,y=1) \cdot P(x=1,y=1) \\
		& = 2 P(a=0,b=0|x=0,y=0) r^2.
	\ea
\ee
This stems from the fact that each event $P(a=0,b=0,x=0,y=0)$ and $P(a=0,b=0,x=1,y=1)$ generates $1$ bit of a key. The equality is a consequence of~\eqref{eq:Hardy_rCond}.

Using \eqref{eq:Hardy_rNonuniform} and \eqref{eq:keyrateR} we calculate that the key rate for the noiseless case is given by
\be
	\nonumber
	2 P(a=0,b=0|x=0,y=0) r^2 \approx 0.06888.
\ee

\subsection{Calculation of the key rates}
\label{sec:keyrate}

We will now describe a method of calculating the key rate achieved by protocols based on the Hardy's paradoxes when we allow a noise to occur.

Let us denote by $n$ the number of iterations with both outcomes equal to $0$, after total number of $N$ iterations. This is on average given by
\be
	\nonumber
	n \approx P(a=0, b=0) \cdot N.
\ee

The malevolent constructor of the device, Eve, proceeds as follows. The iterations can be divided into two groups. For iterations within the first group, Eve makes a guess that the key value is $0$, and for iterations from the second group, she guesses the key value is $1$. We consider the case in which the average values of the Hardy's probabilities,~\eqref{eq:HardyNoisy}, are given by a vector $\mathbf{h}$.

Let us denote the probability that a subsystem belongs to the first group by $p_0$, and let the average values of the Hardy's probabilities from this group be given by $\mathbf{h}_0$. This allows guessing $0$ by Eve with the probability upper bounded by $P_0 \equiv \Gamma_0(\mathbf{h}_0) \geq \frac{1}{2}$ ($P_0 \geq \frac{1}{2}$, since otherwise it would be profitable to guess bit $1$ instead of $0$).

The remaining part of subsystems, which occurs with probability $p_1 = 1 - p_0$, has the Hardy's probabilities given by $\mathbf{h}_1$, and Eve guesses correctly the key value is $1$ with probability not exceeding $P_1 \equiv \Gamma_1(\mathbf{h}_1) \geq \frac{1}{2}$.

\subsubsection{Basic case} 
\label{sec:HardyBasicCase}

\begin{figure}[!htbp]
	\includegraphics[width=0.88\textwidth]{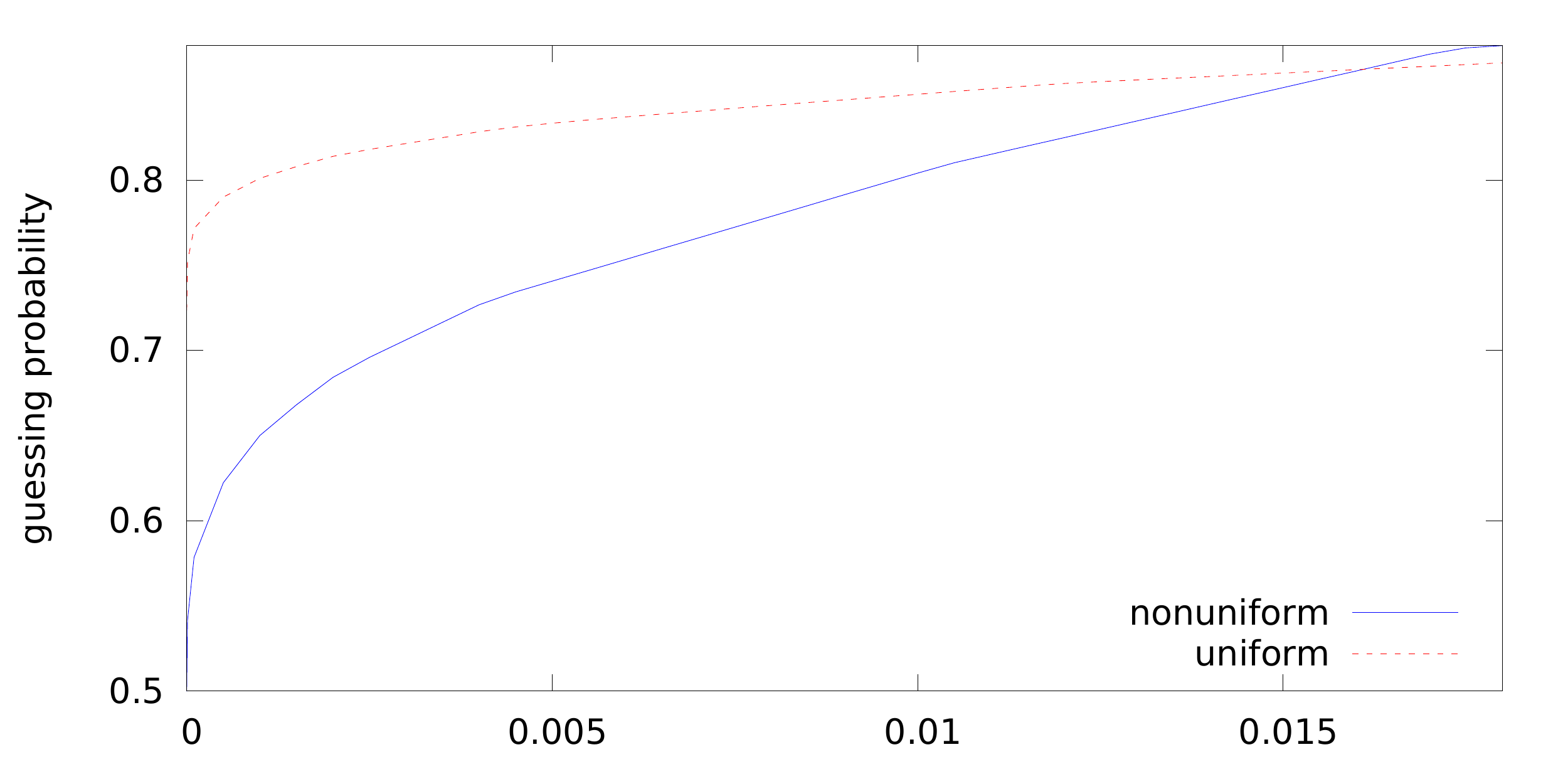}
	\caption[Average guessing probability of a key when no dropping strategy is used.]{\textbf{Average guessing probability of a key when no dropping strategy is used.} This is given by solutions of the problem \eqref{program1}. Cases with uniform and nonuniform distribution of settings are shown.\label{fig:guessingProbabilities}}
\end{figure}

In the simplest case described in sec.~\ref{sec:HardyProtocol}, in which no dropping strategy is used, the best thing Eve may do is to maximize her guessing probability, $P^{(1)}_{guess}(\mathbf{h})$. The relevant upper bound for the average guessing probability is given by the solution of the following problem in variables $p_0, p_1 , \mathbf{h}_0, \mathbf{h}_1$:
\begin{align}
	\label{program1}
	\begin{split}
		\text{maximize } &\null p_0 \Gamma_0(\mathbf{h}_0) + p_1 \Gamma_1(\mathbf{h}_1) \\
		\text{subject to } &\null p_0 \mathbf{h}_0 + p_1 \mathbf{h}_1 = \mathbf{h}, \\
		&\null p_0 + p_1 = 1, \\
		&\null p_0, p_1 \geq 0.
	\end{split}
\end{align}
Surprisingly approximation of this problem may be obtained using LP, see sec.~\ref{sec:gammaLP}. Solutions of this problem for $\mathbf{h}$ given by~\eqref{h_noise} are shown in Fig.~\ref{fig:guessingProbabilities}.

In this case the key rate is given by the following formula, \textit{cf.} \eqref{eq:keyrateMPA}:
\be
	\nonumber
	K_1 = P(a=0, b=0) \left( -\log_2(P^{(1)}_{guess}(\mathbf{h})) - H(x|y) \right).
\ee
The term $P(a=0, b=0)$ stems from the fact that the only round for which $a=b=0$ generates the key. The expression $-\log_2(P^{(1)}_{guess}(\mathbf{h}))$ is the average min-entropy of a bit of key, and $H(x|y)$ is the conditional entropy, which determines the amount of communication needed of reconciliation of the key, as shown in \cite{reconciliation94}. The communicated messages in reconciliation reveal some part of the key to an eavesdropper. Reconciliation is needed when the noise occurs, since then we do not have $P(x \neq y|a=b=0) = 0$.

\begin{figure}[!htbp]
	\includegraphics[width=0.88\textwidth]{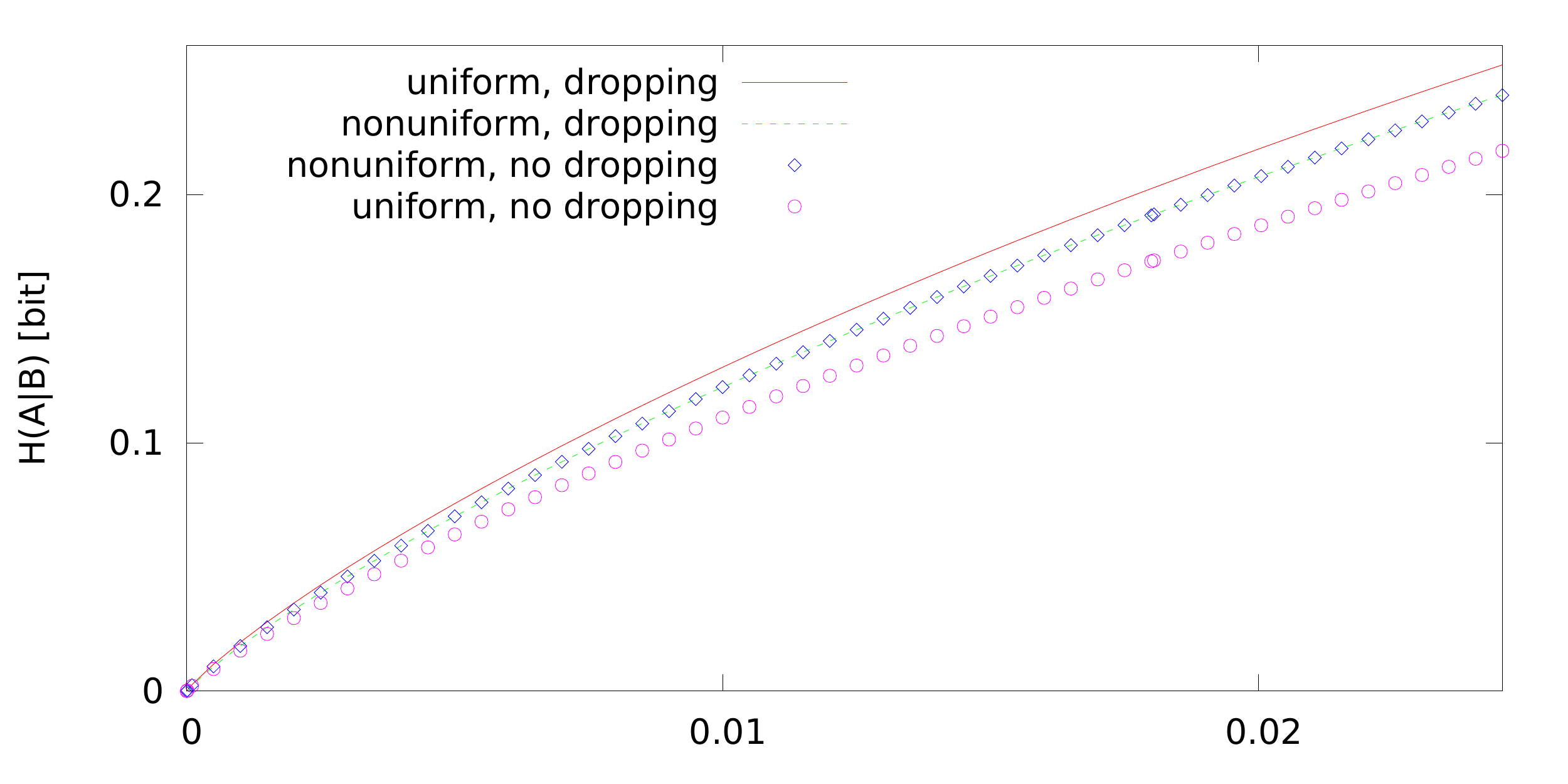}
	\caption[Conditional entropies of settings of Alice and Bob in Hardy's QKD protocol.]{\textbf{Values of conditional entropies of the setting of Alice given the setting of Bob, $H(x|y)$, if outcomes on both sides are $0$}. The cases with uniform and nonuniform distribution of settings, and with and without dropping strategy, are considered. In the case of the nonuniform distribution of settings, the line referring to use of a dropping strategy is slightly above the one without any dropping. Note that the fact that $H(x|y)$ depends on noise is a result of a different dependence of probabilities $P(x,y|a=b=0)$ on the noise for different $x$ and $y$. \label{fig:HABs}}
\end{figure}

Both expressions, $P(a=0, b=0)$ and the conditional entropy $H(x|y)$ can be calculated from the experimental setup\footnote{A brief explanation may be necessary at this point. Even though we are working in DI scheme, we can expect the device to generate a probability distribution obtained with a reasonable setup. Thus a probable probability distribution is given by the one obtained with a noised Hardy's state with proper measurements, which determines the value of $P(a=0, b=0)$ and $H(x|y)$.}, and reflect the amount of communication needed for the key reconciliation. Examples of conditional entropies for different post-processing strategies and distributions of settings are shown in fig.~\ref{fig:HABs}.

\subsubsection{The dropping strategy} 

Let us now consider those iterations in which both published outcomes are $0$. Among them the number of iterations with the setting of Alice equal to $0$ is
\be
	\nonumber
	\left( p_0 P_0 + p_1 (1 - P_1) \right) n \equiv p^\mathcal{A}_0 \cdot n,
\ee
and equal to $1$ is
\be
	\nonumber
	\left( p_0 (1 - P_0) + p_1 P_1 \right) n \equiv p^\mathcal{A}_1 \cdot n.
\ee
In these equations we assume that the device is prepared by an eavesdropper in the way which maximizes his guessing probability. Thus we assume that in those iterations in which he tries to guess the key value to be equal to $0$, the device generates this value with maximal probability, $P_0$; similarly for those iterations in which the eavesdropper guesses the key value $1$, the device generates it with probability $P_1$.

If $p^\mathcal{A}_0 < p^\mathcal{A}_1$, then Alice discards $\left( p^\mathcal{A}_1 - p^\mathcal{A}_0 \right) \cdot n$ part of the iterations having the value $1$. After this she has equal number of iterations with each of the values, namely $p^\mathcal{A}_0 \cdot n$.

For this scenario Eve will on average guess correctly $p_0 P_0 n$ iterations with the value $0$, and $\frac{p^\mathcal{A}_0}{p^\mathcal{A}_1} p_1 P_1 n$ iterations with the value $1$. Thus we get that her guessing probability among those iterations that are not discarded is given by
\be
	\nonumber
	P^{(2)}_{guess} \equiv \frac{1}{2 p^\mathcal{A}_0 n} \left( p_0 P_0 n + \frac{p^\mathcal{A}_0}{p^\mathcal{A}_1} p_1 P_1 n \right) = \frac{1}{2} \left( \frac{p_0 P_0}{p^\mathcal{A}_0} + \frac{p_1 P_1}{p^\mathcal{A}_1} \right).
\ee
The case with $p^\mathcal{A}_0 > p^\mathcal{A}_1$ gives exactly the same formula.

In order to calculate the bound on the guessing probability $P^{(2)}_{guess}(\mathbf{h})$ we use the following problem in variables $p_0, p_1, \mathbf{h}_0, \mathbf{h}_1$:
\begin{align}
	\label{program2}
	\begin{split}
		\text{maximize } &\null \frac{1}{2} \left( \frac{1}{p^\mathcal{A}_0} p_0 \Gamma_0(\mathbf{h}_0) + \frac{1}{p^\mathcal{A}_1} p_1 \Gamma_1(\mathbf{h}_1) \right) \\
		\text{subject to } &\null p_0 \mathbf{h}_0 + p_1 \mathbf{h}_1 = \mathbf{h}, \\
		&\null p_0 + p_1 = 1, \\
		&\null p_0, p_1 \geq 0.
	\end{split}
\end{align}
As in the previous case, the values
\be
	\nonumber
	\ba
		& p^\mathcal{A}_0 = P(x=0|a=0,b=0) \text{, and} \\
		& p^\mathcal{A}_1 = P(x=1|a=0,b=0)
	\ea
\ee
are calculated from the setup.

The key rate is now given by the following formula:
\be
	\nonumber
	\ba
		K_2 = & P(a=0, b=0) \cdot \left( 2 \min \left( p^\mathcal{A}_0,p^\mathcal{A}_1 \right) \right) \cdot \\
		& \cdot \left( -\log_2 \left( P^{(2)}_{guess}(\mathbf{h}) \right) - H(x|y,\text{dropping}) \right).
	\ea
\ee

Both $P(a=0,b=0)$ and the conditional entropy $H(x|y,\text{dropping})$ can be calculated directly from the setup, \textit{i.e.} the state and measurements expected to be used (see the footnote in sec.~\ref{sec:HardyBasicCase}).

\subsection{Biased sources of randomness}
\label{sec:biased}\index{probability distribution!biased}

Let us assume that the average probability distribution of the source of randomness is given by~\eqref{perfectDistribution}. Still, in each iteration, the probability distribution is biased in a way known to an eavesdropper. For the sake of simplicity we consider only \textbf{biases} obtained by changing the parameters $p_{A}$ and $p_{B}$ to $p_{A} \pm \epsilon$ and $p_{B} \pm \epsilon$, respectively, for given $\epsilon$. This is justified by Proposition~1 of \cite{Amp6}. This gives four possible biased distributions of settings, denoted $\left(\mathbb{P}_{biased,i}(x,y)\right)_{i=1,2,3,4}$. The average distribution in~\eqref{perfectDistribution} can be obtained only if the proportions of all biased distributions are equal.

If we know only the average distribution given by~\eqref{perfectDistribution}, then for iterations with a particular biased distribution $\mathbb{P}_{biased,i}(x,y)$, the observed conditional probabilities are under- or overestimated, \textit{viz.}
\be
	\label{underOverEstimated}
	P_{observed}(a,b|x,y) = P(a,b|x,y) \frac{P_{biased,i}(x,y)}{P_{unbiased}(x,y)},
\ee
\textit{cf.}~Eq.~\eqref{eq:IEst}.

Let us consider the noiseless case with nonuniform distribution of settings, given by~\eqref{eq:Hardy2q} with
\be
	P(a=0,b=0|x=0,y=0) = q = \frac{5 \sqrt{5} - 11}{2}, \nonumber
\ee
and thus with
\be
	P(a=0,b=0|x=1,y=1) = \tilde{q} = \sqrt{5} - 2.
\ee
For a given biased distribution $\mathbb{P}_{biased,i}(x,y)$, the probability that the generated bit of the key in a particular iteration is $0$, denoted by $P_{i,key=0}$, is given by the formula
\be
	\nonumber
	P_{i,key=0} \equiv \frac{q P_{biased,i}(0,0)}{q P_{biased,i}(0,0) + \tilde{q} P_{biased,i}(1,1)},
\ee
\textit{cf.}~Eq.~\eqref{guessingBayes0}.

Since the eavesdropper tries to guess the more probable key value, the guessing probability is given by the formula
\be
	\nonumber
	P_{guess,i} = \max(P_{i,key=0}, 1 - P_{i,key=0}).
\ee
In order to obtain the average guessing probability, this expression has to be averaged over all four possible biased probability distributions, \textit{viz.}
\be
	\nonumber
	\sum_{i=1,2,3,4} \frac{1}{4} P_{guess,i}.
\ee
Using this we can get that for $\epsilon \approx \frac{1}{10}$ the Hardy protocol is still able to work. Thus the proposed protocol is able to work even in cases when the only available source of randomness is biased. Up to our knowledge this is the first known protocol able to work with such a high bias \cite{Amp6}.

\subsection{Explanation of the problems in the protocol}

Here we give more details regarding the optimization problems we were using in this section for the analysis of the Hardy's protocol.

\subsubsection{NPA problem for Hardy's paradox}

Now, we will give an example of a problem occurring in the analysis of the protocol from this section. As a particular case let us take the following SDP problem for maximization of the value of $\nu$:
\begin{align}
	\label{problem-Hardy}
	\begin{split}
		\text{maximize } & \Tr \left( \Gamma M_{\nu} \right) \\
		\text{subject to } &\null \Gamma \in \mathcal{Q}_{1+AB}(2,2|2,2) \\
		&\null \Tr \left( \Gamma M_{P(0,0|0,0)} \right) \geq h_1, \\
		&\null \Tr \left( \Gamma M_{P(0,0|1,0)} \right) \leq h_2, \\
		&\null \Tr \left( \Gamma M_{P(0,0|0,1)} \right) \leq h_3, \\
		&\null \Tr \left( \Gamma M_{P(1,1|1,1)} \right) \leq h_4,
	\end{split}
\end{align}
where
\be
	\nonumber
	M_{\nu} = (1-p_A) \left( p_B M_{P(0,0|1,0)} + (1-p_B) M_{P(0,0|1,1)} \right)
\ee
refers to the probability that both results are $0$ and Alice's setting is $1$ (\textit{cf.}~Eq.~\eqref{guessingBayesY}).

We note that this problem relaxes the equality conditions of~\eqref{eq:HardyNoisy} to inequalities. These are simple for an SDP solver to handle, since equalities are introduced as pairs of opposite inequalities. This SDP problem has been used as one of the performance tests in chapter~\ref{chap:solver} with
\be
	\nonumber
	h_1 = q - \epsilon, h_2=h_3=h_4=\epsilon,
\ee
where $\epsilon \equiv 10^{-8}$, and $p_A$ and $p_B$ are given in~\eqref{eq:Hardy_rNonuniform}.

\subsubsection{Guessing probability as a linear problem}
\label{sec:gammaLP}

Let us consider the problem \eqref{program1}. We will show a method to approximate it with a linear problem. Recall from sec.~\ref{sec:sdpRelax} that $\Gamma_0(\mathbf{h})$ and $\Gamma_1(\mathbf{h})$ are concave hulls of functions $\tilde{\Gamma}_0(\mathbf{h})$ and $\tilde{\Gamma}_1(\mathbf{h})$. The latter pair of functions is calculated for given $\mathbf{h}$ with a problem similar to the problem~\eqref{problem-Hardy}. The value of $\Gamma_i(\mathbf{h})$ for given $\mathbf{h}$ is calculated as a maximum over all convex combinations of $\tilde{\Gamma}_i(\cdot)$ calculated for all possible points.

For obvious reasons, we can calculate $\tilde{\Gamma}_i(\cdot)$ only for a finite set of $\kappa$ points, $\{\mathbf{h}^{(k)}_i\}_k$, as stated\footnote{We have chosen an irregular set with more points near the values of particular importance, \textit{i.e.} $0$ and $q=\frac{5 \sqrt{5} - 11}{2}$. More precisely we have chosen
\be
	\nonumber
	\ba
		h_1 \in \{ & 0.0000001, 0.000001, 0.0000025, 0.000005, 0.00001, 0.000025, 0.00005, \\
		& 0.0001, 0.00025, 0.0005, 0.001, 0.003, 0.007, q - 0.0000001, \\
		& q - 0.000001, q - 0.00001, q, q + 0.00001, q + 0.000001, q + 0.0000001\},
	\ea
\ee
\be
	\nonumber
	\ba
		h_2 \in \{ & 0.0000001, 0.000001, 0.0000025, 0.000005, 0.00001, 0.000051, 0.00011, 0.00051, \\
		& 0.0011, 0.0031, 0.0071, q - 0.0000001, q - 0.00001, q, q + 0.00001, \\
		& q + 0.0000001, 0.15, 0.2, 0.25, 0.3\},
	\ea
\ee
\be
	\nonumber
	\ba
		h_3 \in \{ & 0.0000001, 0.000001, 0.0000025, 0.000005, 0.00001, 0.000052, 0.00012, 0.00052, \\
		& 0.0012, 0.005, q - 0.0000001, q - 0.00001, q, q + 0.00001, q + 0.0000001, \\
		& 0.2, 0.4, 0.6, 0.8, 1\},
	\ea
\ee
and
\be
	\nonumber
	\ba
		h_4 \in \{ & 0.0000001, 0.000001, 0.00001, 0.00013, 0.0013, 0.0033, 0.0073, q - 0.00001, \\
		& q, q + 0.00001, 0.15, 0.2, 0.3, 0.4, 0.5, 0.6, 0.7, 0.8, 0.9, 1\}.
	\ea
\ee
Each set has $20$ elements. Thus the calculation of the grid requires the calculation of $160000$ point being solutions of SDP problems. As shown in chapter~\ref{chap:solver}, the calculation of the grid with SDPT3 solver takes about 35h, with SeDuMi solver about 17h, and about 4h with PMSdp.} in sec.~\ref{sec:sdpRelax}, and thus get only an approximation of $\Gamma_i(\mathbf{h})$. Taking this into account we formulate an approximation of the problem \eqref{program1} in variable $x \in \mathbb{R}^{2 \kappa}$ as a primal LP, \eqref{LP-primal} with
\be
	\nonumber
	A^T = 
	\begin{bmatrix}
	\cdots & (\mathbf{h}^{(k)}_0)_1 & \cdots & (\mathbf{h}^{(k)}_1)_1 & \cdots \\
	\cdots & (\mathbf{h}^{(k)}_0)_2 & \cdots & (\mathbf{h}^{(k)}_1)_2 & \cdots \\
	\cdots & (\mathbf{h}^{(k)}_0)_3 & \cdots & (\mathbf{h}^{(k)}_1)_3 & \cdots \\
	\cdots & (\mathbf{h}^{(k)}_0)_4 & \cdots & (\mathbf{h}^{(k)}_1)_4 & \cdots \\
	\cdots & 1 & \cdots & 1 & \cdots
	\end{bmatrix},
\ee
with $b = [h_1, h_2, h_3, h_4, 1]^T$, and
\be
	\nonumber
	c = -[\cdots \tilde{\Gamma}_0(\mathbf{h}^{(k)}_0) \cdots \tilde{\Gamma}_1(\mathbf{h}^{(k)}_1) \cdots].
\ee
The value of $x_k$ for $k \in \{1, \dots, \kappa\}$ determines the share of a point with the value $\mathbf{h}^{(k)}_0$ (see \eqref{eq:HardyNoisy}) and the eavesdropper guessing $0$. The value of $x_k$ for $k \in \{\kappa + 1, \dots, 2 \kappa\}$ determines the share of a point with the value $\mathbf{h}^{(k - \kappa)}_1$ (see \eqref{eq:HardyNoisy}) and the eavesdropper guessing $1$. First four constraints mean that the average observed value in \eqref{eq:HardyNoisy} are given by $h_1$, $h_2$, $h_3$ and $h_4$. The last row of $A$ and $b$ is the normalization of convex coefficients constraint. This normalization includes the constraint $p_0 + p_1 = 1$. The average guessing probability is obtained by averaging the guessing probability over all $\mathbf{h}^{(k)}_{i=0,1}$ and guesses $0$ and $1$.

\section{Randomness expansion with two devices}
\label{sec:expansion}\index{device-independent}\index{randomness!expansion}

In this section we will describe properties of several protocols for DI-QRNG introduced by us in \cite{LubiePlacki}. Each of the protocols is able to certify two bits of randomness in each iteration in the noiseless case. Their efficiency is investigated in the presence of white noise with the NPA method.

One of our results is that for a different level of noise different protocols generate most randomness. For this reason the vendor of DI-QRNG should use different protocols depending on the amount of noise he expects to occur. Another of our results is that some of the protocols can be improved with additional observation of some CHSH\index{Bell operator!CHSH} expression.

As presented in chapter~\ref{chap:DI}, the laws of quantum mechanics give to an honest vendor a way to convince his potential customers that his device generates reliable randomness. With the help of some protocol, \textit{e.g.} based on Bell operators, he can allow them to check with some statistical tests that the device indeed produces certain amount of randomness, regardless of the way it has been constructed. However, he still needs to decide which protocol to use. We consider the scenario with a device consisting of three parts, \textit{viz.} a source of entangled states and two measurement apparatuses.

We note that all protocols are possible to be physically realized with devices obeying the laws of quantum physics. A prominent example of experiments involving similar protocols was conducted by Pironio \textit{et al.} \cite{RNGCBT}.

\subsection{Randomness certification protocols}

In all the protocols described below we assume that Alice's and Bob's devices are separated during the measurements. This holds, \textit{e.g.} if they are distant to each other. If we want the randomness not only to be fair, which is the case for example in gambling and system modeling purposes, but also to be confidential, \textit{e.g.} for cryptography or authentication, we also have to assume that the untrusted device does not communicate with the world outside. Without this assumption even a fair RNG may send the results to the adversary.

The general scheme is as follows. We consider a Bell operator, \textit{resp.} a set of operators. Next we choose a pair of settings, $x_0$ for Alice and $y_0$ for Bob which will be used to generate random numbers. We perform a series of iterations. In majority of the iterations we use settings $x_0$ and $y_0$, but for a randomly chosen iterations we use any other random settings instead. These iterations are used to estimate the values of operators used to certify the randomness.

As mentioned in chapter~\ref{chap:DI}, since for this purpose some initial randomness is needed, protocols described below are \textbf{randomness expanders}\index{randomness!expansion}\footnote{Note that in the case when we are interested not only in generating strings of that with some randomness, but also in extracting shorter random strings of a randomness of a better quality, then we need even more initial randomness for a randomness extraction, see~sec.~\ref{sec:extractors}.}.

To analyze the protocol we plot the min-entropy given some values of the operators. The plots are obtained using the NPA.

\subsection{Investigated protocols}

Below we present several different randomness certification protocols. Recall that $\eta = 1 - p$ is the noise parameter.

First, we consider protocols based on a violation of a single Bell inequality. For this purpose we use inequalities from the Braunstein-Caves family, CHSH, modified CHSH, $I_1$ and $I_2$ (see~secs~\ref{prot:BC}, \ref{sec:CHSH} and \ref{prot:other} in chapter~\ref{chap:basicsQI}).

Afterward we consider two more complicated protocols. First of them is based on a decomposition of CHSH operator into two parts, each having independent constraints. Second of them is T3C operator (see~sec.~\ref{sec:T3}) with two additional CHSH constraints.

Finally, in sec.~\ref{sec:CHSHcond} we show that imposing an additional CHSH condition in general gives a way to improve a protocol.

For the sake of simplicity of the analysis, we assume in these protocols which are composed of more than one operator that in each iteration all of the operators are affected by Eve to the same extent. In general one should consider the case when in some iterations one of the operators has a higher value than the average, whereas another one has a lower value than the average. We gave an example of such analysis previously in sec.~\ref{sec:HardyQKD}.

\subsubsection{Bell inequalities as protocols}
\label{sec:BIprotocols}

As first Bell operators we consider BC3, BC5 and BC7 from Braunstein-Caves\index{Bell operator!Braunstein-Caves}\index{randomness!expansion!Braunstein-Caves} family. As a pair of settings used for randomness generation with BCn operator we use $x_0=1$ and $y_0=\frac{3}{2} + \frac{n}{2}$. The certified min-entropy is calculated as logarithm of the solution of the following problem:
\begin{align}
	\label{problem-BCn}
	\begin{split}
		\text{maximize } & \max_{a,b \in \{1, \cdots, n\}} \left\{ \Tr \left( \Gamma M_{P \left( a,b|1,\frac{3}{2} + \frac{n}{2} \right)} \right) \right\} \\
		\text{subject to } &\null \Gamma \in \mathcal{Q}_{1+AB}(2,2|3,3) \\
		&\null \Tr \left( \Gamma M_{BCn} \right) \geq c_n \\
	\end{split}
\end{align}
where $M_{BCn}$ is the matrix constructed for the $BCn$ operator, and
\be
	\nonumber
	c_n = 2 p \cdot n \cdot \cos \left( \frac{\pi}{2 \cdot n} \right).
\ee

Note that this is not an SDP formulation. In order to solve this problem we have to use $n^2$ different SDPs, one for each possible pair of $a$ and $b$, and afterward take the maximum of these results.

For our numerical tests in chapter~\ref{chap:solver} we take $n=3$, $a=b=1$, and $p = 0.99$.

\begin{figure}[htb]
	\centering
		\includegraphics[width=0.88\textwidth]{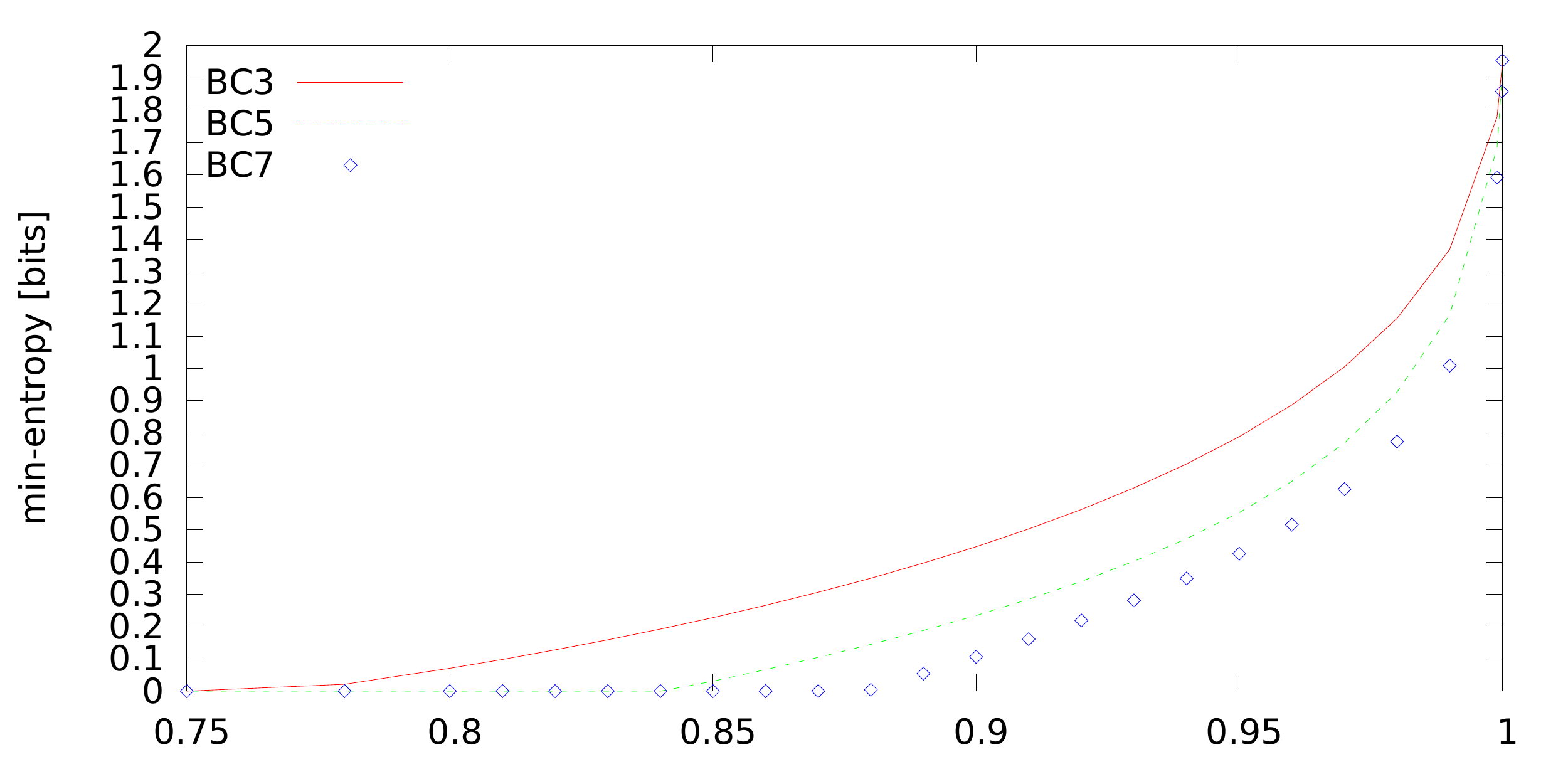}
	\caption{Lower bounds on min-entropy for protocols using Braunstein-Caves operator.}
	\label{fig:BC357}
\end{figure}

The results for protocols using Braunstein-Caves operators, as a function of noise, are shown in fig.~\ref{fig:BC357}. For all levels of noise the simplest operator, $BC_{3}$ (see~Eq.~\ref{BC3BI}) gives the highest min-entropy.

As the next Bell operator we use modified CHSH (see~sec.~\ref{prot:modCHSH}). We examine the min-entropy with the pair of settings $x_0=y_0=1$.

In the last example of protocols we take Bell operators defined above as $I_1$ and $I_2$, see~sec.~\ref{prot:other}. The pair of measurement settings for which the min-entropy is calculated are $x_0=y_0=1$ for both protocols.

\begin{figure}[htb]
	\centering
		\includegraphics[width=0.88\textwidth]{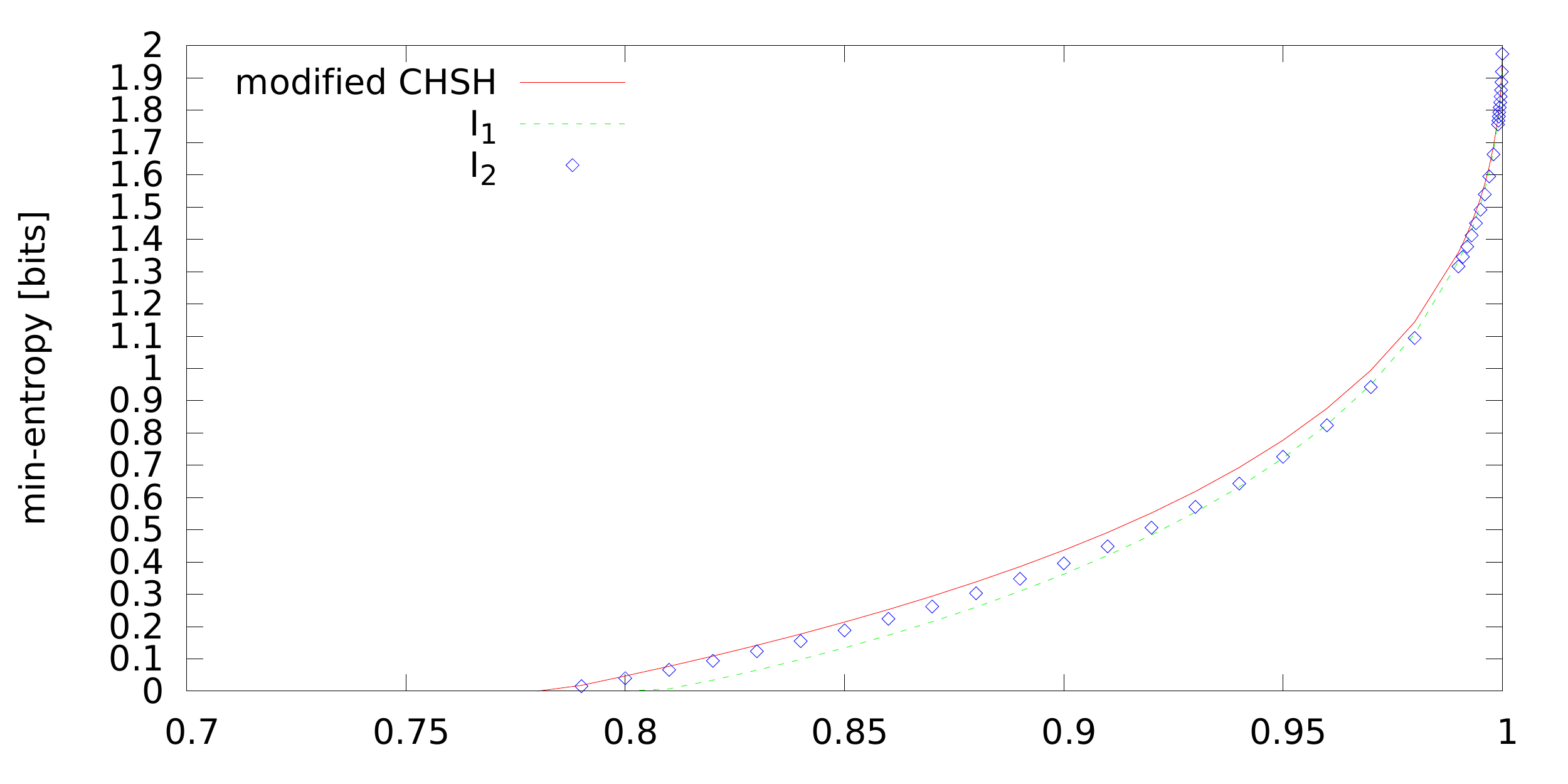}
	\caption[Min-entropies of modified CHSH, $I_1$ and $I_2$ protocols.]{Lower bounds on min-entropies for randomness certification protocols using modified CHSH, $I_1$ and $I_2$, as a function of noise.\label{fig:random}}
\end{figure}

The results for protocols using modified CHSH\index{Bell operator!modified CHSH}\index{randomness!expansion!}, $I_1$ and $I_2$ are shown in fig.~\ref{fig:random}.

\subsubsection{E0 and E1}
\label{prot:E0E1}\index{randomness!expansion!E0E1}

In this case instead of taking only a single operator corresponding to some Bell inequality, we use more than one for randomness certification.

Let us consider the following decomposition of a single CHSH operator into two operators:
\be
	\label{Es}
	\ba
		E_0 & = C(1, 1) + C(1, 2), \\
		E_1 & = C(2, 1) - C(2, 2).
	\ea
\ee
Let us note that this certificate requires both Alice's and Bob's parts of the device to have only $2$ possible measurement settings. It is the smallest requirement among all presented protocols. In this work we refer to this protocol as E0E1.

Using the Uffink's inequality \cite{Uffinka} we obtain that the maximal values of pairs of operators from~\eqref{Es} lie on a circle of radius $2$. Taking into account symmetries of these operators, their maximal values obtainable in quantum mechanics may be parametrized in the following way:
\be
	\label{EsMax}
	\ba
		& E_{0, max}(\phi) = 2 \cos(\phi) \\
		& E_{1, max}(\phi) = 2 \sin(\phi)
	\ea
\ee
where $\phi \in \left[0, \frac{\pi}{2}\right]$. A device is tuned by the constructor to produce probabilities leading to a particular value of $\phi$. Since we are dealing with a correlation form of operators if the noise $\eta$ occurs then the limiting values in~\eqref{EsMax} have to be multiplied by $1-\eta=p$.

Below we use the NPA to determine the min-entropy for measurement settings $x_0=2$ and $y_0=1$ as a function of $\phi$ for maximal values of the operators given in~\eqref{Es} for different values of noise. Afterward we find the optimal angle $\phi$ as a function of noise. To certify the randomness we assume that simultaneously $E_{0} \geq 2 \cdot p \cdot \cos(\phi)$ and $E_{1} \geq 2 \cdot p \cdot \sin(\phi)$. This way we have a family of protocols parametrized by a continuous variable $\phi$.

\begin{figure}[htb]
	\centering
		\includegraphics[width=0.88\textwidth]{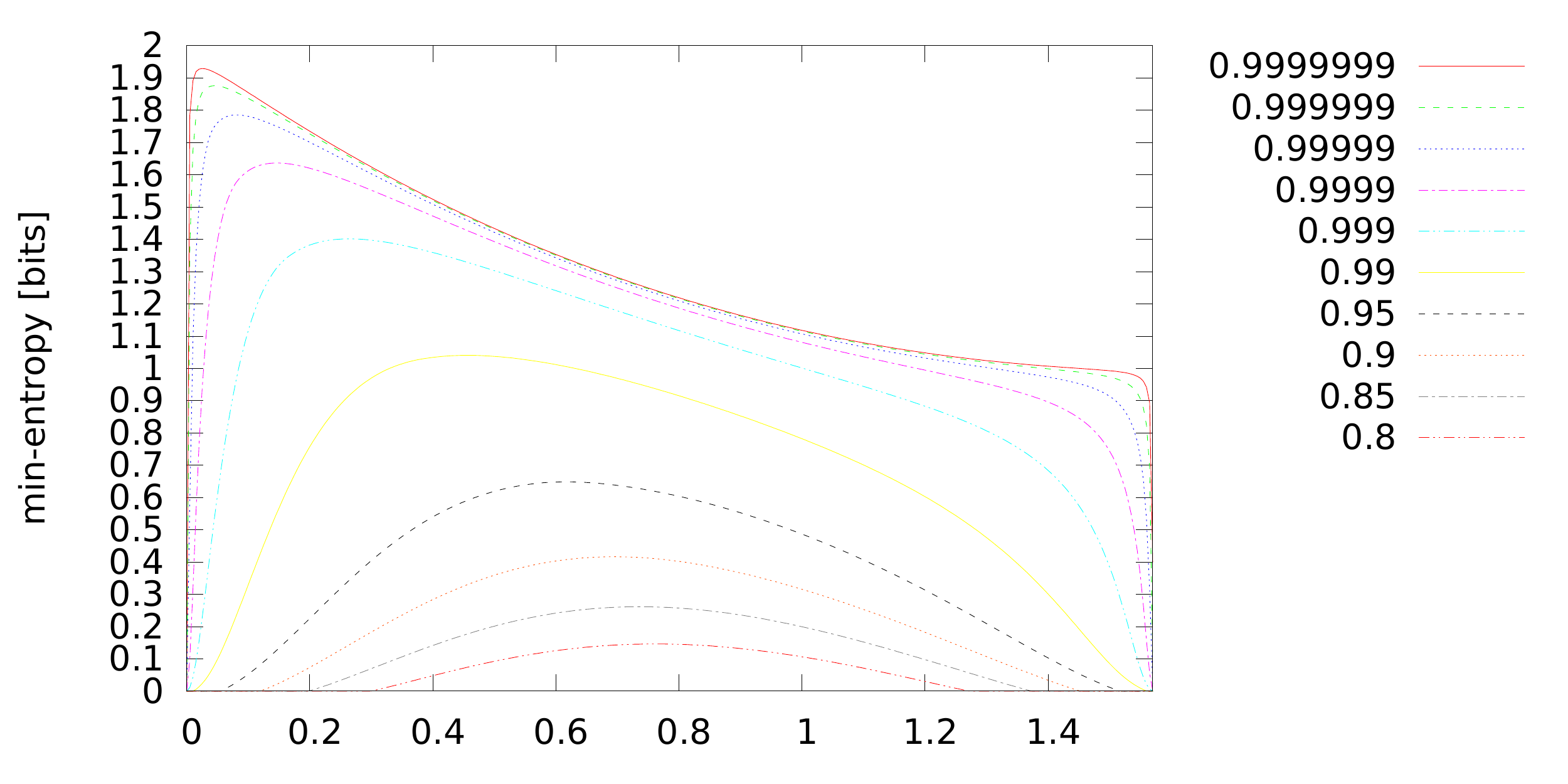}
	\caption[Min-entropy of E0E1 protocol.]{\textbf{Lower bound on min-entropy for E0E1 protocol} as a function of $\phi$ (see~Eq.~\ref{EsMax}) for different values of noise.\label{fig:E0E1}}
\end{figure}

The optimization problem is formulated as
\begin{align}
	\label{problem-E0E1}
	\begin{split}
		\text{maximize } & \max_{a,b \in \{1,2\}} \left\{ \Tr \left( \Gamma M_{P(a,b|2,1)} \right) \right\} \\
		\text{subject to } &\null \Gamma \in \mathcal{Q}_{1+AB}(2,2|2,2) \\
		&\null \Tr \left( \Gamma M_{E0} \right) \geq e_0, \\
		&\null \Tr \left( \Gamma M_{E1} \right) \geq e_1
	\end{split}
\end{align}
where
\be
	\nonumber
	\ba
		M_{E0} \equiv & 4 \left( M_{P(1,1|1,1)} + M_{P(1,1|1,2)} - M_{P_A(1|1)} \right) + \\
		& + 2 \left( -M_{P_B(1|1)} - M_{P_B(1|2)} + M_1 \right),
	\ea
\ee
and
\be
	\nonumber
	M_{E1} \equiv 4 \left( M_{P(1,1|2,1)} - M_{P(1,1|2,2)} \right) + 2 \left( M_{P_B(1|2)} - M_{P_B(1|1)} \right),
\ee
and
\be
	\nonumber
	\begin{aligned}
		& e_0 \equiv 2 p \cos{\phi}, \\
		& e_1 \equiv 2 p \sin{\phi}. \\
	\end{aligned}
\ee
In fig.~\ref{fig:E0E1} the results for this protocol are shown for different noises. As one may expect, for parameter $\phi$ equal $0$ and $\frac{\pi}{2}$ the min-entropy is $0$, since in this case the possible values of the operators from~\eqref{Es} in quantum and classical cases are the same. Thus the behavior of the device can be implemented locally and gives no warranty of the randomness. Another result is that the optimal angle between average values of operators from~\eqref{Es} depends on the noise parameter $p$. This dependence is shown in tab.~\ref{tab:E0E1OptAngle}.

\begin{table}[htb]
	\centering
		\begin{tabular}{|r|l|}
			\hline
			p & $\phi$  \\ \hline
			$0.9999999$ & $0.0252$ \\ \hline
			$0.999999$ & $0.0452$ \\ \hline
			$0.99999$ & $0.0811$ \\ \hline
			$0.9999$ & $0.1460$ \\ \hline
			$0.999$ & $0.2638$ \\ \hline
			$0.99$ & $0.4562$ \\ \hline
			$0.95$ & $0.6179$ \\ \hline
			$0.9$ & $0.6948$ \\ \hline
			$0.85$ & $0.7357$ \\ \hline
			$0.8$ & $0.7617$ \\ \hline
		\end{tabular}
	\caption{Optimal value of $\phi$ depending on noise.}
	\label{tab:E0E1OptAngle}
\end{table}

In our performance tests we use $a=b=1$, $\phi = \frac{\pi}{4}$ and $p=1$. Note that this is on the boundary of the set $\gls{quantumSet}$. For this reason this problem is expected to possibly cause problems with feasibility.

\subsubsection{T3 with an additional condition}
\label{prot:T3C}\index{randomness!expansion!T3C}

Let us consider a scenario in which Alice has $4$ possible measurement settings and Bob has $3$, each having $2$ possible outcomes. In \cite{HWL13} the T3 operator was used. Now let us take two additional Bell operators, which are identical to CHSH\index{Bell operator!CHSH} with certain choices of settings, \textit{viz.}:
\begin{subequations}
\label{CHSHs}
 \begin{align}
  CHSH_{1} & = C(0, 1) + C(2, 1) + C(0, 2) - C(2, 2) \label{CHSH1} \\
  CHSH_{2} & = C(1, 1) + C(3, 1) + C(1, 2) - C(3, 2) \label{CHSH2}
 \end{align}
\end{subequations}

The Tsirelson bound of the operator T3 is $4 \sqrt{3} \approx 6.928$, and for operators from~\eqref{CHSHs} is $2 \cdot \sqrt{2} \approx 2.82$, the same as for the standard CHSH.

If we impose on the device a condition that both operators from~\eqref{CHSHs} achieve the value of at least $0 \leq C \leq 2 \sqrt{2}$, then the maximal value of T3 is a function of $C$, $T3_{max} = T3_{max}(C)$. We require the device to obtain this maximal value. In the case when we cope with noise or imperfections of the device, where $0 < p < 1$, then we have to multiply values of T3 and CHSH by $p$. We call this protocol T3C.

Assuming these conditions we determine the lower bound on the min-entropy as a function of $C$ with maximal possible value of T3 for Alice's setting $x_0=1$ and Bob's $y_0=3$.

The min-entropy is evaluated with the following problem
\begin{align}
	\label{problem-T3C}
	\begin{split}
		\text{maximize } & \max_{a,b \in \{1,2\}} \left\{ \Tr \left( \Gamma M_{P(a,b|1,3)} \right) \right\} \\
		\text{subject to } &\null \Gamma \in \mathcal{Q}_{1+AB}(2,2|4,3) \\
		&\null \Tr \left( \Gamma M_{T3C} \right) \geq c_0, \\
		&\null \Tr \left( \Gamma M_{CHSH1} \right) \geq c_1, \\
		&\null \Tr \left( \Gamma M_{CHSH2} \right) \geq c_2,
	\end{split}
\end{align}
where
\be
	\nonumber
	M_{CHSH1} \equiv 4 \left( M_{P(1,1|1,1)} + M_{P(1,1|3,1)} + M_{P(1,1|1,2)} + M_{P(1,1|3,2)} \right)
\ee
and $M_{T3C}$ and $M_{CHSH2}$ are defined in an analogous way as in previous problems.

For the numerical tests in chapter~\ref{chap:solver} we take $a=b=1$, and
\be
	\nonumber
	\begin{aligned}
		& c_0 \equiv 4 p \sqrt{3}, \\
		& c_1 \equiv 2.3 p, \\
		& c_2 \equiv 2.3 p, \\
		& p = 0.99.
	\end{aligned}
\ee
Nonetheless, the values possible to obtain for $c_1$ and $c_2$ are a little bit higher that $2.3$. We have chosen this value so that the results are easier to reproduce.

\begin{figure}[htb]
	\centering
		\includegraphics[width=0.88\textwidth]{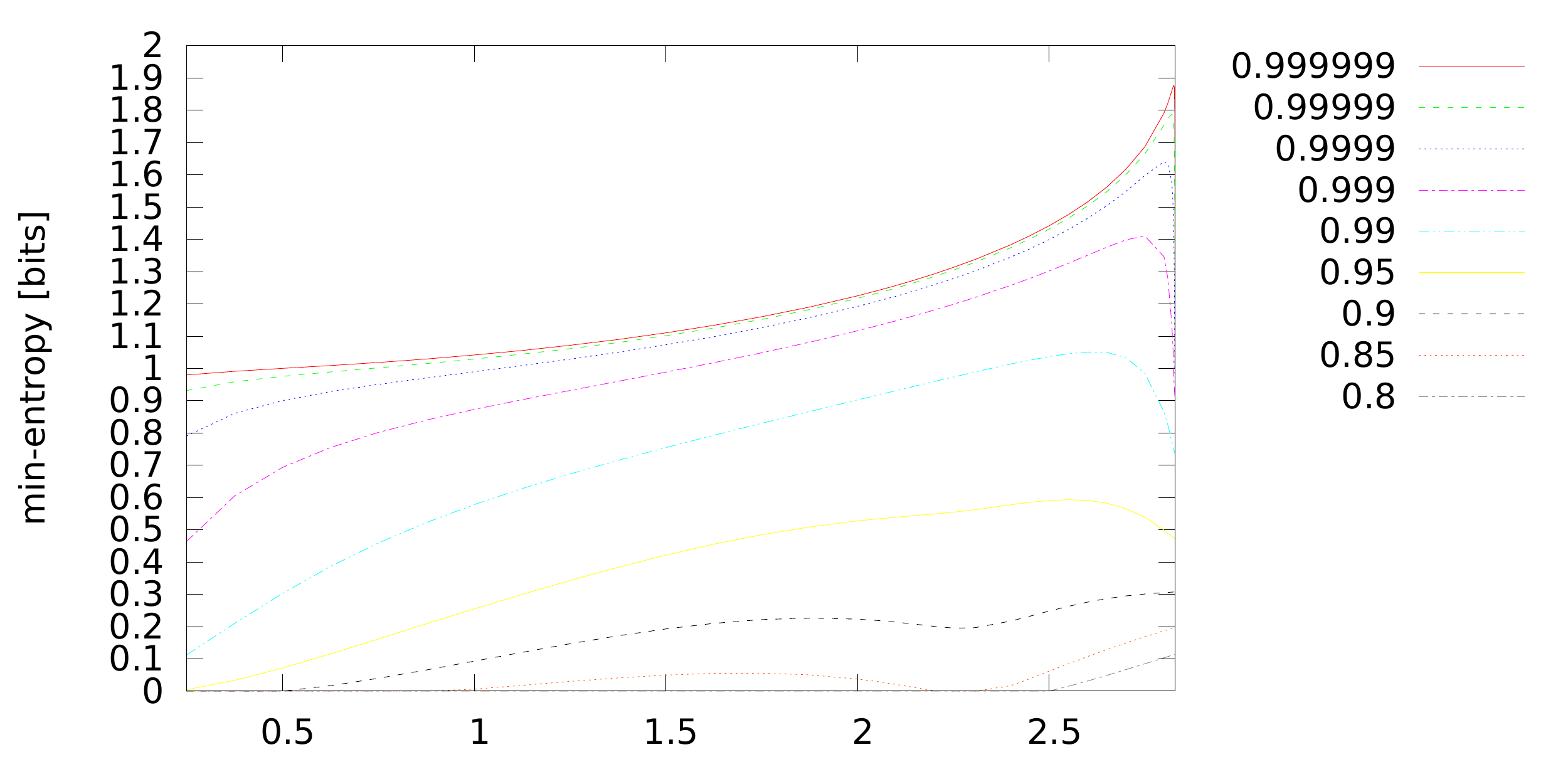}
	\caption[Min-entropy of T3C protocol.]{\textbf{Lower bound on min-entropy for protocol T3C} as a function of $C$ for different values of noise.\label{fig:T3C}}
\end{figure}

Fig.~\ref{fig:T3C} shows the results for this protocol. It is worth noticing that in the case without noise as parameter $C$ approaches its maximal value $2 \sqrt{2}$, then the min-entropy tends to $2$. The min-entropy strongly depends on $C$. This dependence is shown in the tab.~\ref{tab:OptC}.

\begin{table}[htb]
	\centering
		\begin{tabular}{|r|c|c|c|c|}
			\hline
			p & $0.999999$ & $0.99999$ & $0.9999$ & $0.999$ \\ \hline
			$\frac{C}{p}$ & $2.826$ & $2.82$ & $2.8$ & $2.75$ \\ \hline \hline
			p & $0.99$ & $0.95$ & $0.9$ & $0.8$ \\ \hline
			$\frac{C}{p}$ & $2.6$ & $2.55$ & $2.828$ & $2.828$ \\ \hline
		\end{tabular}
	\caption{Optimal value of parameter $C$ for protocol T3C depending on noise.}
	\label{tab:OptC}
\end{table}

Considering the bound on the value of the operator $T3$, it is maximal for $C \approx 2.3094116$. For high noises ($p < 0.95$) min-entropy has a local minimum near this point.

Tab.~\ref{tab:T3vsT3C} compares the amount of certified randomness by T3C and by T3 without additional CHSH conditions.
\begin{table}[htb]
	\centering
		\begin{tabular}{|r|c|c|}
			\hline
			p & $T3$ & T3C \\ \hline
			$0.99999$ & $1.3294$ & $1.7871$ \\ \hline
			$0.999$ & $1.2171$ & $1.4101$ \\ \hline
			$0.95$ & $0.55873$ & $0.5931$ \\ \hline
			$0.9$ & $0.19515$ & $0.3072$ \\ \hline
			$0.8$ & $0$ & $0.1136$ \\ \hline
		\end{tabular}
	\caption{Comparison of min-entropies certified by $T3$ alone and with two additional CHSH conditions.}
	\label{tab:T3vsT3C}
\end{table}

\subsubsection{Improving the modified CHSH protocol with a CHSH constraint}
\label{sec:CHSHcond}\index{Bell operator!CHSH}

Let us note that even though CHSH inequality is not able to certify two bits of randomness per iteration even in the noiseless case \cite{RNGCBT}, it is quite efficient for $p \leq 0.9$, see~tab.~\ref{tab:chsh}, being able to certify more randomness than most of the protocols presented above. Thus if we impose an additional condition for the CHSH operator we may hope to improve some of the above protocols.

\begin{table}[htb]
	\centering
		\begin{tabular}{|r|c|c|}
			\hline
			p & global & local \\ \hline
			$0.99999$ & $1.21757$ & $0.99090$ \\ \hline
			$0.999$ & $1.12231$ & $0.91155$ \\ \hline
			$0.95$ & $0.58411$ & $0.47234$ \\ \hline
			$0.9$ & $0.37757$ & $0.30718$ \\ \hline
			$0.8$ & $0.13510$ & $0.11362$ \\ \hline
		\end{tabular}
	\caption[Randomness certified by CHSH inequality for different noises.]{\textbf{Randomness certified by CHSH inequality for different noises.} The global randomness is the randomness, that is contained in a pair of bits, where one is the outcome of Alice, and the other is the outcome of Bob. The local randomness is the one of one bit of outcome for one of the parties. These data are the same for any possible pair of settings $x_0$ and $y_0$ used for generation. Randomness is given in terms of min-entropy}
	\label{tab:chsh}
\end{table}

Let us consider the modified CHSH protocol augmented with the ``pure'' CHSH. Combining modCHSH with a condition
\be
	\nonumber
	C(1, 2) + C(1, 3) + C(2, 2) - C(2, 3) \geq p \cdot 2 \cdot \sqrt{2}
\ee
gives the results shown in tab.~\ref{tab:modCHSHPlus}. This result shows that for low noises the improved modified CHSH protocol is almost as efficient as the most efficient single-operator protocol described above. One should note that the former requires fewer measurement settings. What is more, for high degrees of noises this protocol is very close to the efficiency of the protocol E0E1. Finally, for intermediate amounts of noise it is the best one.

\begin{table}[htb]
	\centering
		\begin{tabular}{|r|c|c|}
			\hline
			$p$ & modified CHSH & Improved modified CHSH \\ \hline
			$0.99999$ & $1.9764$ & $1.9764$ \\ \hline
			$0.999$ & $1.7751$ & $1.7751$ \\ \hline
			$0.95$ & $0.7775$ & $0.78024$ \\ \hline
			$0.9$ & $0.4365$ & $0.45443$ \\ \hline
			$0.8$ & $0.0468$ & $0.1342$ \\ \hline
		\end{tabular}
	\caption[Comparison of modified CHSH protocols]{Comparison of modified CHSH protocols with and without the additional CHSH constraint in terms of certified global min-entropy.}
	\label{tab:modCHSHPlus}
\end{table}

We refer interested readers to our paper \cite{LubiePlacki} for more technical details.

\subsection{Numerical results with semi-definite problems}
\label{sec:results}

The first interesting fact is that among the protocols under investigation there is no single optimal protocol.

What is even more remarkable is that the three best ones fall into three distinct categories. One of these protocols uses a single Bell operator BC3, another one is a combination of two Bell operators, modified CHSH and CHSH, whereas E0E1protocol is not even composed of Bell operators\footnote{Neither $E0$ nor $E1$ is a Bell operator, since the maximal value of each of them treated separately is the same in the classical and quantum case.}. This proves that there is more than one place to look for optimal protocols.

For $p > 0.9$ the protocol based on BC3 operator certifies the largest amount of randomness. On the other hand, for high noises ($p \approx 0.8$) the largest min-entropy is obtained using the protocol E0E1. The main disadvantage of this protocol is the necessity to chose the angle $\phi$ parameter individually for each level of noise.

The protocol T3C has the same drawback as E0E1, namely it requires the constructor to choose its parameter, $C$, depending on the noise. Similarly it gives good results for high noises, $p \approx 0.8$, but not as good as E0E1. It requires also more measurement settings.

The amount of achieved randomness for small noises, \textit{i.e.} $p \geq 0.9$, by protocols using modified CHSH, $I_{1}$ and $I_{2}$, is slightly smaller that the randomness from the protocol BC3. On the other hand, the protocol based on modified CHSH operator requires less measurement settings than BC3. In the case of intermediate noise, $0.85 \leq p \leq 0.92$, this protocol with additional CHSH condition certifies most randomness. It requires $2$ binary measurements for Alice and $3$ for Bob.

Protocols with more complicated Bell operators, $I_1$ and $I_2$, slightly differ depending on the amount of noise. For $p \geq 0.999$ the protocol $I_1$ gives more randomness than $I_{2}$ (\ref{eq:I2}), while for $p \leq 0.999$ $I_{2}$ gives more randomness than $I_{1}$.

\begin{figure}[htb]
	\centering
		\includegraphics[width=0.88\textwidth]{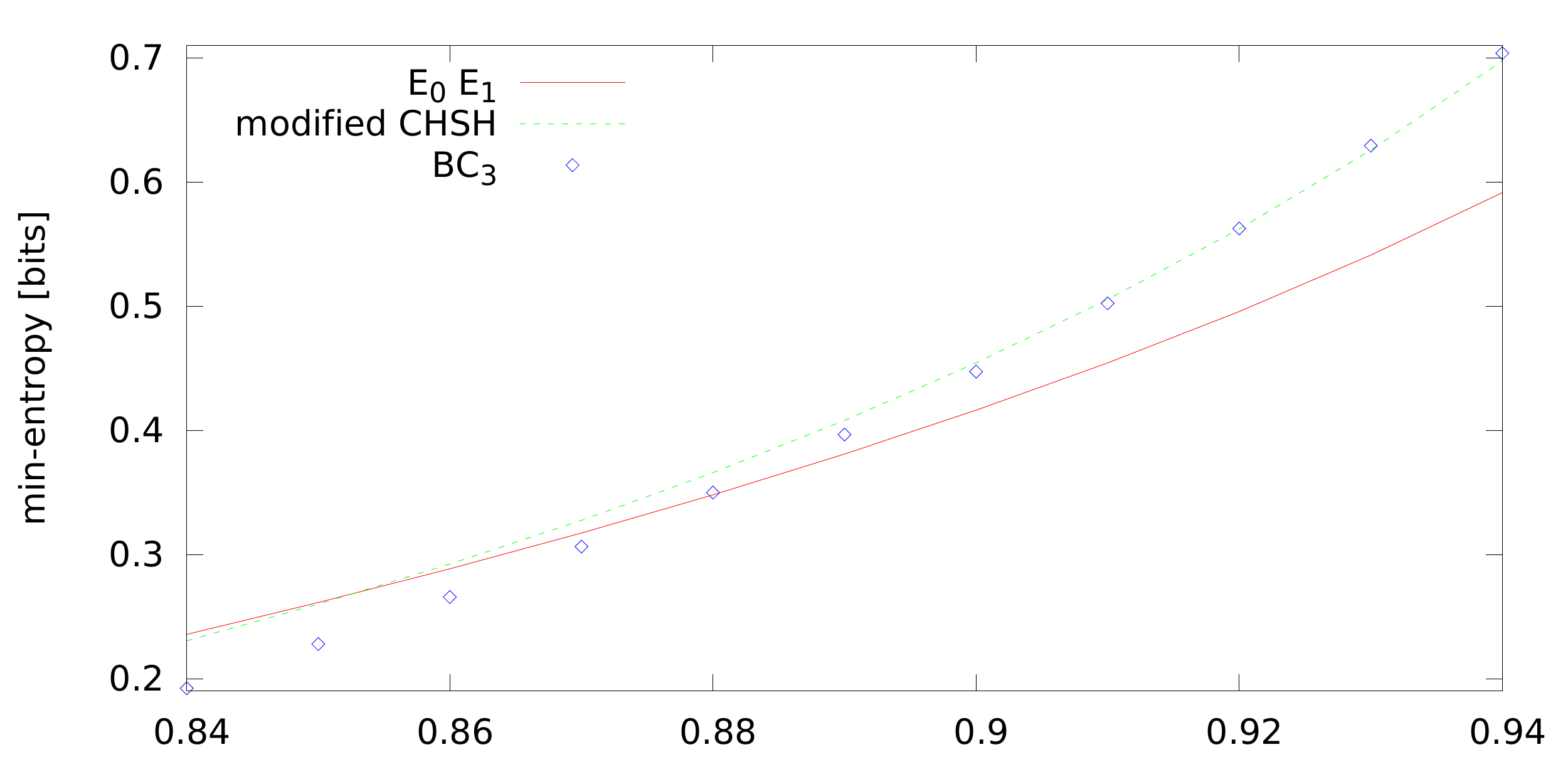}
	\caption{Comparison of three most efficient of investigated protocols for $0.84 \leq p \leq 0.94$.\label{fig:BestChartLarge}}
\end{figure}

All the above protocols are compared for selected values of $p$ in tables~\ref{tab:compare1} and \ref{tab:compare2}. A comparison of the three protocols which are in our opinion the most interesting is shown in fig.~\ref{fig:BestChartLarge}. We conclude that the vendor of QRNGs should choose one of these three protocols, depending on the amount of noise he is expecting the device to have to cope with.

\begin{table}[htb]
	\centering
		\begin{tabular}{|r|c|c|c|c|c|}
			\hline
			$p$ & BC3 & BC5 & BC7 & E0E1 & T3C \\ \hline
			$0.99999$ & $1.9769$ & $1.9656$ & $1.9537$ & $1.7854$ & $1.7871$ \\ \hline
			$0.999$ & $1.7792$ & $1.6841$ & $1.5917$ & $1.4013$ & $1.4101$ \\ \hline
			$0.95$ & $0.7885$ & $0.5534$ & $0.4258$ & $0.6484$ & $0.5931$ \\ \hline
			$0.9$ & $0.4474$ & $0.2342$ & $0.1064$ & $0.4163$ & $0.3072$ \\ \hline
			$0.8$ & $0.0709$ & $0.0000$ & $0.0000$ & $0.1461$ & $0.1136$ \\ \hline
		\end{tabular}
	\caption{Comparison of protocols from Braunstein-Caves family, E0E1, and T3C for different values of noise in terms of certified global min-entropy. In case of E0E1 protocol we take values for optimal angle parameter.}
	\label{tab:compare1}
\end{table}

\begin{table}[htb]
	\centering
		\begin{tabular}{|r|c|c|c|c|c|}
			\hline
			p & modified CHSH & $I_{1}$ & $I_{2}$ \\ \hline
			$0.99999$ & $1.9764$ & $1.9753$ & $1.9742$ \\ \hline
			$0.999$ & $1.7751$ & $1.7649$ & $1.7558$ \\ \hline
			$0.95$ & $0.78024$ & $0.7219$ & $0.7262$ \\ \hline
			$0.9$ & $0.45443$ & $0.3625$ & $0.3959$ \\ \hline
			$0.8$ & $0.1342$  & $0.0000$ & $0.0398$ \\ \hline
		\end{tabular}
	\caption{Comparison in terms of certified global min-entropy of protocols with improved modified CHSH, $I_1$ and $I_2$ for different noises.}
	\label{tab:compare2}
\end{table}

\section{Semi-device-independent scenario in NPA}
\label{sec:sdi}\index{semi-device-independent}\index{dimension witness}\index{randomness!expansion}

As mentioned previously, in this section we describe our method which we introduced and developed in papers \cite{HWL13,HWL14}, and which allows to analyze SDI scenarios with the methods of SDP, \textit{via} a reduction to an DI problem modeled in the NPA. We use this method to establish the robustness of a few SDI-QRNGs. With this method it is possible to impose an arbitrary restriction on the dimension of the communicated quantum system.

We also introduce the procedure of reduction of a dimension witness. Using this it is possible to obtain a new dimension witness. We show that such a reduced witness is able to certify in SDI at least as much randomness as the initial one.

We note that a physical implementation of a protocol obtained with the proposed method has been realized by Ahrens \textit{et al.} \cite{M-expdimwit}.

Recently two other methods which allow to perform a similar task of formulation SDP relaxations of SDI problems have been introduced in \cite{NTV} and \cite{MiguelVertesi}. The drawback of the former is that the size of the SDP problems is growing exponentially with the communicated dimension. In the latter method this is not observed. Both methods are closely related to the NPA method.

As mentioned above, a common method for the certification of QRNGs is to perform a Bell experiment as self-testing. This requires the device to be based on measurements on entangled states. On the other hand, the creation of an entangled system is relatively more difficult than the preparation of a simple quantum state. For this reason it is preferred to use prepare and measure protocols. On the other hand, the topic of Bell inequalities is by far better developed in the literature than the topic of dimension witnesses \cite{HWL11,DimWit,PB11,DW4}. Thus, it is useful to have a method which allows to develop devices of the first kind with the help of the well established knowledge about the devices of the second type.

One of the topics of this section are relations between Bell inequalities and dimension witnesses. One of such relations deals with converting a DI protocol to SDI, and \textit{vice versa} in such a way, that the second protocol inherits some properties of the first one.

\subsection{Properties of probability distributions}

Let us give the following lemma stating some properties of joint probability distributions which we will use further in this work.
\begin{lem}
	\label{probImply}
	For $a, b \in \{0, 1\}$ (binary outcomes) and arbitrary $x, y$, let us assume the no-signaling principle (see the definition~\ref{def:no-signal}). Then we have the following implications:
	\begin{enumerate}
		\item If $P(a,b|x,y)=P(\neg a, \neg b|x,y)$ holds, then we have $P(a|x)=\frac{1}{2}$ and
			\be
				\nonumber
				P(b|a,x,y)+P(b|\neg a,x,y)=1.
			\ee
		
		\item If $P(b|a,x,y)+P(b|\neg a,x,y)=1$ holds, then we have
			\be
				\nonumber
				P(b|a,x,y)=P(\neg b|\neg a,x,y).
			\ee
		
		\item If $P(a|x)=\frac{1}{2}$ and $P(b|a,x,y)=P(\neg b|\neg a,x,y)$ hold, then we have
			\be
				\nonumber
				P(a,b|x,y)=P(\neg a,\neg b|x,y).
			\ee
	\end{enumerate}
\end{lem}

We have proved this lemma in \cite{HWL14}.

\subsection{Analysis of dimension witnesses}

We start the discussion of SDI in the NPA with the statement of some properties of dimension witnesses which we introduced in \cite{HWL14}.

\subsubsection{Zero-summing dimension witnesses}

In our paper \cite{HWL14} we have introduced the following definition of a particular form of dimension witnesses.
\begin{definition}
    \label{symDWdef}
    A dimension witness $W$ of the form of~\eqref{DW} with the set of Alice's settings $\bar{X}$ of even size, $\bar{B} = \{0,1\}$ and $C_W=0$, is \textbf{symmetric}\index{dimension witness!symmetric}, if there exists a bijection $\phi: \bar{X} \rightarrow \bar{X}$ with
		\be
			\nonumber
			\forall_x \phi(x) \neq x,
		\ee
		and
		\be
			\nonumber
			\beta_{b,x,y} = -\beta_{b,\phi(x),y} = -\beta_{\neg b,x,y}.
		\ee

    For a set $\bar{\chi} \subset \bar{X}$ we define
    \be
			\nonumber
			W_{\bar{\chi}} \equiv \sum_{b \in \{0,1\}} \sum_{x \in \bar{\chi}} \sum_{y \in \bar{Y}} \beta_{b,x,y} P(b|x,y).
    \ee

    A set $\bar{\chi} \subset \bar{X}$ satisfying $\bar{\chi} \cap \phi(\bar{\chi}) = \emptyset$, and $\bar{\chi} \cup \phi(\bar{\chi}) = \bar{X}$ is called a \textbf{half} of $\bar{X}$.

    If a set $\bar{\chi}$ is a half, then $W_{\bar{\chi}}$ is called a dimension witness \textbf{reduced} with respect to $\bar{\chi}$. $\phi$ and $\bar{\chi}$ may be omitted if it is obvious which automorphism or set is considered.
\end{definition}

Note that it follows that every symmetric dimension witness is also a binary zero-summing\index{dimension witness!zero-summing}\index{dimension witness!binary} dimension witness. What is more, symmetric dimension witnesses are linear combinations of expressions of the form
\be
	\label{eq:symDW_omega}
	\omega(x, y) \equiv \frac{1}{2} \left( D(x,y) - D(\phi(x),y) \right),
\ee
where
\be
	\label{eq:symDW_D}
	D(x,y) \equiv P(0|x,y) - P(1|x,y).
\ee

In sec.~\ref{sec:symDW}, we investigate the properties of symmetric dimension witnesses in more details.

\subsubsection{Properties of dimension witnesses}
\label{sec:propDW}

Further we will use the following theorem \cite{HWL14}:
\begin{trm}
	\label{dwLemma}
	Let $\gls{HilbertSpace}$ be a Hilbert space of a dimension $2$.
	
	Let $\bar{X}$, $\bar{Y}$, $\{\beta_{b,x,y}\}_{b \in \{0,1\}, x \in \bar{X}, y \in \bar{Y}}$ be sets, and $C_W > 0$. Let $W$ be a binary dimension witness defined by these sets.

	Let $\{\rho_x\}_{x \in \bar{X}} \equiv \mathcal{S}$ be a set of states on $\gls{HilbertSpace}$.
	
	Let $\left\{ \{M^0_y,M^1_y\} \right\}_{y \in Y} \equiv \mathcal{M}$ be a set of binary POVM on $\gls{HilbertSpace}$.
	
	For these sets of states and POVMs let us define
	\be
		\nonumber
		s \equiv W[\mathbb{P}_2 (\mathcal{S},\mathcal{M})].
	\ee

	The following implications hold:
	\begin{enumerate}
	
		\item If $\forall_{y \in \bar{Y}} \sum_x \beta_{0,x,y} = \sum_x \beta_{1,x,y}$, then there exists a set of binary POVM on $\gls{HilbertSpace}$, $\tilde{\mathcal{M}} \equiv \left\{ \{\tilde{M}^0_y,\tilde{M}^1_y\} \right\}_{y \in \bar{Y}}$, such that
			\be
				\nonumber
				\forall_{y,b} \Tr \tilde{M}^b_y = 1,
			\ee
			and
			\be
				\nonumber
				W[\mathbb{P}_2 (\mathcal{S},\tilde{\mathcal{M}})] = s.
			\ee
			
		\item If $\sum_{b,x,y} \beta_{b,x,y} = 0$, and $\forall_{y,b} \Tr M^b_y = 1$, then for
			\be
				\nonumber
				\tilde{\mathcal{S}} \equiv \{\idOp - \rho_x\}_{x \in \bar{X}},
			\ee
			which is a set of states on $\gls{HilbertSpace}$, we have
			\be
				\nonumber
				W[\mathbb{P}_2 (\tilde{\mathcal{S}},\mathcal{M})] = -s.
			\ee
		
		\item If $\forall_{y,b} \Tr M^b_y = 1$, then there exists a set of projective measurements, $\tilde{\mathcal{M}} \equiv \left\{ \{\Pi^0_y,\Pi^1_y\} \right\}_{y \in \bar{Y}}$ with $\forall_{b \in \bar{B}, y \in \bar{Y}} \Tr \left( \Pi^b_y \right) = 1$, such that
			\be
				\nonumber
				W[\mathbb{P}_2 (\mathcal{S},\tilde{\mathcal{M}})] \geq s.
			\ee
	\end{enumerate}
\end{trm}

The first statement in this theorem states that in the dimension $2$ the condition that all measurement operators have trace $1$ is not restrictive with regard to the set of values possible to attain.

The second statement gives sufficient conditions under which an operation of negation of all states gives the same value of a dimension witness but with opposite sign.

The third statement, which may be used to complement the first, says that under certain conditions it is not restrictive to use only projective measurements in case when the values possible to be attained are considered.

Here we will prove only the first of these implications. Interested readers are referred to our paper \cite{HWL14} for a complete proof of other statements.
\begin{proof}
	Let us take $y \in \bar{Y}$. Let 
	\be
		\nonumber
		\ba
			& c_y = \frac{1}{2} \left(1 - \Tr(M^0_y)\right), \\
			& \tilde{M}^0_y = M^0_y + c_y \idOp, \\
			& \tilde{M}^1_y = M^1_y - c_y \idOp.
		\ea
	\ee

	We will prove that
	\be
		\nonumber
		\forall_{y,b} \tilde{M}^b_y \succeq 0.
	\ee
	
	Obviously we have from the completeness of measurements
	\be
		\tilde{M}^0_y + \tilde{M}^1_y = M^0_y + M^1_y = \idOp. \nonumber
	\ee
	
	There exists an orthonormal basis $\{\ket{0_y},\ket{1_y}\}$ in which
	\be
		\nonumber
		M^0_y = v_0 \proj{0_y} + v_1 \proj{1_y},
	\ee
	and
	\be
		\nonumber
		M^1_y = (1-v_0) \proj{0_y} + (1-v_1) \proj{1_y},
	\ee
	where $v_0,v_1 \in [0,1]$.
	
	We have $c_y=\frac{1}{2}(1-v_0-v_1)$, and $\idOp = \proj{0_y}+\proj{1_y}$. Thus
	\be
		\nonumber
		\tilde{M}^0_y = \frac{1}{2}(1+v_0-v_1)\proj{0} + \frac{1}{2}(1-v_0+v_1)\proj{1}.
	\ee
	Since $1+v_0-v_1\geq 0$ and $1-v_0+v_1 \geq 0$, we have $\tilde{M}^0_y \succeq 0$, and $\Tr \tilde{M}^0_y = 1$. Similarly, we check that $\tilde{M}^1_y \succeq 0$ and $\Tr \tilde{M}^1_y = 1$.

	Repeating this construction for all $y \in \bar{Y}$, we obtain a set of POVM,
	\be
		\nonumber
		\tilde{\mathcal{M}} \equiv \left\{ \{\tilde{M}^0_y,\tilde{M}^1_y\} \right\}_{y \in \bar{Y}}.
	\ee

	We have
	\be
		\nonumber
		\Tr (\rho_x \tilde{M}^b_y) = P(b|x,y) + (-1)^b \cdot c_y,
	\ee
	and from this it follows that
	\be
		\nonumber
		\ba
			W [\mathbb{P}_d(\mathcal{S},\tilde{\mathcal{M}})] = & \sum_{b,x,y} \beta_{b,x,y} \Tr (\rho_x \tilde{M}^b_y) = \\
			& s + \sum_y c_y \left( \sum_x \beta_{0,x,y} - \sum_x \beta_{1,x,y} \right) = s.
		\ea
	\ee
	
	This finishes the proof of the first implication.
\end{proof}

\subsection{Relation between Bell operators and dimension witnesses}

If we start with a DI-QRNG protocol based on some Bell operator, and construct out of it a prepare and measure SDI protocol, we hope it to certify a reasonable amount of min-entropy which is somehow related to the efficiency of the DI protocol. This is a useful case, as there exist many randomness expansion protocols based on Bell operators \cite{RNGCBT,CK11,MP13}. We have given a few examples of such protocols in sec.~\ref{sec:expansion}.

Another situation occurs when we have some SDI-QRNG protocol and want to employ the well developed SDP methods to analyze it. Here we present a way to obtain a new Bell operator with the property that the DI protocol using it certifies at most as much randomness as the SDI protocol.

\subsubsection{Conversion from a dimension witness to a Bell operator}
\label{sec:DWtoBI}

Let us consider a device $\textbf{D0}$ obtained from an untrusted vendor. The device consists of two parts, which we consider as black boxes, one used by Alice, and the other used by Bob. The only parameter of the device which we are able to verify is the dimension of the message send from one part of it to the other one. We make also a standard assumption that the device cannot communicate with the world outside the laboratory.

Similarly as in a DI case, let $\bar{X}$ be the set of settings among which Alice can choose. But in contrast to our previous cases, now these settings do not relate to any measurement producing an outcome to Alice. Instead, the box is supposed to produce and emit one of the quantum states of the dimension $d$. Let us denote the prepared states by $\{\rho_{x}\}_{x \in \bar{X}}$. The states are unknown to us.

Bob is also given a part of the device with settings labeled as $\bar{Y}$. Now these are the measurements settings performing one of the measurements given by POVMs from the following set	$\left\{ \{M^b_y\}_{b \in \bar{B}} \right\}_{y \in \bar{Y}}$. We do not know anything about those measurements apart from the fact that they are performed on the state emitted by the part of the device belonging to Alice.

Since we know the dimension of the communicated states, this is obviously an SDI scenario. Let us denote the conditional probability of obtaining the outcome $b \in \bar{B}$ when the chosen settings are $x \in \bar{X}$ and $y \in \bar{Y}$, by $P_{\textbf{D0}}(b|x,y)$.

Now, suppose that we are given a dimension witness $W$ of the form given in~\eqref{DW} above. Let us assume that this dimension witness achieves in the experiments on the device $\textbf{D0}$ on average the value $W_0$. 

Although the device $\textbf{D0}$ is not trusted, we may still \textit{think} about another device, denoted here by $\textbf{D1}$.

The device $\textbf{D1}$ also consists of two parts, with buttons labeled by the same sets $\bar{X}$ and $\bar{Y}$ on the side of Alice and Bob respectively. In this device we assume that both parts share a singlet state of the dimension $d$ (see sec.~\ref{sec:Hilbert}). The part on the side of Alice performs some projective measurement with outcome $0$ (if the projection succeeds\index{projection succeeded}) or $1$ (otherwise), depending on the chosen input $x \in \bar{X}$. We also assume that the measurement projects Alice's part of the singlet on the state $\rho_{x}$ that is the same as the relevant state from the device $\textbf{D0}$.

If the projection has succeeded, which happens with the probability $\frac{1}{d}$, then the device returns $a = 0$ and changes the state on the Alice's side into the state $\rho_{x}$, otherwise it returns $a = 1$. Since the shared state is a singlet, this measurement prepares the same $d$-dimensional state on the side of Bob. Afterward he performs the same POVM $\{M^b_y\}_{b \in \bar{B}}$ as the device $\textbf{D0}$ would perform, and finally he returns the outcome $b \in \bar{B}$. We do not have to analyze the case in which $a = 1$, since probabilities $P_{\textbf{D1}}(1,b|x,y)$ are not used in our relaxation.

Let us denote the probability that Alice gets the outcome $a$ with the setting $x$, and simultaneously Bob gets the outcome $b$ with the setting $y$ by $P_{\textbf{D1}}(a,b|x,y)$. It is easy to see, that
\be
	\nonumber
	P_{\textbf{D0}}(b|x,y) = d \cdot P_{\textbf{D1}}(0,b|x,y).
\ee

Let us consider a third device, $\textbf{D2}$, with the same interface as $\textbf{D1}$. Now the conditions on the internal working are relaxed, namely we do not assume anything about the performed measurements. Now the parts of Alice and Bob are allowed to an arbitrary state $\rho$ of any dimension. For this device we give an additional constraint
\be
	\nonumber
	\forall_{x \in X} P_{\textbf{D2}}(0|x) = \frac{1}{d},
\ee
where $P_{\textbf{D2}}(a|x)$ is the probability of getting the outcome $a$ by Alice with the setting $x$ with the device $\textbf{D2}$. We note that the description of this device is DI. Thus we may employ SDP with the NPA method to model its behavior, \textit{i.e.} the probability distributions of the device $\textbf{D2}$.

For the third device let us denote the probability of getting the outcomes $a$ and $b$ with a given pair of settings $x$ and $y$ for Alice and Bob, respectively, as $P_{\textbf{D2}}(a,b|x,y)$.

It is easy to see that all the conditional probability distributions which are possible to be obtained by the device $\textbf{D1}$, and equivalently by the device $\textbf{D0}$, are also possible to be obtained with the device $\textbf{D2}$. What is more, since the device $\textbf{D2}$ is a relaxed version of the initial device $\textbf{D0}$, if both of them have the same value of the relevant security parameters, then the certified amount of min-entropy generated by the device $\textbf{D2}$ gives a lower bound of the min-entropy certified to be generated by the device $\textbf{D0}$.

This way we have proved the following theorem.
\begin{trm}
    \label{DWtoBItheorem}
    Let $A = \{0,1\}$, $B = \bar{B}$, $X = \bar{X}$ and $Y = \bar{Y}$ be sets. Let us take $s \in \mathbb{R}$, $d \geq 2$, a Bell operator $B$ of the form given by~\eqref{BI}, and a dimension witness $W$ of the form given by~\eqref{DW}.
		
		Let us assume that in the definition of $W$ we have
		\be
			\nonumber
			\ba
				& \beta_{b,x,y} = d \cdot \alpha_{0,b,x,y}, \\
				& \alpha_{1,b,x,y} = 0, \\
				& C_I = C_W.
			\ea
		\ee

    Let $\mathcal{P}_{d,SDI}(s)$ be a subset of $\gls{PpmdBXY}$ with $d \geq 2$ satisfying the condition that
		\be
			\nonumber
			\forall_{\mathbb{P}(\bar{B}|\bar{X},\bar{Y}) \in \gls{PpmdBXY}} W[\mathbb{P}(\bar{B}|\bar{X},\bar{Y})] = s,
		\ee
		\textit{cf.}~the definition \ref{probSDI}.
		
		Let us define
		\be
			\nonumber
			\ba
				\mathcal{P}_{DI}(s) \equiv & \left\{ \mathbb{P}_{SDI}(B|X,Y): \exists \mathbb{P}_{DI}(\{0,1\},B|X,Y) \text{ such that } \right. \\
				& \forall_x P_{A,DI}(0|x)=\frac{1}{d}, B[\mathbb{P}_{DI}(\{0,1\},B|X,Y)]=s, \\
				& \left. P_{SDI}(b|x,y)=d \cdot P_{DI}(0,b|x,y) \right\}.
			\ea
		\ee

    Then $\mathcal{P}_{d,SDI}(s) \subseteq \mathcal{P}_{DI}(s)$.
\end{trm}

The theorem states that if a Bell operator and a dimension witness are related in the way given above, then we can use the set of probabilities allowed with the former in DI as a relaxation of the set of probabilities allowed with the latter in SDI. Fig.~\ref{fig:CGLMP} shows an example of application of the theorem \ref{DWtoBItheorem}.

We have thus developed a relation between Bell operators and dimension witnesses stating that the amount of randomness certified by a Bell operator lower-bounds the amount of randomness certified by the relevant dimension witness.

Recall that one of key features of the set of DI joint probability distributions is that it can be efficiently approximated using the NPA hierarchy. A useful property of the method is that we obtain a bound for any dimension of the communicated system changing only a value of the linear bound.

We conclude that using
\be
	\nonumber
	P(b|x,y) = d \cdot P(0,b|x,y),
\ee
we get that the certified min-entropy of SDI protocol is lower-bounded by the one of the DI protocol minus $\log_2{d}$.

\subsubsection{Conversion from a Bell operator to a dimension witness}
\label{sec:BItoDW}

In \cite{HWL13} we have proposed a heuristic method for obtaining a dimension witness from a Bell operator. Here we only briefly sketch the idea. We rewrite DI joint probabilities as
\be
	\label{eq:PabxyToPbaxy}
	P(a,b|x,y)=P(b|a,x,y) \cdot P(a|x).
\ee
If we now define $\bar{x} \equiv (a,x)$, and $\beta_{b,\bar{x},y} \equiv \beta_{b,(a,x),y} \equiv \alpha_{a,b,x,y} \cdot P(a|x)$, then we can take $\bar{B} = B$, $\bar{X} = A \times X$, and $\bar{Y} = Y$, and create a new dimension witness.

We stress that this method is heuristic. We do not state any general relation between the min-entropy which is possible to be certified with an SDI protocol obtained this way, and with the initial DI protocol. We refer interested readers to \cite{HWL13} and \cite{HWL14} for more details. Examples of such protocols, using witnesses T2, T3, BC3 and modified CHSH, are described below in the section \ref{sec:explicitExamples}. Starting with a Bell operator of the form
\be
	\label{symBI}
	\ba
		\sum_{x,y} & \dot{\alpha}_{x,y} \left( \frac{1}{2 p_{0,x}} \left( P(0,0|x,y) - P(0,1|x,y) \right) \right.\\
		+ &\left. \frac{1}{2 p_{1,x}} \left( P(1,1|x,y) - P(1,0|x,y) \right) \right),
	\ea
\ee
where $p_{0,x}+p_{1,x}=1$, we get a symmetric dimension witness of the form given in~\eqref{DW}, where
\be
	\nonumber
	\ba
		& \beta_{0,x,y} = \dot{\alpha}_{x,y}, \\
		& \beta_{1,x,y} = -\dot{\alpha}_{x,y}.
	\ea
\ee
To obtain a new SDI protocol, we first assume that $a$ is an initial DI scenario chosen randomly by Alice, with the distribution defined by $P(a|x) \equiv p_{a,x}$. Then we rewrite the probabilities in a form in Eq.~\ref{eq:PabxyToPbaxy}. Finally, to move from probabilities $\mathbb{P}(A,B|X,Y)$ to $\mathbb{P}(B|X,Y)$, we replace $P(a|x)$ with constant values, and obtain an operator of a form of a dimension witness.

\subsection[Binary zero-summing dimension witnesses]{Binary zero-summing dimension witnesses in dimension two}
\label{sec:binaryDW}\index{dimension witness!zero-summing}\index{dimension witness!binary}

Now, we will show that for the class of binary zero-summing dimension witnesses (see the definition \ref{def:binaryZerosummming} in chapter~\ref{chap:basicsQI}) it is possible to obtain a tighter SDP relaxations with the NPA, if the scenario with a Hilbert space of dimension $2$ is considered. We will now restrict our considerations to this case.

We show that for binary zero-summing dimension witnesses in dimension two it is possible to obtain tighter bounds in comparison with the one obtained using theorem~\ref{DWtoBItheorem}. Moreover, we show that with a symmetric dimension witness we can perform an operation of reduction which reduces the number of settings used by Alice by half, and thus simplify the scenario, while retaining the ability in certification of randomness.

For the purpose of this considerations we define a negation of two dimensional state as $\neg \rho \equiv \idOp - \rho$.

\subsubsection{Determinisic and projective strategies}

We start with a binary zero-summing dimension witness
\be
	\nonumber
	W = W(\{0,1\},\bar{X},\bar{Y},\{\beta_{b,x,y}\},C_W).
\ee
The witness is used to certify the randomness generated when Alice uses her preparation setting $x_0 \in \bar{X}$, and Bob uses the measurement setting $y_0 \in \bar{Y}$.

Let $\{\rho_x\}_{x \in \bar{X}}$ be the set of states, and $\left\{ \{M^0_y,M^1_y\} \right\}_{y \in \bar{Y}}$ be the measurements maximizing the guessing probability of the generated bits.

Let us note that the value of binary zero-summing dimension witness does not change if for arbitrary $y \in \bar{Y}$, the measurement is changed to
\be
	\nonumber
	\{M^0_y + c \idOp, M^1_y - c \idOp\},
\ee
where $c$ is such that the spectrum of the operators remains in the range $[0,1]$.

From the fact that the potential adversary wants to make the probability of one of the outcomes of the measurement $y_0$ as large as it is possible, we conclude that the form of the measurements which maximizes his guessing probability is
\be
	\label{optMeasY0}
	\begin{array}{l l}
		\begin{bmatrix}
			1 & 0 \\
			0 & 1 - \delta
		\end{bmatrix},
		& \quad
		\begin{bmatrix}
			0 & 0 \\
			0 & \delta
		\end{bmatrix}.
	\end{array}
\ee

Let us note that by the theorem \ref{dwLemma}.1 and \ref{dwLemma}.3 it is not restrictive for the vendor to use only projectors of trace $1$ for the measurements different than $y_0$.

The strategy of using a measurement of the form of~\eqref{optMeasY0} for the setting $y_0$, and projectors of trace $1$ for all remaining measurements, is equivalent to using the mixture of the following two strategies:
\begin{itemize}
	\item P - a pure strategy: a projective measurement of trace $1$ is used as the measurement selected by the setting $y_0$.
	\item D - a deterministic strategy: measurement $y_0$ returns always the same value $0$ or $1$.
\end{itemize}
For the setting $y_0$ in fraction $\delta$ of the cases, the strategy P is used, and in the remaining cases the strategy D. For the remaining measurements the same projective measurements of trace $1$ are used in both cases.

The guessing probability for the strategy D is $1$, and for the strategy P it is $P_{guess}$. This way we get that the average guessing probability is
\be
	\label{deltaAffine}
	(1 - \delta) + \delta \cdot P_{guess}.
\ee

In the case of a zero-summing dimension witness with the deterministic strategy, measurements with the setting $y_0$ give no contribution to the value of the witness. Thus the certification of the randomness with the dimension witness
\be
	\nonumber
  W = W(\{0,1\},\bar{X},\bar{Y},\{\beta_{b,x,y}\},C_W)
\ee
when the vendor of the device uses the mixed strategy is, after applying certain affine transformation,~\eqref{deltaAffine}, equivalent to the certification with a dimension witness
\be
	\nonumber
   W_{(\delta,y_0)}(\{0,1\},\bar{X},\bar{Y},\{\tilde{\beta}_{b,x,y}\},C_W)
\ee
with $\tilde{\beta}_{b,x,y}$ defined in~\eqref{deltaBeta} below, and the strategy P, where the guessing probability of Eve is given by~\eqref{deltaAffine}.

Since the vendor may choose any $\delta \in [0,1]$ which allows to observe the required value of the dimension witness $W$, when calculating lower-bound on the certified min-entropy, the worst case should be considered for each situation.

This way, we have proved the following
\begin{lem}
	Let $W = W(\{0,1\},\bar{X},\bar{Y},\{\beta_{b,x,y}\},C_W)$ be a binary zero-summing dimension witness, $x_0 \in \bar{X}$, and $y_0 \in \bar{Y}$. Let
	\be
		\nonumber
		W_{(\delta,y_0)} = W_{(\delta,y_0)}(\{0,1\},\bar{X},\bar{Y},\{\tilde{\beta}_{b,x,y}\},C_W)
	\ee
	be a dimension witness, where
	\be
		\label{deltaBeta}
		\tilde{\beta}_{b,x,y} = \left\{
		\begin{array}{l l}
			\beta_{b,x,y} & \quad \text{if $y \neq y_0$}\\
			\delta \cdot \beta_{b,x,y} & \quad \text{if $y = y_0$}
		\end{array} \right. .
	\ee
	Then
	\be
		\nonumber
		P_{guess}^{cert} (W,x_0,y_0,s,2) = \max_{\delta \in [0,1]} \left( (1 - \delta) + \delta \cdot P_{guess}^{cert(P)}(W_{(\delta,y_0)},x_0,y_0,s) \right)
	\ee
	where
	\be
		\nonumber
		P_{guess}^{cert}(W,x_0,y_0,s,2),
	\ee
	and
	\be
		\nonumber
		P_{guess}^{cert(P)}(W,x_0,y_0,s)
	\ee
	are defined in~\eqref{PguessCert} and~\eqref{PguessCertP}.
\end{lem}

\subsubsection{Tighter bounds from the NPA}
\label{sec:tighter}

The consequence of restricting the vendor to the dimension $2$ and measurements of trace $1$ is that the following holds for all $x \in \bar{X}$ and $y \in \bar{Y}$, and for any $b \in \{0,1\}$:
\be
	\label{negRho}
	\ba
		\Tr & (\neg \rho_x M^{\neg b}_y) = \Tr((\idOp - \rho_x) (\idOp - M^b_y)) = \\
		& 1 - \Tr(M^b_y) + \Tr(\rho_x M^b_y) = \Tr(\rho_x M^b_y) = P(b|x,y).
	\ea
\ee
This relation allows to refine the relaxation given in the section \ref{sec:DWtoBI}.

Now we consider a device $\textbf{D1}^{\prime}$ that models the strategy P by sharing the singlet state of dimension $2$, projecting on pure states $\{\rho_x\}_{x \in \bar{X}}$ on the side of Alice, and measuring on the side of Bob with projective measurements of trace $1$, $\left\{ \{M^0_y,M^1_y\} \right\}_{y \in \bar{Y}}$. In contrast to the device $\textbf{D1}$ if the projection on a state $\rho_x$ for any $x \in \bar{X}$ fails, then the prepared state is $\neg \rho_x$. It is easy to see that, by~\eqref{negRho}, for all $a, b \in \{0,1\}$, $x \in X$, and $y \in Y$ the probabilities obtained in this device are constrained by the relation
\be
	\label{negAnegB}
	P(a,b|x,y) = P(\neg a, \neg b|x,y).
\ee
A further relaxation, analogous to the one leading from the device $\textbf{D1}$ to the device $\textbf{D2}$, allows to obtain a device $\textbf{D2}^{\prime}$ satisfying the relation \eqref{negAnegB} which can be modeled by SDP.

Thus we have shown that for the case when the communicated state has dimension $2$ and $\bar{B}=\{0,1\}$ we have a relaxation given by the device $\textbf{D2}^{\prime}$ satisfying~\eqref{negAnegB}. In this case we have the following relation
\be
	\nonumber
	P(b|x,y) \stackrel{\text{is relaxed by}}{\longrightarrow} P_{\textbf{D2}^{\prime}}(0,b|x,y) + P_{\textbf{D2}^{\prime}}(1, \neg b|x,y)
\ee
from the construction of the device $\textbf{D2}^{\prime}$. From this it follows that~\eqref{eq:symDW_D}, $D(x,y) = P(0|x,y)-P(1|x,y)$, is relaxed by
\be
	\nonumber
	\ba
		& P_{\textbf{D2}^{\prime}}(0,0|x,y) + P_{\textbf{D2}^{\prime}}(1,1|x,y) - P_{\textbf{D2}^{\prime}}(0,1|x,y) - P_{\textbf{D2}^{\prime}}(1,0|x,y) \\
		& = C_{\textbf{D2}^{\prime}}(x,y).
	\ea
\ee
We can write it symbolically
\be
	\label{eq:DtoC}
	D(x,y) \stackrel{\text{is relaxed by}}{\longrightarrow} C_{\textbf{D2}^{\prime}}(x,y).
\ee
Thus we finally obtain
\be
	\label{eq:omegaToCC}
	\omega(x, y) \stackrel{\text{is relaxed by}}{\longrightarrow} \frac{1}{2} \left( C_{\textbf{D2}^{\prime}}(x,y) - C_{\textbf{D2}^{\prime}}(\phi(x),y) \right),
\ee
where we have $\omega(x, y)=\frac{1}{2} \left( D(x,y) - D(\phi(x),y) \right)$, as defined in~\eqref{eq:symDW_omega}.

We summarize these considerations in the following
\begin{trm}
	\label{ZeroSumTrm}
	Let $X = \bar{X}$ and $Y = \bar{Y}$ be sets.
	
	Let us take $s \in \mathbb{R}$, a Bell operator $B$ of the form given in~\eqref{BI}, and a binary zero-summing dimension witness $W$ of the form given in~\eqref{DW}, satisfying
	\be
		\nonumber
		\beta_{b,x,y} = \alpha_{0,b,x,y} = \alpha_{1,\neg b,x,y}.
	\ee

	Let $\mathcal{P}_{SDI}^{(P)}(s)$ be a subset of $\gls{PPXY}$ containing those probabilities
	\be
		\nonumber
		\mathbb{P}_2 (\{0,1\}|\bar{X},\bar{Y})
	\ee
	which satisfies $W[\mathbb{P}_2] = s$.
	
	Let us define
	\be
		\nonumber
		\ba
			\mathcal{P}_{DI,cond}(s) \equiv & \left\{ \mathbb{P}_{SDI}(B|X,Y): \exists \mathbb{P}_{DI}(\{0,1\},\{0,1\}|X,Y) \text{ such that } \right. \\
			& \forall_{a,b,x,y} P_{DI}(a,b|x,y) = P_{DI}(\neg a, \neg b|x,y), \\
			& B[\mathbb{P}_{DI}(\{0,1\},\{0,1\}|X,Y)]=s, \\
			& \left. P_{SDI}(b|x,y) = P_{DI}(0,b|x,y) + P_{DI}(1,\neg b|x,y) \right\}.
		\ea
	\ee

	Then $\mathcal{P}_{SDI}^{(P)}(s) \subseteq \mathcal{P}_{DI,cond}(s)$.
\end{trm}

Similarly to the theorem \ref{DWtoBItheorem}, this theorem shows how to use the set of probabilities allowed in DI as a relaxation of the relevant set in SDI, this time for a more restricted case.

From the lemma~\ref{probImply} we get that the condition
\be
	\nonumber
	P(a,b|x,y)=P(\neg a, \neg b|x,y)
\ee
is more restrictive than $P(a|x)=\frac{1}{2}$. In consequence
\be
	\nonumber
	\forall_s \mathcal{P}_{SDI}^{(P)}(s) \subseteq \mathcal{P}_{DI,cond}(s) \subseteq \mathcal{P}_{DI}(s),
\ee
where the sets are defined in theorems \ref{DWtoBItheorem} and \ref{ZeroSumTrm}. Thus the theorem~\ref{ZeroSumTrm} \textit{refines} the results of the theorem~\ref{DWtoBItheorem} for the case of binary zero-summing dimension witnesses.

Secs~\ref{sec:PStrategyLower} and~\ref{sec:PStrategyLowerUpper} show examples of lower and upper bounds for min-entropy certified when the untrusted vendor uses the strategy P. In secs~\ref{sec:MixedStrategyLowerDelta} and \ref{sec:MixedStrategyLower} lower bounds for the certified min-entropy in case when the untrusted vendor uses the mixed strategy are presented. All lower bounds are calculated using the NPA.

The upper-bounds are obtained by finding explicit representations of states and measurements with \textsc{fminunc} subroutine in OCTAVE. This optimization has been carried over pure states and projective measurements and, in contrast to SDP, is not guaranteed to reach global minimum.

Interestingly, in all protocols considered in sec.~\ref{sec:MixedStrategyLower}, it is optimal for the adversary to use $\delta = 1$, meaning that using the mixed strategy gives no gain in comparison with the strategy P.

\subsubsection{Symmetric dimension witnesses}
\label{sec:symDW}\index{dimension witness!zero-summing}\index{dimension witness!binary}\index{dimension witness!symmetric}

If a dimension witness is symmetric, then there is a way to reduce the size of the set of settings of Alice, $\bar{X}$, whilst the obtained dimension witness can certify at least the same amount of randomness as the initial one. Recall that symmetric dimension witnesses are special cases of binary zero-summing dimension witnesses.

The following theorem is an immediate result of the theorem~\ref{ZeroSumTrm} and the theorem~\ref{dwLemma},~p.2:
\begin{trm}
	\label{symAddCond}
	Let $\Pi$ be an SDI protocol with quantum communication limited to dimension $2$ with a symmetric dimension witness $W$ satisfying
	\begin{itemize}
		\item the value of the security parameter is equal to $s$,
		\item the protocol certifies the randomness $r$, and
		\item a pure strategy P is used.
	\end{itemize}
	Then the same value of the security parameter $s$ is still possible to be attained and certifies at least the same randomness $r$, if we impose an additional condition
	\be
		\nonumber
		\rho_x = \idOp - \rho_{\phi(x)} = \neg \rho_{\phi(x)},
	\ee
	implying
	\be
		\label{eq:symCond}
		P(b|x,y)=P(\neg b|\phi(x),y).
	\ee
\end{trm}

Roughly speaking, the theorem~\ref{symAddCond} states that symmetric dimension witnesses have some degree of freedom that does not increase the range of values possible to be attained, but allows adversary to ``distribute'' the average value of the witness among the states to mislead about the reliability of the device. The method proposed shows a way to remove this freedom.

Let us define $\phi((a,x)) \equiv (\neg a,x)$ and a set
\be
	\nonumber
	\chi \equiv \{(0,x): x \in X\} \subseteq \{0,1\} \times X.
\ee

We will now employ the heuristic method of conversion from Bell operators to dimension witnesses, sec.~\ref{sec:BItoDW}, for the case of correlation form of Bell operators. These operators can be expressed as in~\eqref{symBI} with $p_{a,x} = \frac{1}{2}$ and $\dot{\alpha}_{x,y} = \alpha_{x,y}$. Then $C(x,y)$ turns into
\be
	\nonumber
	\ba
		& \frac{1}{2} \left( P(0|(0,x),y)+P(1|(1,x),y) -P(1|(0,x),y)-P(0|(1,x),y) \right) = \\
		& \frac{1}{2} \left( D((0,x),y) - D((1,x),y) \right) = \omega(x, y)\\
		& = P(0|(0,x),y) - P(0|(1,x),y),
	\ea
\ee
and thus $\bar{X} = \{0,1\} \times X$. We know that a dimension witness which is a linear combination of expressions $\omega(x, y)$ is symmetric. This means that it is always possible to obtain a symmetric dimension witness from a correlation Bell operator \textit{via}:
\be
	\label{Ctoomega}
	C(x,y) \stackrel{\text{using heuristic conversion}}{\longrightarrow} \omega(x, y).
\ee

In order to apply theorem~\ref{symAddCond} we first assume~\eqref{eq:symCond}, \textit{i.e.} that
\be
	\nonumber
	P(b|(a,x),y)=P(\neg b|(\neg a,x),y),
\ee
and that $W$ is a symmetric DW with
\be
	\nonumber
	\bar{X}=\{0,1\} \times \bar{\chi}
\ee
for some set $\bar{\chi}$. The former assumption is justified below, but first we state its consequence. From the assumptions it follows that
\be
	\label{eq:omegaToD_0}
	\ba
		\forall_{x \in \bar{\chi}, y \in \bar{Y}} \quad & \omega(x, y) = P(0|(0,x),y) - P(0|(1,x),y) \\
		& = P(0|(0,x),y) - P(1|(0,x),y) = D((0,x),y).
	\ea
\ee
Let us now consider once again the P-M device $\textbf{D0}$ and assume that the set of settings of Alice is $\bar{X}=\{0,1\} \times \bar{\chi}$, and $W$ is used for certification. Let $\phi((a,x))=(\neg a,x)$ for $(a,x) \in \bar{X}$. We now introduce a device $\textbf{D1}^{\prime\prime}$ which prepares the same states as $\textbf{D0}$ for settings in $\{0\} \times \bar{\chi}$, and states $\rho_{(1,x)} = \neg \rho_{(0,x)}$ for settings in $\{1\} \times \bar{\chi}$. By theorem~\ref{symAddCond} this device can attain the same value $s$ of $W$ as the device $\textbf{D0}$ (this is obtained when states and measurements were chosen properly, which is always possible), and certifies at least the same amount of randomness. But from~\eqref{eq:omegaToD_0} it follows that with the dimension witness $W_{\bar{\chi}}$ (see definition~\ref{symDWdef}) the device $\textbf{D1}^{\prime\prime}$ will certify exactly the same amount of randomness as with $W$. On the other hand, since for Alice preparations for settings from $\{1\} \times \bar{\chi}$ does not influence the value of $W_{\bar{\chi}}$, we can consider another, \textit{reduced}, device, $\textbf{D2}^{\prime\prime}$, which has only half of the settings compared to devices $\textbf{D0}$ and $\textbf{D1}^{\prime\prime}$, and using the dimension witness $W_{\bar{\chi}}$ certifies at least as much randomness as the initial device $\textbf{D0}$ with $W$.

If we identify the settings $(0,x) \in \{0\} \times \bar{\chi}$ in the initial device, with settings $x \in \bar{\chi}$ of the reduced device we get
\be
	\label{eq:omegaToD}
	\forall_{x \in \bar{\chi}} \omega(x, y) \stackrel{\text{reduction}}{\longrightarrow} D(x,y).
\ee
instead of $\omega(x, y)$ from~\eqref{Ctoomega}. This is an example of the \textit{reduction}. We note that using this method the number of states used by Alice is halved.

\subsection[Dimension witnesses summary]{Summary of relations between Bell operators and dimension witnesses}

To summarize, we have the following conversion:
\begin{itemize}
	\item arbitrary DW $\rightarrow$ BI,~\eqref{BI}, by theorem~\ref{DWtoBItheorem}
	\item binary zero-summing DW $\rightarrow$ BI,~\eqref{BI}, tighter min-entropy bound, by theorem~\ref{ZeroSumTrm}
	\item correlation BI with $A=\{0,1\}$ $\rightarrow$ symmetric DW with $\bar{X}=\{0,1\} \times X$ using $C(x,y) \rightarrow \omega(x,y)$ by~\eqref{Ctoomega}
	\item symmetric DW with $\bar{X}=\{0,1\} \times X$ $\rightarrow$ reduced DW with $\bar{X}=X$ using $\omega(x,y) \rightarrow D(x,y)$ by~\eqref{eq:omegaToD}
	\item symmetric DW $\rightarrow$ correlation BI with $X=\bar{X}$ by~\eqref{eq:omegaToCC}
\end{itemize}

This gives us two chains of transformations.

The first is given by~\eqref{eq:DtoC}, \eqref{Ctoomega}, and~\eqref{eq:omegaToCC}:
\be
	\nonumber
	\ba
		& D(x,y) \stackrel{\text{is relaxed by}}{\longrightarrow} C(x,y) \stackrel{\text{using heuristic conversion}}{\longrightarrow} \\
		& \omega(x, y) \stackrel{\text{is relaxed by}}{\longrightarrow} \frac{1}{2} \left( C(x,y) - C(\phi(x),y) \right).
	\ea
\ee

The second one,~\eqref{Ctoomega}, \eqref{eq:omegaToD}, and~\eqref{eq:DtoC}, gives the following one-to-one relation between correlation Bell operators, symmetric dimension witnesses and reduced dimension witnesses
\be
	\nonumber
	\ba
		& C(x,y) \stackrel{\text{using heuristic conversion}}{\longrightarrow} \omega(x, y) \stackrel{\text{reduction}}{\longrightarrow} \\
		& D(x,y) \stackrel{\text{is relaxed by}}{\longrightarrow} C(x,y).
	\ea
\ee

\subsection{Examples of conversion between DI and SDI protocols}
\label{sec:explicitExamples}

In this section we give several examples of applications of the methods presented above. All figures are plotted with respect to the noise parameter $p$. On the plots $-$ denotes the logical negation of a bit.

\subsubsection{Tn as dimension witnesses}

In \cite{HWL12} a family of dimension witnesses based on quantum random access codes (\textit{QRAC}s) was introduced. In ($n$ to $1$)-QRAC Alice encodes $n$ bits by sending one of the $2^n$ states to Bob who tries to guess one of the bits performing one of $n$ measurements. These dimension witnesses are symmetric.

The symmetric dimension witness of the ($2$ to $1$)-QRAC can be written in the form
\be
	\label{T2DWfull}
	\ba
		\omega(1, 1) + \omega(1, 2) + \omega(2, 1) - \omega(2, 2).
	\ea
\ee
The reduced form of this dimension witness is thus
\be
	\label{T2DW}
	\ba
		D(1, 1) + D(1, 2) + D(2, 1) - D(2, 2).
	\ea
\ee
The randomness certified by these two dimension witnesses is lower-bounded by the values obtained with two Bell operators stated in~\eqref{T2BIfull} and~\eqref{T2BI}, respectively:
\begin{subequations}
	\be
		\label{T2BIfull}
		\ba
			& \frac{1}{2} \left( C(1, 1) + C(1, 2) + C(2, 1) - C(2, 2) \right. \\
			& \left. + C(3, 1) + C(3, 2) + C(4, 1) - C(4, 2) \right),
		\ea
	\ee
	and for the dimension witness from the equation (\ref{T2DW}),
	\be
		\label{T2BI}
		\ba
			T2 \equiv C(1, 1) + C(1, 2) + C(2, 1) - C(2, 2).
		\ea
	\ee
\end{subequations}
The operator defined in~\eqref{T2BI} is exactly the CHSH Bell operator. Lower bounds for this case are shown in sec.~\ref{sec:robustSdi} in figs~\ref{fig:T2_cert_P}, \ref{fig:T2_NPA_P}, \ref{fig:T2_NPA_delta} and \ref{fig:T2_cert}.

Let us now consider the next QRAC, \textit{i.e.} the ($3$ to $1$)-QRAC. This is given by the equation
\be
	\label{T3DWfull}
	\sum_{x \in \bar{X}, y \in \bar{Y}} (-1)^{x_y} P(0|x,y),
\ee
with $\bar{X} = \{000, \ldots, 111\}$ and $\bar{Y} = \{0,1,2\}$.

Taking $\phi(x)=\neg x$, where negation is bit-wise, $\bar{X} \equiv \{000,001,010,011\}$ and $\bar{Y} \equiv \bar{Y}$, the following reduced dimension witness is obtained
\be
	\label{T3DW}
	\ba
		& P(0|000,0)+P(0|001,0)+P(0|010,0)+P(0|011,0) \\
		& +P(0|000,1) +P(0|000,2) +P(0|001,1) -P(0|001,2) \\
		& -P(0|010,1) +P(0|010,2) -P(0|011,1) -P(0|011,2).
	\ea
\ee
Now, we can use this reduced dimension witness and employ the above method,~\eqref{eq:DtoC}, to get the Bell operator $T3$ described above in sec.~\ref{sec:T3}, in~\eqref{T3BI}. On the other hand, if we use the dimension witness without reduction, namely in a form defined in~\eqref{T3DWfull}, we obtain a larger Bell operator on $\mathbb{P}(\{0,1\},\{0,1\}|\{1,\dots,8\},\{1,2,3\})$ with similar properties
	\be
	\label{T3BIfull}
	\ba
		T3^{\prime} & \equiv \frac{1}{2} \left( C(1,1)-C(5,1)+C(2,1)-C(6,1)+C(3,1)-C(7,1) \right. \\
		&  +C(4,1)-C(8,1)+C(1,2)-C(5,2) +C(1,3)-C(5,3)  \\
		& +C(2,2)-C(6,2)-C(2,3)+C(6,3)-C(3,2)+C(7,2) \\
		& \left. +C(3,3)-C(7,3)-C(4,2)+C(8,2) -C(4,3)+C(8,3) \right).
	\ea
\ee
We may calculate a lower-bound on $H_{\infty}^{cert}(T3,x_0,y_0,s,2)$, the certified min-entropy, with $x_0 = 1$, $y_0 = 1$ using the theorem~\ref{ZeroSumTrm} and executing a series of SDP problems for a grid of values of $\delta \in [0,1]$. Fig.~\ref{fig:T3_NPA_P} shows the min-entropies certified with the Bell inequality $T3$ for different additional conditions. NPA lower-bounds on the certified min-entropy obtained by theorem \ref{DWtoBItheorem} with additional condition $P(a,b|x,y)=P(\neg a,\neg b|x,y)$ assuming that the strategy P is used are shown in fig.~\ref{fig:T3_cert_P}, and in figs~\ref{fig:T3_NPA_delta} and \ref{fig:T3_cert} for the mixed strategy.

\subsubsection{BC3 as a dimension witness}
\index{Bell operator!Braunstein-Caves}

We will now consider a dimension witness constructed with the above method out of the BC3 Bell operator. First, we construct a symmetric dimension witness with six settings on the side of Alice given by the following operator on $\mathbb{P}(\{0,1\}|\{0,1\} \times \{1,2,3\}, \{1,2,3\})$:
\be
	\label{eq:BC3DWfull}
	\ba
		& P(0|(0,1),1) - P(0|(1,1),1) + P(0|(0,1),2) - P(0|(1,1),2) \\
		& + P(0|(0,2),2) - P(0|(1,2),2) + P(0|(0,2),3) - P(0|(1,2),3) \\
		& + P(0|(0,3),3) - P(0|(1,3),3)	- P(0|(0,3),1) + P(0|(1,3),1).
	\ea
\ee
Using the method of reduction, we may transform this symmetric dimension witness with six preparation states to a reduced one with only three states. Let us take
\be
	\nonumber
	\ba
		& \phi((a,x)) \equiv (\neg a,x), \\
		& \chi \equiv \{(0,x): x \in \{1,2,3\}\}  \subseteq \{0,1\} \times \{1,2,3\}.
	\ea
\ee
Then we obtain the following dimension witness:
\be
	\nonumber
	\ba
		2 ( & P(0|(0,1),1) + P(0|(0,1),2) + P(0|(0,2),2) + \\
		& + P(0|(0,2),3) + P(0|(0,3),3) - P(0|(0,3),1) ) - 4.
	\ea
\ee
If we identify settings $(0,x)$ with $x$, we get the reduced dimension witness given
\be
	\label{eq:BC3DWreduced}
	\ba
		2 ( & P(0|1,1) + P(0|1,2) + P(0|2,2) + \\
		& + P(0|2,3) + P(0|3,3) - P(0|3,1) ) - 4.
	\ea
\ee

We now do the opposite way, and starting with the dimension witness given by~\eqref{eq:BC3DWfull} we use the theorem~\ref{ZeroSumTrm} to get the following Bell operator on $\mathbb{P}(\{0,1\},\{0,1\}|\{1,\dots,6\},\{1,2,3\})$ for the calculation of the lower bound on min-entropy generated by the protocol using the witness:
\be
	\label{BC3BIfull}
	\ba
		& \frac{1}{2} \left( C(1, 1) + C(1, 2) + C(2, 2) + C(2, 3) + C(3, 3) - C(3, 1) \right. \\
		&\left. + C(4, 1) + C(4, 2) + C(5, 2) + C(5, 3) + C(6, 3) - C(6, 1) \right),
	\ea
\ee
where we have changed the labels of the settings of Alice in the following way: $(0, 1) \equiv 1$, $(0, 2) \equiv 2$, $(0, 3) \equiv 3$, $(1, 1) \equiv 4$, $(1, 2) \equiv 5$, and $(1, 3) \equiv 6$.

If we start with the reduced dimension witness from~\eqref{eq:BC3DWfull} and assume $P(a,b|x,y)=P(\neg a, \neg b|x,y)$, we get that the lower-bounding Bell operator is exactly the initial BC3.

The min-entropies certified with BC3 for different additional conditions are plotted in fig.~\ref{fig:BC3_NPA_P}, \textit{cf.}~fig.~\ref{fig:BC357} in sec.~\ref{sec:expansion}. Lower-bounds on the min-entropy certified in the SDI protocol, calculated using the NPA and the theorem \ref{DWtoBItheorem} with additional condition $P(a,b|x,y)=P(\neg a,\neg b|x,y)$ are plotted for strategy P in fig.~\ref{fig:BC3_cert_P} and for mixed strategy in figs~\ref{fig:BC3_NPA_delta} and \ref{fig:BC3_cert}.

\subsubsection{Modified CHSH as a dimension witness}

Taking into account that the DI-QRNG protocol with modified CHSH, Eq.\eqref{modCHSH}, is very robust in certifying the randomness (see~sec.~\ref{sec:BIprotocols}), the dimension witness with randomness lower-bounded by it may also be expected to be robust. Lower-bounds for the discussed DI and SDI protocols are plotted in figs~\ref{fig:modCHSH_cert_P}, \ref{fig:modCHSH_NPA_P}, \ref{fig:modCHSH_NPA_delta}, and \ref{fig:modCHSH_cert} in sec.~\ref{sec:robustSdi}.

Choosing $P(a|x)=\frac{1}{2}$ and applying the heuristic method from sec.~\ref{sec:BItoDW} to the modified CHSH operator, we get the following symmetric dimension witness
\be
	\label{modCHSHDWfull}
	\omega(1, 2) + \omega(1, 3) + \omega(2, 1) + \omega(2, 2) - \omega(2, 3).
\ee
The reduced dimension witness created with this operator is
\be
	\label{modCHSHDW}
	D(1, 2) + D(1, 3)  + D(2, 1) + D(2, 2) - D(2, 3),
\ee
and acts on $\mathbb{P}(\{0,1\}|\{1,2\},\{1,2,3\})$.

The min-entropy of the protocol using the dimension witness defined in~\eqref{modCHSHDWfull} is lower-bounded by the following Bell operator
\be
	\label{modCHSHBIfull}
	\ba
		& \frac{1}{2} \left( C(1, 2) + C(1, 3) + C(2, 1) + C(2, 2) - C(2, 3) \right. \\
		& \left. + C(3, 2) + C(3, 3) + C(4, 1) + C(4, 2) - C(4, 3) \right)
	\ea
\ee
defined on probability distributions
\be
	\nonumber
	 \mathbb{P}(\{0,1\},\{0,1\}|\{1,2,3,4\},\{1,2,3\}).
\ee

\subsubsection{CGLMP dimension witness}
\label{sec:CGLMPdw}\index{Bell operator!CGLMP}

\begin{figure}[!htbp]
	\includegraphics[width=0.88\textwidth]{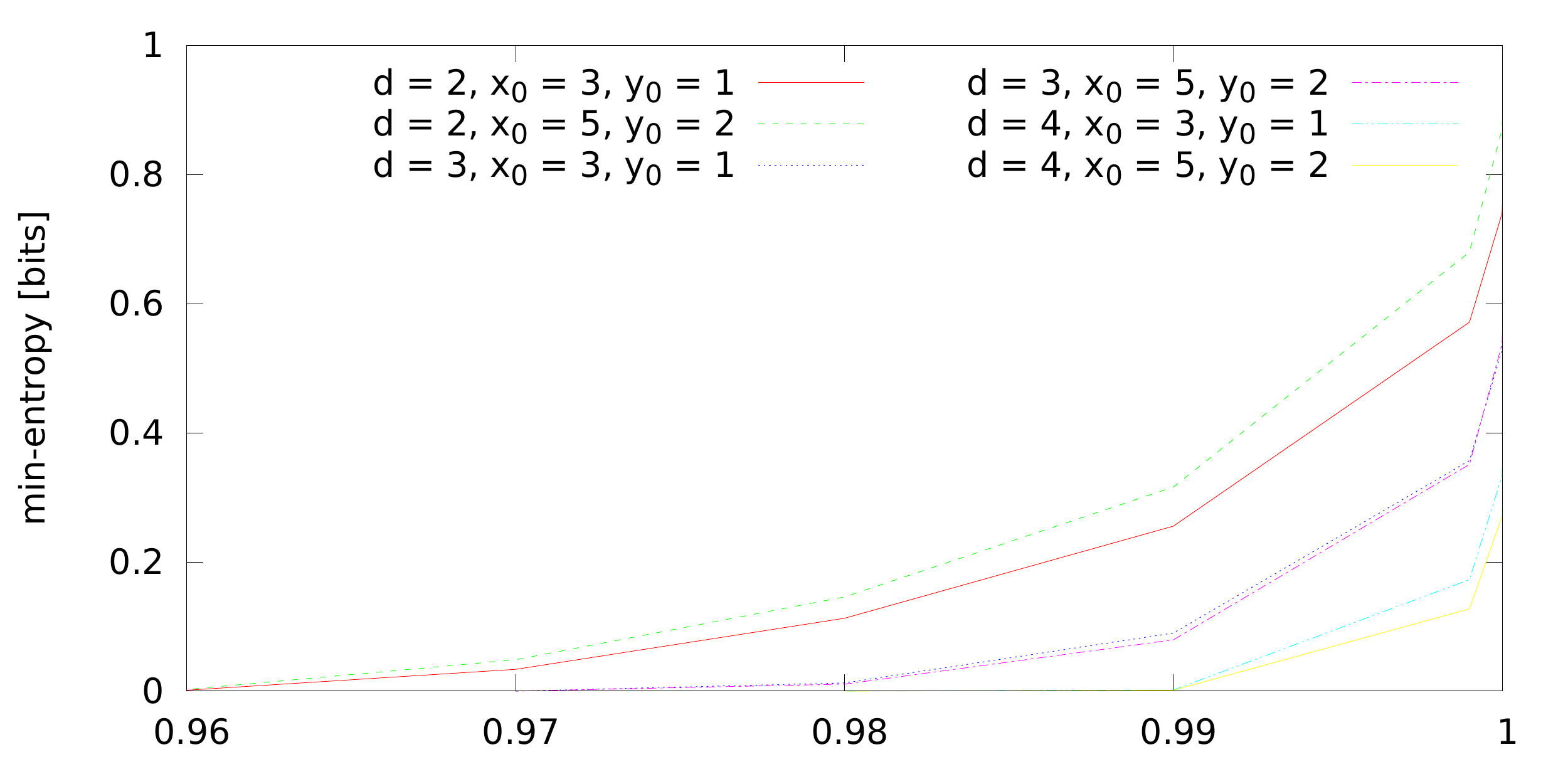}
	\caption[Lower-bounds on the certified randomness for the CGLMP protocol.]{\textbf{Lower-bounds on the certified randomness for the CGLMP protocol}, namely using the dimension witness defined in~\eqref{CGLMPdw} for different values of the dimension $d$ as a function of noise parameter $p$.\label{fig:CGLMP}}
\end{figure}

We will now use the heuristic method mentioned above in order to convert the CGLMP Bell operator defined in~\eqref{eq:CGLMP} in order to obtain the following CGLMP dimension witness:
\be
	\label{CGLMPdw}
	\ba
		&  P(0|1,1)-P(2|1,1)+P(0|1,2)-P(2|1,2)-P(0|2,1)+P(1|2,1) \\
		& -P(0|2,2)+P(1|2,2)-P(1|3,1)+P(2|3,1)-P(1|3,2)+P(2|3,2) \\
		& -P(0|4,1)+P(1|4,1)+P(0|4,2)-P(2|4,2)-P(1|5,1)+P(2|5,1) \\
		& -P(0|5,2)+P(1|5,2)+P(0|6,1)-P(2|6,1)-P(1|6,2)+P(2|6,2).
	\ea
\ee
on
\be
	\nonumber
	\mathbb{P}(\{0,1,2\}|\{1,\dots,6\},\{1,2\}).
\ee
Applying the method from sec.~\ref{sec:DWtoBI}, we get the following Bell operator
\be
	\nonumber
	\ba
		&  P(0,0|1,1)-P(0,2|1,1)+P(0,0|1,2)-P(0,2|1,2)-P(0,0|2,1) \\
		& +P(0,1|2,1)-P(0,0|2,2)+P(0,1|2,2)-P(0,1|3,1)+P(0,2|3,1) \\
		& -P(0,1|3,2)+P(0,2|3,2)-P(0,0|4,1)+P(0,1|4,1)+P(0,0|4,2) \\
		& -P(0,2|4,2)-P(0,1|5,1)+P(0,2|5,1)-P(0,0|5,2)+P(0,1|5,2) \\
		& +P(0,0|6,1)-P(0,2|6,1)-P(0,1|6,2)+P(0,2|6,2)
	\ea
\ee
on $\mathbb{P}(\{0,1\},\{0,1,2\}|\{1,\dots,6\},\{1,2\})$. This expression can be analyzed with the NPA. The certified randomness for CGLMP and different values of $d$ are shown in fig.~\ref{fig:CGLMP}.

\subsection{Investigation of semi-device-independent protocols}
\label{sec:robustSdi}

Now we will use the method of SDP in order to investigate the properties of the SDI-QRNG protocols considered above.

\subsubsection{Tightness of bounds}
\label{sec:PStrategyLowerUpper}

We start with the illustration of how the lower bounds are close to the actual values.

\begin{figure}[!htbp]
	\includegraphics[width=0.88\textwidth]{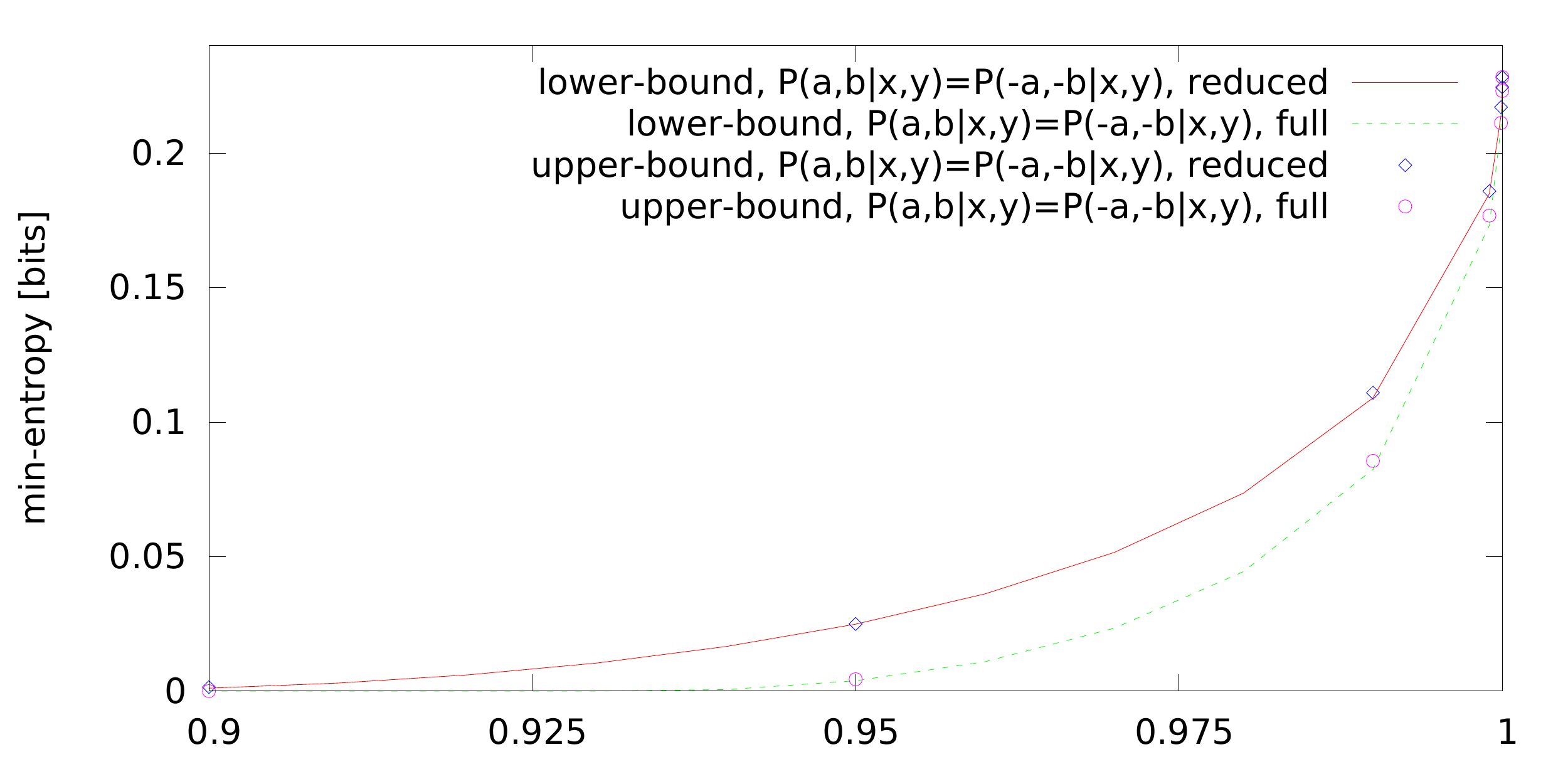}
	\caption[The randomness of T2 SDI-QRNG protocol.]{Numerical lower-bounds via SDP and upper-bounds on the randomness certified for the strategy P with the reduced and full dimension witness T2, see~\eqref{T2DWfull} and~\eqref{T2DW}.\label{fig:T2_cert_P}}
\end{figure}

\begin{figure}[!htbp]
	\includegraphics[width=0.88\textwidth]{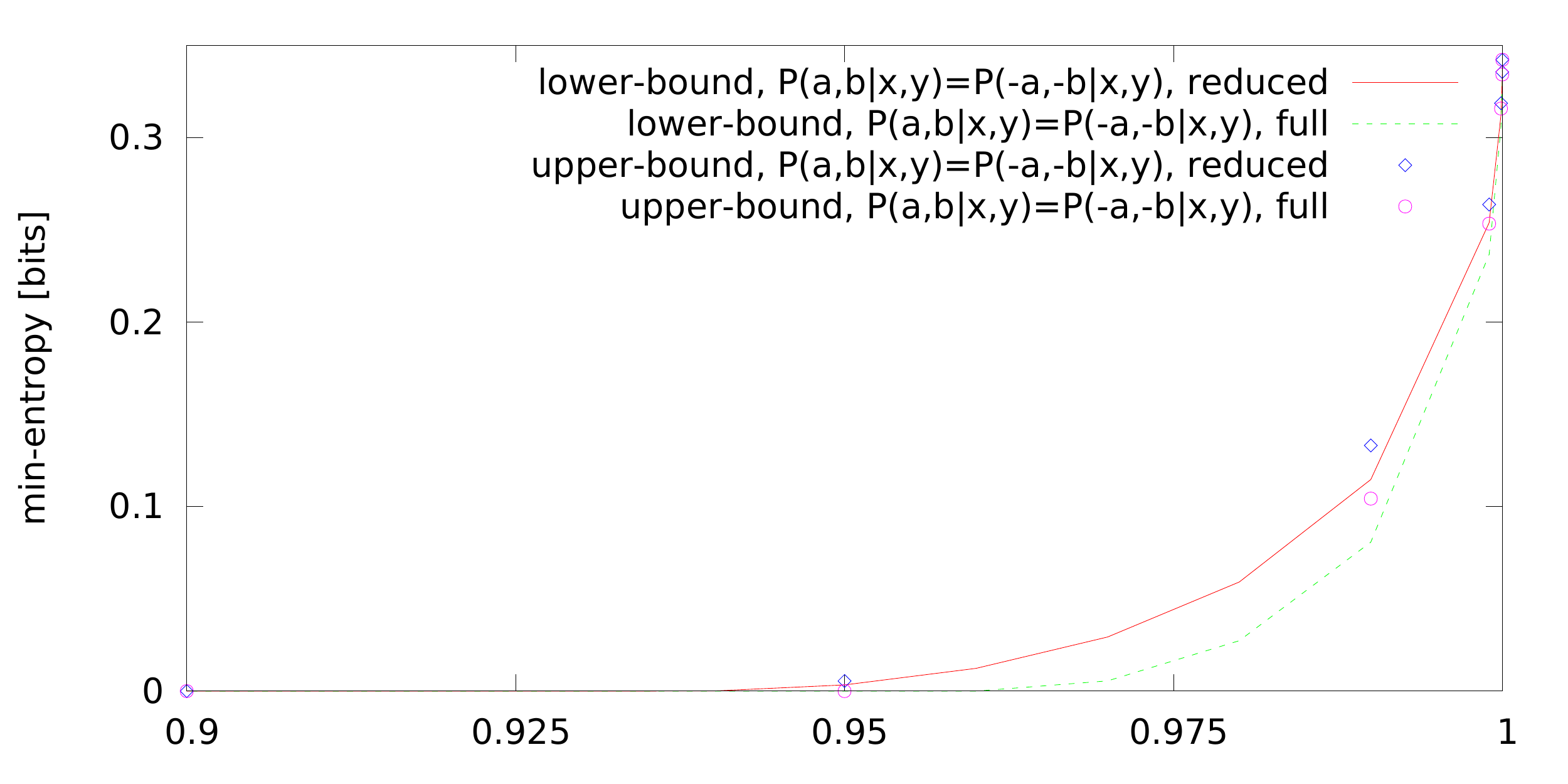}
	\caption[The randomness of T3 SDI-QRNG protocol.]{Numerical lower-bounds via SDP and upper-bounds on the randomness certified for the strategy P with the reduced and full dimension witness T3, see~\eqref{T3DWfull} and~\eqref{T3DW}.\label{fig:T3_cert_P}}
\end{figure}

\begin{figure}[!htbp]
	\includegraphics[width=0.88\textwidth]{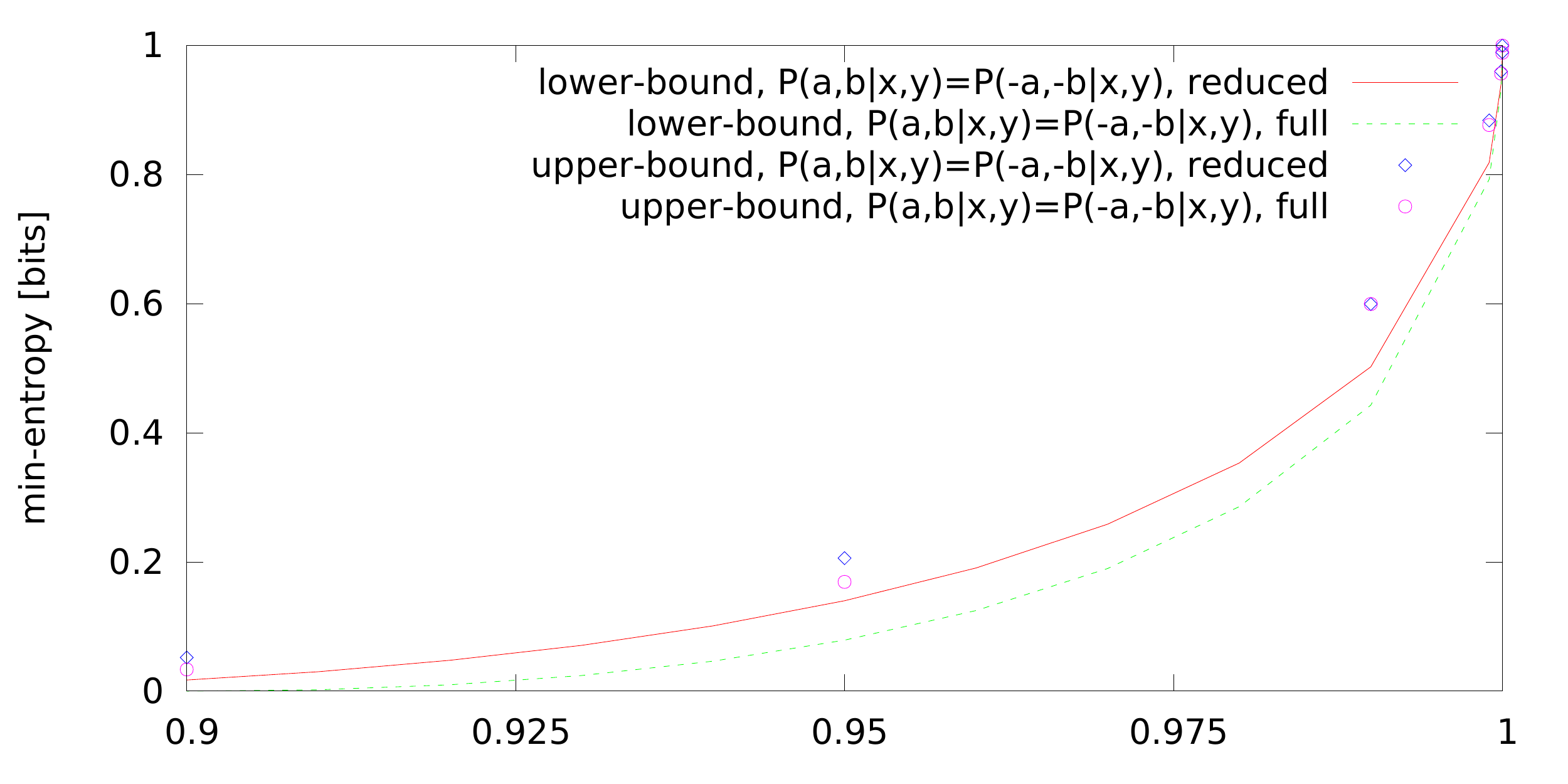}
	\caption[The randomness of BC3 SDI-QRNG protocol.]{Numerical lower-bounds via SDP and upper-bounds on the randomness certified for the strategy P with the reduced and full dimension witness BC3.\label{fig:BC3_cert_P}}
\end{figure}

\begin{figure}[!htbp]
	\includegraphics[width=0.88\textwidth]{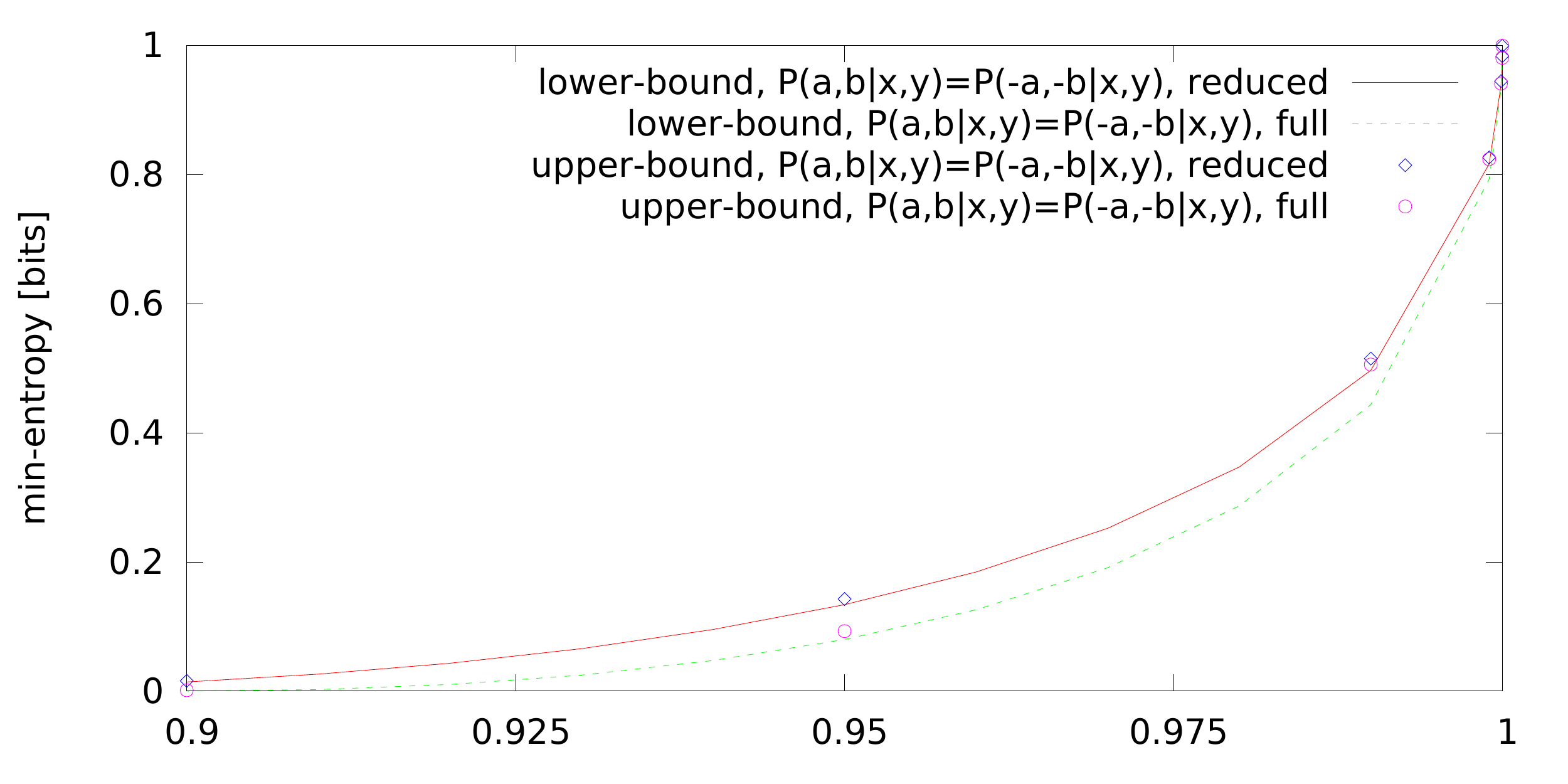}
	\caption[The randomness of modified CHSH SDI-QRNG protocol.]{Numerical lower-bounds via SDP and upper-bounds on the randomness certified for the strategy P with the reduced and full dimension witness obtained from the modified CHSH Bell operator, see~\eqref{modCHSHDWfull} and~\eqref{modCHSHDW}.\label{fig:modCHSH_cert_P}}
\end{figure}

Figs~\ref{fig:T2_cert_P}, \ref{fig:T3_cert_P}, \ref{fig:BC3_cert_P}, and~\ref{fig:modCHSH_cert_P} show numerical lower bounds and upper bounds on the randomness certified for the strategy P. The lower bounds are obtained with SDP in Almost Quantum\index{distribution!Almost Quantum} level, and the upper bound by finding an explicit pure quantum states and rank-1 measurements parametrized by~\eqref{eq:Bloch} and optimized with \textsc{fminunc} subroutine in OCTAVE. We observe an advantage for the reduced versions.

\subsubsection{Relation between families of conditions}
\label{sec:PStrategyLower}

Now, we will consider lower bounds from the NPA method on min-entropy certified by different Bell operators for different levels of noise. The situation differs from the standard quantum randomness expansion considered in sec.~\ref{sec:expansion} in that we take into account additional constraints of the probability distributions, not only the value of a certain Bell operator. These considerations illustrate the relations summarized in lemma~\ref{probImply}. 

\begin{figure}[!htbp]
	\includegraphics[width=0.88\textwidth]{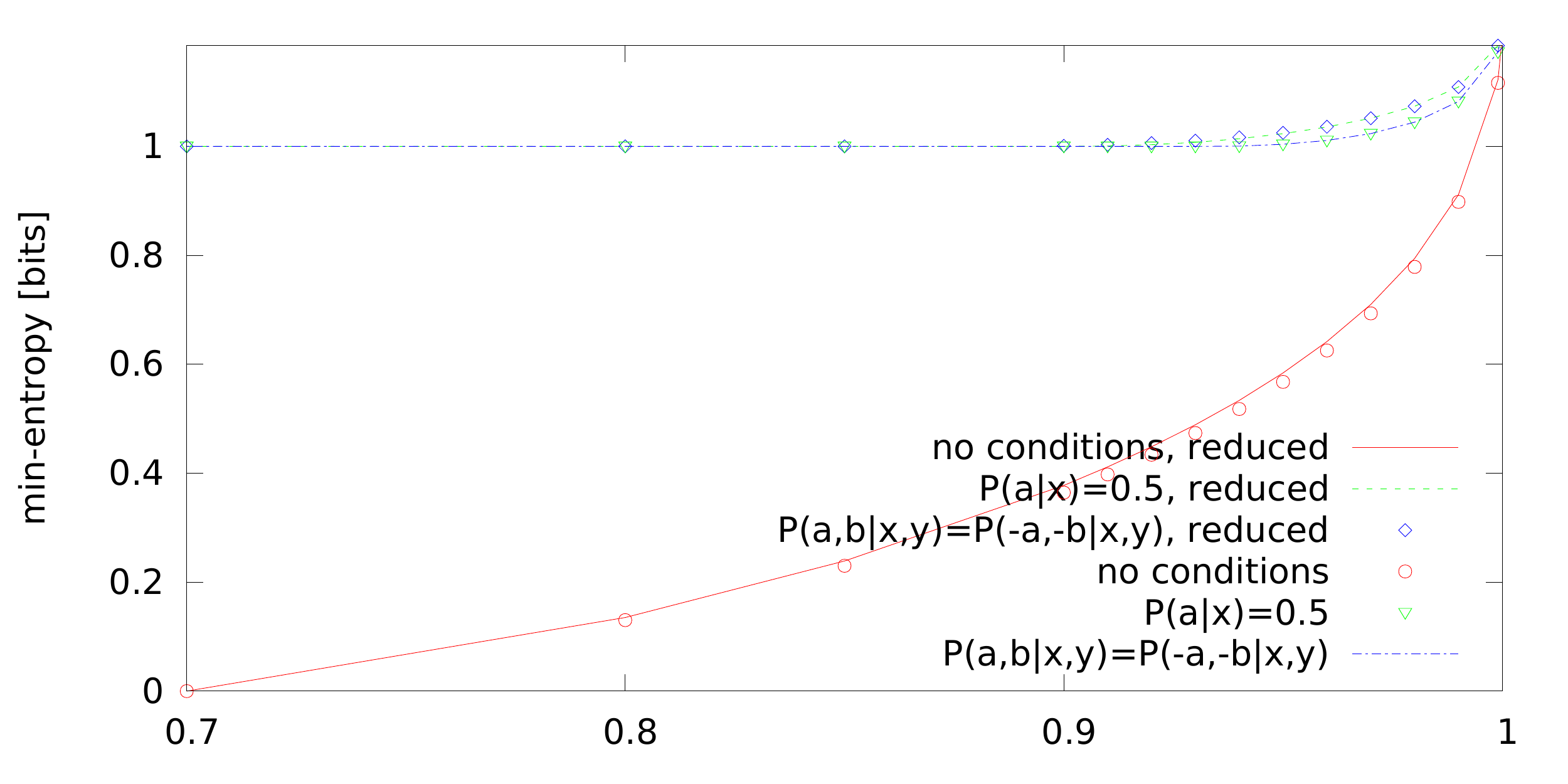}
	\caption[T2 DI-QRNG protocol with additional constraints.]{Lower bounds \textit{via} SDP on min-entropy certified by T2 DI-QRNG protocol for different levels of noise and different additional constraints. The full Bell operator is given by~\eqref{T2BIfull}, and the reduced Bell operator is given in~\eqref{T2BI}.\label{fig:T2_NPA_P}}
\end{figure}

\begin{figure}[!htbp]
	\includegraphics[width=0.88\textwidth]{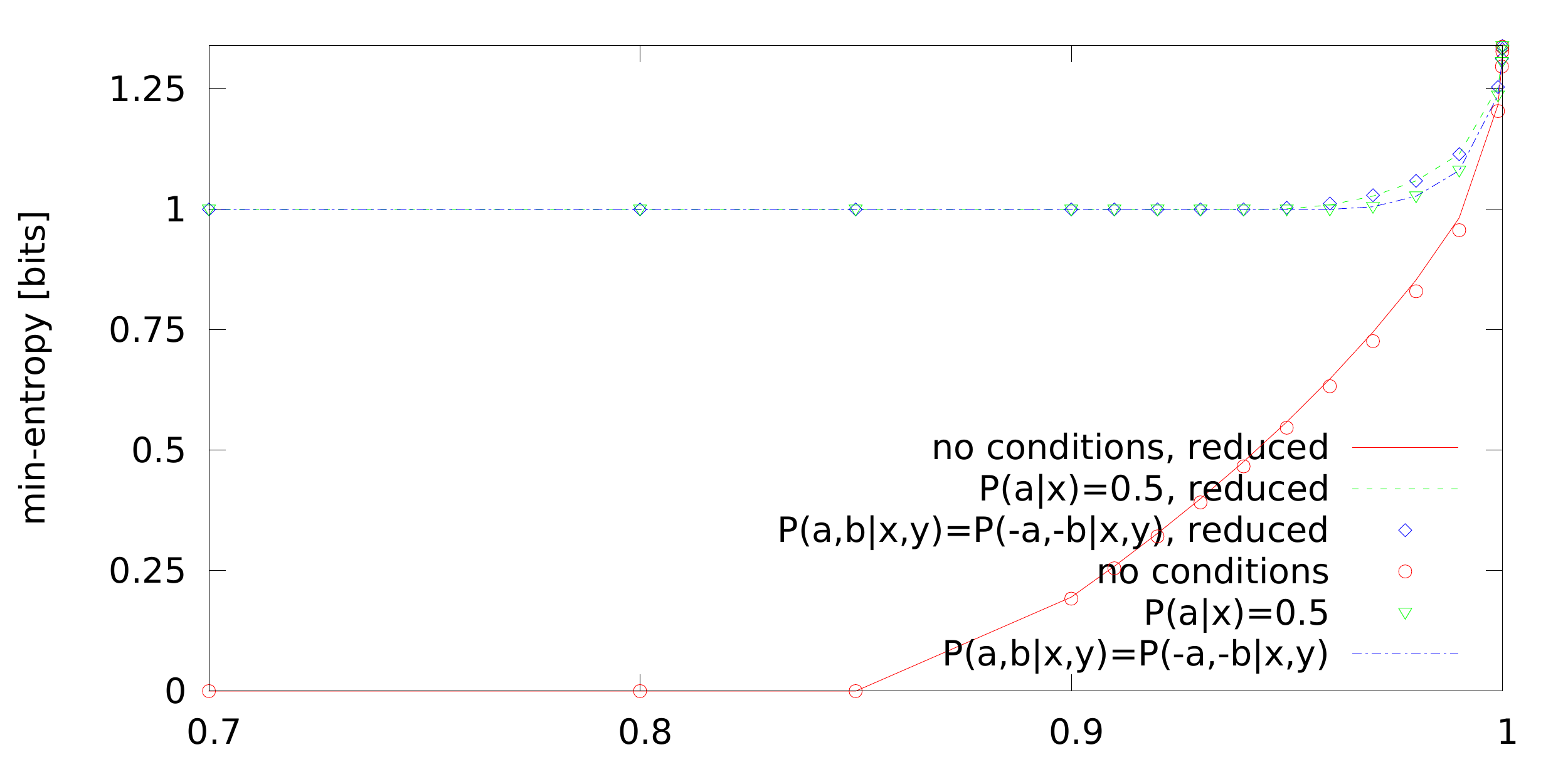}
	\caption[T3 DI-QRNG protocol with additional constraints.]{Lower bounds \textit{via} SDP on min-entropy certified by T3 DI-QRNG protocol for different levels of noise and different additional constraints. The full Bell operator is given by~\eqref{T3BIfull}, and the reduced Bell operator is given in~\eqref{T3BI}.\label{fig:T3_NPA_P}}
\end{figure}

\begin{figure}[!htbp]
	\includegraphics[width=0.88\textwidth]{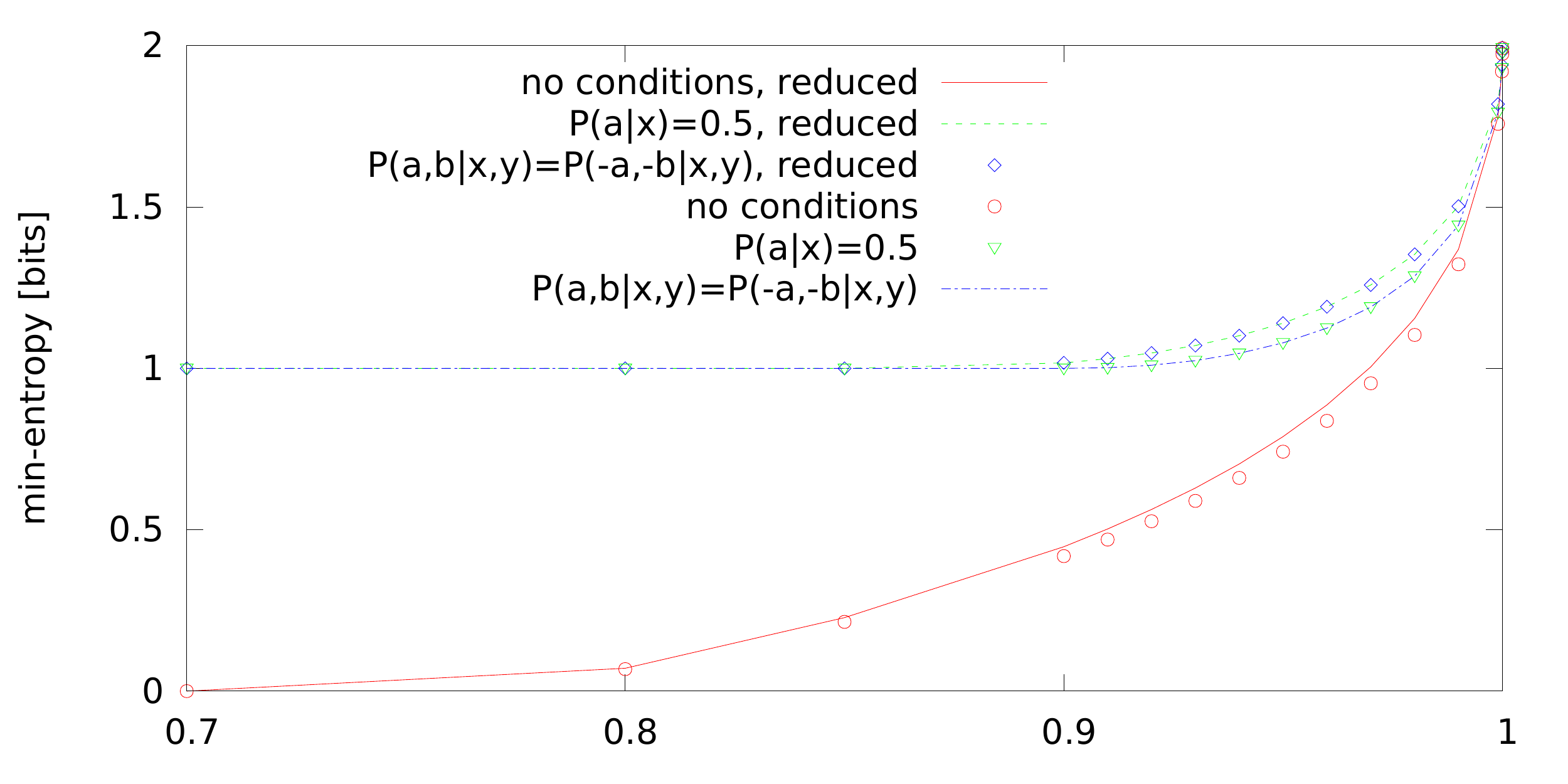}
	\caption[BC3 DI-QRNG protocol with additional constraints.]{Lower bounds \textit{via} SDP on min-entropy certified by BC3 DI-QRNG protocol for different levels of noise and different additional constraints. The full Bell operator is given by~\eqref{BC3BIfull}, and the reduced Bell operator is given in~\eqref{BC3BI}.\label{fig:BC3_NPA_P}}
\end{figure}

\begin{figure}[!htbp]
	\includegraphics[width=0.88\textwidth]{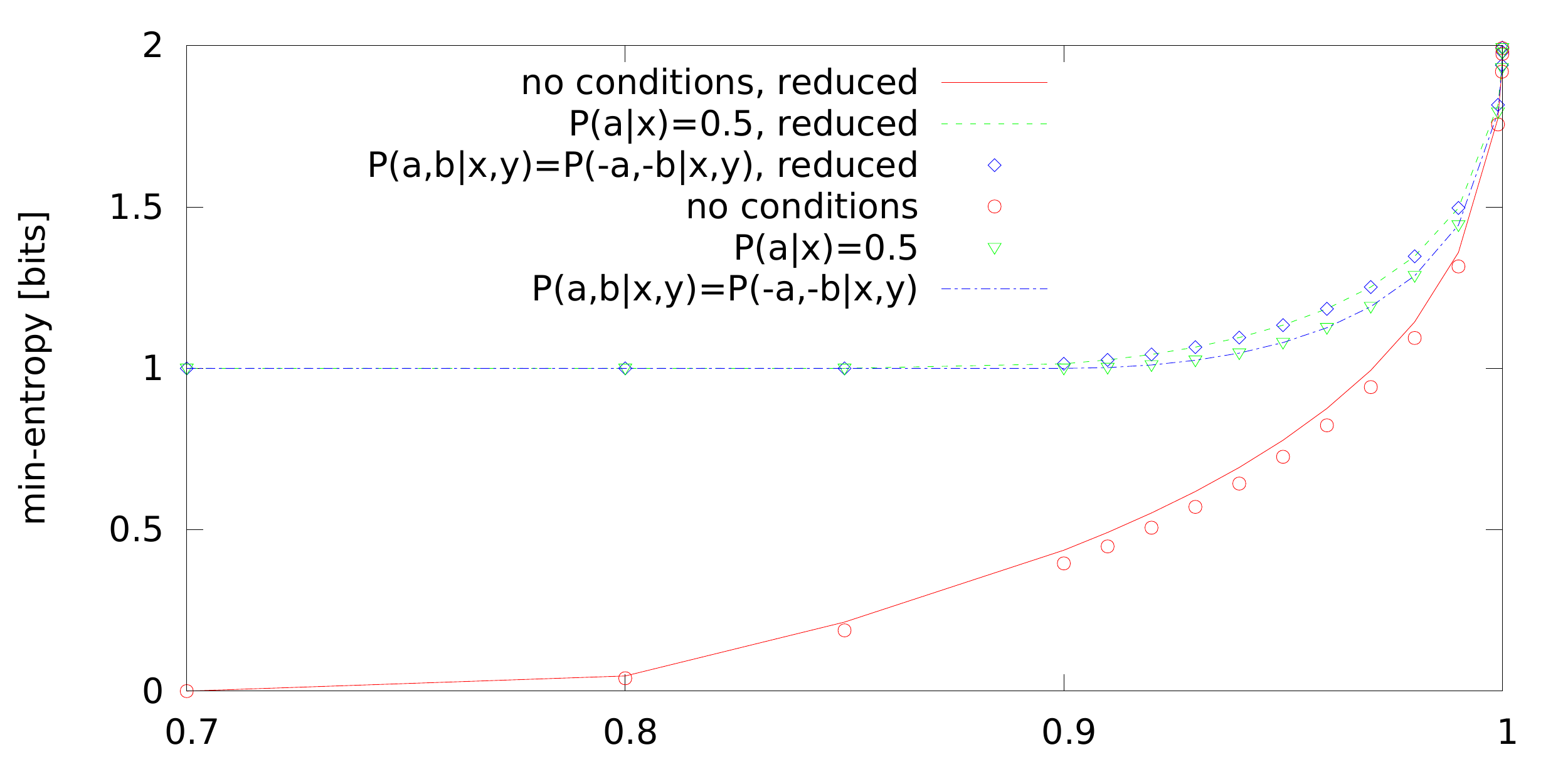}
	\caption[Modified CHSH DI-QRNG protocol with additional constraints.]{Lower bounds \textit{via} SDP on min-entropy certified by modified CHSH DI-QRNG protocol for different levels of noise and different additional constraints. The full Bell operator is given by~\eqref{modCHSHBIfull}, and the reduced Bell operator is given in~\eqref{modCHSH}.\label{fig:modCHSH_NPA_P}}
\end{figure}

In figs~\ref{fig:T2_NPA_P}, \ref{fig:T3_NPA_P}, \ref{fig:BC3_NPA_P}, and \ref{fig:modCHSH_NPA_P}, the \textit{reduced} Bell operators give lower bounds on the relevant reduced symmetric dimension witnesses for the strategy P.  We observe that using reduced dimension witnesses provides an advantage in the terms of certifiable randomness.  The reduced Bell operators are given by~\eqref{T2BI}, \eqref{T3BI}, \eqref{BC3BI}, and~\eqref{modCHSH}, while Bell operators that may be used for lower bounding the randomness of full symmetric dimension witnesses are given by~\eqref{T2BIfull}, \eqref{T3BIfull}, \eqref{BC3BIfull}, and~\eqref{modCHSHBIfull}. These refers to T2, Fig.~\ref{fig:T2_NPA_P}, T3, Fig.~\ref{fig:T3_NPA_P}, BC3, Fig.~\ref{fig:BC3_NPA_P}, and modified CHSH, Fig.~\ref{fig:modCHSH_NPA_P}, cases respectively.

\subsubsection{Mixed P and D strategy}
\label{sec:MixedStrategyLowerDelta}

\begin{figure}[!htbp]
	\includegraphics[width=0.88\textwidth]{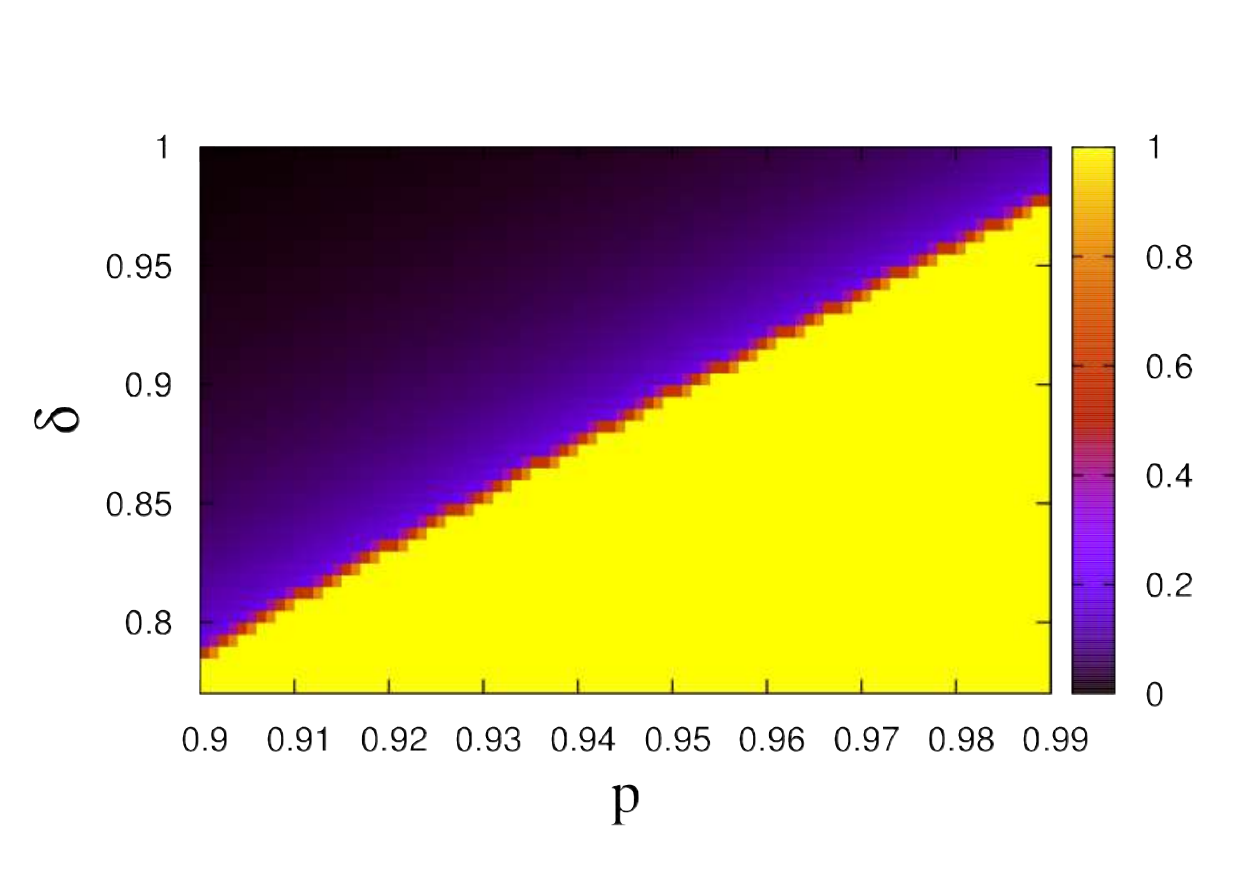}
	\caption[Mixed strategy for reduced T2 SDI-QRNG protocol.]{The efficiency of mixed strategies P and D for different values of $\delta$ in~\eqref{deltaAffine} for reduced T2 SDI-QRNG protocol.\label{fig:T2_NPA_delta}}
\end{figure}

\begin{figure}[!htbp]
	\includegraphics[width=0.88\textwidth]{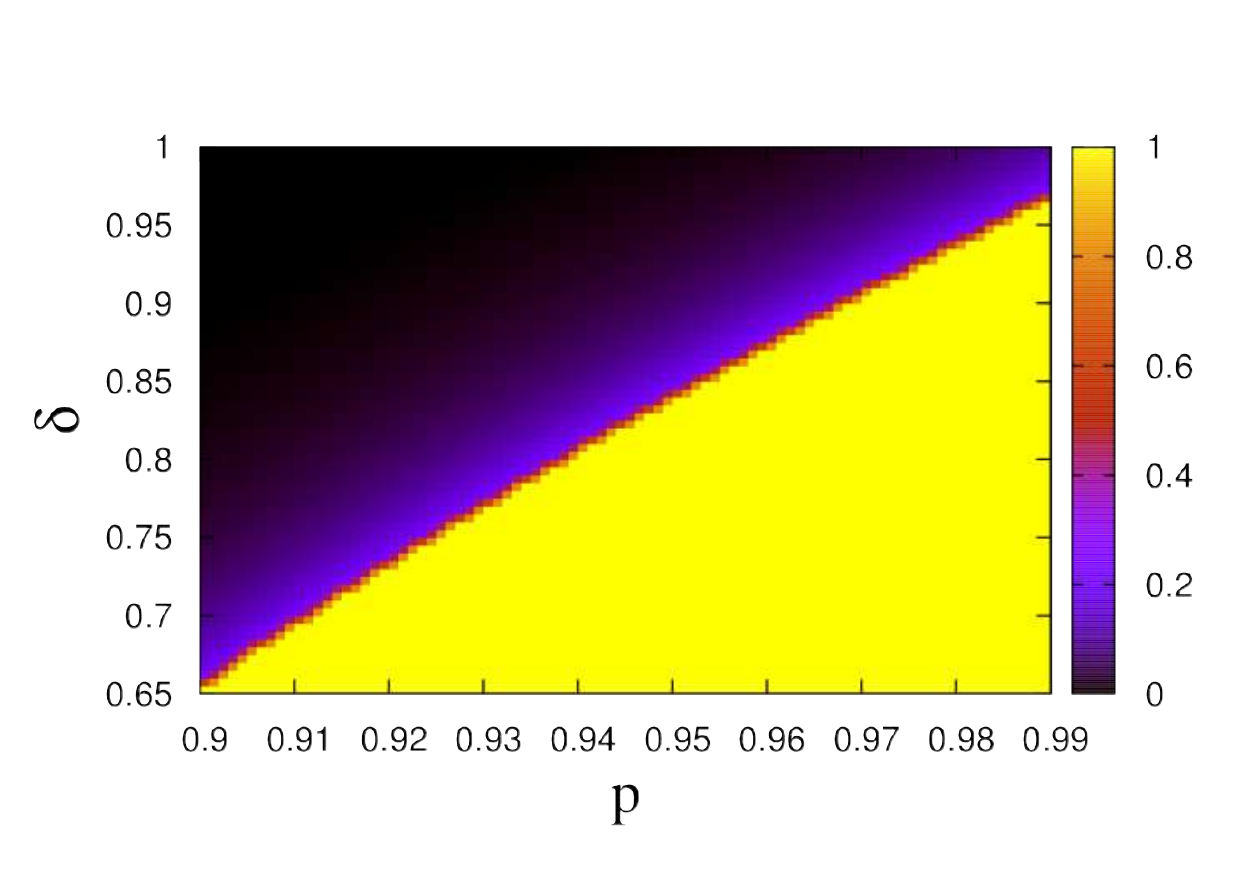}
	\caption[Mixed strategy for reduced T3 SDI-QRNG protocol.]{The efficiency of mixed strategies P and D for different values of $\delta$ in~\eqref{deltaAffine} for reduced T3 SDI-QRNG protocol.\label{fig:T3_NPA_delta}}
\end{figure}

\begin{figure}[!htbp]
	\includegraphics[width=0.88\textwidth]{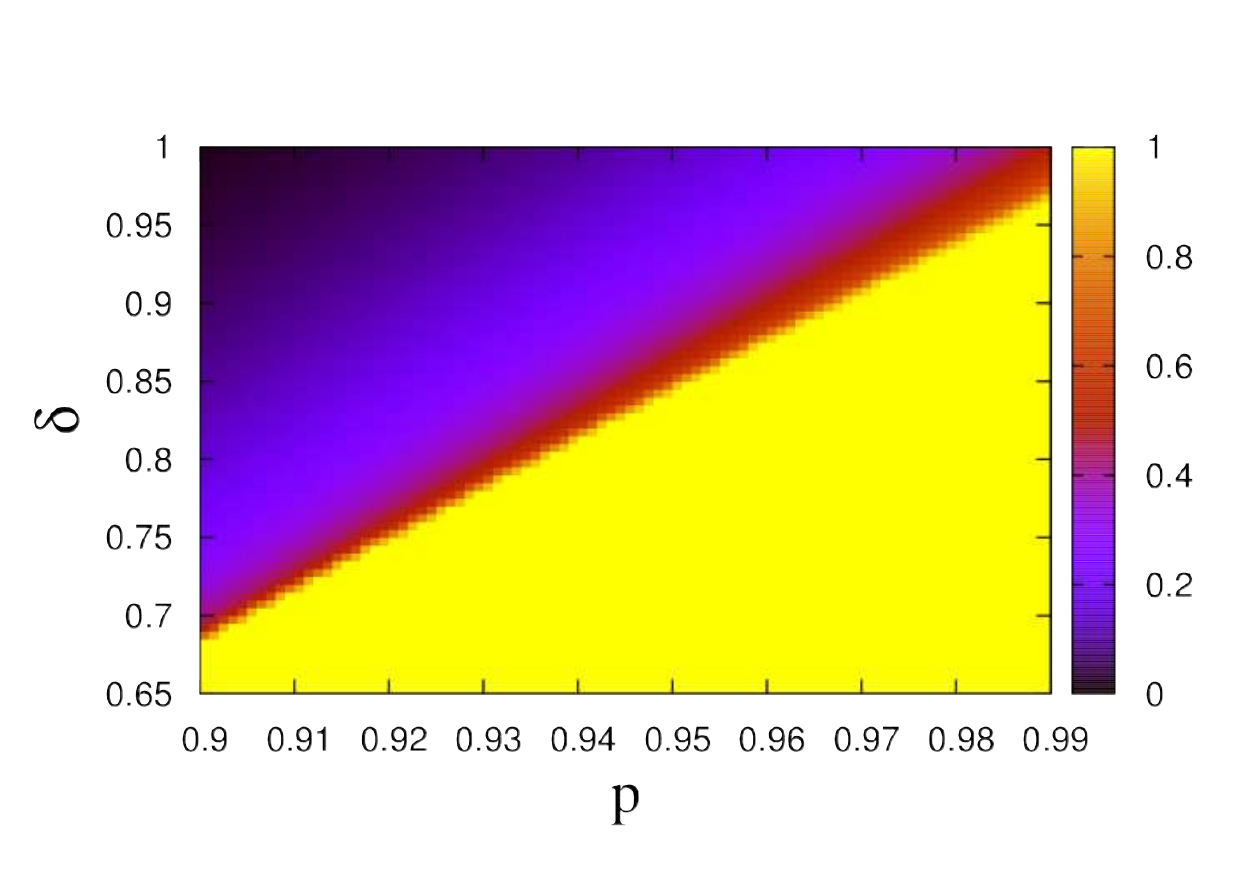}
	\caption[Mixed strategy for reduced BC3 SDI-QRNG protocol.]{The efficiency of mixed strategies P and D for different values of $\delta$ in~\eqref{deltaAffine} for reduced BC3 SDI-QRNG protocol.\label{fig:BC3_NPA_delta}}
\end{figure}

\begin{figure}[!htbp]
	\includegraphics[width=0.88\textwidth]{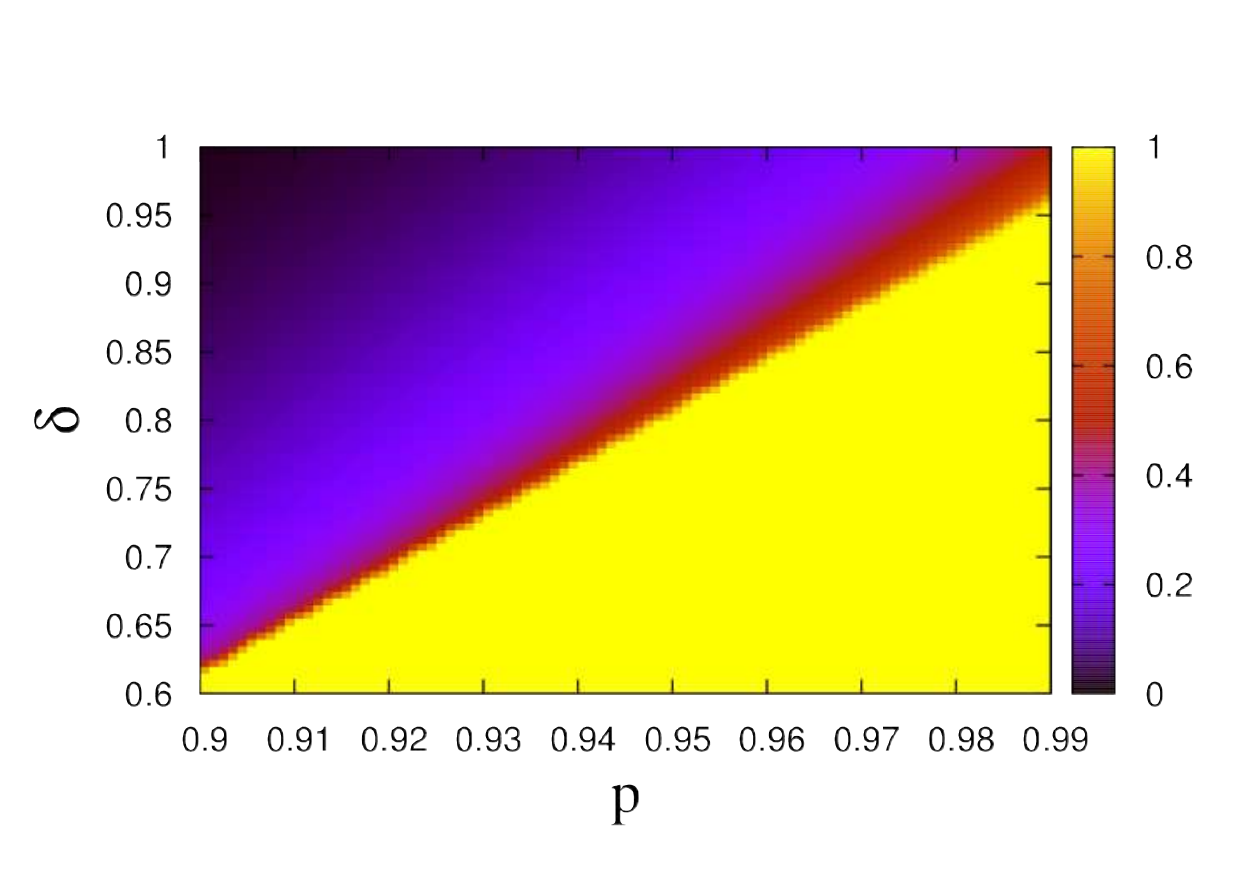}
	\caption[Mixed strategy for reduced modified CHSH SDI-QRNG protocol.]{The efficiency of mixed strategies P and D for different values of $\delta$ in~\eqref{deltaAffine} for reduced modified CHSH SDI-QRNG protocol.\label{fig:modCHSH_NPA_delta}}
\end{figure}

Here we show lower bounds obtained using the NPA on the certified randomness in SDI-QRNG with reduced dimension witnesses when the malevolent constructor of the device uses the mixed strategy with different values of the parameter $\delta$, see sec.~\ref{sec:binaryDW}. If a certain value of a dimension witness is impossible to be achieved with a given $\delta$, then, since the eavesdropper cannot mislead us this way, the value $1$ is put on the figures. The cases of T2, Fig.~\ref{fig:T2_NPA_delta}, T3, Fig.~\ref{fig:T3_NPA_delta}, BC3, Fig.~\ref{fig:BC3_NPA_delta}, and modified modCHSH, Fig.~\ref{fig:modCHSH_NPA_delta}, are shown on these plots.

\subsubsection{The robustness of semi-device-independent protocols}
\label{sec:MixedStrategyLower}

\begin{figure}[!htbp]
	\includegraphics[width=0.88\textwidth]{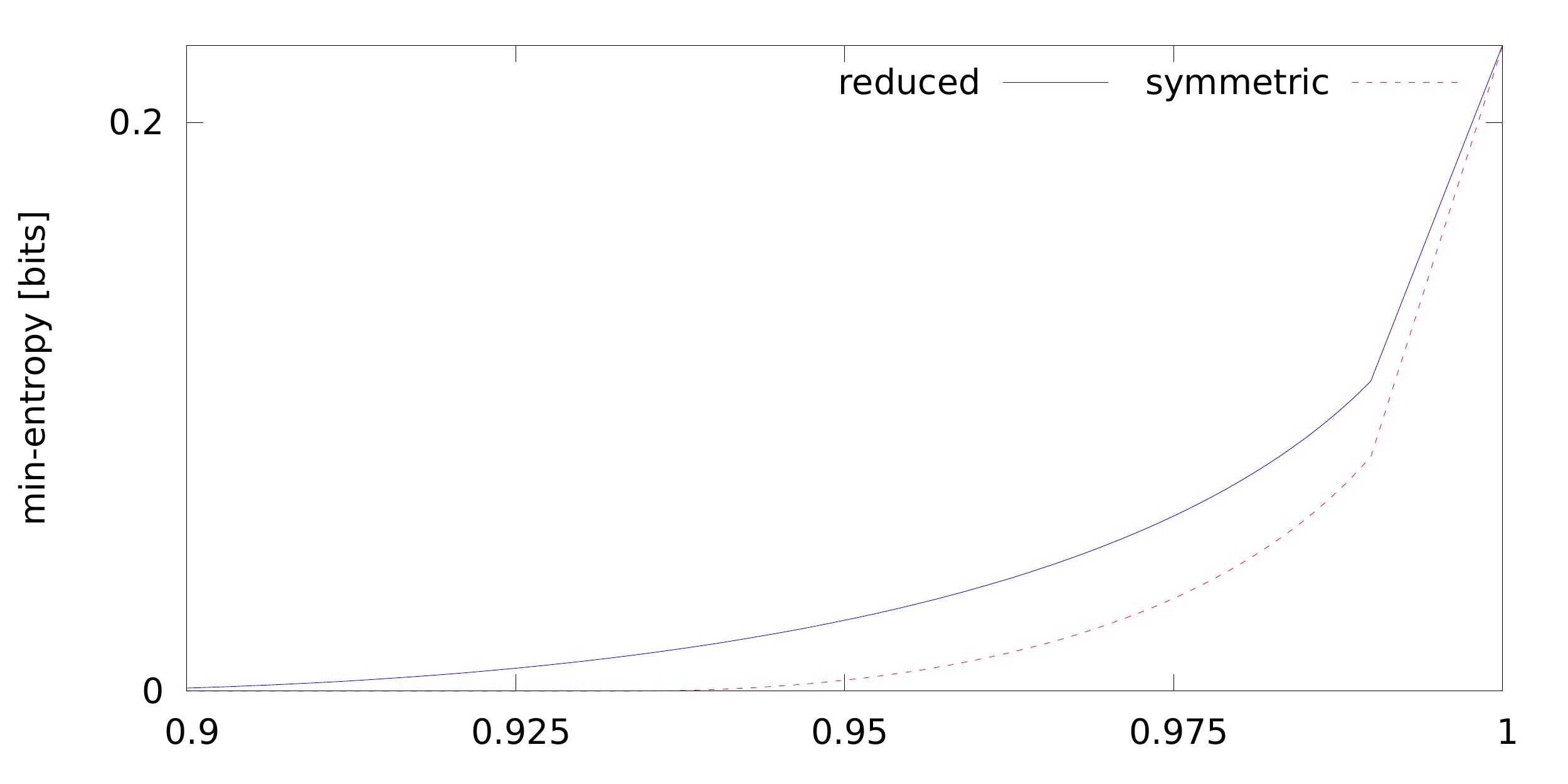}
	\caption[The robustness of T2 SDI-QRNG protocol.]{The robustness of the reduced T2 SDI-QRNG protocol with the worst case value of $\delta$ in~\eqref{deltaAffine}.\label{fig:T2_cert}}
\end{figure}

\begin{figure}[!htbp]
	\includegraphics[width=0.88\textwidth]{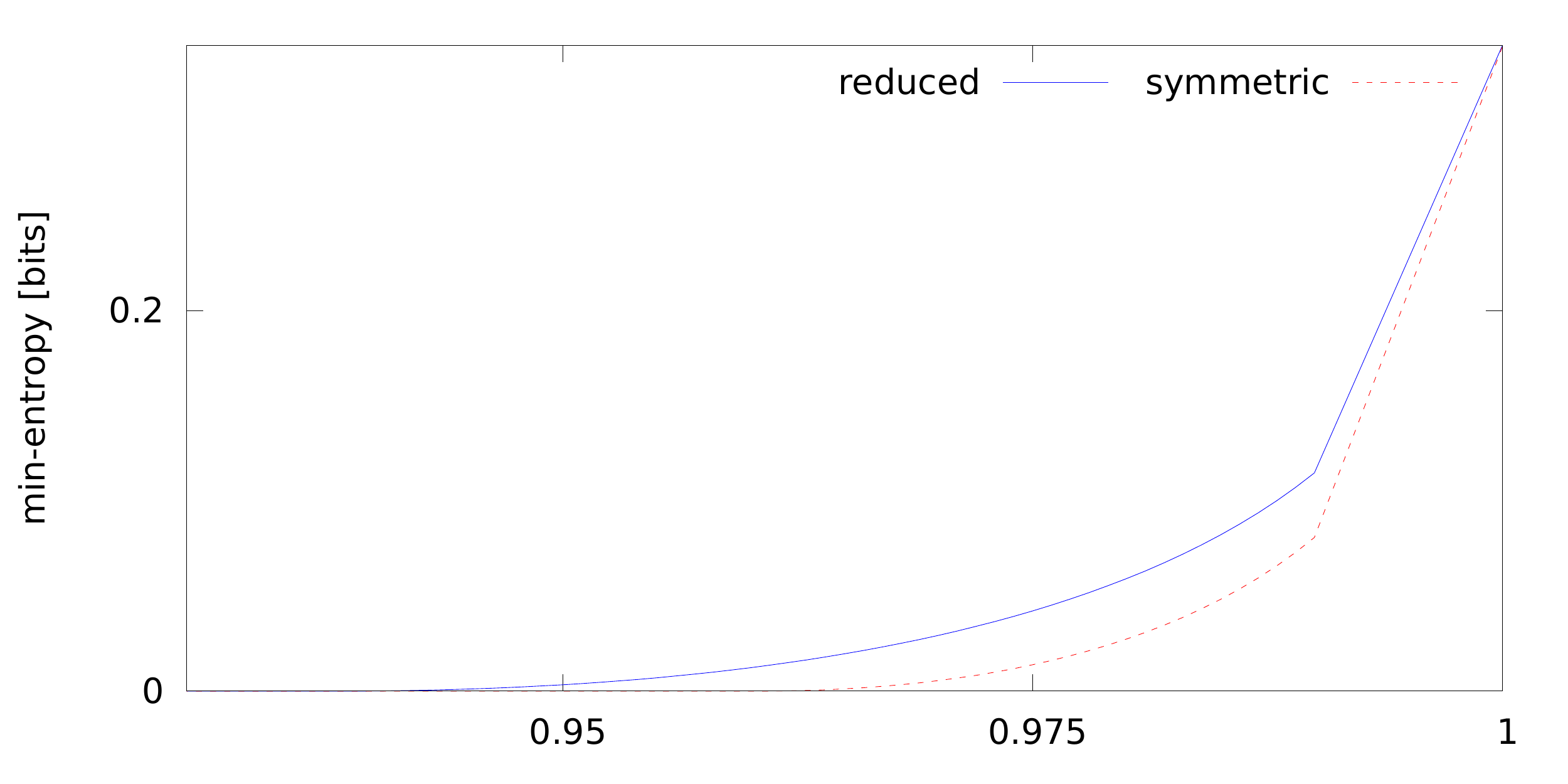}
	\caption[The robustness of T3 SDI-QRNG protocol.]{The robustness of the reduced T3 SDI-QRNG protocol with the worst case value of $\delta$ in~\eqref{deltaAffine}.\label{fig:T3_cert}}
\end{figure}

\begin{figure}[!htbp]
	\includegraphics[width=0.88\textwidth]{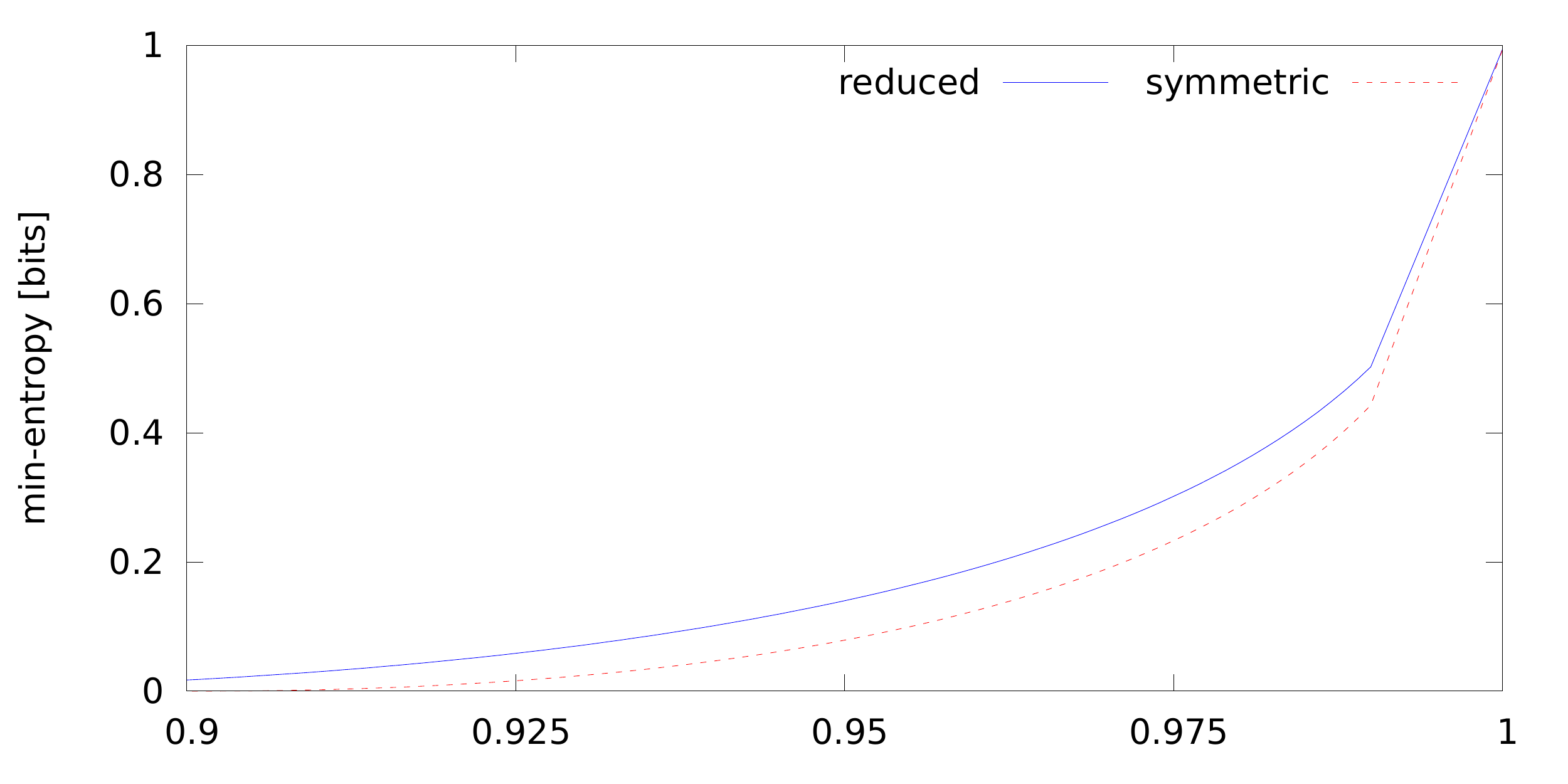}
	\caption[The robustness of BC3 SDI-QRNG protocol.]{The robustness of the reduced BC3 SDI-QRNG protocol with the worst case value of $\delta$ in~\eqref{deltaAffine}.\label{fig:BC3_cert}}
\end{figure}

\begin{figure}[!htbp]
	\includegraphics[width=0.88\textwidth]{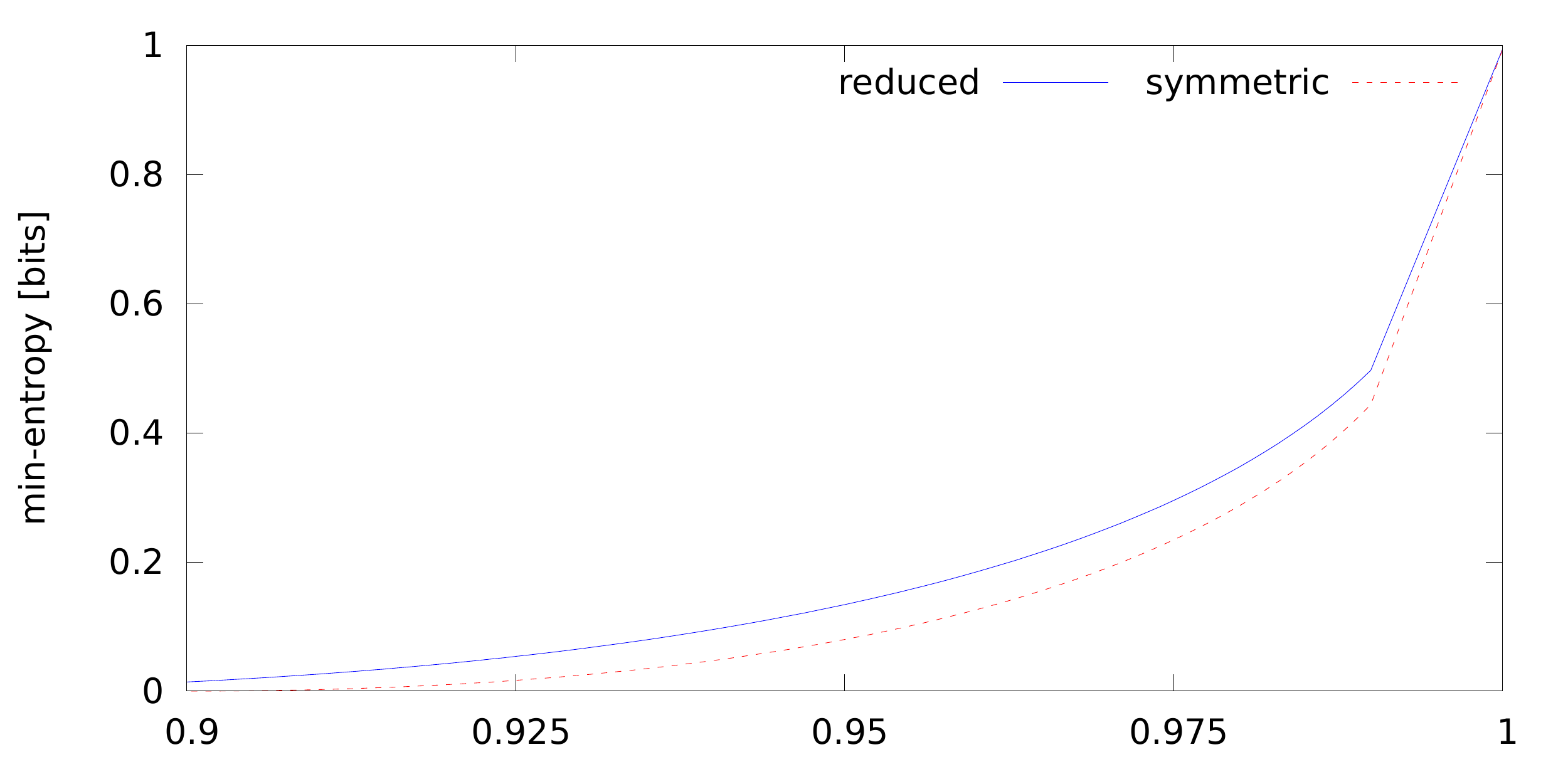}
	\caption[The robustness of modified CHSH SDI-QRNG protocol.]{The robustness of the reduced modified CHSH SDI-QRNG protocol with the worst case value of $\delta$ in~\eqref{deltaAffine}.\label{fig:modCHSH_cert}}
\end{figure}

Now we will give the lower bounds obtained using the NPA on the certified randomness for both reduced and symmetric dimension witnesses in the case when the malevolent vendor uses the mixed strategy with the optimal (from his point of view) value of the parameter $\delta$. This occurs to be $1$ in all cases, which supports a conjecture that there is no gain when a mixed strategy is used. The plots show the cases referring to T2, Fig.~\ref{fig:T2_cert}, T3, Fig.~\ref{fig:T3_cert}, BC3, Fig.~\ref{fig:BC3_cert}), and modified CHSH, Fig.~\ref{fig:modCHSH_cert}.

\section[Randomness amplification with three devices]{Amplification of arbitrarily weak Santha-Vazirani source with three devices}
\label{sec:amplification}\index{device-independent}\index{randomness!amplification}

This section deals with the quantum answer to the question about the possibility of amplification of arbitrary free randomness stated in sec.~\ref{sec:randAmplify}. We show that this task is possible if we are using quantum resources.

Our amplification protocol involves a Bell experiment with the Mermin game\index{Bell operator!Mermin}, see~sec.~\ref{sec:Mermin}. We introduced this protocol in the paper \cite{MP13}. We refer interested readers to our paper for a detailed analysis of physical aspects of the protocol.

We show that the Mermin game has a very important property that for any quality of the SV source of randomness the violation required to amplify it is possible to be observed using quantum resources. Thus, our protocol, in opposition to the protocol proposed by Gallego \textit{et al.} in \cite{GMTDAA12}, can find practical application, since the latter needs a maximal noiseless violation.

The majority of papers on the randomness amplification \cite{Amp1,GMTDAA12,Amp3,Amp4,Amp5,Amp6} assume only the no-signaling principle. Since our task here is not to study the foundations of the quantum theory, but to provide a protocol which can be used in practice, we assume the laws of quantum mechanics. We note that the amplification task differs from cryptographic scenario in the point that, after the amplification process, the generated bits do not have to be kept secret. In this case we are interested only in their indeterminacy, not privacy.

\subsection{Description of the scenario}

In opposition to the DI-QRNGs and SDI-QRNGs considered above, if we are to use $\epsilon$-free string of bits as a source of settings, we cannot assume that the distribution of inputs is unbiased. We discussed this issue in sec.~\ref{sec:biased} in the analysis of DI-QKD using the Hardy's paradox. We assume that the value of $\epsilon$ is known.

The proposed protocol of amplification of weak randomness considers two types of devices, namely the $\epsilon$-free Santha-Vazirani source of randomness, and quantum boxes shared among three honest parties. Since we are working in a DI scheme, we do not assume anything about internal construction of quantum boxes.

The task of randomness amplification is performed in competition with a malevolent vendor of those devices, Eve. She is able to produce a random variable $e$ which is to some extent correlated with those devices. We refer to $e$ as the \textbf{knowledge} of Eve. The task of amplification is now to generate a final bit $y$ that is $\epsilon'$-free with respect to Eve, or precisely
\be
	\nonumber
	\frac{1}{2}-\epsilon' \leq P(y|e) \leq \frac{1}{2} + \epsilon'.
\ee

The procedure runs in a sequence of iterations. With the SV source of randomness the parties are able to produce an arbitrary large sequence of bits, $\mathbf{s}$. Each party gets one bit as their input, and produces one bit of output. For iteration $i$ we denote by $x_i$, $y_i$ and $z_i$ the settings of each party, and by $a_i$, $b_i$ and $c_i$ their outcomes. Since in Mermin game, see sec.~\ref{sec:Mermin}, we have $x_i \oplus y_i \oplus z_i = 1$, we assume that only $x_i$ and $y_i$ are randomized, and $z_i$ is calculated as $x_i \oplus y_i \oplus 1$.

Without going into details, the crucial part is the calculation of randomness of local outputs of any of the three involved parties, say Alice. Below we show that this randomness is good enough to allow, after post-processing, the amplification of the initial randomness.

\subsection{Calculation of the randomness with semi-definite programming}

We will use SDP relaxation given by the NPA method. This time the scenario is more complicated, since it involves not two, but three parties. As mentioned before, the NPA can be extended to such a case. Nonetheless, finding lower bounds on the quality of randomness generated by playing a biased non-local game is non-trivial.

For any given distribution $p(x,y|e)$ the NPA can be used to efficiently find a lower bound on this randomness. The problem with our case is that this probability distribution is not only unknown to us, but it can also be different in each iteration. The only thing that we know about $p(x,y|e)$ is that both bits $x$ and $y$ come from an $\epsilon$-free source (with known $\epsilon$).

Let us write the success probability of the biased Mermin game in the following form
\be
	\label{eq:MerminBiased}
	\sum_{x,y} p(x,y|e) P(a \oplus b \oplus c = xy|x,y,x \oplus y \oplus 1,e).
\ee
The term $P(a \oplus b \oplus c = xy|x,y,x \oplus y \oplus 1,e)$ is defined directly by the description of the quantum box used in the protocol, where $P(a,b,c|x,y,z,e)$ is the probability of outcomes $a$, $b$ and $c$ when the settings are $x$, $y$ and $z$, and $e$ is the knowledge of Eve. We define
\be
	\nonumber
	P_A(a|x,y,z,e) \equiv \sum_{b,c} P(a,b,c|x,y,z,e)
\ee
as the marginal probability of Alice.

Let us consider the following set of constraints on the probabilities for some $\epsilon \in [0,\frac{1}{2})$ and $P_s \in [0,1]$:
\be
	\label{eq:cond}
	\ba
		\sum_{x,y} p(x,y|e) P(a \oplus b \oplus c = xy|x,y,x \oplus y \oplus 1,e) \geq P_s, & \\
		\frac{1}{2} - \epsilon \leq p(x|e) \leq \frac{1}{2} + \epsilon, & \\
		\frac{1}{2} - \epsilon \leq p(y|x,e) \leq \frac{1}{2} + \epsilon, & \\
		\mathbb{P}(\{0,1\},\{0,1\},\{0,1\}|\{0,1\},\{0,1\},\{0,1\},\{e\}) \text{ is quantum for all $e$}. &
	\ea
\ee
$P_s$ may be understood as a \textit{lower} bound on the success probability of the biased Mermin game, \eqref{eq:MerminBiased}, known (\textit{e.g.} from an experiment) to the user of the device. We want to find a function $P_{max}(\epsilon,P_s)$ being the solution of the following optimization problem in variables $P(a,b,c|x,y,z,e)$:
\begin{align}
	\label{problem-Pmax}
	\begin{split}
		\text{maximize } & \max_{x,y,i, \in \{0,1\}, e} P_A(i|x,y,x \oplus y \oplus 1,e) \\
		\text{subject to } &\null \text{the conditions from~\eqref{eq:cond} are fulfilled.}
	\end{split}
\end{align}
This problem cannot be formulated as an SDP problem using the NPA, since the conditions from~\eqref{eq:cond} are not linear in the variables. In order to overcome this difficulty let us consider what the success probability of the unbiased game, (\textit{i.e.} with $\epsilon = 0$) would be if the same states and measurements are used in it.

Let us denote the quantum probability distribution which solves the problem~\eqref{problem-Pmax} as
\be
	\nonumber
	\mathbb{P}_{opt}(\{0,1\},\{0,1\},\{0,1\}|\{0,1\},\{0,1\},\{0,1\},\{e\}).
\ee
The success probability, $P_s^{ub}$, of the unbiased game with quantum box defined by the distribution $\mathbb{P}_{opt}$ is given by the following expression
\be
	\nonumber
	P_s^{ub} \equiv \frac{1}{4} \sum_{x,y} P_{opt}(a \oplus b \oplus c = xy|x,y,x \oplus y \oplus 1,e).
\ee

Let us define
\be
	\nonumber
	P_m \equiv \min_{x,y} P_{opt}(a \oplus b \oplus c = xy|x,y,x \oplus y \oplus 1,e).
\ee
Obviously we have
\be
	\nonumber
	P_s^{ub}\geq P_m.
\ee
The success probability, $P_{success}$, of the biased Mermin game, \eqref{eq:MerminBiased}, for a \textit{given} value of $P_m$ is as large as possible when the probability $p(x,y|e)$ in front of $P_m$ is as small as possible. Since we know that this probability is generated by $\epsilon$-free Santha-Vazirani source, the smallest probability of a pair of outcomes is $\left(\frac{1}{2}-\epsilon\right)^2$. Moreover, $P_{opt}(a \oplus b \oplus c = xy|x,y,x \oplus y \oplus 1,e) \leq 1$ for all $x$ and $y$. Recall that $P_s$ is a lower bound on $P_{success}$. From these considerations we obtain
\be
	\nonumber
	\ba
	P_s \leq & P_{success} \leq (\frac{1}{2}+\epsilon)^2 + 2 (\frac{1}{2}+\epsilon) (\frac{1}{2}-\epsilon) + (\frac{1}{2}-\epsilon)^2 P_m = \\
	& 1-\left(\frac{1}{2}-\epsilon\right)^2+\left(\frac{1}{2}-\epsilon\right)^2 P_m \leq 1- \left(\frac{1}{2}-\epsilon\right)^2\left(1-P_s^{ub}\right)
	\ea
\ee
and, in consequence,
\be
	\label{est}
	P_s^{ub} \geq 1 - \frac{1 - P_s}{\left( \frac{1}{2} - \epsilon \right)^2}.
\ee

We will now find a way to relax the problem \eqref{problem-Pmax} to a form possible to be stated as an NPA problem. This will allow us to calculate an upper bound on $P_{max}(\epsilon,P_s)$. The above considerations show that from \eqref{eq:cond} it follows that there exists a probability distribution
\be
	\nonumber
	\mathbb{P}(\{0,1\},\{0,1\},\{0,1\}|\{0,1\},\{0,1\},\{0,1\},\{e\})
\ee
satisfying conditions:
\be
	\label{eq:PmaxEpsPs}
	\ba
		\frac{1}{4} \sum_{x,y} P(a \oplus b \oplus c = xy|x,y,x \oplus y \oplus 1,e) \geq 1 - \frac{1 - P_s}{\left( \frac{1}{2} - \epsilon \right)^2}, \\
		\mathbb{P}(\{0,1\},\{0,1\},\{0,1\}|\{0,1\},\{0,1\},\{0,1\},\{e\}) \text{ is quantum for all $e$}.
	\ea
\ee
Thus the conditions from~\eqref{eq:cond} imply \eqref{eq:PmaxEpsPs}, and thus the value of the solution on the following maximization problem in variables $P(a,b,c|x,y,z,e)$ gives an upper bound on the value of the solution of the problem~\eqref{problem-Pmax}:
\begin{align}
	\nonumber
	\begin{split}
		\text{maximize } & \max_{x,y,i, \in \{0,1\}, e} \max P_A(i|x,y,x \oplus y \oplus 1,e) \\
		\text{subject to } &\null \text{the conditions from~\eqref{eq:PmaxEpsPs} are fulfilled}.
	\end{split}
\end{align}
It is crucial for practical implementations that the value of $P_s$ can be experimentally estimated.

\begin{figure}[!htbp]
	\center
		\resizebox{8cm}{!}{\includegraphics{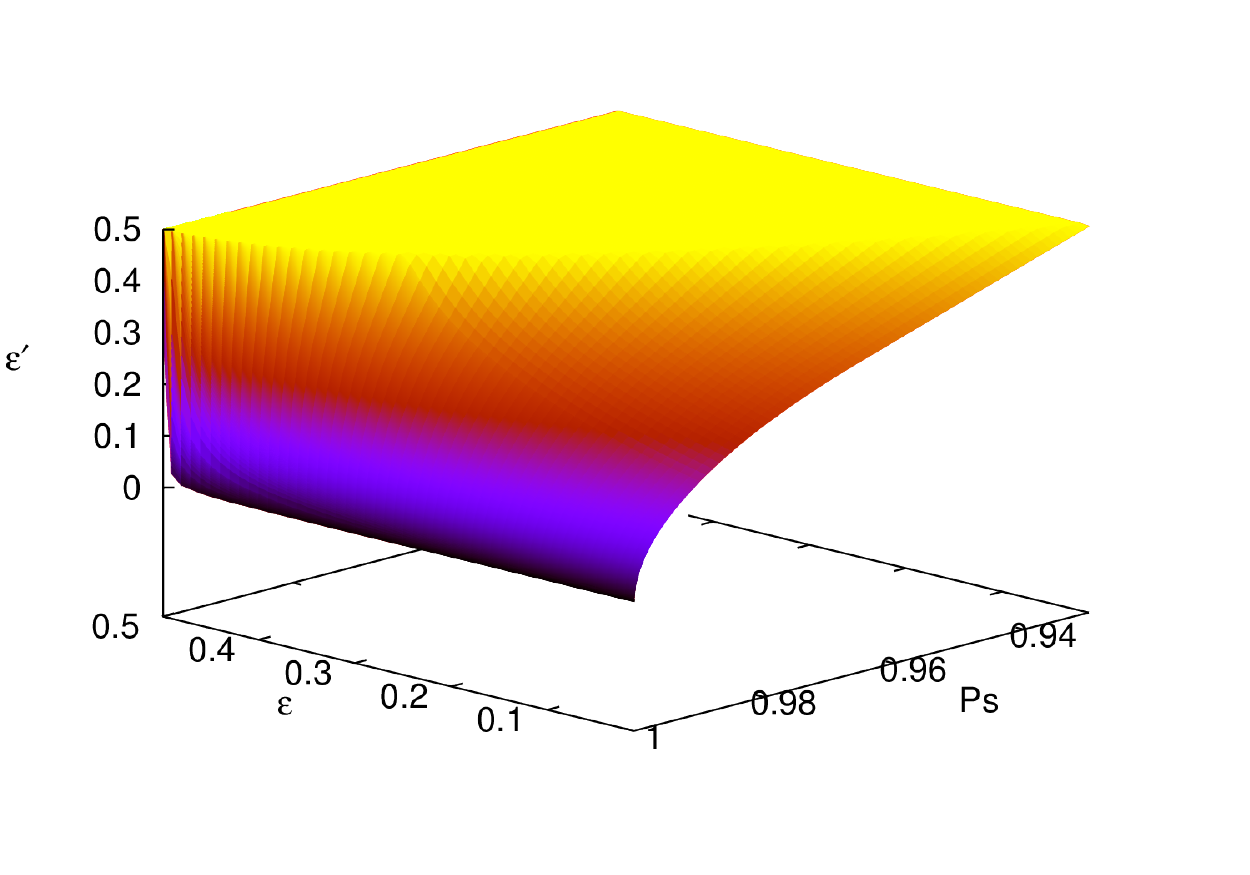}}
	\caption[Maximal bias of the measurement outcome in the Mermin game.]{\textbf{Maximal bias of the measurement outcome} as a function $g(\epsilon,P_s)$ of the weakness of randomness and success probability of winning in the Mermin game. This gives the critical values of the success probabilities required to take inputs from $\epsilon$-free source and get outcomes with bias less than $\epsilon'$.\label{fig:Pmax}}
\end{figure}

We calculated the upper bounds on $P_{max}(\epsilon,P_s)$ with the NPA with the hierarchy level
\be
	\nonumber
	\mathcal{Q}_{1+AB+AC+BC}(2,2,2|2,2,2).
\ee
From an upper bound on $P_{max}(\epsilon,P_s)$ it is straightforward to calculate an upper bound on the bias of the output of Alice, $\epsilon' = P_{max}(\epsilon,P_s) - \frac{1}{2}$. Let us denote this upper bound on the bias by $g(\epsilon,P_s)$. This function is shown in fig.~\ref{fig:Pmax}.

\begin{figure}[!htbp]
	\center
		\resizebox{8cm}{!}{\includegraphics{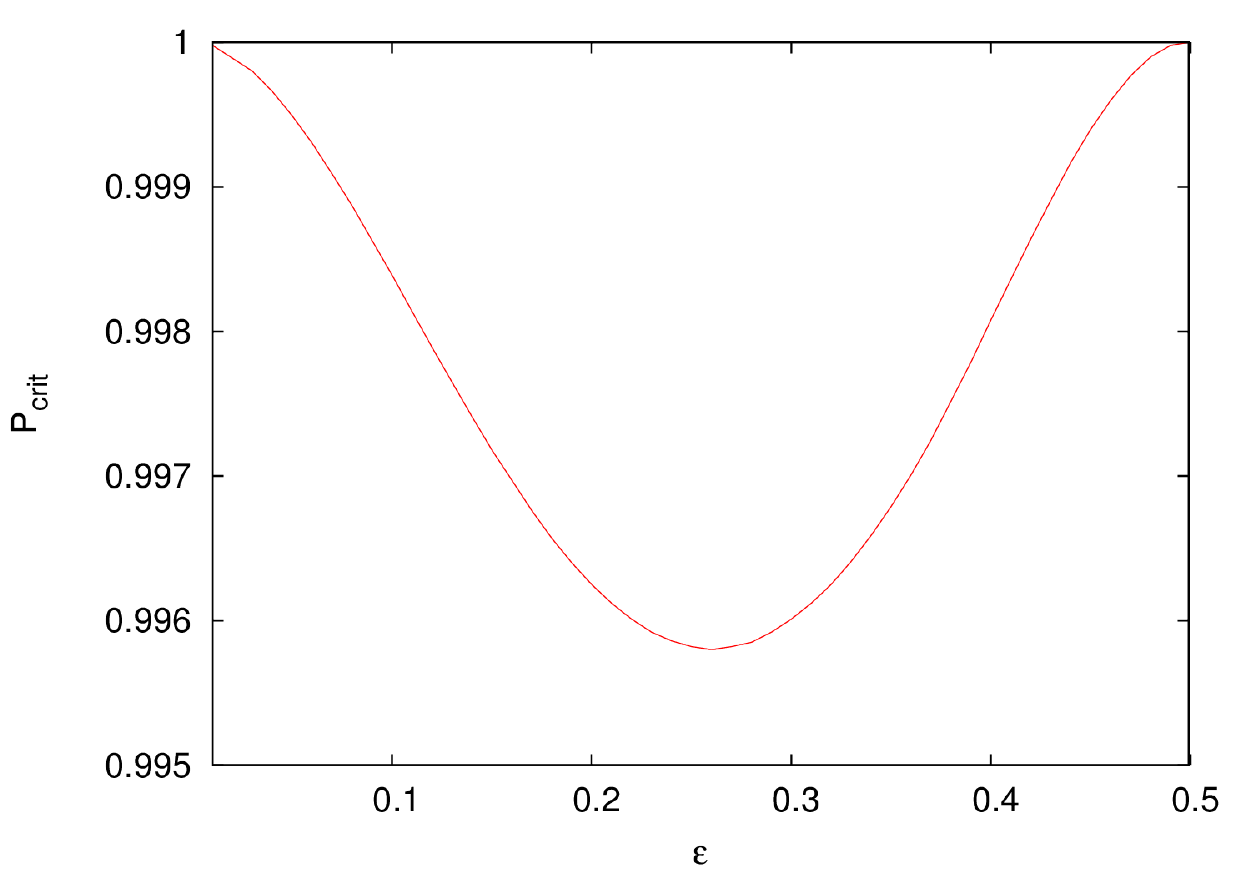}}
	\caption[Winning probabilities in the Mermin game sufficient for randomness amplification.]{\textbf{Winning probabilities in the Mermin game sufficient for randomness amplification} as a function $P_{crit}(\epsilon)$ of the bias of the source of randomness. If $P_s$ is above the plot it means that the bias of the final bit $y$ is lower than that of the source. Note that higher values of $P_s$ are required for extremal initial $\epsilon$'s. This is because if $\epsilon$ is large the initial quantity of randomness is bad which makes the amplification difficult. On the other hand, if $\epsilon$ is already low then to amplify it we need to obtain an even more random bit which is again difficult.\label{fig:Pcrit}}
\end{figure}

Our main result is that for any $0 < \epsilon, \eta < \frac{1}{2}$ there exists $P_s < 1$ such that $g(\epsilon,P_s) \leq \eta$. To be more precise, we have numerically checked only $\epsilon \leq 0.499$, and conjecture this is true for all $\epsilon < \frac{1}{2}$. The critical value of $P_{crit}(\epsilon)$ such that for all $P_s > P_{crit}(\epsilon)$ we have $g(\epsilon,P_s) < \epsilon$ is plotted in fig.~\ref{fig:Pcrit}.

\subsection{The randomness amplification protocol}

If the success probability was the same for each iteration, this would complete the task, since it would be enough to perform an experiment with an average success probability lower than $\epsilon$ and take the outcomes of one of the parties as our final sequence. Then $\epsilon' = g(\epsilon,P_s)$. Yet, we cannot assume this, since the success probability may differ between iterations. For this reason we have to perform some post-processing of the outcomes.

This result is somehow technical, and not directly connected to the topic of this work, \textit{i.e.} SDP in the analysis of quantum protocols. We refer interested readers to our paper \cite{MP13} for a complete proof that the presented protocol is working properly. We mention that the crucial part of the proof uses the following lemma by Chor and Goldreich \cite{chor}:
\begin{lem}
	 Let us define an $(N,b)$-source which produces $N$ bits
	\be
		\nonumber
		s_1,\dots,s_N
	\ee
	such that $P(s_1,\dots,s_N|k) \leq 2^{-b}$. Consider two independent sources $(N,b_1)$ and $(N,b_2)$ producing bits $h_1,\dots,h_N$ and $a_1,\dots,a_N$ respectively. The independence condition is defined by
	\be
		\nonumber
		P(h_1,\dots,h_N,a_1,\dots,a_N|k)=P(h_1,\dots,h_N|k)\times P(a_1,\dots,a_N|k).
	\ee
	Then, the inner product $y=\bigoplus_{i=1}^N h_i \cdot a_i$ is an $\epsilon'$-free bit, that is $\frac{1}{2}-\epsilon' \leq P(y|k)\leq\frac{1}{2}+ \epsilon'$, if
	\be
		\nonumber
		b_1+b_2 \geq N+2+2\log_2\frac{1}{\epsilon'}.
	\ee
\end{lem}

We finish by stating the following amplification protocol. Let $N$ denote the number of iterations we perform.

We begin with $\epsilon$-free Santha-Vazirani source of randomness and tri-partite quantum box, as described above. Then we use the SV source to generate a sequence $\mathbf{s}$ of $2 N$ bits to choose measurement settings for subsequent iterations, and $N$ bits $\mathbf{h} = (h_1, \cdots, h_N)$ which will be needed to perform \textit{hashing}\index{hashing}. We assume that the SV source and the quantum box are correlated only through the variable $e$.

After these initial steps we perform $N$ iterations to collect a sequence of outcomes of the first party, $\mathbf{a} = (a_1, \cdots, a_N)$.

When the statistics are collected, we estimate the average success probability. Let us denote the estimated (\textit{i.e.} directly observed in this particular sequence of iterations) average success probability by $P_{est}$. Let us define\footnote{The formula defined in~\eqref{eq:rpN} is used in an estimation method using Azuma-Hoeffding theorem \cite{Hoeffding63,Azuma67}.}
\be
	\label{eq:rpN}
	r(q, N) \equiv \sqrt{-\ln(1 - q)} N^{-\frac{1}{2}}.
\ee
At this stage we check the following condition:
\be
	\label{eq:condGeps}
	\eta_{est} \equiv g \left(\epsilon, P_{\text{est}}-r(q, N) \right) < \frac{1}{1 + 2 \epsilon} - \frac{1}{2}.
\ee
Such a value can always be obtained in quantum mechanical experiments for $N$ large enough. One may check that from \eqref{eq:condGeps} it follows that
\be
	\nonumber
	\kappa \equiv 1 - \log_2 \left( \left( \frac{1}{2} + \eta_{est} \right)^{-1} \left( \frac{1}{2} + \epsilon \right)^{-1} \right) < 0.
\ee

If the condition \eqref{eq:condGeps} is satisfied, then we perform the hashing by calculating
\be
	\label{eq:hashing}
	y = \bigoplus_{i=1}^N h_i a_i.
\ee
The resulting bit $y$ is then $\epsilon'$-free with $\epsilon' = 2^{2 + \kappa N}$. It is easy to see that the resulting $\epsilon'$ can be made arbitrarily close to $0$ if $N$ is large enough.

If the condition \eqref{eq:condGeps} is not satisfied, then we abort the protocol. Note that the larger $\epsilon$ the condition is more restrictive.

The reliability of this protocol is stated in the following theorem:
\begin{trm}
	For any $0 < \epsilon' < \epsilon < \frac{1}{2}$ and $0 \leq q < 1$ there exists $N$ such that in the protocol presented above the bit $y$ is $\epsilon'$-free with the probability at least $q$ when the condition \eqref{eq:condGeps} is satisfied.
\end{trm}
For the proof we refer interested readers to our work, \cite{MP13}. Note that the protocol can still fail to amplify with probability $1-p$ although the condition \eqref{eq:condGeps} is satisfied.

\chapter{Interior point methods}
\label{chap:IPM}\index{interior point methods}

In this chapter we provide an overview of issues related to interior point methods (\acrshort{IPM}s). We start with the historical overview of the development of these methods, and discuss basic notions related to them. Then we move to a discussion of the interior point algorithm in different variants.

In this chapter we widely use the notation stated in~\eqref{eq:vecMat} by interchanging matrix ($X$ or $Z$) and vector ($x$ or $z$) forms of variables.

\section[Overview of interior point methods]{An overview of interior point methods for semi-definite programming}

In this section we give a short overview of the development of interior point methods. A more detailed treatment on this topic may be found in \cite{Terlaky96,FM00,Wright05,Gondzio12}, and in \cite{RTV06} for the case of LP.

Nowadays it is difficult to imagine the field of optimization without interior point algorithms. Nonetheless they had not played an important role before 1984 when a groundbreaking paper by Karmarkar \cite{Karmarkar84} was published. The revelation of the potential of IPM has been called a \textit{revolution in optimization} \cite{Wright05}, and IPM is mentioned among the most important algorithms of all the time \cite{Kubale}. Moreover, before 1984 there had been almost no connection between linear and nonlinear programming \cite{Wright05}. IPM allowed to introduce a unified framework for analysis of both linear and nonlinear problems \cite{NN92}.

\subsection{A brief history of optimization}

The modern history of linear optimization started during World War II, when the problem of optimal resource management became crucial. The problem was solved in 1947 by George Dantzig \cite{Dantzig90}, when he proposed the simplex method\index{simplex method}. Although the algorithm had exponential worst case complexity, it performed very efficiently in practical problems. Roughly speaking, the idea of the simplex method is to start at some vertex of the relevant convex polytope, and subsequently move over to its extreme point. The problem with this method is that there exist cases where the algorithm visits every vertex of the feasible region, leading to exponential worst case complexity, like \textit{e.g.} in the Klee-Minty problem formulated in 1970 \cite{KleeMinty70}.

During 1960s and 1970s the importance of the topic of computational complexity increased, and thus scientists became concerned with finding polynomial algorithms for different problems, since it was believed that an efficient solution has to be polynomial.

The first polynomial method for LP was \textit{ellipsoid method}\index{ellipsoid method} invented by Khachian in 1979 \cite{Khachian79}. The problem is that this method is very slow in practical problems. Thus, before 1984 there had existed two major methods for LP: one with the exponential worst case complexity but efficient in practice, and an inefficient one with a polynomial complexity.

During 1960s in the field of nonlinear programming there was a practice to convert constrained problems to unconstrained ones using the so-called \textit{barrier methods} \cite{FiaccoMcCormick90}. By introducing a barrier function it is possible to define a so-called \textit{central path}, which is traversed using Newton's method. Nevertheless, the popularity of barrier methods decreased in 1970s.

The IPM revolution began when Karmarkar announced a method for LP of polynomial complexity, which in contrast to the method of Khachian performed well in practical cases. In the following year it was shown \cite{GMSTW86} that the method was equivalent to a logarithmic barrier method applied to LP. Moreover, it was shown that IPM for LP is not worst than the simplex method in terms of efficiency in practical problems.

The origin of the barrier methods from nonlinear optimization suggested that IPM should be possible to be applicable to more general problems than LP. An extremely important result of invention of IPM is the unification of LP and nonlinear programming.

The next mile-stone in the history of IPM is the independent result of Alizadeh \cite{A91} Nesterov and Nemirovskii \cite{NN92,NN94} of late 1980s, who extended the scope of the method to a family of other convex optimization problems. Nesterov and Nemirovskii showed that the key to applying IPM to convex problems was the knowledge of a barrier which had a special property of self-concordance. In addition, if the method is to find a practical application, the first and second derivatives of the barrier have to be easily computable. The theory developed by Nesterov and Nemirovskii was used by Vandenberghe and Boyd \cite{Boyd95}. It allowed them to apply the Gonzaga and Todd's LP method \cite{GT92} to SDP problems.

In short, \textit{self-concordant barrier}\index{self-concordant barrier} is a smooth convex function defined on the interior of the relevant set. It tends to $\infty$ as the boundary is approached, and together with its derivatives it satisfies certain Lipschitz continuity conditions. They showed that IPM can be applied to any set for which it is possible to formulate such a barrier. Fortunately, a relatively easily computable self-concordant barrier is known for the case of SDP, namely
\be
	\label{eq:sdp-barrier}
	F(X) \equiv -\ln \det X,
\ee
for $X \succ 0$. A detailed modern introduction to self-concordant functions is contained in \cite{N04}. In their further works, Nesterov and Todd \cite{NT97,NT98} introduced and developed a notion of self-scaled barriers, which allow IPM to take longer steps than simple self-concordant barriers.

Another important development in LP which was further applied in SDP is the predictor-corrector\index{predictor-corrector} (\acrshort{P-C}) scheme \cite{Mehrotra92}. This topic will be described in more details further in sec.~\ref{sec:PC}.

We have briefly mentioned the barrier IPM for SDP. From the point of view of this work, the most important variant of IPM is the so-called \textit{primal-dual} interior point method. This method was introduced by Kojima \textit{et al.} in \cite{KMY89}, and allowed for implementation on efficient LP solvers by Lustig \textit{et al.} \cite{LMS91,LMS92,LMS94a,LMS94b}.

One of the most fascinating differences between IPM for LP and SDP is the problem of \textit{search directions}. We elucidate this topic in sec.~\ref{sec:searchDir}. In short the problem arises from the fact in LP the variables are scalars which commute, but in SDP the variables are symmetric matrices which in general do not commute.

\subsection{Self-dual embedding}
\label{sec:embedding}\index{self-dual embedding}

The self-dual embedding method for LP has been introduced by Ye \textit{et al.} \cite{Ye94,Ye96}, and later generalized to other problems \cite{LSZ00}. We briefly describe this method since it is used in SeDuMi\index{SeDuMi} solver \cite{Sturm02,SeDuMi}, which is one of two popular solvers we use in this work as a reference.

The self-dual embedding is one of the ways to overcome the difficulty with finding the initial solution for an SDP solver. The problem arises when a variant of IPM which requires all iterates to be feasible is to be implemented. Since this method is not used by the solver implemented in this work, we only sketch it to allow understanding how SeDuMi solver works. We refer interested readers to the description given in \cite{Sturm02}.

Roughly speaking, the idea is to \textit{embed} the SDP problem into a larger SDP problem, for which it is trivial to find an initially feasible solution. This is done by introducing additional variables $x_{e},y_{e},z_{e} \in \mathbb{R}_{+}$, whose values can be calculated for some initial iterates $X^{(0)}$, $y^{(0)}$ and $Z^{(0)}$. Then it is possible to construct a larger problem which has a solution allowing to approximate the solution of the initial problem. This is done by re-normalizing the solution of the larger problem, namely
\be
	\nonumber
	\frac{1}{x_{e}^{*}} (X^{*},y^{*},Z^{*})
\ee
gives an approximate solution of the initial problem. Here $(X^{*},y^{*},Z^{*})$ are the values obtained for the optimum of the larger problem.

\subsection{Semi-definite programming solvers}

We will briefly discuss modern implementations of SDP solvers. We refer readers to \cite{Mittelmann12} for a comprehensive overview.

A standard choice for an SDP solver among the NPA community seems \cite{NPA07,NPA08,MiguelVertesi} to be the SeDuMi\index{SeDuMi} solver \cite{SeDuMi,Sturm02} created by Sturm and, after his death, developed and maintained by Imre Polik. As mentioned above, this solver implements self-dual embedding IPM.

Another SDP solver of particular interest is SDPT3\index{SDPT3} solver \cite{TTT03,TTT12} implemented by Toh, Todd and Tutuncu. It implements infeasible primal-dual IPM with NT and HKM search directions (see sec.~\ref{sec:searchDir} for the definition of the search directions).

SeDuMi and SDPT3 are described in details in sec.~\ref{sec:parameters}, where their performance is analyzed. Other examples of SDP solvers include CSDP \cite{CSDP} by Borchers, DSDP \cite{DSDP}, SDPA \cite{SDPA}.

The choice of SeDuMi and SDPT3 as reference solvers further in this paper is motivated in the case of SeDuMi by its popularity, and in the case of SDPT3 by its scientific importance reflected by the number of papers referring to its mechanisms, and the number of their citations. At the moment of writing this work, among the mentioned solver, these two were most cited.

\section[Formulation of interior point method]{Formulation of the primal-dual interior point algorithm}
\label{sec:IPMalg}

In this section we give an overview of working of a predictor-corrector path-following interior point algorithm for SDP. For the sake of simplicity we assume that the problem on the input of the algorithm has a feasible optimal solution for which the strong duality condition (see~Eq.~\ref{eq:strongDuality}) holds. We expect the algorithm to output a solution which is (at least) close to be feasible, and for which the gap, see sec.~\ref{sec:duality}, between primal and dual solutions do not exceed some threshold.

Recall from sec.~\ref{sec:SDPformulation} that an SDP problem is defined by a set of symmetric matrices $A_1, \ldots A_m \in \mathbb{R}^{n \times n}$, $b \in \mathbb{R}^{m}$, and $C \in \mathbb{R}^{n \times n}$. Below we fix the meaning of \gls{n} and \gls{m}, so that they refer to the definition of an SDP problem.

A crucial notion for path-following IPMs is the \textbf{central path}\index{interior point methods!central path}. This is a set of solutions, $(X,y,Z)$, parametrized by a non-negative variable $\nu$, which satisfy (\textit{cf.}~Eqs~\eqref{SDP-primal} and~\eqref{SDP-dual})
\be
	\nonumber
	\ba
		& X, Z \succeq 0, \\
		& \Tr(A_i X) = b_i, \text{ for } i = 1, \cdots, m \\
		& C - \sum_{i=1}^{m} y_i A_i = Z, \\
		& X Z = \nu \idOp,
	\ea
\ee
or, equivalently (see~\eqref{eq:mathcalA})
\begin{subequations}
	\label{eq:optimalConds}
	\be
		\begin{aligned}
			\label{eq:primalFeas}
			\mathcal{A}^T x = b \\
		\end{aligned}
	\ee
	\be
		\begin{aligned}
			\label{eq:dualFeas}
			c - \mathcal{A} y = z, \\
		\end{aligned}
	\ee
	\be
		\begin{aligned}
			\label{eq:centralPath}
			X Z = \nu \idOp,
		\end{aligned}
	\ee
\end{subequations}
together with $X, Z \succeq 0$.

From the strong duality,~\eqref{eq:strongDuality}, we get that the optimal solution,
\be
	\nonumber
	(X^{*}, y^{*}, Z^{*}),
\ee
has the property that it is on the central path at the point with $\nu = 0$. The idea of path-following IPM methods is to keep the iterates possibly close to some point at the central path, with $\nu$ subsequently decreasing.

Using the conditions from~\eqref{eq:primalFeas} and~\eqref{eq:dualFeas} we get that the Newton step $(\Delta X, \Delta y, \Delta Z)$ is defined by the following conditions
\be
	\label{eq:rprd}
	\ba
		\mathcal{A}^T \Delta x = r_p, \\
		\mathcal{A} \Delta y + \Delta z = r_d.
	\ea
\ee
If the method assumes that $r_p=r_d=0$, then it is called \textbf{feasible}\index{interior point methods!feasible} IPM, and \textbf{infeasible}\index{interior point methods!infeasible} otherwise.

The most problematic is the condition imposed on the Newton step from the condition in \eqref{eq:centralPath}. As mentioned above, the problem arises from the fact that the matrices $X^{(i)}$ and $Z^{(i)}$ (where $i$ stands for the current iteration) possibly do not commute. For this reason the following form of conditions on the target of the Newton step is imposed:
\be
	\label{eq:NewtonSymmetrization}
	\Theta_{\nu}(X, Z) = \zeroOp \in \mathbb{R}^{n \times n},
\ee
where $\Theta_{\nu}(X, Z)$ is some symmetrization of $X Z - \nu \idOp$ (\textit{cf.}~Eq.~\eqref{eq:centralPath}) satisfying $\Theta_{\nu}(X, Z) = \Theta_{\nu}(X, Z)^T$. The problem of choice of $\Theta_{\nu}(X, Z)$ is discussed further in sec.~\ref{sec:searchDir}. We also define 
\be
	\label{eq:NewtonSymmetrizationVector}
	\theta_{\nu}(x, z) \equiv \vec(\Theta_{\nu}(X, Z)) \in \mathbb{R}^{n^2}.
\ee

\subsection{The Newton step}
\label{sec:NewtonStep}\index{Newton"'s method}

From~\eqref{eq:NewtonSymmetrization} we have the following constraint for the Newton step:
\be
	\nonumber
	\mathcal{E} \Delta x + \mathcal{F} \Delta z  = r_c,
\ee
where
\begin{subequations}
	\be
		\label{eq:E}
		\gls{mathcalE} \equiv \partial_x \theta_{\nu}(x, z) \in \mathbb{R}^{n^2 \times n^2},
	\ee
	\be
		\label{eq:F}
		\gls{mathcalF} \equiv \partial_z \theta_{\nu}(x, z) \in \mathbb{R}^{n^2 \times n^2},
	\ee
	\be
		\label{eq:Rc}
		\ba
			& R_c \equiv -\Theta_{\nu}(X, Z) \in \mathbb{R}^{n \times n}, \\
			& r_c \equiv \vec(R_c)  \in \mathbb{R}^{n^2}.
		\ea
	\ee
\end{subequations}
This is a classical notation used in \cite{TTT98,AHO98,T99,TTT99,TTT03}. This together with conditions~\eqref{eq:rprd} gives the following formula for the Newton step:
\be
	\label{eq:NewtonStep}
	\begin{bmatrix}
		\mathcal{A}^T & \zeroOp & \zeroOp \\
		\zeroOp & \mathcal{A} & \idOp \\
		\mathcal{E} & \zeroOp & \mathcal{F}
	\end{bmatrix}
	\begin{bmatrix}
		\Delta x \\
		\Delta y \\
		\Delta z
	\end{bmatrix}
	=
	\begin{bmatrix}
		r_p \\
		r_d \\
		r_c.
	\end{bmatrix}.
\ee
If $X, Z \succ 0$ and the so-called Monteiro-Zhang symmetrization is used, \eqref{eq:MZsymmetrization} in sec.~\ref{sec:searchDir}, then this system has a unique solution \cite{SSK98,TTT98}. The size of the matrix occurring on left-hand-side (LHS) of~\eqref{eq:NewtonStep} is $(2 n^2 + m) \times (2 n^2 + m)$.

A way of solving this equation with the Schur complement\index{Schur complement} method is described in sec.~\ref{sec:SchurIPM}. By now it is enough to say that it allows to reduce the size of the equation to $m \times m$. Although this reduction allows to save much of computations, but the creation of this complement is still one of the most expensive operations in SDP iterations. Methods of computing it are discussed in more detail in sec.~\ref{sec:Schur}.

In the simplest case, when no \acrshort{P-C} is used, in each iteration one determines the Newton direction solving~\eqref{eq:NewtonStep}. Afterward one needs to choose the step-length, \textit{i.e.} a pair of constants, $\alpha,\beta \in (0,1]$ (we keep here the notation of \cite{TTT98,TTT12}), such that
\be
	\label{eq:stepPSDcond}
	\ba
		X + \alpha \Delta X \succeq 0, \\
		Z + \beta \Delta Z \succeq 0.
	\ea
\ee
One may also impose additional constraint, \textit{e.g.} concerning the distance from the central path. We note here that the Newton's method is used only to establish the direction, but the iterative solver does not necessary take the full Newton step. The issue of computing the step-length is discussed in sec.~\ref{sec:step-lengths}.

\subsection{Step-lengths and the convergence of interior point methods}
\label{sec:step-lengths}\index{interior point methods!convergence}

We will here describe methods of evaluation the step-length taken within an iteration after the step direction has been evaluated.

Some implementations of IPM solvers take advantage of the so-called \textit{central path neighborhood}\index{interior point method!central path!neighborhood}, which formalizes some notion of a distance of a solution $(X,y,Z)$, not necessarily feasible, from the central path. Examples of neighborhoods are given in \cite{Sturm95,Sturm99,Sturm02}. Using suitable neighborhoods many authors were able to prove some convergence results of particular variants of IPM. The algorithms differ not only in the choice of a neighborhood, but also in the length of the steps, $\alpha$ and $\beta$, taken within it.

The bounds on the number of iterations are usually expressed for a given neighborhood in terms of $n$ being the size of the matrices describing the problem, and $\epsilon$ expressing the required limit on the gap between primal and dual solutions. In \cite{Sturm99} the bound $O \left( \sqrt{n} \log{\frac{1}{\epsilon}} \right)$ for a particular choice of neighborhood is obtained. Similar bounds have been obtained by other authors with different choices of neighborhood (see, \textit{e.g.} \cite{T99} for overview). For the proof of convergence the algorithms also require the starting point to be contained within the neighborhood.

The topic of efficient evaluation of the step-length is covered in more details in \cite{Toh02}. In particular, one may show that $t$ being the solution of the following problem defined by $M$ and $\Delta M$
\begin{align}
	\nonumber
	\begin{split}
		\text{maximize } &\null t \\
		\text{subject to } &\null M + t \Delta M \succeq 0
	\end{split}
\end{align}
is given by the formula
\be
	\label{eq:step-length}
	\max \left( \eig(C^{-T} \Delta M C^{-1}) \right),
\ee
where $C$ is the Cholesky decomposition of $M$. This gives a method for computing the step-length, $\alpha$ and $\beta$. We have $\alpha = \min(t, 1)$ with $M=X$ and $\Delta M = \Delta X$, and similarly for $\beta$. In the case of OCTAVE \cite{octave} the $\eig$ function is computed with \textsc{DSYEV} subroutine from LAPACK \cite{LAPACK,Golub96}.

We stress that since the proofs of convergence of IPM usually assume that the steps are contained in some neighborhood of the central path, the cost of dropping this assumption in some implementation is that there is no warranty that the algorithm converges. Nonetheless, the cost of usage of the neighborhood is the time of its computation, and the decrease of step-lengths. For this reasons some solvers, \textit{e.g.} SDPT3 \cite{TTT03}, and the solver proposed in chapter~\ref{chap:solver}, PMSdp, decide not to impose any neighborhood restrictions on the iterates.

We briefly mention other reasons why a practical implementation of IPM may not find a solution. We distinguish the \textbf{failure}\index{interior point method!failure} and the situation when the number of iteration has \textbf{exceeded}\index{interior point method!iteration exceeded} some limit imposed by a user or implementation. In the latter case it still may be possible that after some number of iterations the specific problem would have converged. We observe that a typical reason of failure is during the Schur complement matrix decomposition. This may occur when the iterates approach the optimal solution where the primal variable and the dual slack are close to be orthogonal, $\Tr(XZ) \approx 0$. In this case the Schur complement may become ill-conditioned.

\subsection{Predictor-corrector method}
\label{sec:PC}\index{predictor-corrector}

The tests described in the introduction to chapter~\ref{chap:solver} (see also appendices \ref{app:SDPT3_SeDuMi}, and~\ref{app:NPAsolver}) indicates that the usage of the P-C allows a significant performance improvement for most of the considered cases. The cost P-C is that in each iteration one has to solve the equation for Newton step \eqref{eq:NewtonStep} twice. Nonetheless the LHS remains unchanged within the single iteration, and thus if the Schur complement\index{Schur complement} method or matrix factorization is used, then it is enough to run it only once per iteration.

The first part of P-C is the calculation of the predictor direction (sometimes called an \textit{affine scaling direction} \cite{Sturm02}), which is an aggressive strategy for moving along the central path. As the target for Newton step one sets $\nu = 0$ in~\eqref{eq:NewtonSymmetrization}, meaning that the predictor step tries to get close to the optimal solution. Let $(\delta X, \delta y, \delta Z)$ denote the predictor step with the step-length given by $\alpha_P$ and $\beta_P$.

The predictor step is not actually performed, but used to derive a second-order correction for the corrector (or \textit{centering direction}). Let us define \cite{TTT98,TTT12})
\be
	\label{eq:sigma}
	\sigma \equiv \left( \frac{\Tr \left( (X + \alpha_P \delta X) (Z + \beta_P \delta Z) \right)}{\Tr(X Z)} \right)^{\text{expon\_used}},
\ee
where $\text{expon\_used}$ is a variable whose value is determined with some algorithm. (We follow here the naming convention of \cite{TTT12} to use the name $\text{expon\_used}$ for this variable, and $\text{expon}$ for the related parameter of an algorithm. See sec.~\ref{sec:SDPT3} for more details.) Let us also define
\be
	\label{eq:mu}
	\mu \equiv \frac{1}{n} \Tr(X Z).
\ee

In the corrector part of the iteration one sets
\be
	\label{eq:Rc-corr}
	R_{c,\text{corr}} \equiv -\Theta_{(\sigma \mu)}(X, Z) + F(X, Z, \delta X, \delta Z)
\ee
(instead of~\eqref{eq:Rc}) and solves~\eqref{eq:NewtonStep} for the second time, with a different right-hand-side (RHS). Here $F(X, Z, \delta X, \delta Z)$ is some expression for the second order correction (see~sec.~\ref{sec:implementation} for more details).

Let us denote the correctors direction as $(\Delta X, \Delta y, \Delta Z)$, and let the step-length be given by $\alpha_C$ and $\beta_C$. In the subsequent iterate the IPM algorithm sets
\be
	\begin{aligned}
		X & := X + \alpha_C \Delta X, \\
		y & := y + \beta_C \Delta y, \\
		Z & := Z + \beta_C \Delta Z.
	\end{aligned}
\ee

The algorithm stops when residual norms, $\epsilon_P$ and $\epsilon_D$, and the gap $|Tr(X Z)|$, see sec.~\ref{sec:duality}, are all less than the specified threshold.

\section{Search directions in interior point methods}
\label{sec:searchDir}\index{interior point methods!search directions}

As mentioned previously, there are many possible symmetrizations of the expression $X Z - \nu \idOp$. Here we describe three historically most important of them, \textit{viz.} AHO, HKM and NT. Most of the modern SDP solvers in the overview by Mittelmann \cite{Mittelmann12} use either HKM or NT. We refer readers to \cite{T99} for a detailed survey of the topic of search directions.

\subsection{AHO and HKM search directions}

Let us consider the condition in~\eqref{eq:centralPath}, namely	$X Z = \nu \idOp$. Taking the step gives
\be
	\nonumber
	(X + \Delta X) ( Z + \Delta Z) = \nu \idOp
\ee
or ignoring the second order term $\Delta X \Delta Z$
\be
	\nonumber
	\Delta X Z + X \Delta Z = \nu \idOp - X Z.
\ee
We require the steps $\Delta X$ and $\Delta Z$ to be symmetric. From the second equation in \eqref{eq:rprd} it is clear that $\Delta Z$ is always symmetric (up to numerical precision). Nonetheless, it may not be true for $\Delta X$. For this reason there is a need for a symmetrization of the equation \eqref{eq:centralPath}.

In \cite{AHO98} Alizadeh, Haeberly and Overton (\acrshort{AHO}\index{interior point methods!search directions!AHO}) introduced the following natural symmetrization of~\eqref{eq:centralPath}
\be
	\nonumber
	\Theta_{\nu}^{AHO}(X, Z) \equiv \frac{1}{2} (X Z + Z X) = \nu \idOp.
\ee
Although historically important, this search direction is not currently used by majority of SDP solvers, \textit{e.g.} it has been removed from the recent implementations of SDPT3\index{SDPT3} \cite{TTT12}.

The \acrshort{HKM}\index{interior point methods!search directions!HKM} direction has been introduced independently by Helmberg \textit{et al.} \cite{HRVW96}, Kojima \textit{et al.} \cite{KSH97} and Monteiro \cite{M97}. This has a primal 
\be
	\nonumber
	\Theta_{\nu}^{HKM,primal}(X, Z) \equiv Z^{\frac{1}{2}} X Z^{\frac{1}{2}} = \nu \idOp,
\ee
and a dual 
\be
	\nonumber
	\Theta_{\nu}^{HKM,dual}(X, Z) \equiv X^{\frac{1}{2}} Z X^{\frac{1}{2}} = \nu \idOp
\ee
versions. This method is used by many modern solvers. Further in this work, in chapter~\ref{chap:solver}, we perform the performance profiling of this method.

We briefly mention that many other directions exist \cite{T99}, including for example the $XZ$\index{interior point methods!search directions!XZ} direction \cite{SP97}.

\subsection{Nesterov-Todd search direction}
\label{sec:NT}\index{interior point methods!search directions!NT}

The method developed by Nesterov and Todd (\acrshort{NT}) \cite{NT97,NT98} is of a particular interest, since the implementation we propose in chapter~\ref{chap:solver} uses this direction. Its direct definition is more involved than in the cases of AHO and HKM.

In their works Nesterov and Todd state the existence of the so-called \textbf{scaling point}\index{scaling point} $W$, which is defined in terms of the self-scaled barrier of a given convex set. Without going into details, their result states in particular that there exists a unique PSD matrix $W$ such that
\be
	\label{eq:WXWZ}
	W^{-1} X W^{-1} = Z.
\ee
Obviously we have equivalently
\be
	\label{eq:WZWX}
	W Z W = X.
\ee

It can be shown with elementary operations that
\be
	\label{eq:W}
	W = X^{\frac{1}{2}} \left( X^{\frac{1}{2}} Z X^{\frac{1}{2}} \right)^{-\frac{1}{2}} X^{\frac{1}{2}} = Z^{-\frac{1}{2}} \left( Z^{\frac{1}{2}} X Z^{\frac{1}{2}} \right)^{\frac{1}{2}} Z^{-\frac{1}{2}}.
\ee
Note that evaluation of a square root of a matrix is a relatively expensive operation. Fortunately, as it is described later, there is no need to perform this calculation when implementing the method.

The matrix $W$ allows the following symmetrization of~\eqref{eq:centralPath}:
\be
	\nonumber
	\frac{1}{2} \left( W^{-\frac{1}{2}} X Z + Z X W^{-\frac{1}{2}} \right) = \nu W^{-1},
\ee
or, taking $V \equiv W^{\frac{1}{2}} Z W^{\frac{1}{2}} = W^{-\frac{1}{2}} X W^{-\frac{1}{2}}$,
\be
	\nonumber
	\Theta_{\nu}^{NT}(X, Z) \equiv V^2 = \nu \idOp.
\ee
The latter approach of the symmetrization is used in \cite{Sturm99}. This form, called the v-space approach, or v-linearization\index{interior point methods!search directions!v-linearization}, gives the same search direction as the original NT direction. We mention it, since one of the investigated solvers, SeDuMi\index{SeDuMi} \cite{SeDuMi}, uses this nomenclature.

A key reference for issues related to implementation of the NT search direction is \cite{TTT98}.

\subsection{Monteiro-Zhang family of search directions}
\index{interior point methods!search directions!Monteiro-Zhang}

Let us describe the following linear transformation introduced by Monteiro and Zhang \cite{Z98,MZ98,M98} in 1998, which allows the formulation of the Monteiro-Zhang family of search directions. This family includes the three mentioned directions, AHO, HKM and NT. This linear transformation is given by the following formula
\be
	\label{eq:MZoperator}
	\gls{HpM} \equiv \frac{1}{2} \left( P M P^{-1} + P^{-T} M^T P^T \right),
\ee
where $P$ is an invertible matrix. Now we take
\be
	\label{eq:MZsymmetrization}
	\Theta_{\nu}(X,Z) \equiv H_P(X Z) - \nu \idOp.
\ee

From~\eqref{eq:E} and~\eqref{eq:F} we get
\be
	\nonumber
	\Mat \left( \mathcal{E} \Delta x \right) = \frac{1}{2} \left( P \Delta X Z P^{-1} + P^{-T} Z \Delta X P^T \right)
\ee
and
\be
	\nonumber
	\Mat \left( \mathcal{F} \Delta z \right) = \frac{1}{2} \left( P X \Delta Z P^{-1} + P^{-T} \Delta Z X P^T \right),
\ee
or equivalently
\be
	\label{eq:EKron}
	\mathcal{E} = P \otimes_S \left( P^{-T} Z \right) \in \mathbb{R}^{n^2 \times n^2},
\ee
and
\be
	\label{eq:FKron}
	\mathcal{F} = (PX) \otimes_S P^{-T} \in \mathbb{R}^{n^2 \times n^2},
\ee
where $\otimes_S$ is the \textit{symmetrized Kronecker product} of two matrices, defined in~\eqref{eq:symKron}. The operators $\mathcal{E}$ and $\mathcal{F}$ act on the vector form of the step, $\Delta x$ and $\Delta z$. Further in this work we will use this Kronecker product form.

One may check that using different choices of the invertible matrix $P$, it is possible to obtain each of the discussed search directions, namely with $P = \idOp$ we get AHO\index{interior point methods!search directions!AHO} search direction, with $P=Z^{\frac{1}{2}}$ and $P=X^{-\frac{1}{2}}$ the primal and dual HKM\index{interior point methods!search directions!HKM} search directions.

One can show that if we take an invertible matrix $P$ with a property that $P^T P = W^{-1}$, then we obtain the NT\index{interior point methods!search directions!NT} search direction, see sec.~\ref{sec:NT}. In sec.~\ref{sec:factorW}, a method of evaluation of a matrix $G$ such that
\be
	\label{eq:GGTW}
	G G^T = W
\ee
is described. Thus we can take $P = G^{-1}$.

\section{The Schur complement equation}
\label{sec:SchurIPM}\index{Schur complement}

The Schur complement method is a well known method \cite{Schur06} which allows to reduce a large system of equations to a smaller one, involving only a subset of variables. For example, a matrix
\be
	\nonumber
	M = 
	\begin{bmatrix}
		A && B \\
		C && D
	\end{bmatrix}
\ee
has the Schur complement given by $A - B D^{-1} C$. The method requires the submatrix $D$ to be invertible. $D$ also needs to be well-conditioned in order to give accurate results under finite-precision arithmetic. The method allows us to solve the equation
\be
	\nonumber
	M \begin{bmatrix} x_1 \\ x_2 \end{bmatrix} = \begin{bmatrix} y_1 \\ y_2 \end{bmatrix}
\ee
solving two simpler equations:
\be
	\nonumber
	\ba
		(A - B D^{-1} C) x_1 = y_1 - B D^{-1} y_2, \text{ and afterward} \\
		C x_1 + D x_2 = y_2.
	\ea
\ee
The reason why this method is useful is that one usually needs $O(n^3)$ operations to solve linear equation with $n$ variables, and thus it is profitable to decompose the initial equation into two smaller equations.

If we reorder the variables in~\eqref{eq:NewtonStep}, then we can obtain the following Schur complement equation for $\Delta y$ \cite{TTT98}:
\be
	\label{eq:SchurComplement}
	\left( \mathcal{A}^T \mathcal{E}^{-1}\mathcal{F} \mathcal{A} \right) \Delta y = r_p + \mathcal{A}^T \left( \mathcal{E}^{-1} \mathcal{F} r_d - \mathcal{E}^{-1} r_c \right).
\ee

After evaluation of $\Delta y$ it is possible to get the values of $\Delta X$ and $\Delta Z$. Namely we have
\be
	\label{eq:DeltaZ}
	\Delta Z = R_d - \Mat \left( \mathcal{A} \Delta y \right),
\ee
where $R_d = \Mat{r_d}$, and
\be
	\label{eq:DeltaX}
	\Delta X = \Mat \left( \mathcal{E}^{-1} (r_c - \mathcal{F} \Delta z) \right).
\ee
Thus we are able to decompose the procedure of solving of~\eqref{eq:NewtonStep} into three steps. The advantage is the following. Using Schur complement method we need to solve a smaller system of equation. The solution of a linear system of $k$ equations requires $O(k^3)$ floating point operations (FlOps). In the initial system we have $k=2 n^2 + m$, with $n > m$, and thus this requires $O(n^6)$ FlOps to get the solution. On the other hand, the Schur complement equation has size $m$, and requires only $O(m^3)$ FlOps. The calculation of~\eqref{eq:DeltaZ} and~\eqref{eq:DeltaX} are simple matrix-matrix and matrix-vector calculations on matrices of size $n \times n$ described in sec.~\ref{sec:implementation}.

\chapter[SDP solver for the NPA]{Semi-definite programming solver for Navascues-Pironio-Ac\'in problems}
\label{chap:solver}\index{interior point methods}

In this chapter we propose a special purpose SDP solver, further referred to as PMSdp, intended to act with NPA problems with the performance better than general solvers. There are a few methods of improving the performance of an SDP solver if it is supposed to solve a limited set of problems.

The simplest way is to optimize the code by making some simplifications, \textit{e.g.} there is no need to diverge the problem depending upon conditions, if these can be assumed to be fulfilled. The results presented in sec.~\ref{sec:profile} suggest that this factor can have a very large impact on the performance.

Another obvious method to improve the performance of an SDP solver is to choose the proper algorithm. Further in sec.~\ref{sec:parameters} some conclusions regarding the selection of the interior point method, including search direction and algorithm parameters, are made. In this section we also give an overview of the implementation of the three considered solvers, SDPT3, SeDuMi and PMSdp.

A more involved way to ameliorate a solver is to take advantage of the structure of the problem. This includes general techniques, like exploiting the sparsity of matrices in \cite{Fujisawa97}, or reorganization of computations \cite{TTT99}. The method proposed in sec.~\ref{sec:Schur} deals with the utilization of the NPA matrix structure in the calculation of the Schur complement\index{Schur complement} matrix which is considered to take the most computational effort in IPM.

The topic of warm-start strategies in the context of IPM has been widely studied in the case of linear programming. In the case where series of similar problems of the same size are to be solved, one may use one of the iterates of already solved instances as a starting point for another instance \cite{warmStart02,Forsgren06,warmStart08}. An overview of warm start strategies in the context of combinatorial optimization problems for SDP is given in \cite{warmStart12}, nonetheless up to our knowledge the problem for the general case has not been solved. Our method of choosing an initial solution for NPA problems is presented in sec.~\ref{sec:warmStart}.

The obstacles which occur in practical IPM solvers cover such problems like ill-conditioning of the factorized matrices, or deterioration of step lengths. Both problems can be overcome to some extent by the application of adequate perturbations to the current solution. Such a method is used in, \textit{e.g.} SDPT3\index{SDPT3} \cite{TTT12}. The cost of the perturbation is a loss of feasibility of the solution in the current iterate. Different strategies for choosing the criterion when to apply a perturbation are discussed in Sec.~\ref{sec:perturb}.

We decided to divide the course of the discussion into separate, relatively independent, topics. We hope that this form will be more useful for readers interested in implementing an IPM solver. Although PMSdp works with mixed LP-SDP problems (\textit{i.e.} SDP problems with additional LP constraints), in implementation discussion we concentrate on the SDP part, as this is more specific for the considered issues.

\SkipTocEntry\section*{Test specification}

We have performed a series of performance tests using SeDuMi and SDPT3 solvers executed for different sets of parameters. The results are contained in the appendix~\ref{app:SDPT3_SeDuMi} and discussed among the sections below.

Let $n$ denote the size of the SDP variable $\Gamma$ in the considered problem, and thus the length of the sequence used to formulate the considered hierarchy level. Let $m$ denote the number of linear constraint matrices\index{linear constraint matrices} of the problem, \textit{i.e.} the dimension of the dual variable.

Since this work concentrates on applications of the NPA method, we have prepared a set of test cases which are examples of applications described in chapter~\ref{chap:quantumProtocols}. These are given in problems~\eqref{problem-CHSH}, \eqref{problem-I3322}, \eqref{problem-Hardy}, \eqref{problem-BCn}, \eqref{problem-E0E1} and~\eqref{problem-T3C}. We consider the cases regarding CHSH and I3322 Bell operators, Hardy QKD and E0E1 protocol as smaller, whereas these regarding protocols from BCn family and T3C as larger. The problems are summarized in tab.~\ref{tab:problemsSizes}. In sec.~\ref{sec:parameters} we discuss also cases with higher hierarchy levels for problems of maximization of CHSH an I3322.

\begin{table}[htbp]
	\caption[Sizes of NPA test cases.]{\textbf{NPA test cases.} The sizes of mixed linear and SDP parts of test cases. The $n_L$ column refers to the number of linear constraints (each linear constraint introduces one additional linear variable to the problem), and $n$ column to the size of an SDP variable (Note that in sec.~\ref{sec:mixed} we have denoted the size of SDP variable with $n_S$ instead of $n$.). The column $m$ refers to the dimension of the dual variable $y$. The last column refers to the equation in the text in which the relevant problem has been stated. All problems are formulated in the Almost Quantum level of the NPA hierarchy.}
	\begin{tabular}{|l|r|r|r|l|}
		\hline
		 & \multicolumn{1}{l|}{$n_L$} & \multicolumn{1}{l|}{$n$ (SDP)} & \multicolumn{1}{l|}{$m$} & Eq. \\ \hline
		CHSH & 0 & 9 & 16 & \eqref{problem-CHSH} \\ \hline
		I3322 & 0 & 16 & 57 & \eqref{problem-I3322} \\ \hline
		E0E1 & 2 & 9 & 16 & \eqref{problem-E0E1} \\ \hline
		Hardy & 4 & 9 & 16 & \eqref{problem-Hardy} \\ \hline
		BC3 & 1 & 16 & 57 & \eqref{problem-BCn} \\ \hline
		T3C & 3 & 20 & 94 & \eqref{problem-T3C} \\ \hline
		BC5 & 1 & 36 & 355 & \eqref{problem-BCn} \\ \hline
		BC7 & 1 & 64 & 1281 & \eqref{problem-BCn} \\ \hline
	\end{tabular}
	\label{tab:problemsSizes}
\end{table}

\SkipTocEntry\section*{Technology and test environment}

The three solvers under comparison are toolboxes written as M-files in a high-level language MATLAB with parts implemented in C and C++. The latter are compiled to a binary form as MEX files with GCC 4.8.4 compilers.

MATLAB is an interpreted fourth-generation programming language used for scientific and engineering purposes. The M-files are either script files or function files. The \acrshort{MEX} files, or MATLAB Executable files, are binary executables callable from MATLAB. These files can be compiled from C, C++ or FORTRAN. The MEX files are dynamically linked subroutines loaded and executed by the MATLAB interpreter. Each MEX file contains only one subroutine which can be called in the same way as built-in MATLAB subroutines. It is a common practice in the field of High Performance Computing to write the parts which are crucial from the point of view of efficiency as MEX files.

All tests have been performed on a machine equipped with Intel Core 2 Quad CPU Q9400@2.66GHz x 4, 15.5GiB RAM under Linux Ubuntu 14.04 in OCTAVE\footnote{OCTAVE is an open source software with API compatible with MATLAB.} 3.8.1 \cite{octave}. The times in tables are the execution times averaged over $100$ executions.
The considered solvers do not use any randomization, and thus the number of iterations does not change between different executions\footnote{The only place in which we use randomization is in creation of a warm-start solution for PMSdp as described in sec.~\ref{sec:warmStartNPA}. Nonetheless the seed is constant, and thus the results are the same for different execution.}.

\section{Algorithm selection and specification of the solvers}
\label{sec:parameters}\index{interior point methods!search directions}

We will now investigate which parameters are optimal for the problems from the NPA method. We consider here two variants of interior point methods, an infeasible primal-dual interior point method implemented in SDPT3 and PMSdp solvers, and a self-dual embedding interior point method implemented in SeDuMi.

All three solvers can work with the mixed cone, which include problems having many linear and SDP variables\footnote{SeDuMi and SDPT3 allows the optimization over the so-called Lorentz, or second order, convex cone (which is a particular case of SDP) which is beyond the scope of this work.}. All of them are MATLAB packages and implement parts of their subroutines as MEX files compiled to a binary code to improve the performance.

As we have mentioned in sec.~\ref{sec:step-lengths}, if step lengths are taken within some specified neighborhood of the central path, then one may prove that a solver will always give a feasible solution $(X,y,Z)$ with $\Tr(X Z) \leq \epsilon$ within a number of steps polynomial in $n$ and $\log{\frac{1}{\epsilon}}$, if the solution exists. For example the neighborhood used by SeDuMi guarantees convergence in $O \left( \sqrt{n} \log{\frac{1}{\epsilon}} \right)$ steps. On the other hand, both SDPT3 and PMSdp do not restrict their steps to any neighborhood. In consequence these solvers are not guaranteed to converge. The reason not to use neighborhood is that the solver takes larger steps, and thus possibly achieves the result faster.

\subsection[SDPT3 solver]{SDPT3 as an example of implementation of primal-dual infeasible interior point method}
\label{sec:SDPT3}\index{SDPT3}\index{interior point methods!infeasible}

SDPT3 is an example of an SDP solver which does not employ the self-dual embedding techniques. Optionally it allows a user to specify a warm-start solution. The solver performs infeasible iterations till the desired feasibility and gap is achieved. The default stopping condition is that both these values do not exceed $10^{-8}$. Under certain conditions this solver performs some perturbations on the iterates; the algorithm is very complicated, and includes many special cases.

The solver implements HKM\index{interior point methods!search directions!HKM} and NT\index{interior point methods!search directions!NT} search directions. To calculate the scaling point the solver used the method from~\eqref{eq:GRUD}.

This solver calculates the Schur complement with the method described in sec.~\ref{sec:Fujisawa}. It uses \textsc{chol} function from the API of MATLAB to factorize the Schur complement matrix. The factors are used twice: first in the predictor and then in the corrector step.

To calculate step lengths SDPT3 uses the Lanczos method of finding the largest eigenvalues \cite{Toh02} in~\eqref{eq:step-length}.
The steps are not constrained to stay within any neighborhood.

Recall from the~\eqref{eq:mu}, that we define
\be
	\nonumber
	\mu \equiv \frac{1}{n} \Tr(X Z).
\ee
One of the input parameters of SDPT3 is $\text{expon}$ used to calculate the value of $\text{expon\_used}$ in~\eqref{eq:sigma} with the following algorithm:
\begin{algorithmic}
	\State $\text{step\_pred} \gets \min(\alpha_P, \beta_P)$
	\If {$\mu > 10^{-6}$}
		\If {$\text{step\_pred} < \frac{\sqrt{3}}{3}$}
			\State $\text{expon\_used} \gets 1$
		\Else
			\State $\text{expon\_used} \gets \max(\text{expon}, 3 \cdot \text{step\_pred}^2)$
		\EndIf
	\Else
		\State $\text{expon\_used} \gets \max(1, \min(\text{expon}, 3 \cdot \text{step\_pred}^2))$
	\EndIf
\end{algorithmic}

The appendix~\ref{app:SDPT3_SeDuMi} contains results of performance tests of SDPT3 solver on test cases specified above. We discuss these results below.

\subsubsection{Almost Quantum level}
\index{distribution!Almost Quantum}

The general conclusion one can draw from this performance data is the following. The HKM search direction seems to be slightly more efficient than the NT search direction for small problems, whereas the latter achieves a better performance for large problems. 

For the majority of test cases it is profitable to use the predictor-corrector scheme. Surprisingly this is not true for two test cases, \textit{viz.} E0E1 and Hardy. In these two cases still the predictor-corrector scheme reduced the number of iterations needed to obtain the final result, but the execution time increased. What is specific to these cases is that they involve more linear constraints than other problems. It is important to note that both cases were not able to attain the threshold of $10^{-8}$, which had to be increased to $10^{-5}$.

It is difficult to draw a conclusion with regard to the optimal value of the parameter $\text{expon}$. For the smaller problems and BC3 and T3C the optimum seems to be narrow at about $1.5$ or $2$, whereas BC5 and BC7 perform better for values $1$ or less.

To sum up, the only clear conclusion is that in most cases it is profitable to use predictor-corrector method\index{predictor-corrector}, whereas the impact of other factors is not unequivocal.

\subsubsection{High hierarchy levels}

We draw more conclusions for the higher hierarchy levels. In these cases the NT search direction performs much better than HKM. The former attains the desired quality of the solution in about $1$-$3$ iterations less that the latter and requires by $10$-$20\%$ less CPU time. It is highly profitable to use the predictor-corrector scheme. This allows to reduce the iteration number and CPU time by about $30\%$.

Note that the level $\mathcal{Q}_4$ for I3322 HKM without P-C failed to give the solution. On the other hand, levels $\mathcal{Q}_{10}$ and $\mathcal{Q}_{13}$ failed in most cases. This supports the conclusion that the method used by SDPT3 can handle NPA problems with $n$ at most equal to about $250$. The results also suggest that the value $m$ is less important.

The reason why higher hierarchy levels fail to converge may be the fact that for each NPA problem there exists a level of the hierarchy from which the rank of the $\Gamma$ matrix does not increase \cite{NPA08}. This leads to strongly ill-conditioned iterates near the optimal solution, when the hierarchy level is increased.

\subsection[SeDuMi solver]{SeDuMi as an example of implementation of self-dual embedding interior point method}
\index{SeDuMi}\index{self-dual embedding}

In the appendix~\ref{app:SDPT3_SeDuMi} the results of performance tests are shown. The SeDuMi solver is an example of solver which uses the self-dual embedding method (see~sec.~\ref{sec:embedding}).

SeDuMi allows to use one of the following algorithms: a first-order wide region method, a centering-predictor-corrector algorithm with v\hyp{}linearization\index{interior point methods!search directions!v-linearization}\footnote{Note that the term v-linearization refers in the case of SeDuMi to the NT direction \cite{Sturm99}.}, and a centering-predictor-corrector algorithm with xz-linearization\index{interior point methods!search directions!XZ}.

There are three possibilities of step-length differentiation implemented in SeDuMi, namely no differentiation, an algorithm with primal/dual step length differentiation, and with an adaptive heuristic to control step differentiation. The algorithm takes two parameters $\theta$ and $\beta$, which are wide region \cite{Sturm95} and neighborhood\index{interior point method!central path!neighborhood} parameters. The detailed discussion of these notions is beyond the scope of this work. We refer interested readers to \cite{SeDuMi} for details (see~sec.~\ref{sec:step-lengths} for a brief overview). A strong dependence on the algorithm parameters is observed.

\subsubsection{Almost Quantum level}
\index{distribution!Almost Quantum}

For almost all cases we observed that the optimal selection of parameters is $\theta \approx 0.7$ and $\beta \approx 0.9$. What is more, for small values of these parameters many problems were not able to converge within the iteration limit $150$.

It is difficult to draw a conclusion with regards to the optimal step-length differentiation method. In tab.~\ref{tab:optStepLenghtSeDuMi} the optimal choices of each problem are given. The only conclusion is that it is profitable to use some step differentiation in SeDuMi.

\begin{table}[htbp]
	\caption[Optimal step-length differentiation strategies]{\textbf{Optimal step-length differentiation strategies} of each of the testing problems for $\theta = 0.7$ and $\beta = 0.9$. The value 0 in the second column refers to the algorithm without differentiation, the value 1 means the primal/dual step length differentiation, and the value 2 means the adaptive heuristic to control step differentiation. The values in brackets mean that the choice is optimal in terms of the number of iterations, but on the testing machine it was not optimal in terms of CPU time.}
	\small
	\begin{tabular}{|l|r|r|r|r|r|r|r|r|}
		\hline
		step length diff. & \multicolumn{1}{l|}{CHSH} & \multicolumn{1}{l|}{I3322} & \multicolumn{1}{l|}{E0E1} & \multicolumn{1}{l|}{Hardy} & \multicolumn{1}{l|}{BC3} & \multicolumn{1}{l|}{T3C} & \multicolumn{1}{l|}{BC5} & \multicolumn{1}{l|}{BC7} \\ \hline
		Wide-region & 2 & 2 & 1 & (1) & 2 & 0 & 1 & 1 \\ \hline
		V-linearization & 0 & 1 & 2 & 1 & 0 & 2 & 1 & 1 \\ \hline
		XZ-linearization & 2 & 1 & 2 & (1) & 2 & 2 & 1 & 1 \\ \hline
	\end{tabular}
	\label{tab:optStepLenghtSeDuMi}
\end{table}

The conclusion that the impact of step-length differentiations is not unequivocal is supported by the results contained in tab.~\ref{tab:optimalAlgorithmsSeDuMi}. This table shows the optimal linearization and step-length differentiation method for each problem case.

From tab.~\ref{tab:optimalAlgorithmsSeDuMi} it is clear that the v-linearization is the optimal choice of the algorithm for Almost Quantum SDP problems.

\begin{table}[htbp]
	\small
	\caption[Optimal algorithms for the NPA test cases.]{\textbf{Optimal algorithms.} The optimal choice of linearization and step-length differentiation method for test problems. The value 0 in the second column refers to the algorithm without differentiation, the value 1 means the primal/dual step length differentiation, and the value 2 means the adaptive heuristic to control step differentiation. The optimality is in terms of the number of iteration (and in most cases agrees with the optimal CPU time on the testing machine).}
	\small
	\begin{tabular}{|l|l|l|l|l|l|l|l|l|}
		\hline
		 & CHSH & I3322 & E0E1 & Hardy & BC3 & T3C & BC5 & BC7 \\ \hline
		linearization & V & V & V & V & V & V or XZ & V & V \\ \hline
		step length diff. & \multicolumn{1}{l|}{1} & \multicolumn{1}{l|}{1} & \multicolumn{1}{l|}{0} & \multicolumn{1}{l|}{1} & \multicolumn{1}{l|}{0} & \multicolumn{1}{l|}{2} & 0 or 2 & 0 or 1 \\ \hline
	\end{tabular}
	\label{tab:optimalAlgorithmsSeDuMi}
\end{table}

\subsubsection{High hierarchy levels}

In high hierarchy levels the optimal parameters in most cases are $\theta \approx 1$ and $\beta \approx 0.9$ for the wide-region algorithm.

In the case of v-linearization algorithm the role of $\beta$ is less important if $\theta$ is not greater than about $0.25$. Otherwise the method fails for higher hierarchy levels.

For all high hierarchy level test cases the v-linearization method needed the smaller number of iterations and CPU time, in all cases the wide region algorithm was the least efficient choice. Moreover, the latter failed in all tests in $\mathcal{Q}_4$ with I3322 test case. Nonetheless, one should note that also v-linearization failed in some cases with large values of $\theta$.

Similarly as in the case of Almost Quantum level, it is difficult to draw a conclusion regarding the preferred step-length differentiation.

\subsection{PMSdp solver}

We decided to implement the primal-dual infeasible IPM, since there is no known method of finding a warm-start solution for the primal variable $X$ (in the dual formulation of the NPA). The solver always uses the warm-start technique described in sec.~\ref{sec:warmStartNPA} for the dual variable $y$.

The above results regarding optimal parameter selection for SDPT3 and SeDuMi suggest that for the NPA problems the NT\index{interior point methods!search directions!NT} search direction (or v-linearization in the nomenclature of SeDuMi) achieves the best performance. For this reason we have chosen PMSdp to implement this search direction. We use~\eqref{eq:GRUD} to calculate the scaling point.

Our implementation uses \textsc{eig} function from the API of MATLAB to calculate the step-length using~\eqref{eq:step-length}. We follow SDPT3 in the algorithm of calculating $\sigma$ for the corrector step,~\eqref{eq:sigma}.

We use a method from sec.~\ref{sec:sparsityNPA} to calculate the Schur complement. This method is designed for the case of NPA problems. To factorize the Schur complement we use \textsc{chol} function from the API of MATLAB.

The stopping criterion for a success is when both primal and dual infeasibilites, and the gap, are less than a threshold specified in the invocation. The default value for both these thresholds is $10^{-8}$.

More implementation details are described in sec.~\ref{sec:implementation}.

\section[Schur complement]{The computation of the Schur complement and the structure of NPA problems}
\label{sec:Schur}\index{Schur complement}

The calculation of the Schur complement matrix (see~secs~\ref{sec:NewtonStep} and \ref{sec:SchurIPM}) is often considered to have the most significant computation cost when solving an SDP problem \cite{Fujisawa97,TTT99,TTT03,Sturm02,TTT12}. For this reason it is the most promising field for inventing a solution tailored for a particular need.

We start this section with a discussion of the structure of NPA matrices, \textit{i.e.} matrices occurring in the NPA method. We then move to methods of calculation of the Schur complement matrix.

We give a discussion of two general approaches. These approaches take the advantage of sparsity occurring in many SDP problems. Then a method which uses the structure of NPA problems is introduced.

\subsection{NPA problem matrices}
\label{sec:NPAmatrices}\index{$\Gamma$ matrix}

We will now describe some obvious properties of the NPA matrices.

Recall the constraints defining the NPA matrices given in~\eqref{eq:gammaConds}. Let $\tilde{\mathcal{S}}$ be the set of indices of the given $\Gamma$ matrix of size $|\tilde{\mathcal{S}}| = n$. In the dual formulation of a problem defined by the above condition, $m$ is the number of algebraically different (up to transposition) concatenations $\{\tilde{O}_k\}_{k \in \{1, \dots, m\}}$ of sequences of operators\index{sequence of operators} of the form
\be
	\label{eq:concatOperators}
	O_i^{\dagger} O_j \equiv \tilde{O}_k.
\ee
From this we have that the $\Gamma$ matrix is a linear combination of $m$ matrices $A_i$, and each of these matrices has entries from sets $\{0,-1\}$, so that each $-A_i$ becomes a binary matrix\footnote{Note the minus sign in front of $A_i$. It is placed to confirm the form of dual SDP problems,~\eqref{SDP-dual}.} (see~appendix~\ref{app:exampleNPA}).

Another property of NPA matrices for a given problem is the fact that these matrices are orthogonal. Moreover, each entry $(k,l)$, $k,l \in \{1, \cdots, n\}$ is different than $0$ in at most one matrix.

As shown in tabs~\ref{tab:CHSHsizes} and \ref{tab:I3322sizes} the value of $m$ grows significantly with the hierarchy level, but slower than $n^2$, and the NPA linear constraint matrices\index{linear constraint matrix} $\{A_i\}$ may be considered as sparse.

\begin{table}[htbp]
	\caption[Sizes of different hierarchy levels for I3322.]{\textbf{Sizes of different hierarchy levels for I3322} problem (\textit{i.e.} the scenario with binary outcomes, and two parties with three settings each).}
	\begin{tabular}{|l|r|r|r||r|}
		\hline
		hierarchy level & \multicolumn{1}{l|}{n} & \multicolumn{1}{l|}{m} & \multicolumn{1}{l|}{average density} & \multicolumn{1}{l|}{m-primal} \\ \hline
		$\mathcal{Q}_2$ & 28 & 154 & 0.196 & 629 \\ \hline
		$\mathcal{Q}_3$ & 88 & 868 & 0.112 & 6875 \\ \hline
		$\mathcal{Q}_4$ & 244 & 4492 & 0.075 & 55043 \\ \hline
		$\mathcal{Q}_5$ & 628 & 22180 & 0.056 & 372203 \\ \hline
	\end{tabular}
	\label{tab:I3322sizes}
\end{table}

The above considerations were taken under the assumption that the NPA problem is formulated in the dual form,~\eqref{SDP-dual}. The reason for this choice of formulation is that it has a much smaller value of $m$ compared to primal formulation, see tabs~\ref{tab:CHSHsizes} and \ref{tab:I3322sizes}.

\subsection{Formula for the Schur complement for the NT direction}

Recall from the~\eqref{eq:SchurComplement} that the Schur complement is given by the following formula
\be
	\nonumber
	\mathcal{A}^T \mathcal{E}^{-1}\mathcal{F} \mathcal{A}.
\ee

First, let us show that for $\mathcal{E}$ and $\mathcal{F}$ defined by~\eqref{eq:EKron} and~\eqref{eq:FKron} for the NT direction we have \cite{TTT98}
\be
	\label{eq:eInvF}
	\mathcal{E}^{-1} \mathcal{F} = W \otimes_S W.
\ee
In derivation we use the equalities $W^{-1} X W^{-1} = Z$ and $G G^T = W$ (see~\eqref{eq:WXWZ} and~\eqref{eq:GGTW}).
\be
	\nonumber
	\begin{aligned}
		\mathcal{E} \left( W \otimes_S W \right) &= \left( G^{-1} W \right) \otimes_S \left( G^T Z W \right) = \left( G^{-1} W \right) \otimes_S \left( G^T W^{-1} X \right) = \\
		& \left( G^{-1} G G^T \right) \otimes_S \left( G^T G^{-T} X \right) = G^T \otimes_S \left( G^{-1} X \right) = \\
		& \left( G^{-1} X \right) \otimes_S G^T = \mathcal{F}.
	\end{aligned}
\ee

Thus the Schur complement $B$ for the NT direction is given by
\be
	\nonumber
	B \equiv \mathcal{A}^T W \otimes_S W \mathcal{A}.
\ee
From this we easily get
\be
	\nonumber
	B_{i,j} = \Tr (W A_i W A_j).
\ee

\subsection{Exploiting sparsity in Fujisawa \textit{et al.}}
\label{sec:Fujisawa}

In their paper, Fujisawa \textit{et al.} \cite{Fujisawa97} consider methods of exploiting the sparsity of the problem matrices.

First, they perform reordering of $\{A_i\}$ matrices. Let $f_i$ denote the number of nonzero elements of matrix $A_i$. They prove that for the methods used in the paper, the most effective is a reordering which has the property that $f_i$ occur in a non-increasing order. Using the symmetry of the Schur complement $B$ only entries $B_{i,j}$ for $j \geq i$ are evaluated directly.

Fujisawa \textit{et al.} consider three different methods for computation of the element $B_{i,j}$. The choice of the method depends on the sparsity of matrices $A_i$ and $A_j$.

For this work, the most interesting is the method when $A_i$ and $A_j$ are very sparse. The method of computing the Schur complement which we use in PMSdp is a modification of this case. The relevant formula is
\be
	\label{eq:Fujisawa}
	B_{i,j} = \sum_{a=1}^n \sum_{b=1}^n \sum_{c=1}^n \sum_{d=1}^n (A_i)_{c,d} W_{c,a} W_{d,b} (A_j)_{a,b}.
\ee
Calculation of this formula requires $(2 f_i + 1) f_j$ multiplications.

\subsection{Exploiting the structure of NPA problems}
\label{sec:sparsityNPA}

We will now introduce our method of exploiting the structure of NPA matrices to compute the Schur complement.

The presented method was proposed with a premise that it is possible to use the specific properties of NPA matrices described in sec.~\ref{sec:NPAmatrices} in order to design a proper data structure which is suitable for calculations of the Schur complement, and to implement a simple and fast kernel for this task.

The obvious choice for a binary sparse matrix storage format is to use one of the existing formats, \textit{e.g.} CRS (Compressed Row Storage) or COO (COOrdinate list) \cite{Saad03}. The former uses less memory, and simplifies certain operations, but for the task of fast iteration we decided to choose the latter one which contains coordinates of all nonzero element of a matrix in an explicit form.

Since the matrices under consideration are binary, there is no need to store the values on the elements. Thus a matrix $A_i$ is stored as an array of integer values, expressing row and column coordinates of all nonzero elements. The kernel accesses these elements in a linear way, so to promote local memory accesses we encode row and column coordinates alternately.

The implementations of SeDuMi and SDPT3 reduces the number of calculations by using a symmetric representation of matrices, which reduced the number of elements to be stored in a problem matrix by a factor of two.
This method stores only a half of the matrix with off-diagonal elements multiplied by $\sqrt{2}$ \cite{TTT98}. For this reason we decided not to take the advantage of the symmetry, since it would require a different treatment of diagonal and off-diagonal elements, which complicates the kernel.

The kernel implements the following algorithm. Let
\be
	\nonumber
	L_i = \left( (a_1,b_1), \cdots, (a_{f_i},b_{f_i}) \right)
\ee
be the array of elements of the matrix $A_i$, and
\be
	\nonumber
	L_j = \left( (c_1,d_1), \cdots, (c_{f_j},d_{f_j}) \right).
\ee

Using the formula by Fujisawa \textit{et al.},~\eqref{eq:Fujisawa} with NPA matrices, we have
\be
	\label{eq:FujisawaNPA}
	B_{i,j} = \Tr(W A_i W A_j) = \sum_{k=1}^{f_i} \sum_{l=1}^{f_j} W_{c_l,a_k} W_{d_l,b_k}.
\ee

We store the scaling point symmetric matrix $W$ as a one-dimensional array $\tilde{W}$ with subsequent elements. It is easy to see that, indexing from $0$, we have
\be
	\nonumber
	W_{i,j} = W_{j,i} = \tilde{W}[n \cdot i + j].
\ee

Thus, using the formula in~\eqref{eq:FujisawaNPA} we get
\be
	\nonumber
	B_{i,j} = \sum_{k=1}^{f_i} \sum_{l=1}^{f_j} \tilde{W}[n \cdot a_k + c_l] \cdot \tilde{W}[n \cdot b_k + d_l].
\ee

Let us consider the expression in the inner sum, namely
\be
	\label{eq:sk}
	s[k] \equiv \sum_{l=1}^{f_j} \tilde{W}[n \cdot a_k + c_l] \cdot \tilde{W}[n \cdot b_k + d_l].
\ee
If we use the symmetry of $A_i$, we can reduce the number of calculations at a cost of two additional condition expressions in the external loop. The value of \eqref{eq:sk} is calculated only if $a_k \leq b_k$. The values are summed with the weight $1$ if $a_k = b_k$, and with the weight $2$ if $a_k < b_k$.

The evaluation of $n \cdot a_k$ and $n \cdot b_k$ is performed in the external loop. Thus each iteration requires one floating point operation, two integer additions, two pointer iterations and two random memory accesses to $\tilde{W}$. The testing machine with Intel Core 2 Quad CPU Q9400 has $6$ MB of L2 cache which allows to store a matrix of sizes up to about $880 \times 880$.

\section{Start strategies}
\label{sec:warmStart}

We start the discussion of warm-start strategies for SDP-IPM with a description of a standard way of choosing the initial solution which does not take into account the problem formulation, \textit{i.e.} the so-called cold-start strategy used by SDP solvers.

\subsection{Cold-start in semi-definite programming solvers}
\label{sec:coldStart}

The investigated solver SeDuMi\index{SeDuMi} starts with some solution within some neighborhood of the central path\index{interior point method!central path!neighborhood} \cite{Sturm02} using self\hyp{}embedding \index{self-dual embedding} technique (see~sec.~\ref{sec:embedding}). The initial values of primal and dual SDP variable, $X^{(0)}$ and $Z^{(0)}$, are identity matrices, and $y^{(0)}$ the zero vector of the relevant dimension. SeDuMi applies the self-embedding to construct a primal and dual feasible solution for a larger problem, which will be used to approximate the solution of the stated problem.

As a general rule for cold-start, in \cite{Fujisawa97} it has been observed that it is desirable for the initial iterate to have the magnitude of at least the same order as the optimal solution. \cite{TTT12} discusses the following method of cold-start used in SDPT3\index{SDPT3}:
\be
	\label{eq:initIdIterate}
	\ba
		& X^{(0)} \equiv \xi \idOp, \\
		& Z^{(0)} \equiv  \eta \idOp,
	\ea
\ee
and $y^{(0)}$ is the zero vector of the relevant dimension. $\xi$ and $\eta$ are given by the following formulas
\be
	\nonumber
	\ba
		& \xi \equiv \max \left( 10, \sqrt{n}, n \max_{i = \{1, \cdots, m\}} \frac{1+|b_i|_F}{1+|A_i|_F} \right), \\
		& \eta \equiv \max \left( 10, \sqrt{n}, |C|_F, \max_{i = \{1, \cdots, m\}} |A_i|_F \right),
	\ea
\ee
where the matrix norm $|\cdot|_F$ is the Frobenius norm.

\subsection{Warm-start strategy for NPA problems} 
\label{sec:warmStartNPA}

We will now introduce the proposed method of creating a starting point for an NPA problem.

Our aim is to construct an initial $\Gamma$ matrix\footnote{A related issue of construction of a basis of moment matrices is discussed in \cite{MiguelVertesi}.}, \textit{i.e.} dual variables $y_0$ and $Z_0$. We start with a method of construction for $y_0$.

A simple way to construct a $\Gamma$ matrix is to find some \textbf{representation}\index{$\Gamma$ matrix!representation} including a state $\ket{\psi}$ (\textit{cf.}~Eq.~\eqref{eq:P-NPA}), and measurement operators for the number of settings and outcomes specified by the considered setup. Let $d$ denote the dimension of this representation.

A natural choice for the state is the maximally entangled state of the dimension $d$. The motivation is that for many Bell inequalities their maximal violation is achieved with this state\footnote{For almost forty years it was believed that the maximally entangled state reveals the most no-local behavior. Ultimately it turned out that this is not true, see~ \cite{Scarani07} for a detailed discussion.}. In case of maximally entangled qubits (\textit{i.e.} for $d=2$) shared by two parties this state is represented by the following density matrix
\be
	\label{eq:singlet2d}
	\begin{bmatrix}
		\frac{1}{2} & 0 & 0 & \frac{1}{2} \\
		0 & 0 & 0 & 0 \\
		0 & 0 & 0 & 0 \\
		\frac{1}{2} & 0 & 0 & \frac{1}{2}
	\end{bmatrix},
\ee
or $\ket{\psi} = \frac{\sqrt{2}}{2}(\ket{0} \otimes \ket{0} + \ket{1} \otimes \ket{1})$.

The second step is a construction of measurement operators. For each measurement setting of a party, these are represented by random orthogonal vectors of dimension $d$. With such a representation it is easy to construct sequences of measurement operators needed for a given NPA hierarchy level. In order to improve the performance, we store an array containing results of scalar products of the vectors representing measurements.

At this point it is easy to calculate a vector $\tilde{y}$ for a given state and measurements. The values of entries occur directly in the $\Gamma$ matrix, and can be evaluated with $\tilde{y}_k = \bra{\psi} \tilde{O}_k \ket{\psi}$, where $\tilde{O}_k$ is the representation of the relevant sequence of measurement operators, see~\eqref{eq:concatOperators}.

The representation of dimension $d$ leads to the $\Gamma$ matrix of rank $d^2$. Since the iterates have to be PSD matrices, such a matrix is not an appropriate starting point for IPM. On the other hand, since the representation is generated randomly, and the convex combination of certificates is still a certificate, we may repeat this procedure to obtain a certificate with a higher rank.

The more matrices in superposition are used, the more calculations are needed. On the other hand, the starting point can be reused for other optimization problems with the same setup and hierarchy level. We observe that the superposition of about $25 n$ different $\tilde{y}$ (obtained for random representations) is a good compromise between the time of creation and the quality of a warm-start point.

\begin{table}[htbp]
	\caption[The time needed to create a warm-start solution.]{\textbf{The time needed to create a warm-start solution} for different problems with superposition of $25 n$ certificates.}
	\begin{tabular}{|l|l|l|l|l|}
		\hline
		problem & CHSH & I3322 & E0E1 & Hardy \\ \hline
		creation time of $y_0$ & \multicolumn{1}{r|}{0.008} & \multicolumn{1}{r|}{0.017} & \multicolumn{1}{r|}{0.010} & \multicolumn{1}{r|}{0.009} \\ \hline \hline
		problem & BC3 & T3C & BC5 & BC7 \\ \hline
		creation time of $y_0$ & \multicolumn{1}{r|}{0.020} & \multicolumn{1}{r|}{0.027} & \multicolumn{1}{r|}{0.089} & \multicolumn{1}{r|}{0.408} \\ \hline
	\end{tabular}
	\label{tab:warmStartCreation}
\end{table}

Since we observed that for high hierarchy levels it is impossible to obtain a full rank matrix, regardless of the number of certificates under superposition (\textit{cf.}~sec.~4.2 in \cite{NPA08}), in order to get an initial PSD iterate, we add the expression $0.00000001 \cdot \idOp$ to the certificate\footnote{The cost of this is that we lose the dual feasibility, and thus we have to use a primal-dual infeasible IPM\index{interior point methods!infeasible}.}.

As the primal solution we take a simplified version of the expression in~\eqref{eq:initIdIterate}, namely we take $\xi = \max \left(10, \sqrt{n}\right)$.

\subsection{Warm-start in SDPT3 solver}
\index{SDPT3}

The SDPT3 solver has an option of supplying the initial iterate. We have performed tests regarding the impact of the warm-start strategy introduced in sec.~\ref{sec:warmStartNPA} on the number of iterations and CPU time needed by SDPT3. The results are shown in appendix~\ref{app:initSDPT3}.

The tests show that the warm-start strategy allows to improve both performance indicators in cases with no or with one linear condition. In those cases it allowed to gain a reduction by up to $5$ iterations. The strategy seems to perform best in the case of the NT direction, especially with P-C.

The method failed for E0E1 and T3C test cases with the HKM direction without P-C.

Thus we draw a conclusion that the proposed warm-start strategy is beneficial in a typical optimization scenario with simple constraints. The probable reason why it is less efficient with many linear constraints is the following. The creation of the initial solution takes into account only the NPA linear constraints for SDP part, and thus the solutions can be still far from dual feasibility in the LP cone.

\section{Methods of perturbation}
\label{sec:perturb}

In this section we discuss methods of performing perturbations on the iterates. The perturbation means that under certain conditions the current iterates are slightly changed. The reason for performing such operations is to prevent the solver from stalling in a point close to the boundary of the PSD cone. The main premise of perturbation strategies is not to reduce the number of iterations or CPU time, but to avoid failures.

We consider seven perturbation strategies described below. First, we specify for each strategy why we expected it to work for some encountered cases. Afterward we discuss the efficiency or inefficiency of each of them. We decided to describe also those strategies which failed to meet their expectations, so that this experience can be used for further improvements.

The expressions $\epsilon_P$ and $\epsilon_D$ are the primal and dual infeasibility norms, calculated using~\eqref{eq:rpNorm} and~\eqref{eq:rdNorm} from sec.~\ref{sec:SDPformulation}. Below $t_p$ means the threshold for the acceptance of primal and dual infeasibility. $t_g$ is the threshold for the acceptance of the gap.

\SkipTocEntry\subsection*{Strategy 1}

The first perturbation strategy perturbs the primal iterate if the primal infeasibility is significantly lower that the gap, and similarly for the dual iterate.

\begin{algorithmic}
	\If {$\text{gap} > 100 \cdot \epsilon_P $}
		\State $X \gets X + 0.01 \cdot t_p \cdot \idOp$
	\EndIf
	\If {$\epsilon_P > 100 \cdot \epsilon_D$}
		\State $Z \gets Z + 0.1 \cdot t_p \cdot \idOp$
	\EndIf
\end{algorithmic}

The motivation for this strategy is to balance the improvements of the gap and feasibility. We observed that the size of the gap is in most cases the most problematic, namely the iterates relatively fast reach the feasibility threshold, and most of the iterations are used to decrease the gap.

\SkipTocEntry\subsection*{Strategy 2}

Let us introduce variables $\text{notPertPrimal}$ and $\text{notPertDual}$ initialized with values $0$. They represent counters since the last perturbation of either primal or dual iterate.

\begin{algorithmic}
	\If {$\text{gap} > 100 \cdot \epsilon_P $ and $\text{notPertPrimal} > 1$}
		\State $X \gets X + 0.01 \cdot t_p \cdot \idOp$
		\State $\text{notPertPrimal} \gets 0$
	\Else
		\State $\text{notPertPrimal} \gets \text{notPertPrimal} + 1$
	\EndIf
	\If {$\epsilon_P > 100 \cdot \epsilon_D$ and $\text{notPertDual} > 1$}
		\State $Z \gets Z + 0.1 \cdot t_p \cdot \idOp$
		\State $\text{notPertDual} \gets 0$
	\Else
		\State $\text{notPertDual} \gets \text{notPertDual} + 1$
	\EndIf
\end{algorithmic}

The only difference in comparison with the previous strategy is that it prevents performing a perturbation ``too often''.

\SkipTocEntry\subsection*{Strategy 3}

Let $\alpha_C$ and $\beta_C$ denote the step-length of the corrector phase (see sec.~\ref{sec:PC}).

\begin{algorithmic}
	\If {$\alpha_C > 0.9$}
		\State $X \gets X + 0.01 \cdot t_p \cdot \idOp$
	\EndIf
	\If {$\beta_C > 0.9$}
		\State $Z \gets Z + 0.1 \cdot t_p \cdot \idOp$
	\EndIf
\end{algorithmic}

This strategy is intended to use these values as a heuristic that near the current iterate the solver will not have any problems with improving the primal and dual feasibility, therefore it may be profitable to change the point with hope to improve the $\text{gap}$ at the cost of feasibility.

\SkipTocEntry\subsection*{Strategy 4}

This strategy is an opposition of the previous strategy. The motivation is to avoid stalling at a point in which the step-lengths are low.

\begin{algorithmic}
	\If {$\alpha_C < 0.1$}
		\State $X \gets X + \epsilon_P \cdot \idOp$
	\EndIf
	\If {$\beta_C < 0.1$}
		\State $Z \gets Z + 10 \cdot \epsilon_D \cdot \idOp$
	\EndIf
\end{algorithmic}

In this strategy the size of the perturbation depends on the current primal or dual feasibility. The reason for this choice is that such a perturbation will keep the infeasibility at the same order, and thus will not disrupt the current solution significantly. The dual perturbation is much stronger. This is motivated by an observation that in most cases it is easier for the solver to improve the dual feasibility.

\SkipTocEntry\subsection*{Strategy 5}

This strategy also depends on current feasibility. The motivation of this strategy is that if the primal or dual solution is by far better than the threshold imposed by the stopping criteria, then a small perturbation can potentially help to improve the step-length while keeping the feasibility at the desired level.

\begin{algorithmic}
	\If {$\epsilon_P < 0.01 \cdot t_p$}
		\State $X \gets X + 0.0002 \cdot \epsilon_P \cdot \idOp$
	\EndIf
	\If {$\epsilon_D < 0.01 \cdot t_p$}
		\State $Z \gets Z + 0.001 \cdot \epsilon_D \cdot \idOp$
	\EndIf
\end{algorithmic}

\SkipTocEntry\subsection*{Strategy 6}

The most complicated of the considered strategies performs a perturbation at most only at either primal or dual iterate.

\begin{algorithmic}
	\If {$\epsilon_P > 100 \cdot |r_d|$}
		\State $Z \gets Z + 0.000000001 \cdot \idOp$
	\ElsIf {$\text{gap} > 100 \cdot \epsilon_D$}
		\State $Z \gets Z + 0.000000001 \cdot \idOp$
	\ElsIf {$\epsilon_P > 0.1$ and $\text{gap} > 0.1$}
		\State $X \gets X + 0.01 \cdot \idOp$
	\ElsIf {$\text{gap} > 1000 \cdot \epsilon_P$}
		\State $X \gets X + 0.00000001 \cdot \idOp$
	\EndIf
\end{algorithmic}

This method prefers a small perturbation on the dual iterate. The motivation is that the dual infeasibility is easier to be reduced, whereas the dual solution approaches the boundary when getting close to the optimal solution. The aim of this small perturbation is to prevent the matrices from becoming highly ill-conditioned (we observed the condition number at order of $10^{19}$ close to the optimal solution for some problems).

If this way of perturbation is not used, then perturbation on the primal solution is considered. If both the gap and primal infeasibility are high, we expect a primal perturbation not to harm the primal solution, but on the other hand, we expect the perturbation to allow longer steps in further iterations.

In the case when the primal infeasibility is relatively small compared to the gap, then it is expected that keeping the primal infeasibility small is less profitable than a potential longer step-length in further iterations.

\SkipTocEntry\subsection*{Strategy 7}

The last simple strategy is simply the third conditional of the previous strategy.

\begin{algorithmic}
	\If {$\epsilon_P > 0.1$ and $\text{gap} > 0.1$}
		\State $X \gets X + 0.01 \cdot \idOp$
	\EndIf
\end{algorithmic}

\subsection{Discussion of perturbation strategies}

As noted in the beginning of this section, the aim of perturbation strategies is not to improve performance, but to prevent failures. This is confirmed by the results in appendix~\ref{app:NPAsolver}, \textit{viz.} in all cases in which the solver did not fail, the most efficient was the solution without perturbations.

Test cases CHSH, BC3, T3C, BC5 and BC7 succeeded with all of the perturbation strategies. Thus, we should look for a perturbation strategy which is able to solve all cases with performance as good as possible.

Strategies 1 and 6 seem to be most reliable for cases E0E1 and Hardy, these are the only strategies which were able to converge in both cases.

In almost all cases strategy 3 was not able to converge. Thus, we conclude that one should not perturb iterates when the step-length is large.

The only case in which strategy 2 is better than strategy 1 is I3322 with $\text{expon}=3$, for which the latter failed. Thus, it seems that there is no need to limit the frequency of perturbations.

In all the cases strategy 5 performs better than strategy 4. Strategy 5 performs also better than strategy 7 in all but CHSH with $\text{expon}=4$. Moreover, it is able to attain the solution in cases when strategy 7 fails.

Let us concentrate on the cases with $\text{expon}=1$, see~tabs~\ref{tab:perturbSmall} and \ref{tab:perturbLarge}. In sec.~\ref{sec:profile} we consider this case in more details. In this case strategy 1 is able to converge in all considered situations. Unfortunately, when other strategies produce results, they are in most cases better in terms of performance.

\begin{table}[htbp]
	\caption[The number of iterations and CPU time for PMSdp (1).]{\textbf{The number of iterations and CPU time} for $\text{expon}=1$ and different perturbation strategies for smaller problems with PMSdp. Averaged over 100 executions.}
	\begin{tabular}{|r|r|r|r|r|c|r|r|r|}
		\hline
		\multicolumn{1}{|l|}{} & \multicolumn{ 2}{c|}{CHSH} & \multicolumn{ 2}{c|}{I3322} & \multicolumn{ 2}{c|}{E0E1} & \multicolumn{ 2}{c|}{Hardy} \\ \hline
		\multicolumn{1}{|l|}{pertubStrat} & \multicolumn{1}{l|}{iter.} & \multicolumn{1}{l|}{time [s]} & \multicolumn{1}{l|}{iter.} & \multicolumn{1}{l|}{time [s]} & \multicolumn{1}{l|}{iter.} & \multicolumn{1}{l|}{time [s]} & \multicolumn{1}{l|}{iter.} & \multicolumn{1}{l|}{time [s]} \\ \hline
		0 & 10 & 0.026 & 11 & 0.037 & \multicolumn{ 2}{c|}{failed} & \multicolumn{ 2}{c|}{failed} \\ \hline
		1 & 11 & 0.028 & 11 & 0.038 & \multicolumn{1}{r|}{33} & 0.091 & 26 & 0.074 \\ \hline
		2 & 11 & 0.028 & 11 & 0.038 & \multicolumn{ 2}{c|}{exceeded} & 26 & 0.074 \\ \hline
		3 & \multicolumn{ 2}{c|}{exceeded} & \multicolumn{ 2}{c|}{failed} & \multicolumn{ 2}{c|}{failed} & 26 & 0.073 \\ \hline
		4 & 10 & 0.026 & 11 & 0.037 & \multicolumn{ 2}{c|}{failed} & \multicolumn{ 2}{c|}{failed} \\ \hline
		5 & 10 & 0.026 & 11 & 0.038 & \multicolumn{ 2}{c|}{exceeded} & 27 & 0.07423 \\ \hline
		6 & 12 & 0.030 & 12 & 0.041 & \multicolumn{1}{r|}{31} & 0.082 & 25 & 0.068 \\ \hline
		7 & 10 & 0.02645 & 11 & 0.037 & \multicolumn{ 2}{c|}{failed} & \multicolumn{ 2}{c|}{failed} \\ \hline
	\end{tabular}
	\label{tab:perturbSmall}
\end{table}

\begin{table}[htbp]
	\caption[The number of iterations and CPU time for PMSdp (2).]{\textbf{The number of iterations and CPU time} for $\text{expon}=1$ and different perturbation strategies for larger problems with PMSdp. Averaged over 100 executions.}
	\begin{tabular}{|r|r|r|r|r|r|r|r|r|}
		\hline
		\multicolumn{1}{|l|}{} & \multicolumn{ 2}{c|}{BC3} & \multicolumn{ 2}{c|}{T3C} & \multicolumn{ 2}{c|}{BC5} & \multicolumn{ 2}{c|}{BC7} \\ \hline
		\multicolumn{1}{|l|}{pertubStrat} & \multicolumn{1}{l|}{iter.} & \multicolumn{1}{l|}{time [s]} & \multicolumn{1}{l|}{iter.} & \multicolumn{1}{l|}{time [s]} & \multicolumn{1}{l|}{iter.} & \multicolumn{1}{l|}{time [s]} & \multicolumn{1}{l|}{iter.} & \multicolumn{1}{l|}{time [s]} \\ \hline
		0 & 13 & 0.04534 & 18 & 0.08065 & 15 & 0.22247 & 18 & 3.43399 \\ \hline
		1 & 13 & 0.04557 & 18 & 0.0808 & 16 & 0.23832 & 19 & 3.61991 \\ \hline
		2 & 13 & 0.04557 & 18 & 0.08079 & 16 & 0.23811 & 19 & 3.61916 \\ \hline
		3 & \multicolumn{ 2}{c|}{exceeded} & \multicolumn{ 2}{c|}{exceeded} & \multicolumn{ 2}{c|}{exceeded} & \multicolumn{ 2}{c|}{exceeded} \\ \hline
		4 & 13 & 0.04567 & 19 & 0.08585 & 19 & 0.28443 & 23 & 4.37437 \\ \hline
		5 & 13 & 0.04686 & 18 & 0.08079 & 15 & 0.22411 & 18 & 3.41851 \\ \hline
		6 & \multicolumn{ 2}{c|}{exceeded} & \multicolumn{ 2}{c|}{exceeded} & \multicolumn{ 2}{c|}{exceeded} & \multicolumn{ 2}{c|}{exceeded} \\ \hline
		7 & 13 & 0.04574 & 18 & 0.08124 & 15 & 0.22404 & 18 & 3.41714 \\ \hline
	\end{tabular}
	\label{tab:perturbLarge}
\end{table}

\section{Implementation issues}
\label{sec:implementation}

In this section we discuss the implementation of PMSdp. The solver uses the NT search direction. We start with a brief overview of the main subroutines of the solver, and then discuss the most important implementation issues of the NT search direction.

The first element of the calculation of the NT direction is the \textit{scaling} operation. The issues related to this topic are discussed in sec.~\ref{sec:factorW}. We note that our implementation calculates scaling matrices $W$, $D$, $G$ and $G^{-1}$ with the singular value decomposition (\acrshort{SVD}) using~\eqref{eq:GLVD}.

The scaling matrices and the SDP problem definition are stored in the dedicated format. We use the method described in sec.~\ref{sec:sparsityNPA} in order to calculate the Schur complement matrix, which is crucial for determining the Newton steps. The Schur complement is afterward factorized using Cholesky factorization function \textsc{chol}. The factorized form is used twice to solve two systems with different RHS.

\subsection{PMSdp architecture overview}

The main subroutine of the PMSdp solver is the function \textsc{solver\_NT\_PC} which implements the NT search direction with the predictor-corrector scheme. This function takes as parameters structures stating the NPA problem as well as parameters regarding the IPM algorithm, $t_p$, $t_g$, $\text{expon}$ and the perturbation strategy, as defined in sec.~\ref{sec:perturb}.

The function \textsc{solver\_NT\_PC} contains the main loop of the solver. The loop continues until the condition
\be
	\label{eq:mainLoopCond}
	\epsilon_P < t_p \wedge \epsilon_D < t_p \wedge \text{gap} < t_g
\ee
is satisfied (\textit{i.e.} both primal and dual infeasibility norms are below the value $t_p$, and the duality gap is less than $t_g$), or until the maximal number of iterations, $100$, is exceeded.

The main loop consists of the following steps:
\begin{enumerate}
	\item Calculate scaling matrices $W$, $D$, $G$ and $G^{-1}$ with a function \textsc{Scaling} implementing the method stated in \eqref{eq:GLVD} below.
	\item Calculate residuals, \eqref{eq:rp} and \eqref{eq:rd} with a function \textsc{Residuals}.
	\item Calculate the LHS of \eqref{eq:NewtonStep} using the method described in sec.~\ref{sec:sparsityNPA} with a function \textsc{LHS} executing a MEX subroutine \textsc{AWWA}.
	\item Perform the Cholesky factorization using the built-in function \textsc{chol}.
	\item State the predictor equation with a function \textsc{PredictorEqs}.
	\item Calculate the Newton step, \eqref{eq:NewtonStep}, for the predictor with a function \textsc{NTDirection}. The function \textsc{NTDirection} solves the Newton step equation with a MEX subroutine \textsc{mextriang}.
	\item Calculate step lengths $\alpha_P$ and $\beta_P$ for the predictor step with a function \textsc{StepLengths} using \eqref{eq:step-length}. The built-in function \textsc{chol} is used for this purpose.
	\item State the corrector equation with a function \textsc{CorrectorEqs}.
	\item Calculate the Newton step for the corrector with the function \textsc{NTDirection}.
	\item Calculate step lengths $\alpha_C$ and $\beta_C$ for the corrector with the function \textsc{StepLengths}.
	\item Update the iterates $X^{(i+1)}$, $y^{(i+1)}$ and $Z^{(i+1)}$ with the corrector step.
	\item Calculate the norms $\epsilon_P$ and $\epsilon_D$, and the gap with a function \textsc{EvalNorms}.
	\item If the condition in \eqref{eq:mainLoopCond} is not satisfied, then perform perturbation on the iterates using one of the strategies described in sec.~\ref{sec:perturb}; otherwise leave the loop.
\end{enumerate}

The function \textsc{solver\_NT\_PC} returns the value of the solution, the final iterates and information about the execution including the number of iteration used to calculate the solution and the following timing information:
\begin{itemize}
	\item the total time of the \textsc{Scaling} function in all iterations,
	\item the total time of the formulation of the LHS in all iterations,
	\item the total time of the factorization of the LHS in all iterations,
	\item the total time of the predictor step calculation in all iterations,
	\item the total time of the corrector step calculation in all iterations, and
	\item the total time of execution.
\end{itemize}

\subsection{Calculation of the search direction}

The NT direction for the step $\Delta y$ is calculated as the solution of the relevant (\textit{i.e.} either predictor or corrector) Schur complement equation. The steps for $\Delta Z$ and $\Delta X$ are calculated with the formulas (\textit{cf.}~Eqs~\eqref{eq:DeltaZ} and~\eqref{eq:DeltaX}, and the implementation of SDPT3)
\be
	\nonumber
	\Delta Z \equiv R_d - \Mat(\mathcal{A} \Delta y)
\ee
and
\be
	\nonumber
	\Delta X \equiv G R_c G^T - W \Delta Z  W^T.
\ee
The latter formula is obtained from \eqref{eq:DeltaX} using \eqref{eq:EKron} and \eqref{eq:eInvF}.

The term $R_c$ is calculated separately for predictor and corrector. For predictor this is given by a simple formula, $-D$. The corrector's $R_c$ calculation is more complicated, as it requires the second order term, see~Eq.~\eqref{eq:Rc-corr}. The first order expression is for this case $\sigma \mu D^{-1} - D$. We follow the Mehrotra \cite{Mehrotra92} method as described in \cite{TTT98}, Eq.~(44).

In the case of predictor, the RHS of the Schur complement of \eqref{eq:NewtonStep} is calculated with the following formula (\textit{cf.}~Eq.~\eqref{eq:SchurComplement} and Eq.(36) in \cite{TTT98}):
\be
	\nonumber
	RHS_{\text{pred}} \equiv b^T + \mathcal{A}^T \cdot \vec(W R_d W).
\ee
The RHS for the case of corrector is calculated with
\be
	\nonumber
	RHS_{\text{corr}} \equiv r_p + \mathcal{A}^T \cdot \vec \left( W R_d W - G R_c G^T \right).
\ee
Since the expression $\vec(W R_d W)$ is used twice, it is stored to avoid the second computation.

After updating the iterates with corrector's steps, the norms are calculated using~\eqref{eq:rpNorm} and~\eqref{eq:rdNorm}. If these are below the defined threshold, then the solver finishes successfully.

\subsection{Factorization of the scaling point}
\label{sec:factorW}\index{interior point methods!search directions!NT}

The paper \cite{TTT98} discusses the method of evaluation of $G$ and $W$ matrices. For the sake of completeness of the description of the implementation of PMSdp, we provide here the description of this method.

In this section $X$ and $Z$ refer to the values of primal and dual SDP variables at the current iteration.

Let $L$ and $R$ be the Cholesky factors of $X$ and $Z$, \textit{viz.} $X = L L^T$ and $Z = R R^T$. Let $U D V^T = R^T L$ be the SVD decomposition of $R^T L$. From this it follows that $U D^2 U^T = R^T X R$ and $V D^2 V^T = L^T Z L$.

Recall the equation for the scaling point, $W$ (Eq.~\eqref{eq:W}):
\be
	\nonumber
	W = X^{\frac{1}{2}} \left( X^{\frac{1}{2}} Z X^{\frac{1}{2}} \right)^{-\frac{1}{2}} X^{\frac{1}{2}} = Z^{-\frac{1}{2}} \left( Z^{\frac{1}{2}} X Z^{\frac{1}{2}} \right)^{\frac{1}{2}} Z^{-\frac{1}{2}}.
\ee

We will prove two formulas for $G$ satisfying~\eqref{eq:GGTW}. These are given by
\be
	\label{eq:GLVD}
	G = L V D^{-\frac{1}{2}},
\ee
and
\be
	\label{eq:GRUD}
	G = R^{-T} U D^{\frac{1}{2}}.
\ee
We leave the proof to sections~\ref{sec:firstG} and \ref{sec:secondG}.

In order to compute $L$ and $R$ matrices, two Cholesky decompositions are needed. Depending on the formula for $G$, either $U$ or $V$ matrix is needed, together with the diagonal matrix $D$. Note that the SVD gives both $U$ and $V$, whereas these matrices can be obtained also with the eigenvalue decomposition of $R^T X R$ and $L^T Z L$, respectively. The eigenvalue decompositions give also the $D$ matrix. The method with eigenvalue decomposition is less stable, but requires less CPU time. 

\subsubsection{The first formula for $G$}
\label{sec:firstG}

We will now prove that the formula in~\eqref{eq:GRUD} satisfies $G G^T = W$.

One may show that $Q \equiv L^{-1} X^{\frac{1}{2}}$ is an orthogonal matrix. We have
\be
	\nonumber
	X^{\frac{1}{2}} Z X^{\frac{1}{2}} = Q^T L^T R R^T L Q = Q^T V D^2 V^T Q.
\ee
From this, we use the fact that $Q$ and $U$ are orthogonal to get
\be
	\nonumber
	\left( X^{\frac{1}{2}} Z X^{\frac{1}{2}} \right)^{-\frac{1}{2}} = Q^T V D^{-1} V^T Q.
\ee
Now, using the formula for $W$ from~\eqref{eq:W}, namely
\be
	\nonumber
	W = X^{\frac{1}{2}} \left( X^{\frac{1}{2}} Z X^{\frac{1}{2}} \right)^{-\frac{1}{2}} X^{\frac{1}{2}},
\ee
we get
\be
	\nonumber
	W = L V D^{-1} V^T L^T = G G^T,
\ee
with $G = L V D^{-\frac{1}{2}}$, which proves~\eqref{eq:GLVD}.

\subsubsection{The second formula for $G$}
\label{sec:secondG}

Here we prove that $G$ given in~\eqref{eq:GLVD} satisfies $G G^T = W$.

It is possible to show that $Q \equiv R^{-1} S^{\frac{1}{2}}$ is an orthogonal matrix. With this we get, similarly to the case in sec.~\ref{sec:firstG}
\be
	\nonumber
	W = Z^{-\frac{1}{2}} \left( Z^{\frac{1}{2}} X Z^{\frac{1}{2}} \right)^{\frac{1}{2}} Z^{-\frac{1}{2}} = R^{-T} U D U^T R^{-1} = G G^T,
\ee
with $G$ given by~\eqref{eq:GRUD}.

\section{Performance analysis}
\label{sec:profile}

In this section we analyze and compare the performance of SDPT3\index{SDPT3}, SeDuMi\index{SeDuMi}, and PMSdp.

\subsection{Test parameters}

For test cases of SeDuMi and SDPT3 we have chosen the parameters which seemed to be optimal separately for each test case. Because of the self-dual embedding method used by SeDuMi it was difficult to modify the code so that it stopped when the desired quality is achieved (\textit{i.e.} the gap and infeasibility are low enough). Instead we run test for SeDuMi, and after obtaining the solution we calculated the relevant norms, see~tab.~\ref{tab:profileParametersSeDuMi}. Afterward these parameters were given to both SDPT3 and PMSdp as tasks to be achieved (roughly speaking both solvers had to be at least as good as SeDuMi).

If one of these solvers was not able to attain a given quality, we decreased it (if it revealed to be better, we noted the better value), see~tabs~\ref{tab:profileParametersSDPT3} and \ref{tab:profileParametersNPA}. Nonetheless, the values were of the same order in all cases, so the performance comparison can be considered as fair.

\begin{table}[htbp]
	\caption{The parameters used by SeDuMi and the quality of the solutions in the profiling test.}
	\begin{tabular}{|l|r|r|r|r|r|r|}
		\hline
		SeDuMi & \multicolumn{1}{l|}{algorithm} & \multicolumn{1}{l|}{step.diff.} & \multicolumn{1}{l|}{$\theta$} & \multicolumn{1}{l|}{$\beta$} & \multicolumn{1}{l|}{$\epsilon_P$} & \multicolumn{1}{l|}{gap} \\ \hline
		CHSH & V & 1 & 0.1 & 0.1 & 4.0E-011 & -9.4E-009 \\ \hline
		I3322 & V & 2 & 0.1 & 0.1 & 6.5E-010 & 2.9E-009 \\ \hline
		E0E1 & XZ & 2 & 0.1 & 0.5 & 1.1E-008 & -1.9E-005 \\ \hline
		Hardy & XZ & 0 & 0.25 & 0.1 & 4.0E-009 & -1.1E-005 \\ \hline
		BC3 & V & 0 & 0.1 & 0.1 & 4.6E-010 & 4.9E-010 \\ \hline
		T3C & XZ & 0 & 0.1 & 0.5 & 9.9E-010 & 3.8E-010 \\ \hline
		BC5 & V & 2 & 0.25 & 0.5 & 2.3E-011 & 1.5E-010 \\ \hline
		BC7 & V & 2 & 0.25 & 0.9 & 1.4E-008 & 3.5E-008 \\ \hline
	\end{tabular}
	\label{tab:profileParametersSeDuMi}
\end{table}

\begin{table}[htbp]
	\caption{The parameters used by SDPT3 and the quality of the solutions in the profiling test.}
	\begin{tabular}{|l|r|r|r|r|}
	\hline
		SDPT3 & \multicolumn{1}{l|}{predcorr} & \multicolumn{1}{l|}{expon} & \multicolumn{1}{l|}{$t_p$} & \multicolumn{1}{l|}{$t_g$} \\ \hline
		CHSH & 1 & 0.1 & 1.00E-011 & 2.00E-010 \\ \hline
		I3322 & 1 & 2 & 3.00E-010 & 2.80E-009 \\ \hline
		E0E1 & 0 & 0 & 1.20E-006 & 1.80E-005 \\ \hline
		Hardy & 0 & 0.1 & 4.40E-006 & 1.00E-005 \\ \hline
		BC3 & 1 & 1.5 & 3.00E-010 & 4.00E-010 \\ \hline
		T3C & 1 & 2 & 1.60E-010 & 2.50E-010 \\ \hline
		BC5 & 1 & 1.5 & 5.00E-010 & 3.00E-010 \\ \hline
		BC7 & 1 & 1.5 & 8.00E-010 & 3.00E-009 \\ \hline
	\end{tabular}
	\label{tab:profileParametersSDPT3}
\end{table}

\begin{table}[htbp]
	\caption{The parameters used by PMSdp and the quality of the solutions in the profiling test.}
	\begin{tabular}{|l|r|r|r|r|}
		\hline
		PMSdp & \multicolumn{1}{l|}{expon} & \multicolumn{1}{l|}{perturb.strat.} & \multicolumn{1}{l|}{$t_p$} & \multicolumn{1}{l|}{$t_g$} \\ \hline
		CHSH & 1 & 1 & 1.00E-011 & 2.00E-010 \\ \hline
		I3322 & 1 & 1 & 3.00E-010 & 2.80E-009 \\ \hline
		E0E1 & 1 & 1 & 1.00E-008 & 1.80E-005 \\ \hline
		Hardy & 1 & 1 & 2.00E-009 & 1.00E-005 \\ \hline
		BC3 & 1 & 1 & 3.00E-010 & 4.00E-010 \\ \hline
		T3C & 1 & 1 & 1.60E-010 & 2.50E-010 \\ \hline
		BC5 & 1 & 1 & 2.00E-011 & 1.40E-010 \\ \hline
		BC7 & 1 & 1 & 1.00E-010 & 3.00E-010 \\ \hline
	\end{tabular}
	\label{tab:profileParametersNPA}
\end{table}

The SDPT3 solver was supplied the same initial solutions as the one with which PMSdp started, namely the one introduced in sec.~\ref{sec:warmStartNPA}. The SeDuMi solver used its cold-start strategy, see sec.~\ref{sec:coldStart}.

\subsection{Comparison of the solvers}

The execution profiles of the test cases are shown in tabs~\ref{tab:profileSeDuMi}, \ref{tab:profileSDPT3} and \ref{tab:profileNPA}. In these profiles we have chosen crucial operations and measured their execution time on a single CPU using functions \textsc{tic} and \textsc{toc} in MATLAB. There is a direct relation between profiled stages of SDPT3 and PMSdp. It is difficult to state such a relation between all stages in SeDuMi and the other two solvers.

The profile stages in SDPT3 and PMSdp are scaling, LHS, the factor, the predictor, and the corrector. The first refers to the creation of all scaling matrices, as described in sec.~\ref{sec:implementation}. LHS is the creation of the Schur complement matrix. The factor time refers to the time spent at Cholesky factorization. The predictor and corrector refer to the operation of solving the Schur equation, determining the Newton direction and step-length.

The last column in lower parts of the tables shows the percentage of the selected stages in the total execution time. This may be considered a measure of the performance gain obtained due to simplifications of the code structure, as mentioned in the introduction to this chapter. As expected, from this point of view, PMSdp has the simplest code, and the major part of it is concentrated in a few crucial operations.

Comparing the execution time, PMSdp outperforms both SeDuMi and SDPT3 in all cases. In some cases it finds a solution of the desired quality more than twenty times faster than SDPT3, and more than 5 times faster than SeDuMi. On the other hand, SeDuMi in all but one cases required the lowest number of iterations to converge. SeDuMi also executed faster than SDPT3, with the exception for the largest test case with BC7.

In most cases of SDPT3, the most time consuming operation was the LHS creation phase. On the other hand, in PMSdp this stage in most cases takes only a small part of the computations. For test cases CHSH, I3322, E0E1, Hardy, BC3 and T3C this stage is performed between $47$ and $208$ times faster. Both implementations use MEX compiled files to calculate the Schur complement. Thus we conclude, that the proposed method of computation serves its purpose.

\begin{table}[htbp]
	\footnotesize
	\caption{The profile of SeDuMi.}
	\begin{tabular}{|l|r|r|r|r|r|r|l|}
		\hline
		SeDuMi & \multicolumn{1}{l|}{build [s]} & \multicolumn{1}{l|}{factor [s]} & \multicolumn{1}{l|}{pred. [s]} & \multicolumn{1}{l|}{corr. [s]} & \multicolumn{1}{l|}{neighborhood [s]} & \multicolumn{1}{l|}{total [s]} & iter. \\ \hline
		CHSH & 0.01 & 0.03 & 0.02 & 0.04 & 0.02 & 0.21 & \multicolumn{1}{r|}{9} \\ \hline
		I3322 & 0.01 & 0.04 & 0.02 & 0.05 & 0.02 & 0.20 & \multicolumn{1}{r|}{12} \\ \hline
		E0E1 & 0.01 & 0.07 & 0.06 & 0.10 & 0.02 & 0.36 & \multicolumn{1}{r|}{23} \\ \hline
		Hardy & 0.01 & 0.09 & 0.08 & 0.11 & 0.03 & 0.42 & \multicolumn{1}{r|}{25} \\ \hline
		BC3 & 0.01 & 0.03 & 0.02 & 0.05 & 0.02 & 0.19 & \multicolumn{1}{r|}{12} \\ \hline
		T3C & 0.02 & 0.05 & 0.03 & 0.07 & 0.02 & 0.27 & \multicolumn{1}{r|}{17} \\ \hline
		BC5 & 0.10 & 0.15 & 0.03 & 0.06 & 0.02 & 0.44 & \multicolumn{1}{r|}{11} \\ \hline
		BC7 & 1.64 & 9.39 & 0.07 & 0.13 & 0.03 & 11.45 & \multicolumn{1}{r|}{12} \\ \hline
		 & \multicolumn{1}{l|}{} & \multicolumn{1}{l|}{} & \multicolumn{1}{l|}{} & \multicolumn{1}{l|}{} & \multicolumn{1}{l|}{} & \multicolumn{1}{l|}{} &  \\ \hline
		CHSH & 4\% & 15\% & 8\% & 18\% & 9\% & 54\% &  \\ \hline
		I3322 & 4\% & 18\% & 11\% & 23\% & 8\% & 64\% &  \\ \hline
		E0E1 & 3\% & 19\% & 16\% & 27\% & 6\% & 71\% &  \\ \hline
		Hardy & 3\% & 22\% & 18\% & 27\% & 6\% & 76\% &  \\ \hline
		BC3 & 5\% & 17\% & 12\% & 24\% & 10\% & 68\% &  \\ \hline
		T3C & 7\% & 18\% & 12\% & 25\% & 7\% & 70\% &  \\ \hline
		BC5 & 23\% & 35\% & 6\% & 13\% & 5\% & 81\% &  \\ \hline
		BC7 & 14\% & 82\% & 1\% & 1\% & 0\% & 98\% &  \\ \hline
	\end{tabular}
	\label{tab:profileSeDuMi}
\end{table}

\begin{table}[htbp]
	\footnotesize
	\caption{The profile of SDPT3.}
	\begin{tabular}{|l|r|r|r|r|r|r|l|}
		\hline
		SDPT3 & \multicolumn{1}{l|}{scaling [s]} & \multicolumn{1}{l|}{LHS [s]} & \multicolumn{1}{l|}{factor [s]} & \multicolumn{1}{l|}{pred. [s]} & \multicolumn{1}{l|}{corr. [s]} & \multicolumn{1}{l|}{total [s]} & iter. \\ \hline
		CHSH & 0.02 & 0.13 & 0.00 & 0.14 & 0.16 & 0.67 & \multicolumn{1}{r|}{17} \\ \hline
		I3322 & 0.02 & 0.46 & 0.00 & 0.12 & 0.14 & 0.92 & \multicolumn{1}{r|}{14} \\ \hline
		E0E1 & 0.03 & 0.18 & 0.00 & 0.21 & 0.00 & 0.71 & \multicolumn{1}{r|}{23} \\ \hline
		Hardy & 0.04 & 0.20 & 0.00 & 0.24 & 0.00 & 0.81 & \multicolumn{1}{r|}{26} \\ \hline
		BC3 & 0.03 & 0.57 & 0.00 & 0.16 & 0.18 & 1.18 & \multicolumn{1}{r|}{17} \\ \hline
		T3C & 0.03 & 1.05 & 0.01 & 0.18 & 0.20 & 1.74 & \multicolumn{1}{r|}{19} \\ \hline
		BC5 & 0.05 & 0.07 & 0.08 & 0.23 & 0.26 & 1.04 & \multicolumn{1}{r|}{20} \\ \hline
		BC7 & 0.10 & 1.10 & 2.79 & 0.67 & 0.66 & 5.99 & \multicolumn{1}{r|}{24} \\ \hline
		 & \multicolumn{1}{l|}{} & \multicolumn{1}{l|}{} & \multicolumn{1}{l|}{} & \multicolumn{1}{l|}{} & \multicolumn{1}{l|}{} & \multicolumn{1}{l|}{} &  \\ \hline
		CHSH & 4\% & 19\% & 0\% & 20\% & 24\% & 67\% &  \\ \hline
		I3322 & 2\% & 50\% & 0\% & 13\% & 15\% & 80\% &  \\ \hline
		E0E1 & 5\% & 25\% & 0\% & 29\% & 0\% & 59\% &  \\ \hline
		Hardy & 5\% & 25\% & 0\% & 29\% & 0\% & 59\% &  \\ \hline
		BC3 & 2\% & 48\% & 0\% & 13\% & 15\% & 79\% &  \\ \hline
		T3C & 2\% & 60\% & 0\% & 10\% & 12\% & 84\% &  \\ \hline
		BC5 & 4\% & 7\% & 7\% & 22\% & 25\% & 66\% &  \\ \hline
		BC7 & 2\% & 18\% & 47\% & 11\% & 11\% & 89\% &  \\ \hline
	\end{tabular}
	\label{tab:profileSDPT3}
\end{table}

\begin{table}[htbp]
	\footnotesize
	\caption{The profile of PMSdp.}
	\begin{tabular}{|l|r|r|r|r|r|r|l|}
		\hline
		PMSdp & \multicolumn{1}{l|}{scaling [s]} & \multicolumn{1}{l|}{LHS [s]} & \multicolumn{1}{l|}{factor [s]} & \multicolumn{1}{l|}{pred. [s]} & \multicolumn{1}{l|}{corr. [s]} & \multicolumn{1}{l|}{total [s]} & iter. \\ \hline
		CHSH & 0.00 & 0.00 & 0.00 & 0.01 & 0.01 & 0.04 & \multicolumn{1}{r|}{14} \\ \hline
		I3322 & 0.01 & 0.00 & 0.00 & 0.01 & 0.01 & 0.04 & \multicolumn{1}{r|}{11} \\ \hline
		E0E1 & 0.01 & 0.00 & 0.00 & 0.02 & 0.03 & 0.09 & \multicolumn{1}{r|}{33} \\ \hline
		Hardy & 0.01 & 0.00 & 0.00 & 0.02 & 0.03 & 0.08 & \multicolumn{1}{r|}{28} \\ \hline
		BC3 & 0.01 & 0.00 & 0.00 & 0.01 & 0.02 & 0.05 & \multicolumn{1}{r|}{15} \\ \hline
		T3C & 0.01 & 0.01 & 0.00 & 0.03 & 0.03 & 0.09 & \multicolumn{1}{r|}{20} \\ \hline
		BC5 & 0.04 & 0.08 & 0.09 & 0.06 & 0.07 & 0.36 & \multicolumn{1}{r|}{23} \\ \hline
		BC7 & 0.12 & 1.05 & 2.41 & 0.20 & 0.21 & 4.00 & \multicolumn{1}{r|}{21} \\ \hline
		 & \multicolumn{1}{l|}{} & \multicolumn{1}{l|}{} & \multicolumn{1}{l|}{} & \multicolumn{1}{l|}{} & \multicolumn{1}{l|}{} & \multicolumn{1}{l|}{} &  \\ \hline
		CHSH & 11\% & 4\% & 1\% & 26\% & 35\% & 78\% &  \\ \hline
		I3322 & 14\% & 5\% & 2\% & 27\% & 34\% & 82\% &  \\ \hline
		E0E1 & 10\% & 4\% & 1\% & 28\% & 36\% & 79\% &  \\ \hline
		Hardy & 10\% & 4\% & 1\% & 28\% & 36\% & 79\% &  \\ \hline
		BC3 & 14\% & 6\% & 2\% & 27\% & 34\% & 82\% &  \\ \hline
		T3C & 14\% & 8\% & 4\% & 27\% & 33\% & 86\% &  \\ \hline
		BC5 & 11\% & 24\% & 25\% & 17\% & 18\% & 95\% &  \\ \hline
		BC7 & 3\% & 26\% & 60\% & 5\% & 5\% & 99\% &  \\ \hline
	\end{tabular}
	\label{tab:profileNPA}
\end{table}

We finish this performance comparison with a brief discussion of one case with a higher hierarchy level. Although in the majority of practical purposes the Almost Quantum and $\mathcal{Q}_2$ levels are enough, it may be interesting to check how the proposed method scales with hierarchy levels, and thus with the size of the problem. As it can be seen from the tables in the appendix~\ref{app:SDPT3_SeDuMi}, SDPT3 is more efficient than SeDuMi for high levels of the NPA hierarchy. For this reason we only consider the profiles of SDPT3 and PMSdp in the comparison. The profiles are shown in tab.~\ref{tab:I3322_Q4}. These profiles were created with $\mathcal{Q}_4$ level of the NPA hierarchy using I3322 Bell operator, see tab.~\ref{tab:I3322sizes}. 

\begin{table}[htbp]
	\caption[PMSdp and SDPT3 with I3322 in $\mathcal{Q}_4$]{Comparison of the profiles of PMSdp and SDPT3 in the hierarchy level $\mathcal{Q}_4$ and I3322 Bell operator.}
	\begin{tabular}{|l|r|r|r|r|r|r|r|}
		\hline
		solver & \multicolumn{1}{l|}{scaling [s]} & \multicolumn{1}{l|}{LHS [s]} & \multicolumn{1}{l|}{factor [s]} & \multicolumn{1}{l|}{pred. [s]} & \multicolumn{1}{l|}{corr. [s]} & \multicolumn{1}{l|}{total [s]} & \multicolumn{1}{l|}{iter.} \\ \hline
		PMSdp & 3.0 & 63.1 & 91.5 & 4.2 & 4.8 & 169.7 & 22 \\ \hline
		SDPT3 & 2.0 & 60.9 & 92.0 & 5.4 & 5.3 & 168.5 & 23 \\ \hline
	\end{tabular}
	\label{tab:I3322_Q4}
\end{table}

These profiles show that in this case SDPT3 achieves a slightly better performance than PMSdp. We conjecture that our implementation of the method of calculation of the Schur complement matrix (the LHS column of profiles) from sec.~\ref{sec:sparsityNPA} scales with the size of the problem worse than the method used by SDPT3. This result is confirmed by the data in tabs~\ref{tab:profileSDPT3} and~\ref{tab:profileNPA} showing that for the largest of considered problems, BC7, PMSdp is only by about $\frac{1}{3}$-rd more efficient than SDPT3, and that the performance gain at the LHS operation is in this case small.

\chapter{Summary}

In this work we have discussed a number of issues concerning the semi-definite optimization with interior point methods with applications to quantum information protocols.

In the field of applications we have presented a new quantum key distribution protocol with novel properties, a few device-independent and semi-device-independent randomness expansion protocols, and answered the question of amplification of arbitrary weak randomness with quantum devices. In theory we developed a method which allows to construct SDP relaxations of SDI problems using the NPA method.

We proposed several methods which can be used in the implementations of solvers dealing with the NPA method of modeling the set of quantum probabilities. These include a warm-start strategies for SDP solvers, a few perturbation methods with evaluation of their efficiency, and a specific-purpose method of implementation of subroutines computing the Schur complements for NPA problems.

\backmatter \appendix

\chapter{Example of a simple NPA matrix}
\label{app:exampleNPA}

This is an example of $\mathcal{A}$ matrix for the dual formulation of $\mathcal{Q}_1(2,2|2,2)$. The LP part is omitted. The matrix is defined by \eqref{eq:mathcalA}, where each column refers to one linear constraint matrix $A_i$, and entry of the dual variable $y_i$. The construction of this matrix is discussed in sec.~\ref{sec:exampleNPA}.

\be
	\nonumber
	\begin{bmatrix}
		0 & 0 & 0 & 0 & 0 & 0 & 0 & 0 & 0 & 0 \\
		-1 & 0 & 0 & 0 & 0 & 0 & 0 & 0 & 0 & 0 \\
		0 & -1 & 0 & 0 & 0 & 0 & 0 & 0 & 0 & 0 \\
		0 & 0 & 0 & -1 & 0 & 0 & 0 & 0 & 0 & 0 \\
		0 & 0 & 0 & 0 & 0 & 0 & -1 & 0 & 0 & 0 \\
		-1 & 0 & 0 & 0 & 0 & 0 & 0 & 0 & 0 & 0 \\
		-1 & 0 & 0 & 0 & 0 & 0 & 0 & 0 & 0 & 0 \\
		0 & 0 & -1 & 0 & 0 & 0 & 0 & 0 & 0 & 0 \\
		0 & 0 & 0 & 0 & -1 & 0 & 0 & 0 & 0 & 0 \\
		0 & 0 & 0 & 0 & 0 & 0 & 0 & -1 & 0 & 0 \\
		0 & -1 & 0 & 0 & 0 & 0 & 0 & 0 & 0 & 0 \\
		0 & 0 & -1 & 0 & 0 & 0 & 0 & 0 & 0 & 0 \\
		0 & -1 & 0 & 0 & 0 & 0 & 0 & 0 & 0 & 0 \\
		0 & 0 & 0 & 0 & 0 & -1 & 0 & 0 & 0 & 0 \\
		0 & 0 & 0 & 0 & 0 & 0 & 0 & 0 & -1 & 0 \\
		0 & 0 & 0 & -1 & 0 & 0 & 0 & 0 & 0 & 0 \\
		0 & 0 & 0 & 0 & -1 & 0 & 0 & 0 & 0 & 0 \\
		0 & 0 & 0 & 0 & 0 & -1 & 0 & 0 & 0 & 0 \\
		0 & 0 & 0 & -1 & 0 & 0 & 0 & 0 & 0 & 0 \\
		0 & 0 & 0 & 0 & 0 & 0 & 0 & 0 & 0 & -1 \\
		0 & 0 & 0 & 0 & 0 & 0 & -1 & 0 & 0 & 0 \\
		0 & 0 & 0 & 0 & 0 & 0 & 0 & -1 & 0 & 0 \\
		0 & 0 & 0 & 0 & 0 & 0 & 0 & 0 & -1 & 0 \\
		0 & 0 & 0 & 0 & 0 & 0 & 0 & 0 & 0 & -1 \\
		0 & 0 & 0 & 0 & 0 & 0 & -1 & 0 & 0 & 0
	\end{bmatrix}
\ee

\chapter{Performance of SDPT3 and SeDuMi}
\label{app:SDPT3_SeDuMi}
\index{SDPT3}\index{SeDuMi}

In this appendix the results of performance tests of SDP solvers SDPT3 and SeDuMi are contained.

The tables referring to SDPT3 contain the numbers of iterations and timing for different test cases depending on three parameters: the search direction (either HKM or NT), the usage or not of predictor-corrector, and the value of $\text{expon}$.

For the case of SeDuMi, the tables refer to the following algorithms: the first-order wide region method, the centering-predictor-corrector algorithm with v-linearization, and the centering-predictor-corrector algorithm with xz\hyp{}linearization. The value 0 in the second column refers to the algorithm without differentiation, the value 1 means the primal/dual step length differentiation, and the value 2 means the adaptive heuristic to control step differentiation. $\theta$ and $\beta$ are wide region and neighborhood parameters.

For each problem set the number of iterations and the execution time is given. The results are discussed in chapter~\ref{chap:solver}.

\begin{table}[htbp]
	\tiny
	\caption[Performance of SDPT3 (1a).]{\textbf{Performance of SDPT3 (1).} The table shows the number of iterations and the execution time on SDPT3 of Almost Quantum for CHSH, I3322, E0E1 and Hardy problem sets. Averaged over 100 executions.}
\begin{tabular}{|l|l|r|r|r|r|r|r|r|r|r|}
		\hline
		 &  & \multicolumn{1}{l|}{} & \multicolumn{ 2}{c|}{CHSH} & \multicolumn{ 2}{l|}{I3322} & \multicolumn{ 2}{l|}{E0E1} & \multicolumn{ 2}{l|}{Hardy} \\ \hline
		HKM/NT & P-C & \multicolumn{1}{l|}{expon} & \multicolumn{1}{l|}{iter.} & \multicolumn{1}{l|}{time [s]} & \multicolumn{1}{l|}{iter.} & \multicolumn{1}{l|}{time [s]} & \multicolumn{1}{l|}{iter.} & \multicolumn{1}{l|}{time [s]} & \multicolumn{1}{l|}{iter.} & \multicolumn{1}{l|}{time [s]} \\ \hline
		HKM & no & 0 & 17 & 0.45 & 21 & 1.12 & 26 & 0.76 & 26 & 0.76 \\ \hline
		HKM & no & 0.01 & 17 & 0.46 & 21 & 1.12 & 26 & 0.76 & 26 & 0.76 \\ \hline
		HKM & no & 0.1 & 17 & 0.46 & 21 & 1.11 & 26 & 0.76 & 26 & 0.76 \\ \hline
		HKM & no & 0.5 & 17 & 0.46 & 21 & 1.11 & 26 & 0.76 & 26 & 0.76 \\ \hline
		HKM & no & 1 & 17 & 0.45 & 21 & 1.11 & 26 & 0.76 & 26 & 0.76 \\ \hline
		HKM & no & 1.5 & 17 & 0.46 & 21 & 1.11 & 26 & 0.76 & 26 & 0.76 \\
\hline
	\end{tabular}
\end{table}

\begin{table}[htbp]
	\tiny
	\caption[Performance of SDPT3 (1b).]{\textbf{Performance of SDPT3 (1).} The table shows the number of iterations and the execution time on SDPT3 of Almost Quantum for CHSH, I3322, E0E1 and Hardy problem sets. Averaged over 100 executions.}
	\begin{tabular}{|l|l|r|r|r|r|r|r|r|r|r|}
		\hline
		 &  & \multicolumn{1}{l|}{} & \multicolumn{ 2}{c|}{CHSH} & \multicolumn{ 2}{l|}{I3322} & \multicolumn{ 2}{l|}{E0E1} & \multicolumn{ 2}{l|}{Hardy} \\ \hline
		HKM/NT & P-C & \multicolumn{1}{l|}{expon} & \multicolumn{1}{l|}{iter.} & \multicolumn{1}{l|}{time [s]} & \multicolumn{1}{l|}{iter.} & \multicolumn{1}{l|}{time [s]} & \multicolumn{1}{l|}{iter.} & \multicolumn{1}{l|}{time [s]} & \multicolumn{1}{l|}{iter.} & \multicolumn{1}{l|}{time [s]} \\ \hline
		NT & no & 0 & 17 & 0.48 & 19 & 1.02 & 26 & 0.79 & 23 & 0.70 \\ \hline
		NT & no & 0.01 & 17 & 0.48 & 19 & 1.02 & 26 & 0.79 & 23 & 0.71 \\ \hline
		NT & no & 0.1 & 17 & 0.49 & 19 & 1.02 & 26 & 0.80 & 23 & 0.70 \\ \hline
		NT & no & 0.5 & 17 & 0.48 & 19 & 1.02 & 26 & 0.79 & 23 & 0.71 \\ \hline
		NT & no & 1 & 17 & 0.48 & 19 & 1.02 & 26 & 0.79 & 23 & 0.70 \\ \hline
		NT & no & 1.5 & 17 & 0.48 & 19 & 1.02 & 26 & 0.79 & 23 & 0.70 \\
\hline
	\end{tabular}
\end{table}

\begin{table}[htbp]
	\tiny
	\caption[Performance of SDPT3 (2a).]{\textbf{Performance of SDPT3 (2).} The table shows the number of iterations and the execution time on SDPT3 of Almost Quantum for BC3, T3C, BC5, and BC7 problem sets. Averaged over 100 executions.}
	\begin{tabular}{|l|l|r|r|r|r|r|r|r|r|r|}
		\hline
		 &  & \multicolumn{1}{l|}{} & \multicolumn{ 2}{c|}{BC3} & \multicolumn{ 2}{l|}{T3C} & \multicolumn{ 2}{l|}{BC5} & \multicolumn{ 2}{l|}{BC7} \\ \hline
		HKM/NT & P-C & \multicolumn{1}{l|}{expon} & \multicolumn{1}{l|}{iter.} & \multicolumn{1}{l|}{time [s]} & \multicolumn{1}{l|}{iter.} & \multicolumn{1}{l|}{time [s]} & \multicolumn{1}{l|}{iter.} & \multicolumn{1}{l|}{time [s]} & \multicolumn{1}{l|}{iter.} & \multicolumn{1}{l|}{time [s]} \\ \hline
		HKM & no & 0 & 21 & 1.17 & 30 & 2.34 & 30 & 1.06 & 36 & 7.71 \\ \hline
		HKM & no & 0.01 & 21 & 1.17 & 30 & 2.34 & 30 & 1.06 & 36 & 7.71 \\ \hline
		HKM & no & 0.1 & 21 & 1.17 & 30 & 2.34 & 30 & 1.06 & 36 & 7.70 \\ \hline
		HKM & no & 0.5 & 21 & 1.16 & 30 & 2.34 & 30 & 1.06 & 36 & 7.71 \\ \hline
		HKM & no & 1 & 21 & 1.16 & 30 & 2.35 & 30 & 1.05 & 36 & 7.73 \\ \hline
		HKM & no & 1.5 & 21 & 1.17 & 30 & 2.34 & 30 & 1.06 & 36 & 7.72 \\
\hline
	\end{tabular}
\end{table}

\begin{table}[htbp]
	\tiny
	\caption[Performance of SDPT3 (2b).]{\textbf{Performance of SDPT3 (2).} The table shows the number of iterations and the execution time on SDPT3 of Almost Quantum for BC3, T3C, BC5, and BC7 problem sets. Averaged over 100 executions.}
	\begin{tabular}{|l|l|r|r|r|r|r|r|r|r|r|}
		\hline
		 &  & \multicolumn{1}{l|}{} & \multicolumn{ 2}{c|}{BC3} & \multicolumn{ 2}{l|}{T3C} & \multicolumn{ 2}{l|}{BC5} & \multicolumn{ 2}{l|}{BC7} \\ \hline
		HKM/NT & P-C & \multicolumn{1}{l|}{expon} & \multicolumn{1}{l|}{iter.} & \multicolumn{1}{l|}{time [s]} & \multicolumn{1}{l|}{iter.} & \multicolumn{1}{l|}{time [s]} & \multicolumn{1}{l|}{iter.} & \multicolumn{1}{l|}{time [s]} & \multicolumn{1}{l|}{iter.} & \multicolumn{1}{l|}{time [s]} \\ \hline
		NT & no & 0 & 19 & 1.06 & 27 & 2.09 & 27 & 1.00 & 36 & 7.64 \\ \hline
		NT & no & 0.01 & 19 & 1.07 & 27 & 2.10 & 27 & 1.00 & 36 & 7.65 \\ \hline
		NT & no & 0.1 & 19 & 1.07 & 27 & 2.10 & 27 & 1.00 & 36 & 7.64 \\ \hline
		NT & no & 0.5 & 19 & 1.07 & 27 & 2.10 & 27 & 1.00 & 36 & 7.65 \\ \hline
		NT & no & 1 & 19 & 1.07 & 27 & 2.09 & 27 & 1.00 & 36 & 7.64 \\ \hline
		NT & no & 1.5 & 19 & 1.07 & 27 & 2.10 & 27 & 1.00 & 36 & 7.64 \\
\hline
	\end{tabular}
\end{table}

\begin{table}[htbp]
	\tiny
	\caption[High NPA hierarchy levels in SDPT3 (1a).]{\textbf{High NPA hierarchy levels in SDPT3 (1).} The table shows the performance of the SDPT3 solver on I3322 problem case in $\mathcal{Q}_2$, $\mathcal{Q}_3$, $\mathcal{Q}_4$ and $\mathcal{Q}_5$. For the level $\mathcal{Q}_5$ the solver was not able to attain the desired quality of solutions.}
	\begin{tabular}{|r|r|r|r|r|r|r|r|r|c|l|}
		\hline
		\multicolumn{1}{|l|}{} & \multicolumn{1}{l|}{} & \multicolumn{1}{l|}{} & \multicolumn{ 2}{c|}{$\mathcal{Q}_2$} & \multicolumn{ 2}{c|}{$\mathcal{Q}_3$} & \multicolumn{ 2}{c|}{$\mathcal{Q}_4$} & \multicolumn{ 2}{c|}{$\mathcal{Q}_5$} \\ \hline
		\multicolumn{1}{|l|}{HKM/NT} & \multicolumn{1}{l|}{predcorr} & \multicolumn{1}{l|}{expon} & \multicolumn{1}{l|}{iter.} & \multicolumn{1}{l|}{time [s]} & \multicolumn{1}{l|}{iter.} & \multicolumn{1}{l|}{time [s]} & \multicolumn{1}{l|}{iter.} & \multicolumn{1}{l|}{time [s]} & \multicolumn{1}{l|}{iter.} & time [s] \\ \hline
		HKM & no & 0.01 & 27 & 2.98 & 36 & 5.80 & \multicolumn{ 2}{c|}{failed} & \multicolumn{ 2}{c|}{failed} \\ \hline
		HKM & no & 0.1 & 27 & 3.01 & 36 & 5.77 & \multicolumn{ 2}{c|}{failed} & \multicolumn{ 2}{c|}{failed} \\ \hline
		HKM & no & 0.5 & 27 & 3.02 & 36 & 5.76 & \multicolumn{ 2}{c|}{failed} & \multicolumn{ 2}{c|}{failed} \\ \hline
		HKM & no & 1 & 27 & 3.02 & 36 & 5.74 & \multicolumn{ 2}{c|}{failed} & \multicolumn{ 2}{c|}{failed} \\ \hline
		HKM & no & 1.5 & 27 & 3.00 & 36 & 5.77 & \multicolumn{ 2}{c|}{failed} & \multicolumn{ 2}{c|}{failed} \\ \hline
		HKM & no & 2 & 27 & 2.98 & 36 & 5.71 & \multicolumn{ 2}{c|}{failed} & \multicolumn{ 2}{c|}{failed} \\
\hline
	\end{tabular}
\end{table}

\begin{table}[htbp]
	\tiny
	\caption[High NPA hierarchy levels in SDPT3 (1b).]{\textbf{High NPA hierarchy levels in SDPT3 (1).} The table shows the performance of the SDPT3 solver on I3322 problem case in $\mathcal{Q}_2$, $\mathcal{Q}_3$, $\mathcal{Q}_4$ and $\mathcal{Q}_5$. For the level $\mathcal{Q}_5$ the solver was not able to attain the desired quality of solutions.}
	\begin{tabular}{|r|r|r|r|r|r|r|r|r|c|l|}
		\hline
		\multicolumn{1}{|l|}{} & \multicolumn{1}{l|}{} & \multicolumn{1}{l|}{} & \multicolumn{ 2}{c|}{$\mathcal{Q}_2$} & \multicolumn{ 2}{c|}{$\mathcal{Q}_3$} & \multicolumn{ 2}{c|}{$\mathcal{Q}_4$} & \multicolumn{ 2}{c|}{$\mathcal{Q}_5$} \\ \hline
		\multicolumn{1}{|l|}{HKM/NT} & \multicolumn{1}{l|}{predcorr} & \multicolumn{1}{l|}{expon} & \multicolumn{1}{l|}{iter.} & \multicolumn{1}{l|}{time [s]} & \multicolumn{1}{l|}{iter.} & \multicolumn{1}{l|}{time [s]} & \multicolumn{1}{l|}{iter.} & \multicolumn{1}{l|}{time [s]} & \multicolumn{1}{l|}{iter.} & time [s] \\ \hline
		NT & no & 0.01 & 25 & 2.67 & 34 & 5.03 & 37 & 263 & \multicolumn{ 2}{c|}{failed} \\ \hline
		NT & no & 0.1 & 25 & 2.68 & 34 & 5.04 & 37 & 262 & \multicolumn{ 2}{c|}{failed} \\ \hline
		NT & no & 0.5 & 25 & 2.69 & 34 & 5.04 & 37 & 263 & \multicolumn{ 2}{c|}{failed} \\ \hline
		NT & no & 1 & 25 & 2.68 & 34 & 5.04 & 37 & 263 & \multicolumn{ 2}{c|}{failed} \\ \hline
		NT & no & 1.5 & 25 & 2.68 & 34 & 5.04 & 37 & 262 & \multicolumn{ 2}{c|}{failed} \\ \hline
		NT & no & 2 & 25 & 2.67 & 34 & 5.05 & 37 & 263 & \multicolumn{ 2}{c|}{failed} \\
\hline
	\end{tabular}
\end{table}

\begin{table}[htbp]
	\tiny
	\caption[High NPA hierarchy levels in SDPT3 (2).]{\textbf{High NPA hierarchy levels in SDPT3 (2).} The table shows the performance of the SDPT3 solver on CHSH problem case in $\mathcal{Q}_5$, $\mathcal{Q}_{10}$, and $\mathcal{Q}_{13}$.}
	\begin{tabular}{|r|r|r|r|r|c|l|c|l|}
		\hline
		\multicolumn{1}{|l|}{} & \multicolumn{1}{l|}{} & \multicolumn{1}{l|}{} & \multicolumn{ 2}{c|}{$\mathcal{Q}_5$} & \multicolumn{ 2}{c|}{$\mathcal{Q}_10$} & \multicolumn{ 2}{c|}{$\mathcal{Q}_{13}$} \\ \hline
		\multicolumn{1}{|l|}{HKM/NT} & \multicolumn{1}{l|}{predcorr} & \multicolumn{1}{l|}{expon} & \multicolumn{1}{l|}{iter.} & \multicolumn{1}{l|}{time [s]} & \multicolumn{1}{l|}{iter.} & time [s] & \multicolumn{1}{l|}{iter.} & time [s] \\ \hline
		HKM & no & 0.01 & 23 & 1.89 & \multicolumn{ 2}{c|}{failed} & \multicolumn{ 2}{c|}{failed} \\ \hline
		HKM & no & 0.1 & 23 & 1.86 & \multicolumn{ 2}{c|}{failed} & \multicolumn{ 2}{c|}{failed} \\ \hline
		HKM & no & 0.5 & 23 & 1.85 & \multicolumn{ 2}{c|}{failed} & \multicolumn{ 2}{c|}{failed} \\ \hline
		HKM & no & 1 & 23 & 1.86 & \multicolumn{ 2}{c|}{failed} & \multicolumn{ 2}{c|}{failed} \\ \hline
		HKM & no & 1.5 & 23 & 1.85 & \multicolumn{ 2}{c|}{failed} & \multicolumn{ 2}{c|}{failed} \\ \hline
		HKM & no & 2 & 23 & 1.84 & \multicolumn{ 2}{c|}{failed} & \multicolumn{ 2}{c|}{failed} \\
\hline
	\end{tabular}
\end{table}

\begin{table}[htbp]
	\tiny
	\caption[The first-order wide region method in SeDuMi with the first-order wide region method algorithm (1).]{\textbf{The first-order wide region method in SeDuMi with the first-order wide region method algorithm (1).} The table shows the number of iterations and the execution time on SeDuMi of Almost Quantum for CHSH, I3322, E0E1 and Hardy problem sets when the first-order wide region method algorithm is used. Averaged over 100 executions.}
	\begin{tabular}{|r|r|r|r|r|r|r|r|r|r|r|}
		\hline
		\multicolumn{1}{|l|}{} & \multicolumn{1}{l|}{} & \multicolumn{1}{l|}{} & \multicolumn{ 2}{c|}{CHSH} & \multicolumn{ 2}{l|}{I3322} & \multicolumn{ 2}{l|}{E0E1} & \multicolumn{ 2}{l|}{Hardy} \\ \hline
		\multicolumn{1}{|l|}{step. diff.} & \multicolumn{1}{l|}{$\theta$} & \multicolumn{1}{l|}{$\beta$} & \multicolumn{1}{l|}{iter.} & \multicolumn{1}{l|}{time [s]} & \multicolumn{1}{l|}{iter.} & \multicolumn{1}{l|}{time [s]} & \multicolumn{1}{l|}{iter.} & \multicolumn{1}{l|}{time [s]} & \multicolumn{1}{l|}{iter.} & \multicolumn{1}{l|}{time [s]} \\ \hline	
		0 & 0.1 & 0.1 & 91 & 1.24 & 87 & 1.32 & \multicolumn{ 2}{c|}{exceeded} & \multicolumn{ 2}{c|}{exceeded} \\ \hline
		0 & 0.1 & 0.5 & 91 & 1.16 & 78 & 1.21 & \multicolumn{ 2}{c|}{exceeded} & \multicolumn{ 2}{c|}{exceeded} \\ \hline
		0 & 0.1 & 0.9 & 91 & 1.17 & 78 & 1.21 & \multicolumn{ 2}{c|}{exceeded} & 144 & 2.29 \\ \hline
		0 & 0.25 & 0.1 & 55 & 0.74 & 58 & 0.94 & \multicolumn{ 2}{c|}{exceeded} & \multicolumn{ 2}{c|}{exceeded} \\ \hline
		0 & 0.25 & 0.5 & 38 & 0.50 & 43 & 0.72 & 136 & 2.06 & 125 & 2.14 \\ \hline
		0 & 0.25 & 0.9 & 38 & 0.54 & 39 & 0.61 & 102 & 1.55 & 91 & 1.58 \\
\hline
	\end{tabular}
\end{table}

\begin{table}[htbp]
	\tiny
	\caption[The first-order wide region method in SeDuMi with the centering-predictor-corrector algorithm with v-linearization (1).]{\textbf{The first-order wide region method in SeDuMi with the centering-predictor-corrector algorithm with v-linearization (1).} The table shows the number of iterations and the execution time on SeDuMi of Almost Quantum for CHSH, I3322, E0E1 and Hardy problem sets when the centering-predictor-corrector algorithm with v-linearization is used. Averaged over 100 executions.}
	\begin{tabular}{|r|r|r|r|r|r|r|r|r|r|r|}
		\hline
		\multicolumn{1}{|l|}{} & \multicolumn{1}{l|}{} & \multicolumn{1}{l|}{} & \multicolumn{ 2}{c|}{CHSH} & \multicolumn{ 2}{l|}{I3322} & \multicolumn{ 2}{l|}{E0E1} & \multicolumn{ 2}{l|}{Hardy} \\ \hline
		\multicolumn{1}{|l|}{step. diff.} & \multicolumn{1}{l|}{$\theta$} & \multicolumn{1}{l|}{$\beta$} & \multicolumn{1}{l|}{iter.} & \multicolumn{1}{l|}{time [s]} & \multicolumn{1}{l|}{iter.} & \multicolumn{1}{l|}{time [s]} & \multicolumn{1}{l|}{iter.} & \multicolumn{1}{l|}{time [s]} & \multicolumn{1}{l|}{iter.} & \multicolumn{1}{l|}{time [s]} \\ \hline
		0 & 0.1 & 0.1 & 10 & 0.14 & 12 & 0.19 & 29 & 0.43 & 27 & 0.42 \\ \hline
		0 & 0.1 & 0.5 & 10 & 0.14 & 12 & 0.19 & 29 & 0.42 & 27 & 0.42 \\ \hline
		0 & 0.1 & 0.9 & 10 & 0.14 & 12 & 0.19 & 29 & 0.43 & 27 & 0.42 \\ \hline
		0 & 0.25 & 0.1 & 10 & 0.14 & 12 & 0.19 & 29 & 0.42 & 28 & 0.44 \\ \hline
		0 & 0.25 & 0.5 & 10 & 0.14 & 12 & 0.19 & 29 & 0.42 & 27 & 0.42 \\ \hline
		0 & 0.25 & 0.9 & 10 & 0.14 & 12 & 0.19 & 29 & 0.43 & 27 & 0.42 \\
\hline
	\end{tabular}
\end{table}

\begin{table}[htbp]
	\tiny
	\caption[The first-order wide region method in SeDuMi with the centering-predictor-corrector algorithm with xz-linearization (1).]{\textbf{The first-order wide region method in SeDuMi with centering-predictor-corrector algorithm with xz-linearization (1).} The table shows the number of iterations and the execution time on SeDuMi of Almost Quantum for CHSH, I3322, E0E1 and Hardy problem sets when the centering-predictor-corrector algorithm with xz-linearization is used. Averaged over 100 executions.}
	\begin{tabular}{|r|r|r|r|r|r|r|r|r|r|r|}
		\hline
		\multicolumn{1}{|l|}{} & \multicolumn{1}{l|}{} & \multicolumn{1}{l|}{} & \multicolumn{ 2}{c|}{CHSH} & \multicolumn{ 2}{l|}{I3322} & \multicolumn{ 2}{l|}{E0E1} & \multicolumn{ 2}{l|}{Hardy} \\ \hline
		\multicolumn{1}{|l|}{step. diff.} & \multicolumn{1}{l|}{$\theta$} & \multicolumn{1}{l|}{$\beta$} & \multicolumn{1}{l|}{iter.} & \multicolumn{1}{l|}{time [s]} & \multicolumn{1}{l|}{iter.} & \multicolumn{1}{l|}{time [s]} & \multicolumn{1}{l|}{iter.} & \multicolumn{1}{l|}{time [s]} & \multicolumn{1}{l|}{iter.} & \multicolumn{1}{l|}{time [s]} \\ \hline
		0 & 0.1 & 0.1 & 14 & 0.20 & 15 & 0.27 & 25 & 0.37 & 25 & 0.41 \\ \hline
		0 & 0.1 & 0.5 & 14 & 0.20 & 15 & 0.28 & 25 & 0.37 & 25 & 0.42 \\ \hline
		0 & 0.1 & 0.9 & 14 & 0.20 & 15 & 0.28 & 25 & 0.37 & 25 & 0.41 \\ \hline
		0 & 0.25 & 0.1 & 14 & 0.20 & 15 & 0.28 & 25 & 0.38 & 25 & 0.41 \\ \hline
		0 & 0.25 & 0.5 & 14 & 0.20 & 15 & 0.28 & 25 & 0.38 & 25 & 0.42 \\ \hline
		0 & 0.25 & 0.9 & 14 & 0.20 & 15 & 0.28 & 25 & 0.38 & 25 & 0.41 \\
\hline
	\end{tabular}
\end{table}

\begin{table}[htbp]
	\tiny
	\caption[The first-order wide region method in SeDuMi with the first-order wide region method algorithm (2).]{\textbf{The first-order wide region method in SeDuMi with the first-order wide region method algorithm (2).} The table shows the number of iterations and the execution time on SeDuMi of Almost Quantum for BC3, T3C, BC5 and BC7 problem sets when the first-order wide region method algorithm is used. Averaged over 100 executions.}
	\begin{tabular}{|r|r|r|r|r|r|r|r|r|r|r|}
		\hline
		\multicolumn{1}{|l|}{} & \multicolumn{1}{l|}{} & \multicolumn{1}{l|}{} & \multicolumn{ 2}{c|}{BC3} & \multicolumn{ 2}{c|}{T3C} & \multicolumn{ 2}{c|}{BC5} & \multicolumn{ 2}{c|}{BC7} \\ \hline
		\multicolumn{1}{|l|}{step. diff.} & \multicolumn{1}{l|}{$\theta$} & \multicolumn{1}{l|}{$\beta$} & \multicolumn{1}{l|}{iter.} & \multicolumn{1}{l|}{time [s]} & \multicolumn{1}{l|}{iter.} & \multicolumn{1}{l|}{time [s]} & \multicolumn{1}{l|}{iter.} & \multicolumn{1}{l|}{time [s]} & \multicolumn{1}{l|}{iter.} & \multicolumn{1}{l|}{time [s]} \\ \hline
		0 & 0.1 & 0.1 & 83 & 1.12 & 135 & 1.96 & 96 & 3.68 & 101 & 100.44 \\ \hline
		0 & 0.1 & 0.5 & 86 & 1.17 & 104 & 1.50 & 80 & 3.03 & 84 & 81.99 \\ \hline
		0 & 0.1 & 0.9 & 86 & 1.17 & 104 & 1.50 & 83 & 3.11 & 88 & 86.31 \\ \hline
		0 & 0.25 & 0.1 & 58 & 0.83 & 90 & 1.36 & 56 & 2.14 & 70 & 68.94 \\ \hline
		0 & 0.25 & 0.5 & 38 & 0.53 & 61 & 0.91 & 41 & 1.56 & 55 & 54.11 \\ \hline
		0 & 0.25 & 0.9 & 37 & 0.49 & 56 & 0.83 & 40 & 1.52 & 49 & 49.16 \\
\hline
	\end{tabular}
\end{table}

\begin{table}[htbp]
	\tiny
	\caption[The first-order wide region method in SeDuMi with the centering-predictor-corrector algorithm with v-linearization (2).]{\textbf{The first-order wide region method in SeDuMi with the centering-predictor-corrector algorithm with v-linearization (2).} The table shows the number of iterations and the execution time on SeDuMi of Almost Quantum for BC3, T3C, BC5 and BC7 problem sets when the centering-predictor-corrector algorithm with v-linearization is used. Averaged over 100 executions.}
	\begin{tabular}{|r|r|r|r|r|r|r|r|r|r|r|}
		\hline
		\multicolumn{1}{|l|}{} & \multicolumn{1}{l|}{} & \multicolumn{1}{l|}{} & \multicolumn{ 2}{c|}{BC3} & \multicolumn{ 2}{c|}{T3C} & \multicolumn{ 2}{c|}{BC5} & \multicolumn{ 2}{c|}{BC7} \\ \hline
		\multicolumn{1}{|l|}{step. diff.} & \multicolumn{1}{l|}{$\theta$} & \multicolumn{1}{l|}{$\beta$} & \multicolumn{1}{l|}{iter.} & \multicolumn{1}{l|}{time [s]} & \multicolumn{1}{l|}{iter.} & \multicolumn{1}{l|}{time [s]} & \multicolumn{1}{l|}{iter.} & \multicolumn{1}{l|}{time [s]} & \multicolumn{1}{l|}{iter.} & \multicolumn{1}{l|}{time [s]} \\ \hline
		0 & 0.1 & 0.1 & 12 & 0.18 & 18 & 0.28 & 11 & 0.44 & 12 & 11.08 \\ \hline
		0 & 0.1 & 0.5 & 12 & 0.19 & 18 & 0.28 & 11 & 0.44 & 12 & 11.06 \\ \hline
		0 & 0.1 & 0.9 & 12 & 0.19 & 18 & 0.29 & 11 & 0.44 & 12 & 11.03 \\ \hline
		0 & 0.25 & 0.1 & 12 & 0.19 & 18 & 0.28 & 11 & 0.44 & 12 & 11.13 \\ \hline
		0 & 0.25 & 0.5 & 12 & 0.18 & 18 & 0.28 & 11 & 0.44 & 12 & 11.16 \\ \hline
		0 & 0.25 & 0.9 & 12 & 0.19 & 18 & 0.28 & 11 & 0.44 & 12 & 11.08 \\
\hline
	\end{tabular}
\end{table}

\begin{table}[htbp]
	\tiny
	\caption[The first-order wide region method in SeDuMi with the centering-predictor-corrector algorithm with xz-linearization (2).]{\textbf{The first-order wide region method in SeDuMi with the centering-predictor-corrector algorithm with xz-linearization (2).} The table shows the number of iterations and the execution time on SeDuMi of Almost Quantum for BC3, T3C, BC5 and BC7 problem sets when the centering-predictor-corrector algorithm with xz-linearization is used. Averaged over 100 executions.}
	\begin{tabular}{|r|r|r|r|r|r|r|r|r|r|r|}
		\hline
		\multicolumn{1}{|l|}{} & \multicolumn{1}{l|}{} & \multicolumn{1}{l|}{} & \multicolumn{ 2}{c|}{BC3} & \multicolumn{ 2}{c|}{T3C} & \multicolumn{ 2}{c|}{BC5} & \multicolumn{ 2}{c|}{BC7} \\ \hline
		\multicolumn{1}{|l|}{step. diff.} & \multicolumn{1}{l|}{$\theta$} & \multicolumn{1}{l|}{$\beta$} & \multicolumn{1}{l|}{iter.} & \multicolumn{1}{l|}{time [s]} & \multicolumn{1}{l|}{iter.} & \multicolumn{1}{l|}{time [s]} & \multicolumn{1}{l|}{iter.} & \multicolumn{1}{l|}{time [s]} & \multicolumn{1}{l|}{iter.} & \multicolumn{1}{l|}{time [s]} \\ \hline
		0 & 0.1 & 0.1 & 15 & 0.24 & 17 & 0.27 & 14 & 0.57 & 16 & 15.44 \\ \hline
		0 & 0.1 & 0.5 & 15 & 0.24 & 17 & 0.27 & 14 & 0.57 & 16 & 15.56 \\ \hline
		0 & 0.1 & 0.9 & 15 & 0.24 & 17 & 0.27 & 14 & 0.57 & 16 & 15.21 \\ \hline
		0 & 0.25 & 0.1 & 14 & 0.22 & 17 & 0.27 & 14 & 0.56 & 16 & 14.93 \\ \hline
		0 & 0.25 & 0.5 & 16 & 0.26 & 17 & 0.27 & 14 & 0.57 & 16 & 15.47 \\ \hline
		0 & 0.25 & 0.9 & 16 & 0.26 & 17 & 0.27 & 14 & 0.57 & 16 & 15.47 \\
\hline
	\end{tabular}
\end{table}

\begin{table}[htbp]
	\tiny
	\caption[High NPA hierarchy levels in SeDuMi (wide region).]{\textbf{High NPA hierarchy levels in SeDuMi (wide region).} The table shows the performance of the SeDuMi solver on CHSH and I3322 problem cases in high NPA hierarchy levels. The execution of I3322 in $\mathcal{Q}_4$ failed in all cases. The first-order wide region method has been used.}
	\begin{tabular}{|r|r|r|r|r|r|r|r|r|r|r|}
		\hline
		\multicolumn{ 1}{|l|}{step. diff.} & \multicolumn{ 1}{l|}{$\theta$} & \multicolumn{ 1}{l|}{$\beta$} & \multicolumn{ 2}{c|}{I3322 in $\mathcal{Q}_2$} & \multicolumn{ 2}{c|}{I3322 in $\mathcal{Q}_3$} & \multicolumn{ 2}{c|}{CHSH in $\mathcal{Q}_5$} & \multicolumn{ 2}{c|}{CHSH in $\mathcal{Q}_{10}$} \\ \cline{ 4- 11}
		\multicolumn{ 1}{|l|}{} & \multicolumn{ 1}{l|}{} & \multicolumn{ 1}{l|}{} & \multicolumn{1}{l|}{iter.} & \multicolumn{1}{l|}{time [s]} & \multicolumn{1}{l|}{iter.} & \multicolumn{1}{l|}{time [s]} & \multicolumn{1}{l|}{iter.} & \multicolumn{1}{l|}{time [s]} & \multicolumn{1}{l|}{iter.} & \multicolumn{1}{l|}{time [s]} \\ \hline
		0 & 0.1 & 0.1 & 97 & 1.85 & 122 & 36.2 & 55 & 1.72 & 69 & 53.0 \\ \hline
		0 & 0.1 & 0.5 & 91 & 1.69 & 97 & 28.4 & 59 & 1.81 & 61 & 46.5 \\ \hline
		0 & 0.1 & 0.9 & 91 & 1.68 & 97 & 28.0 & 59 & 1.80 & 61 & 46.6 \\ \hline
		0 & 0.25 & 0.1 & 78 & 1.50 & 88 & 25.6 & 36 & 1.11 & 37 & 28.3 \\ \hline
		0 & 0.25 & 0.5 & 53 & 0.99 & 69 & 20.2 & 33 & 1.01 & 38 & 29.1 \\ \hline
		0 & 0.25 & 0.9 & 49 & 0.93 & 55 & 16.1 & 33 & 1.03 & 35 & 26.7 \\
\hline
	\end{tabular}
\end{table}

\begin{table}[htbp]
		\tiny
		\caption[High NPA hierarchy levels in SeDuMi.]{\textbf{High NPA hierarchy levels in SeDuMi.} The table shows the performance of the SeDuMi solver on CHSH and I3322 problem cases in high NPA hierarchy levels. The centering-predictor-corrector algorithm with v-linearization has been used.}
	\begin{tabular}{|r|r|r|r|r|r|r|r|r|r|r|r|r|}
		\hline
		\multicolumn{1}{|c|}{} & \multicolumn{1}{l|}{} & \multicolumn{1}{l|}{} & \multicolumn{ 2}{c|}{I3322 in $\mathcal{Q}_2$} & \multicolumn{ 2}{c|}{I3322 in $\mathcal{Q}_3$} & \multicolumn{ 2}{c|}{I3322 in $\mathcal{Q}_4$} & \multicolumn{ 2}{c|}{CHSH in $\mathcal{Q}_5$} & \multicolumn{ 2}{c|}{CHSH in $\mathcal{Q}_{10}$} \\ \hline
		\multicolumn{1}{|l|}{step diff.} & \multicolumn{1}{l|}{$\theta$} & \multicolumn{1}{l|}{$\beta$} & \multicolumn{1}{l|}{iter.} & \multicolumn{1}{l|}{time [s]} & \multicolumn{1}{l|}{iter.} & \multicolumn{1}{l|}{time [s]} & \multicolumn{1}{l|}{iter.} & \multicolumn{1}{l|}{time [s]} & \multicolumn{1}{l|}{iter.} & \multicolumn{1}{l|}{time [s]} & \multicolumn{1}{l|}{iter.} & \multicolumn{1}{l|}{time [s]} \\ \hline
		0 & 0.1 & 0.1 & 14 & 0.29 & 19 & 5.6 & 18 & 869 & 12 & 0.42 & 12 & 9.6 \\ \hline
		0 & 0.1 & 0.5 & 14 & 0.29 & 19 & 5.6 & 18 & 870 & 12 & 0.41 & 12 & 9.6 \\ \hline
		0 & 0.1 & 0.9 & 14 & 0.30 & 19 & 5.5 & 18 & 876 & 12 & 0.41 & 12 & 9.7 \\ \hline
		0 & 0.25 & 0.1 & 14 & 0.29 & 19 & 5.6 & 18 & 873 & 12 & 0.41 & 12 & 9.6 \\ \hline
		0 & 0.25 & 0.5 & 14 & 0.29 & 19 & 5.6 & 18 & 876 & 12 & 0.41 & 12 & 9.6 \\ \hline
		0 & 0.25 & 0.9 & 14 & 0.29 & 19 & 5.5 & 18 & 873 & 12 & 0.41 & 12 & 9.7 \\
\hline
	\end{tabular}
\end{table}

\begin{table}[htbp]
	\tiny
	\caption[High NPA hierarchy levels in SeDuMi (xz-linearization).]{\textbf{High NPA hierarchy levels in SeDuMi (xz-linearization).} The table shows the performance of the SeDuMi solver on CHSH and I3322 problem cases in high NPA hierarchy levels. The centering-predictor-corrector algorithm with xz-linearization has been used. Levels $\mathcal{Q}_2$, $\mathcal{Q}_3$ and $\mathcal{Q}_4$ refer to I3322, and levels $\mathcal{Q}_5$ and $\mathcal{Q}_{10}$ to CHSH.}
	\begin{tabular}{|r|r|r|r|r|r|r|r|r|r|r|r|r|}
		\hline
		\multicolumn{1}{|l|}{} & \multicolumn{1}{l|}{} & \multicolumn{1}{l|}{} & \multicolumn{ 2}{c|}{I3322 in $\mathcal{Q}_2$} & \multicolumn{ 2}{c|}{I3322 in $\mathcal{Q}_3$} & \multicolumn{ 2}{c|}{I3322 in $\mathcal{Q}_4$} & \multicolumn{ 2}{c|}{CHSH in $\mathcal{Q}_5$} & \multicolumn{ 2}{c|}{CHSH in $\mathcal{Q}_{10}$} \\ \hline
		\multicolumn{1}{|l|}{step diff.} & \multicolumn{1}{l|}{$\theta$} & \multicolumn{1}{l|}{$\beta$} & \multicolumn{1}{l|}{iter.} & \multicolumn{1}{l|}{time [s]} & \multicolumn{1}{l|}{iterations} & \multicolumn{1}{l|}{time [s]} & \multicolumn{1}{l|}{iter.} & \multicolumn{1}{l|}{time [s]} & \multicolumn{1}{l|}{iter.} & \multicolumn{1}{l|}{time [s]} & \multicolumn{1}{l|}{iter.} & \multicolumn{1}{l|}{time [s]} \\ \hline
		0 & 0.1 & 0.1 & 15 & 0.31 & 18 & 5.1 & 19 & 916 & 14 & 0.46 & 16 & 13.0 \\ \hline
		0 & 0.1 & 0.5 & 15 & 0.32 & 18 & 5.1 & 19 & 911 & 14 & 0.46 & 16 & 13.0 \\ \hline
		0 & 0.1 & 0.9 & 15 & 0.31 & 18 & 5.1 & 19 & 908 & 14 & 0.46 & 16 & 13.0 \\ \hline
		0 & 0.25 & 0.1 & 15 & 0.31 & 18 & 5.1 & 19 & 905 & 14 & 0.47 & 16 & 13.0 \\ \hline
		0 & 0.25 & 0.5 & 15 & 0.32 & 18 & 5.1 & 19 & 918 & 14 & 0.46 & 16 & 13.0 \\ \hline
		0 & 0.25 & 0.9 & 15 & 0.32 & 18 & 5.1 & 19 & 916 & 14 & 0.46 & 16 & 13.0 \\
\hline
	\end{tabular}
\end{table}

\chapter{Performance of NPA-solver}
\label{app:NPAsolver}


The PMSdp tests are performed with different choices of parameters $\text{expon}$ and perturbation strategies (0 means no perturbations). In all cases the stopping condition was that the gap and infeasibilites are not larger that $10^{-8}$; only for cases E0E1 and Hardy the gap had to be less than $2 \cdot 10^{-5}$. A case is considered to be \textbf{failed} if an error, \textit{e.g.} in Cholesky factorization occurred, and to \textbf{exceed} after reaching more that $100$ of iterations. 

\begin{table}[htbp]
	\tiny
	\caption[Performance of NPA-solver (1a).]{\tiny \textbf{Performance of NPA-solver (1a).} The table shows the performance of the special purpose NPA-solver in CHSH, I3322, E0E1 and Hardy problems in the Almost Quantum level of the NPA hierarchy. Averaged over 100 executions.}
\begin{tabular}{|r|r|r|r|r|r|c|l|r|r|}
		\hline
		\multicolumn{2}{|c|}{parameters} & \multicolumn{2}{c|}{CHSH} & \multicolumn{2}{c|}{I3322} & \multicolumn{2}{c|}{E0E1} & \multicolumn{2}{c|}{Hardy} \\ \hline
		\multicolumn{1}{|l|}{expon} & \multicolumn{1}{l|}{pertub. str.} & \multicolumn{1}{l|}{iter.} & \multicolumn{1}{l|}{time} & \multicolumn{1}{l|}{iter.} & \multicolumn{1}{l|}{time} & \multicolumn{1}{l|}{iter.} & time & \multicolumn{1}{l|}{iter.} & \multicolumn{1}{l|}{time} \\ \hline

		0 & 0 & 10 & 0.03 & 11 & 0.04 & \multicolumn{ 2}{c|}{failed} & \multicolumn{ 2}{c|}{failed} \\ \hline
		0 & 1 & 11 & 0.03 & 11 & 0.04 & \multicolumn{1}{r|}{33} & 0.09 & 26 & 0.07 \\ \hline
		0 & 2 & 11 & 0.03 & 11 & 0.04 & \multicolumn{ 2}{c|}{exceeded} & 26 & 0.07 \\ \hline
		0 & 3 & \multicolumn{ 2}{c|}{exceeded} & \multicolumn{ 2}{c|}{failed} & \multicolumn{ 2}{c|}{failed} & 26 & 0.07 \\ \hline
		0 & 4 & 10 & 0.03 & 11 & 0.04 & \multicolumn{ 2}{c|}{failed} & \multicolumn{ 2}{c|}{failed} \\ \hline
		0 & 5 & 10 & 0.03 & 11 & 0.04 & \multicolumn{ 2}{c|}{exceeded} & 27 & 0.07 \\
\hline
	\end{tabular}
\end{table}

\begin{table}[htbp]
	\caption[Performance of NPA-solver (1b).]{\tiny \textbf{Performance of NPA-solver (1b).} The table shows the performance of the special purpose NPA-solver in CHSH, I3322, E0E1 and Hardy problems in the Almost Quantum level of the NPA hierarchy. Averaged over 100 executions.}
	\begin{tabular}{|r|r|r|r|r|r|c|l|r|r|}
		\hline
		\multicolumn{2}{|c|}{parameters} & \multicolumn{2}{c|}{CHSH} & \multicolumn{2}{c|}{I3322} & \multicolumn{2}{c|}{E0E1} & \multicolumn{2}{c|}{Hardy} \\ \hline
		\multicolumn{1}{|l|}{expon} & \multicolumn{1}{l|}{pertub. str.} & \multicolumn{1}{l|}{iter.} & \multicolumn{1}{l|}{time} & \multicolumn{1}{l|}{iter.} & \multicolumn{1}{l|}{time} & \multicolumn{1}{l|}{iter.} & time & \multicolumn{1}{l|}{iter.} & \multicolumn{1}{l|}{time} \\ \hline

		2.5 & 0 & 11 & 0.03 & 12 & 0.04 & \multicolumn{ 2}{c|}{failed} & \multicolumn{ 2}{c|}{failed} \\ \hline
		2.5 & 1 & 11 & 0.03 & 12 & 0.05 & \multicolumn{1}{r|}{29} & \multicolumn{1}{r|}{0.08} & 26 & 0.07 \\ \hline
		2.5 & 2 & 12 & 0.03 & 13 & 0.05 & \multicolumn{ 2}{c|}{exceeded} & 26 & 0.07 \\ \hline
		2.5 & 3 & \multicolumn{ 2}{c|}{exceeded} & \multicolumn{ 2}{c|}{failed} & \multicolumn{ 2}{c|}{failed} & 26 & 0.07 \\ \hline
		2.5 & 4 & 11 & 0.03 & 12 & 0.05 & \multicolumn{ 2}{c|}{failed} & \multicolumn{ 2}{c|}{failed} \\ \hline
		2.5 & 5 & 11 & 0.03 & 12 & 0.05 & \multicolumn{ 2}{c|}{failed} & 29 & 0.08 \\
\hline
	\end{tabular}
\end{table}

\begin{table}[htbp]
	\tiny
	\caption[Performance of NPA-solver (2a).]{\tiny \textbf{Performance of NPA-solver (2a).} The table shows the performance of the special purpose NPA-solver in BC3, T3C, BC5, and BC7 problems in the Almost Quantum level of the NPA hierarchy. Averaged over 100 executions.}
	\begin{tabular}{|r|r|r|r|r|r|r|r|r|r|}
		\hline
		\multicolumn{2}{|c|}{parameters} & \multicolumn{2}{c|}{BC3} & \multicolumn{2}{c|}{T3C} & \multicolumn{2}{c|}{BC5} & \multicolumn{2}{c|}{BC7} \\ \hline
		\multicolumn{1}{|l|}{expon} & \multicolumn{1}{l|}{pertub. str.} & \multicolumn{1}{l|}{iter.} & \multicolumn{1}{l|}{time} & \multicolumn{1}{l|}{iter.} & \multicolumn{1}{l|}{time} & \multicolumn{1}{l|}{iter.} & time & \multicolumn{1}{l|}{iter.} & \multicolumn{1}{l|}{time} \\ \hline

		0 & 0 & 13 & 0.05 & 18 & 0.09 & 15 & 0.24 & 18 & 3.46 \\ \hline
		0 & 1 & 13 & 0.05 & 18 & 0.08 & 16 & 0.26 & 19 & 3.64 \\ \hline
		0 & 2 & 13 & 0.04 & 18 & 0.08 & 16 & 0.26 & 19 & 3.64 \\ \hline
		0 & 3 & \multicolumn{ 2}{c|}{exceeded} & \multicolumn{ 2}{c|}{exceeded} & \multicolumn{ 2}{c|}{exceeded} & \multicolumn{ 2}{c|}{exceeded} \\ \hline
		0 & 4 & 13 & 0.04 & 19 & 0.09 & 19 & 0.31 & 23 & 4.42 \\ \hline
		0 & 5 & 13 & 0.05 & 18 & 0.08 & 15 & 0.24 & 18 & 3.47 \\
\hline
	\end{tabular}
\end{table}

\begin{table}[htbp]
	\caption[Performance of NPA-solver (2b).]{\tiny \textbf{Performance of NPA-solver (2b).} The table shows the performance of the special purpose NPA-solver in BC3, T3C, BC5, and BC7 problems in the Almost Quantum level of the NPA hierarchy. Averaged over 100 executions.}
	\begin{tabular}{|r|r|r|r|r|r|r|r|r|r|}
		\hline
		\multicolumn{2}{|c|}{parameters} & \multicolumn{2}{c|}{BC3} & \multicolumn{2}{c|}{T3C} & \multicolumn{2}{c|}{BC5} & \multicolumn{2}{c|}{BC7} \\ \hline
		\multicolumn{1}{|l|}{expon} & \multicolumn{1}{l|}{pertub. str.} & \multicolumn{1}{l|}{iter.} & \multicolumn{1}{l|}{time} & \multicolumn{1}{l|}{iter.} & \multicolumn{1}{l|}{time} & \multicolumn{1}{l|}{iter.} & time & \multicolumn{1}{l|}{iter.} & \multicolumn{1}{l|}{time} \\ \hline

		2.5 & 0 & 13 & 0.05 & 17 & 0.08 & 16 & 0.26 & 18 & 3.46 \\ \hline
		2.5 & 1 & 14 & 0.05 & 17 & 0.08 & 16 & 0.25 & 19 & 3.65 \\ \hline
		2.5 & 2 & 14 & 0.05 & 18 & 0.08 & 16 & 0.26 & 18 & 3.54 \\ \hline
		2.5 & 3 & \multicolumn{ 2}{c|}{exceeded} & 27 & 0.12 & \multicolumn{ 2}{c|}{exceeded} & \multicolumn{ 2}{c|}{exceeded} \\ \hline
		2.5 & 4 & 13 & 0.05 & 18 & 0.08 & 18 & 0.29 & 22 & 4.23 \\ \hline
		2.5 & 5 & 13 & 0.05 & 17 & 0.08 & 16 & 0.25 & 18 & 3.46 \\
\hline
	\end{tabular}
\end{table}

\chapter{Impact of initial solution on SDPT3}
\label{app:initSDPT3}

\begin{table}[htbp]
	\index{interior point methods!search directions!HKM}
	\caption[Initial solution for SDPT3 with HKM (I)]{The impact of supplying initial iterates for SDPT3 in CHSH, I3322, E0E1 and Hardy problems in the Almost Quantum level of the NPA hierarchy for HKM search direction.}
\begin{tabular}{|l|r|r|r|r|r|r|r|r|r|}
\hline
\multicolumn{2}{|c|}{parameters} & \multicolumn{2}{c|}{CHSH} & \multicolumn{2}{c|}{I3322} & \multicolumn{2}{c|}{E0E1} & \multicolumn{2}{c|}{Hardy} \\ \hline
PC & \multicolumn{1}{l|}{expon} & \multicolumn{1}{l|}{iter.} & \multicolumn{1}{l|}{time} & \multicolumn{1}{l|}{iter.} & \multicolumn{1}{l|}{time} & \multicolumn{1}{l|}{iter.} & time & \multicolumn{1}{l|}{iter.} & \multicolumn{1}{l|}{time} \\ \hline
no & 0 & -2 & -1\% & -2 & -9\% & \multicolumn{ 2}{c|}{failed} & 4 & 22\% \\ \hline
no & 0.01 & -2 & -12\% & -2 & -10\% & \multicolumn{ 2}{c|}{failed} & 4 & 24\% \\ \hline
no & 0.1 & -2 & -12\% & -2 & -9\% & \multicolumn{ 2}{c|}{failed} & 4 & 20\% \\ \hline
no & 0.5 & -2 & -12\% & -2 & -9\% & \multicolumn{ 2}{c|}{failed} & 4 & 18\% \\ \hline
no & 1 & -2 & -12\% & -2 & -9\% & \multicolumn{ 2}{c|}{failed} & 4 & 17\% \\ \hline
no & 1.5 & -2 & -12\% & -2 & -8\% & \multicolumn{ 2}{c|}{failed} & 4 & 18\% \\
\hline
\end{tabular}
\end{table}

\begin{table}[htbp]
	\index{interior point methods!search directions!HKM}
	\caption[Initial solution for SDPT3 with HKM (II)]{The impact of supplying initial iterates for SDPT3 in BC3, T3C, BC5 and BC7 problems in the Almost Quantum level of the NPA hierarchy for HKM search direction.}
\begin{tabular}{|l|r|r|r|r|r|r|r|r|r|}
\hline
\multicolumn{2}{|c|}{parameters} & \multicolumn{2}{c|}{BC3} & \multicolumn{2}{c|}{T3C} & \multicolumn{2}{c|}{BC5} & \multicolumn{2}{c|}{BC7} \\ \hline
PC & \multicolumn{1}{l|}{expon} & \multicolumn{1}{l|}{iter.} & \multicolumn{1}{l|}{time} & \multicolumn{1}{l|}{iter.} & time & \multicolumn{1}{l|}{iter.} & \multicolumn{1}{l|}{time} & \multicolumn{1}{l|}{iter.} & \multicolumn{1}{l|}{time} \\ \hline
no & 0 & -1 & -4\% & \multicolumn{ 2}{c|}{failed} & -5 & -19\% & -1 & -2\% \\ \hline
no & 0.01 & -1 & -4\% & \multicolumn{ 2}{c|}{failed} & -5 & -19\% & -1 & -2\% \\ \hline
no & 0.1 & -1 & -4\% & \multicolumn{ 2}{c|}{failed} & -5 & -19\% & -1 & -2\% \\ \hline
no & 0.5 & -1 & -3\% & \multicolumn{ 2}{c|}{failed} & -5 & -18\% & -1 & -2\% \\ \hline
no & 1 & -1 & -4\% & \multicolumn{ 2}{c|}{failed} & -5 & -17\% & -1 & -2\% \\ \hline
no & 1.5 & -1 & -5\% & \multicolumn{ 2}{c|}{failed} & -5 & -19\% & -1 & -3\% \\
\hline
\end{tabular}
\end{table}

\begin{table}[htbp]
	\index{interior point methods!search directions!NT}
	\caption[Initial solution for SDPT3 with NT (I)]{The impact of supplying initial iterates for SDPT3 in CHSH, I3322, E0E1 and Hardy problems in the Almost Quantum level of the NPA hierarchy for NT search direction.}
\begin{tabular}{|l|r|r|r|r|r|r|r|r|r|}
\hline
\multicolumn{2}{|c|}{parameters} & \multicolumn{2}{c|}{CHSH} & \multicolumn{2}{c|}{I3322} & \multicolumn{2}{c|}{E0E1} & \multicolumn{2}{c|}{Hardy} \\ \hline
PC & \multicolumn{1}{l|}{expon} & \multicolumn{1}{l|}{iter.} & \multicolumn{1}{l|}{time} & \multicolumn{1}{l|}{iter.} & \multicolumn{1}{l|}{time} & \multicolumn{1}{l|}{iter.} & \multicolumn{1}{l|}{time} & \multicolumn{1}{l|}{iter.} & \multicolumn{1}{l|}{time} \\ \hline
no & 0 & -4 & -22\% & 0 & 0\% & -1 & -4\% & 3 & 14\% \\ \hline
no & 0.01 & -4 & -25\% & 0 & 0\% & -1 & -4\% & 3 & 14\% \\ \hline
no & 0.1 & -4 & -25\% & 0 & 0\% & -1 & -1\% & 3 & 13\% \\ \hline
no & 0.5 & -4 & -25\% & 0 & 0\% & -1 & 3\% & 3 & 13\% \\ \hline
no & 1 & -4 & -25\% & 0 & 0\% & -1 & 2\% & 3 & 14\% \\ \hline
no & 1.5 & -4 & -23\% & 0 & 0\% & -1 & 3\% & 3 & 14\% \\ \hline
no & 2 & -4 & -24\% & 0 & 1\% & -1 & 5\% & 3 & 12\% \\ \hline
no & 2.5 & -4 & -25\% & 0 & 0\% & -1 & 2\% & 3 & 14\% \\ \hline
no & 3 & -4 & -24\% & 0 & 0\% & -1 & 0\% & 3 & 14\% \\ \hline
no & 4 & -4 & -23\% & 0 & 0\% & -1 & -1\% & 3 & 12\% \\ \hline
no & 8 & -4 & -25\% & 0 & 0\% & -1 & -5\% & 3 & 13\% \\ \hline
yes & 0 & -1 & -9\% & -2 & -14\% & -2 & -5\% & -1 & -3\% \\ \hline
yes & 0.01 & -1 & -9\% & -2 & -15\% & -2 & -6\% & -1 & -4\% \\ \hline
yes & 0.1 & -1 & -10\% & -2 & -15\% & -2 & -8\% & -1 & -4\% \\ \hline
yes & 0.5 & -1 & -10\% & -2 & -15\% & -2 & -6\% & -1 & -3\% \\ \hline
yes & 1 & -1 & -9\% & -2 & -15\% & -2 & -7\% & -1 & -6\% \\ \hline
yes & 1.5 & -1 & -10\% & -2 & -14\% & -2 & -8\% & 0 & -2\% \\ \hline
yes & 2 & -1 & -11\% & -1 & -8\% & -1 & -3\% & 0 & -1\% \\ \hline
yes & 2.5 & -1 & -10\% & -1 & -7\% & -1 & -2\% & 0 & 1\% \\ \hline
yes & 3 & 0 & 0\% & -1 & -7\% & 0 & 3\% & 0 & 4\% \\ \hline
yes & 4 & 0 & -1\% & -1 & -7\% & 0 & 4\% & 0 & 0\% \\ \hline
yes & 8 & 0 & -1\% & 0 & -1\% & -1 & -1\% & 0 & 0\% \\ \hline
\end{tabular}
\end{table}

\begin{table}[htbp]
	\index{interior point methods!search directions!NT}
	\caption[Initial solution for SDPT3 with NT (II)]{The impact of supplying initial iterates for SDPT3 in BC3, T3C, BC5 and BC7 problems in the Almost Quantum level of the NPA hierarchy for NT search direction.}
\begin{tabular}{|l|r|r|r|r|r|r|r|r|r|}
\hline
\multicolumn{2}{|c|}{parameters} & \multicolumn{2}{c|}{BC3} & \multicolumn{2}{c|}{T3C} & \multicolumn{2}{c|}{BC5} & \multicolumn{2}{c|}{BC7} \\ \hline
PC & \multicolumn{1}{l|}{expon} & \multicolumn{1}{l|}{iter.} & \multicolumn{1}{l|}{time} & \multicolumn{1}{l|}{iter.} & \multicolumn{1}{l|}{time} & \multicolumn{1}{l|}{iter.} & \multicolumn{1}{l|}{time} & \multicolumn{1}{l|}{iter.} & \multicolumn{1}{l|}{time} \\ \hline
no & 0 & -1 & -4\% & -2 & -7\% & -3 & -12\% & -2 & -4\% \\ \hline
no & 0.01 & -1 & -6\% & -2 & -7\% & -3 & -13\% & -2 & -3\% \\ \hline
no & 0.1 & -1 & -6\% & -2 & -7\% & -3 & -13\% & -2 & -4\% \\ \hline
no & 0.5 & -1 & -5\% & -2 & -7\% & -3 & -12\% & -2 & -5\% \\ \hline
no & 1 & -1 & -5\% & -2 & -7\% & -3 & -11\% & -2 & -5\% \\ \hline
no & 1.5 & -1 & -4\% & -2 & -7\% & -3 & -12\% & -2 & -3\% \\ \hline
no & 2 & -1 & -4\% & -2 & -7\% & -3 & -12\% & -2 & -5\% \\ \hline
no & 2.5 & -1 & -5\% & -2 & -7\% & -3 & -12\% & -2 & -6\% \\ \hline
no & 3 & -1 & -5\% & -2 & -7\% & -3 & -12\% & -2 & -6\% \\ \hline
no & 4 & -1 & -4\% & -2 & -7\% & -3 & -12\% & -2 & -5\% \\ \hline
no & 8 & -1 & -5\% & -2 & -7\% & -3 & -11\% & -2 & -6\% \\ \hline
yes & 0 & -1 & -6\% & -1 & -6\% & -1 & -8\% & -2 & -8\% \\ \hline
yes & 0.01 & -1 & -6\% & -1 & -5\% & -1 & -7\% & -2 & -4\% \\ \hline
yes & 0.1 & -1 & -7\% & -1 & -5\% & -1 & -8\% & -2 & -4\% \\ \hline
yes & 0.5 & -1 & -8\% & -1 & -5\% & -1 & -8\% & -2 & -5\% \\ \hline
yes & 1 & -1 & -7\% & -1 & -5\% & -1 & -7\% & -2 & -3\% \\ \hline
yes & 1.5 & 0 & 2\% & -1 & -6\% & -1 & -6\% & -2 & -2\% \\ \hline
yes & 2 & -1 & -6\% & 1 & 7\% & -1 & -7\% & -1 & 0\% \\ \hline
yes & 2.5 & 0 & 0\% & 0 & 1\% & -1 & -8\% & -2 & -7\% \\ \hline
yes & 3 & 1 & 9\% & 0 & 0\% & -1 & -7\% & -4 & -13\% \\ \hline
yes & 4 & 0 & 1\% & 0 & 0\% & -2 & -12\% & -3 & -12\% \\ \hline
yes & 8 & 0 & 2\% & 0 & 0\% & -2 & -13\% & -4 & -16\% \\ \hline
\end{tabular}
\end{table}

\listoffigures
\listoftables
\printindex

\end{document}